\documentclass[10pt]{article}

\usepackage{graphicx}
\usepackage{amsmath}
\usepackage[numbers]{natbib}
\usepackage{algorithm}
\usepackage{algorithmic}

\usepackage{verbatim}
\usepackage{mathrsfs}
\usepackage{xcolor, tikz, fancybox, morefloats}

\newtheorem{example}{Example}

\definecolor{lightgray}{gray}{0.9}

\newenvironment{evb}{\VerbatimEnvironment%
    \noindent\begin{Sbox}\footnotesize
        \begin{minipage}{\dimexpr\linewidth-8\fboxsep-2\fboxrule}
          \begin{Verbatim}
}{%
      \end{Verbatim}%
      \end{minipage}%
      \end{Sbox}%
      \fcolorbox{black}{lightgray}{\TheSbox}%
}

\begin{document}

\title{A Scalable Thermal Reservoir Simulator for Giant Models on Parallel Computers}

{\author{Hui Liu\thanks{Authors to whom correspondence may be addressed. Email:
hui.sc.liu@gmail.com}, Zhangxin Chen
		\\
		\vspace{6pt} \\
		Dept. of Chemical and Petroleum Engineering, University of Calgary\\
		2500 University Drive NW, Calgary, AB, Canada\\
	}
}

\date{}
\maketitle

\begin{abstract}
This paper introduces the model, numerical methods, algorithms and parallel implementation of a thermal 
    reservoir simulator that designed for numerical simulations of thermal reservoir with multiple components
    in three dimensional domain using distributed-memory parallel computers.
    Its full mathematical model is introduced with correlations for important properties and well modeling.
    Various well constraints, such as fixed bottom hole pressure, fixed oil, water, gas and liquid rates at
    surface condition and reservoir condition,
    constant heat transfer model, convective heat transfer model, heater model (temperature control, rate
    control, dual rate/temperature control), and subcool (steam trap), are introduced in details, including
    their mathematical models and methods.
    Efficient numerical methods (discretization scheme, matrix decoupling methods, and preconditioners),
    parallel computing technologies and implementation details are presented, including option parsing,
    keyword parsing, parallel IO (input and output), data management and visualization.
    The simulator is designed for giant models with billions or even trillions of grid blocks using hundreds
    of thousands of CPUs.
    Numerical experiments show that our results match commercial simulators, which confirms the correctness of
    our methods and implementations. SAGD simulation with 15106 well pairs is also presented to study the
    effectiveness of our numerical methods. Scalability testings demonstrate that our simulator can handle giant
    models with 216 billion grid blocks using 100,800 CPU cores and the simulator has good scalability.
\end{abstract}

\section{Introduction}

Reservoir simulations play critical roles in reservoir management, since it provides one way to examine
production plan and to predict future oil and gas production\cite{dengh}. Simulators have been developed and applied for
decades, such as CMG STARS. They have widely used in reservoir management.
When multiple chemicals are considered in a model or the geological model is complicated, it may take too long
for one simulation, which reduces the productivity of reservoir engineers.
Acceleration of simulations is important to oil and gas industry.

Reservoir simulations have been studied for decades, and various models and methods have been proposed. 
Crookston et al.\cite{Crookston1979} {proposed a simple two-dimensional model to deal with three phases flow} and
to handle vaporization-condensation effects.
Grabowski \cite{Grabowski1979} developed a sequential implicit
method for thermal reservoir model.
A general four-phase multi-component in-situ combustion model was proposed by Coats \cite{Coats1980},
which was improved by Rubin \cite{Rubin1985} that a fully coupled
implicit wellbore model was considered.
Variable substitution \cite{Coats1980} methods and pseudo-equilibrium ratio (PER) methods \cite{Crookston1979} were
designed to discretize the thermal models, while Mifflin et
al\cite{Mifflin1991} suggested to use global variables, pressure, moles and energy as unknowns.
Barua \cite{Barua1989} proposed algorithms to solve the nonlinear equations in
parallel and combined the iterative solutions to linear systems and Quasi-Newton method.
Effective linear solver and preconditioner methods have been
proposed to accelerate the solution of linear systems from reservoir simulations, such as constrained pressure
residual (CPR) methods \cite{CPR-old,CPR-cao}, multi-stage methods \cite{Study-Two-Stage}, multiple level
preconditioners \cite{mlp} and FASP (fast auxiliary space preconditioners) \cite{FASP,FASP2}.
{Chen et al. designed a family of CPR-type preconditioners, such as
CPR-FP, CPR-FPF and CPR-FFPF methods \cite{bos-pc}}, which have been applied to different simulations
\cite{poly-wang,plat-liu,bos-3p-dpdp}.

Parallel computers have more memory and higher performance, which provide excellent approaches to accelerate
reservoir simulations \cite{Dogru2000,Coumou2008, Wu2002,Edwards2012a}.
In the early stage, vectorization techniques in shared-memory machines was widely applied though
it didn't scale very well \cite{Coats1987,Dogru2000}.
Meijerink \cite{Meijerink1991}
developed a black oil simulator using the IMPES method and implemented on a local-memory MIMD computer.
Chien \cite{Chien1997}
applied domain decomposition and MPI on an IBM SP-2 parallel computer.
Wang \cite{Wang1997,Parashar1997} implemented a
fully implicit equation-of-state compositional simulator for distributed-memory parallel computers,
and large-scale reservoir models were simulated \cite{Wang1999}.
Reservoir models with millions of grid blocks on parallel computers were reported\cite{Verdiere1999}.
Killough \cite{Killough1993} reviewed the parallel reservoir models and parallel computing technologies.
Saudi Aramco developed new-generation massively-parallel reservoir simulator
\cite{Al-Shaalan2003,Dogru1999,Dogru2002, Fung2000}, and reservoir models with millions of grid blocks
were studied. 
Zhang et al. developed a scalable general-purpose platform, which has been applied to
reservoir simulations \cite{phg,phg-quad,kwang}.

This paper introduces our work on developing a parallel thermal simulator, including mathematical model,
numerical methods and implementation. The model is introduced in details, and explanations are provided, which
compared with CMG STARS, such as modeling method and its default behaviors. Here are the features of our
methods and simulator:
\begin{enumerate}
    \item An automatic configure script has been developed to detect operating system and compiler options.
        With its help, the simulator can be compiled under any Unix-like systems, Linux systems and Mac OS,
        and any MPI implementations, such as IBM Spectrum MPI, Intel MPI, MPICH, OpenMPI, and MVAPICH.
        The codes are written by C language from scratch, and at this moment, over 80,000 lines of C code
        have been written.

    \item The simulator is designed to work with arbitrary CPU cores (MPI processes), such 1,000,000 cores.

    \item All data types are customized. The integer can be configured as integer (\verb|int|), long integer
        (\verb|long int|) and long long integer (\verb|long long int|). The floating point number could be
        double precision (\verb|double|) and long double (\verb|long double|). MPI support is required to
        handle long double.

    \item The simulator can handle arbitrary grid size, arbitrary oil components (heavy oil and light oil),
        arbitrary gas components (non-condensable gas) and arbitrary wells (injector, producer, and heating wells).
        Only parallel computing resource and MPI compilers can limit the capacity of the simulator. 
        The simulator has tested
        models with hundreds of billions of grid blocks, hundreds of oil and gas components,
        and tens of thousands of wells.

    \item A flexible keyword parsing model has been developed to handle user input. All properties, such as
        rock, water, oil and gas, heat and well, are handled by the keyword parsing module. Arbitrary oil and
        gas components, wells and schedules can be read and parsed.

    \item The K-value method is applied. The gas phase can be treated as ideal gas or non-ideal gas, which is
        controlled by keyword file. If it is non-ideal gas, the RK EOS is employed to handle it.

    \item Effective discretization schemes, multi-stage CPR-type preconditioners, decoupling methods and
        Newton methods have been developed.
    
    \item Techniques for accelerating Newton methods, such as damping, Appleyard method, modified Appleyard
        method and weighted upstream \cite{tomj}, have been developed in the simulator.

    \item Various well controls have been implemented, such as fixed bottom hole pressure, fixed water rate,
        fixed oil rate, fixed gas rate, fixed liquid rate, constant heat transfer model, convective heat
        transfer model, heating well (\verb|HTWELL| in CMG STARS), subcool (steamtrap, rate control,
        temperature control, dual rate/temperature control), and combinations of these controls.
        Their mathematical details are provided.

    \item The well index has several models, which are the same as CMG STARS, including user input and a few
        analytical models. Different well weights, such as unweighted, explicit weight and implicit weight for injector,
        explicit weight and implicit weight for producer, have been developed.

    \item Isenthalpic flash calculations and surface flash calculations are implemented to model injection, production and
        performance report.

    \item Various enthalpy calculation formula have been studied, including gas enthalpy, liquid enthalpy,
        and vaporization enthalpy.

    \item Anaytical formulas and table-based input for relative permeability and viscosity. For table input
        data, we have linear interpolation and cubic monotone interpolation.

    \item Various mixing rules have been developed for conduct (simple and complex), viscosity, and density.
        Different rock modeling, such as bulk constant and rock constant, are implemented as CMG STARS.

\end{enumerate}

The structure of the paper is as follows. In \S \ref{sec-model}, the thermal reservoir model is introduced and
the equations for various properties are presented. 
In \S \ref{sec-num}, numerical methods and parallel computing approaches are proposed. 
In \S \ref{sec-exp}, numerical experiments are carried out to validate our
results against commercial simulator, CMG STARS, and to show the scalability of the parallel thermal
simulator.

\section{Mathematical Model}
\label{sec-model}

Most simulators share the same theory framework\cite{cmg2015starguide,cdong,ckhuang}.
For the sake of completeness, the mathematical model of the thermal simulator is introduced here, and the
models are almost the same as reference \cite{cmg2015starguide,cdong,ckhuang}. 
The content of this section is borrowed from our previous
manuscript \cite{in-situ} and CMG STARS \cite{cmg2015starguide}. 
In reference \cite{in-situ}, the following assumptions were made: water
component exists in water and gas phases, all oil components exist in oil and gas phases, non-condensable
gas components exist in gas phase only, and all three phases co-exist during the entire simulation. In this
paper, different assumptions are made: the water component exists in water and gas phases, heavy oil
components exist in oil phase only, light oil components exist in both oil and gas phases, non-condensable gas
components exist in gas phase. Phase appearance and dis-appearance are allowd. Depending on the input,
arbitrary oil components and non-condensable gas components are allowed. Necessary changes have been made to
address the difference between the in-situ combustion model \cite{in-situ} and the thermal model applied here.

\subsection{Darcy’s Law}

Darcy's law is applied to model the velocity of a fluid phase, which describes the relation among
permeability, viscosity, saturation and pressure difference.
In our thermal model, the water phase ($w$), the oil phase ($o$) and the gas phase ($g$) co-exist
(\cite{chen2007reservoir}),
\begin{gather}
\begin{split}
\vec{u}_w & = -\frac{k_{rw}}{\mu_w} \vec{k} \left( \nabla p_w - \gamma_w \nabla z\right)\\
\vec{u}_o & = -\frac{k_{ro}}{\mu_o} \vec{k} \left( \nabla p_o - \gamma_o \nabla z\right)\\
\vec{u}_g & = -\frac{k_{rg}}{\mu_g} \vec{k} \left( \nabla p_g - \gamma_g \nabla z\right).
\end{split}
\label{eq4.1}
\end{gather}

\subsection{Mass Conservation Equations}

For a multi-phase, multi-component system, $x_{c,\alpha}$ denotes the mole fraction of a component in
the $\alpha$-phase. The mole number of a component in a phase and the total mole number of the phase are
denoted as $n_{c,\alpha}$ and $n_{\alpha}$, respectively. Thus the mole fractions are
\begin{gather}
x_{c,\alpha} = \frac{n_{c,\alpha}}{n_{\alpha}}.
\end{gather}
In the simplest thermal model, water phase has water component only, so $n_{w} = 1$. If the gas phase exists,
it may contains water, light oil and non-condensable gas components.
Since each component may exist in several phases, total mole number of component $c$ is written as 
below (\cite{chen2007reservoir}):
\begin{gather}
\frac{\partial}{\partial t}\left(\phi \Sigma_{\alpha}^{N_\alpha} \rho_\alpha S_\alpha x_{c,\alpha}\right) =
- \nabla \cdot \left( \Sigma_{\alpha}^{N_\alpha} \rho_\alpha S_\alpha \vec{u}_{\alpha} \right)
+ \Sigma_{\alpha}^{N_\alpha} q_{\alpha,well} x_{c,\alpha}.
\label{eq4.2}
\end{gather}

In this equation, it is noticeable that different from other models, the “mass” conserved here is only the
mole number rather than the mass. Also, $\rho_\alpha$ and $q_\alpha$ are the mole density and mole
production/injection of phase $\alpha$.

\subsection{Energy Conservation Equation}

The energy conservation equation for a thermal process (\cite{chen2007reservoir}) is described as:
\begin{gather}
\begin{split}
& \frac{\partial}{\partial t}\left(\phi(\rho_w S_w U_w + \rho_o S_o U_o + \rho_g S_g U_g) + (1-\phi)U_r\right) \\
= \quad & \nabla \cdot \left( K_T \nabla T \right)
- \nabla \cdot \left( \rho_w H_w \vec{u}_w + \rho_o H_o \vec{u}_o + \rho_g H_g \vec{u}_g \right)\\
& + (q_{w,well} H_w + q_{o,well} H_o + q_{g,well} H_g) - Q_{loss},
\end{split}
\label{eq4.3}
\end{gather}
where $U$ denotes the volumetric internal energy.  On the right-hand side, the first term
represents the conduction term. This is caused by a difference in temperature, where the rate of conduction is
constraint by $K_T$, the bulk thermal conductivity. The thermal conductivity here is a combination of liquid,
and rock, where a linear mixing rule is applied (\cite{cmg2015starguide}),
\begin{gather}
K_T = \phi \left[S_w K_w + S_o K_o + S_g K_g \right] + (1-\phi)K_r.
\end{gather}
In the equation, $K_w, K_o, K_g, K_r$ denote thermal conductivities for water phase, oil phase, gas
phase, and rock separately. This rule is also called simple mixing rule in CMG STARS. The complex mixing rule
is also implemented in the simulator, whose details can be read from CMG STARS manual. We should mention that
there are different ways to model rock internal energy: $(1-\phi)U_r$. In above equation, the porosity,
$\phi$, is a function of pressure and temperature, and $U_r$ is a function of temperature, so the rock
internal energy is a function of pressure and temperature. This method assumes the volume of a grid block does
not change. Another way is to assume the rock volume does not change, which uses $(1-\phi_i)U_r$ to model rock
internal energy. $\phi_i$ does not change during the simulation, and this method preserves the rock energy.
The second method is applied as the default method in CMG STARS and our simulator.

A heat loss term to underburden and overburden is also considered, and the semi-analytical method developed 
by Vinsome et al. \cite{hloss-vin} is applied.

\subsection{Capillary Pressure}

A capillary pressure $P_c$ is the pressure difference across the interface between two immiscible
fluids arising from capillary forces,
which are usually functions of saturation,
relationship (\cite{chen2007reservoir}):
\begin{gather}
p_w = p_o - p_{cow}(S_w), \quad p_g = p_o + p_{cog}(S_g).
\label{eq4.5}
\end{gather}

\subsection{Phase Saturation Constraint}
The solid phase is not considered. The water, oil and gas saturations have the following constraint,
\begin{equation}
    S_w + S_o + S_g = 1.
\end{equation}
The gas phase can appear and disappear. The PER (Pseudo-Equilibrium Ratios) method is applied to calculate
K-values of water component and light oil components such that water phase and oil phase do not disappear.
However, the water saturation and oil saturation should be handled carefully when they are too small and the
gas phase exists. The partial derivatives of K-values to saturations must be included when the saturations are
small.

\subsection{Phase Composition Constraints}

A constraint implies that the sum of all the components' mole fractions in a phase adds up to one, which is
usually encountered for in compositional flow (\cite{chen2007reservoir}):
\begin{gather}
\Sigma_{\alpha}^{N_\alpha} x_{c,\alpha} = 1, \quad \alpha = w,o,g.
\label{eq4.6}
\end{gather}
It comes from the total mole number of a given phase that
\begin{gather}
\Sigma_{\alpha}^{N_\alpha} n_{c,\alpha} = n_{\alpha}, \quad \alpha = w,o,g.
\end{gather}

\subsection{Phase Equilibrium Constraints}

In a multi-component system, a K-value (or an equilibrium ratio) is defined as the ratio of the mole fractions
of a component in its distributed two phases:
\begin{gather}
{K}_{c,\alpha_1,\alpha_2} = \frac{x_{c,\alpha_1}}{x_{c,\alpha_2}}.
\end{gather}
In our model, a K-value is a function of pressure and temperature, 
which is calculated from an analytic equation as:
\begin{gather}
K = \left(\frac{{kv}_1}{p} + {kv}_2 p + {kv}_3\right)\exp\left(\frac{{kv}_4}{T-{kv}_5}\right).
\end{gather}
When gas phase exists, calculations of K-values for water, light oil and heavy oil are as follows
(\cite{chen2007reservoir}; \cite{cmg2015starguide}):

\begin{gather}
\begin{split}
&{K}_{W} = K_W(p, T) \\
= & \left(\frac{kv1_{W}}{p} + kv2_{W} \cdot p + kv3_{W}\right)\exp{\left(\frac{kv4_{W}}{T-kv5_{W}}\right)},\\
&{K}_{O,i} = K_O[i](p, T) \\
= & \left(\frac{kv1_{O,i}}{p} + kv2_{O,i} \cdot p + kv3_{O,i}\right)\exp{\left(\frac{kv4_{O,i}}{T-kv5_{O,i}}\right)}.\\
&{K}_{O,i} = K_O[i](p, T) = 0. \\
\end{split}
\end{gather}
The KV1, KV4 and KV5 from CMG STARS are $1.1705e5$ atm, -3816.44, and -227.02 C.

In our thermal model, the calculations of K-values are modified, where 
the PER (Pseudo-Equilibrium Ratios) method(\cite{crookston1979numerical,
abou1985handling}) is applied for water and light oil,
\begin{gather}
{K*}_{W} = K^*_W(p, T) = \left(\frac{S_w}{S_w+n_{cg}}\right)K_W(p, T),\\
{K*}_{O,i} = K^*_O[i](p, T) = \left(\frac{S_o}{S_o+\epsilon}\right) K_O[i](p, T).
\end{gather}
In calculations of pseudo K-values, $\epsilon$ is a small number of the order of $1e-4$.
The water phase and oil phase exist through the entrie simulation. However, the gas phase is allowed to
disappear.
The gas phase mole fraction for the oil components and water component are functions of $p, T,
S_w, S_g$. The mole fraction in the gas phase for gas components are the basic unknowns:
\begin{gather}
y = y (p, T, S_w, S_g).
\end{gather}

\subsection{Phase Changes}
Gas phase is allowed to re-appear and disappear, which has to be checked and determined in each Newton
iteration. The K-value only validates when gas phase exists.

When gas phase exists, its saturation, $S_g$, is positive. If a non-positive gas saturation is detected, then
gas phase disappears. $S_g$ is set to 0.

When gas phase doesn't exist, $S_g$ is 0. If the following relationship is detected,
\begin{gather}
\Sigma_{i} y_{i} > 1,
\label{eq4.6}
\end{gather}
then gas phase re-appears. A small gas saturation is set, such as 1e-3.

A cell type boolean DOF (degrees of freedom, which will be introduced) is applied to store the gas phase status.

\subsection{Compressibility Factor of Real Gas}

In the thermal model, the Redlich-Kwong EOS (\cite{redlich1949thermodynamics}) is used to calculate the Z factor.
\begin{gather}
A = A(p, T) = 0.427480 \left(\frac{p}{p_{crit}}\right) \left(\frac{T_{crit}}{T}\right)^{2.5},\\
B = B(p, T) = 0.086640 \left(\frac{p}{p_{crit}}\right) \left(\frac{T_{crit}}{T}\right).
\end{gather}
In addition, the following mixing method is applied:
\begin{gather}
a = \sum_{i} {y_i T_{crit,i} \sqrt{\frac{T_{crit,i}}{p_{crit,i}}}},\\
b = \sum_{i} {y_i \frac{T_{crit,i}}{p_{crit,i}}},\\
T_{crit} = \left(\frac{a^2}{b}\right)^{\frac{2}{3}},\\
p_{crit} = \frac{T_{crit}}{b}.
\end{gather}
Then, after we have the coefficients A and B, the compressibility factor of real gas satisfies the equation \begin{gather}
Z^3 - Z^2 + (A - B - B^2)Z - AB = 0.
\end{gather}
This equation is cubic. Therefore, there are three roots for the equation. Also, a root might be virtual. In
this case, we choose the biggest real root. With the calculation of all the coefficients, the $Z$ factor is a
function of $p$, $T$, $x_i$ and $y_i$:
\begin{gather}
Z = Z(p, T, x_i, y_i).
\end{gather}

\subsection{Density}

For real gas mixture, the density of the gas phase can be calculated as:
$$\rho_g = \rho_g(p, T, x_i, y_i) = \frac{p}{Z(p, T, x_i, y_i) \cdot R \cdot T}$$

The water phase only contains one water component in this model, so the calculation of the water density is simple:
\begin{gather}
\rho_w = \rho_w(p, T) = \rho_{w,ref} \exp (cp_w (p-p_{ref})-ct1_w (T-T_{ref}) \\
-\frac{ct2_w}{2}(T-T_{ref})^2 + cpt_w (p-p_{ref}) (T-T_{ref}))
\end{gather}
where $\rho_{w,ref}$ is the reference density of the water phase at the reference temperature and pressure.

For oil component $O[i]$ is in the oil phase, the density can be calculated the same:
\begin{gather}
\rho_{O[i]} = \rho_{O[i]}(p, T) = \rho_{{O[i]},ref} \exp (cp_{O[i]} (p-p_{ref})-ct1_{O[i]} (T-T_{ref}) \\
-\frac{ct2_{O[i]}}{2}(T-T_{ref})^2 + cpt_{O[i]} (p-p_{ref}) (T-T_{ref}))
\end{gather}
The density of oil phase, $\rho_o$, which is mixture of multiple oil components, is calculated as:
\begin{gather}
\frac{1}{\rho_o} = \sum^{n_{co}}_i \frac{x_i}{\rho_{O[i]}}.
\end{gather}

\subsection{Viscosity}

There are a few ways to calculate viscosity, such as table input and analytical correlations. For table input
method, interpolations are required to calculate the viscosity of a component or a phase at a given
temperature. In the following, analytical method is introduced for oil, water and gas.

The viscosity of heavy oil is very high, and we assume the viscosity of an oil component is a function of
temperature,
\begin{gather}
\mu_{O[i]} = avisc_{O[i]} \exp{\left(\frac{bvisc_{O[i]}}{T}\right)}.
\end{gather}
The oil phase viscosity is calculated by a logarithmic mixing rule:
\begin{gather}
\ln(\mu_o) = \sum_i^{n_{co}} x[i] \ln (\mu_{O[i]}(T)),
\end{gather}
which is equivalent to,
\begin{gather}
\mu_o = \mu_o (T, x_i) = \exp {\left(\sum_i^{n_{co}} x[i] \ln (\mu_{O[i]}(T))\right) } = 
    \sum_i^{n_{co}} \left(\mu_{O[i]}(T)\right)^{x_i}.
\end{gather}

The water phase has only one component, and its viscosity is calculated as:
\begin{gather}
\mu_w = \mu_w (T) = avisc_w \exp{\left(\frac{bvisc_w}{T}\right)}.
\end{gather}
Another option for water is to use an internal viscosity table as shown by Figure \ref{fig-wat-vis}.
\begin{figure}[!htb]
\centering
\begin{evb}
    5.0         1.5182         125.0       0.2227             500.0       7.2818E-02
    8.0         1.386          150.0       0.1848             600.0       7.2818E-02
    10.0        1.311          175.0       0.1586             700.0       7.2818E-02
    20.0        1.005          200.0       0.1394             800.0       7.2818E-02
    30.0        0.8004         225.0       0.1238
    40.0        0.6543         250.0       0.1117
    50.0        0.5518         275.0       0.1005
    60.0        0.4714         300.0       9.9125E-02
    70.0        0.4066         325.0       8.4075E-02
    80.0        0.3570         350.0       7.7437E-02
    90.0        0.3182         375.0       7.2818E-02
    100.0       0.2828         400.0       7.2818E-02
\end{evb}
\caption{Water viscosity table: temperature (C) vs viscosity (cp)}
\label{fig-wat-vis}
\end{figure}

The gas component viscosity is calculated as, 
\begin{gather}
\mu_{g,i} = \mu_{g,i}(T) = avg_i \cdot T ^ {bvg_i}.
\end{gather}
According to a mixing rule, the mole mass of a component is included:
\begin{gather}
\mu_{g} = \mu_{g}(p, T, S_w, S_g, x_i, y_i) = \frac{\sum_i{\mu_{g,i} \cdot y_i\sqrt{M_i} }}{\sum_i{y_i\sqrt{M_i}}},
\end{gather}
where $M_i$ is molecular weight of $i$-th component.
Another way to calculate gas phase viscosity is to use the following correlation,
\begin{gather}
    \mu_{g} = \mu_{g}(T) = 0.0136 + 3.8 * 10^{-5} * T,
\end{gather}
where $T$ is in degree C. An internal gas phase viscosity is also available as shown by Figure \ref{fig-gas-vis}.
\begin{figure}[!htb]
\centering
\begin{evb}
           10.0       1.3979E-02              200.0       2.1203E-02
           20.0       1.4360E-02              225.0       2.2153E-02
           30.0       1.4740E-02              250.0       2.3103E-02
           40.0       1.5120E-02              275.0       2.4054E-02
           50.0       1.5500E-02              300.0       2.5004E-02
           60.0       1.5880E-02              325.0       2.5955E-02
           70.0       1.6260E-02              350.0       2.6905E-02
           80.0       1.6641E-02              375.0       2.7855E-02
           90.0       1.7021E-02              400.0       2.8806E-02
          100.0       1.7401E-02              500.0       3.2607E-02
          125.0       1.8351E-02              600.0       3.6409E-02
          150.0       1.9302E-02              700.0       4.0210E-02
          175.0       2.0252E-02              800.0       4.4012E-02
\end{evb}
\caption{Gas phase viscosity table: temperature (C) vs viscosity (cp)}
\label{fig-gas-vis}
\end{figure}
\subsection{Porosity}

Porosity is the ratio of the pore volume to the bulk volume in a porous medium, describing the volume
containing fluids. When pressure is high, due to the effort of fluids, pores are also enlarged. For a
non-isothermal model, the porosity is also influenced by temperature.
We define a coefficient as a total compressibility of porosity (\cite{chen2007reservoir}):
\begin{gather}
\phi_c = \phi_c(p, T) = cpor(p-p_{ref}) - ctpor(T-T_{ref}) + cptpor(p-p_{ref})(T-T_{ref}).
\end{gather}
This factor is a function of pressure and temperature. For the calculation of porosity, we have two approaches with this factor:

Linear:
\begin{gather}
\phi = \phi (p, T) = \phi_{ref} \cdot (1 + \phi_c(p, T)).
\end{gather}

Nonlinear:
\begin{gather}
\phi = \phi (p, T) = \phi_{ref} \cdot e^{\phi_c(p, T)}.
\end{gather}
For both two approaches, porosity is a function of pressure and temperature. The porosity may be set to 0 if
it is smaller than certain value, such as 1e-3.

\subsection{Relative Permeabilities}

There are two ways for calculating relative permeabilities. The first one is to use analytical correlations,
and the second one is to use input tables as shown by Figure \ref{rela-swt}. It has four columns, water
saturation $S_w$, $k_{rw}$, $k_{row}$, and capillary pressure. Here capillary pressures are ignored, and all
values are zero.
\begin{figure}[!htb]
\centering
\begin{evb}
swt: 
#   Sw        Krw        Krow         Pcw
    0.45     0.0         0.4
    0.47     0.000056    0.361
    0.50     0.000552    0.30625
    0.55     0.00312     0.225
    0.60     0.00861     0.15625
    0.65     0.01768     0.1
    0.70     0.03088     0.05625
    0.75     0.04871     0.025
    0.77     0.05724     0.016
    0.80     0.07162     0.00625
    0.82     0.08229     0.00225
    0.85     0.1         0.0 
\end{evb}
\caption{Oil-water relative permeability table}
\label{rela-swt}
\end{figure}

The water phase relative permeability, $k_{rw}$, can be obtained with interpolation from oil-water relative
permeability table, which is a function of $S_{w}$ (and temperature):
\begin{gather}
k_{rw} = k_{rw}(S_w).
\end{gather}
The gas phase relative permeability, $k_{rg}$, can be calculated the same from a gas-oil relative permeability
table or liquid-gas relative permeability table, which is a function of $S_{g}$ (and temperature):
\begin{gather}
k_{rg} = k_{rg}(S_g).
\end{gather}
Figure \ref{rela-slt} is a sample liquid-gas relative permeability table, which also has four columns: $S_l =
S_o + S_w = 1 - S_g$,
$k_{rg}$, $k_{rog}$, and capillary pressure. Here the capillary pressure is zero.
\begin{figure}[!htb]
\centering
\begin{evb}
slt:
#   Sl       Krg         Krog           Pcg
    0.45     0.2         0.0
    0.55     0.14202     0.0
    0.57     0.13123     0.00079
    0.60     0.11560     0.00494
    0.62     0.10555     0.00968
    0.65     0.09106     0.01975
    0.67     0.08181     0.02844
    0.70     0.06856     0.04444
    0.72     0.06017     0.05709
    0.75     0.04829     0.07901
    0.77     0.04087     0.09560
    0.80     0.03054     0.12346
    0.83     0.02127     0.15486
    0.85     0.01574     0.17778
    0.87     0.01080     0.20227
    0.90     0.00467     0.24198
    0.92     0.00165     0.27042
    0.94     0.0         0.30044
    1.       0.0         0.4 
\end{evb}
\caption{A liquid-gas relative permeability table}
\label{rela-slt}
\end{figure}

As for the relative permeability of oil $k_{ro}$, there are several models available (\cite{corey1956three};
\cite{naar1961three}, 1961; \cite{stone1970probability}; \cite{delshad1989comparison}). In our model, 
the Stone’s model II method (\cite{stone1973estimation}) is applied:
\begin{gather} k_{ro} = k_{ro}(S_w, S_g)\\ =
    k_{rocw}\left[\left(\frac{k_{row}(S_w)}{k_{rocw}}+k_{rw}(S_w)\right)
    \left(\frac{k_{rog}(S_g)}{k_{rocw}}+k_{rg}(S_g)\right)-k_{rw}(S_w)-k_{rg}(S_g)\right],
\end{gather}
where $k_{rocw}$ is the oil-water two-phase relative permeability to oil at connate water
saturation, $krog$ is the oil-gas two-phase relative permeability to oil, and $krow$ is the oil-water
two-phase relative permeability to oil.
\begin{gather}
k_{rocw} = k_{row}(S_w = S_{wc}) = k_{rog}(S_g = 0).
\end{gather}
$krow$ and $krog$ are interpolated from input tables.

\subsection{Energy}

Enthalpy is a measurement of energy in a thermodynamic system, which is equal to 
the internal energy of the system plus the product of
pressure and volume. There are three ways for calculation of enthalpy: gas-based, liquid based, and simple
Hvap.

\noindent \textbf{Gas-based enthalpy.}
The enthalpy of a gas component is calculated as follows (\cite{cmg2015starguide}):
\begin{gather}
H_{g,i} = H_{g,i}(T) = \int_{T_{ref}}^T {\left(cpg1_i + cpg2_i \cdot t + cpg3_i \cdot t^2 + cpg4_i \cdot
    t^3 + cpg5_i \cdot t^4\right)}dt,
\end{gather}
$cpg1_i$, $cpg2_i$,  $cpg3_i$, $cpg4_i$, and $cpg5_i$ are constants for component $i$.
The gas phase enthalpy can be calculated by a weighted mean with gas mole fractions $y_i$:
\begin{gather}
H_g = H_g (p, T, S_w, S_g, x_i, y_i) = \sum_{i}^{N_c} {y_i H_{g,i}}.
\end{gather}

For the oil and water phases, the heat of vaporization should be considered, which can be calculated by:
\begin{equation}
H_{v,i} = H_{v,i} (T) =\left\{
 \begin{aligned}
     & hvr_i \cdot (T_{crit,i} - T)^{ev_i}, & & T < T_{crit,i}; \\
     & 0,                                   & & T >= T_{crit,i};
\end{aligned}
 \right.
\end{equation}
The enthalpy of a liquid component can be calculated as:
\begin{gather}
H_{i} = H_{i} (T) = H_{g,i} - H_{v,i}.
\end{gather}
where $H_{g,i}$ is the enthalpy of component $i$ in the gas phase.
As a result, for the water phase which only includes one component, the enthalpy is:
\begin{gather}
H_w = H_w (T) = H_{g,W} - H_{v,W}.
\end{gather}
For the oil phase, as a mixture, the enthalpy is:
\begin{gather}
H_o = H_o (p, T, x_i) = \sum_{i}^{n_{co}} {x_i (H_{g, O[i]} - H_{v, O[i]})}.
\end{gather}

\noindent \textbf{Liquid-based enthalpy.}
The enthalpy of a liquid component is calculated as follows (\cite{cmg2015starguide}):
\begin{gather}
H_{l,i} = H_{l,i}(T) = \int_{T_{ref}}^T {\left(cpl1_i + cpl2_i \cdot t + cpl3_i \cdot t^2 + cpl4_i \cdot
    t^3 + cpl5_i \cdot t^4\right)}dt,
\end{gather}
$cpl1_i$, $cpl2_i$,  $cpl3_i$, $cpl4_i$, and $cpl5_i$ are constants for component $i$.
The enthalpy of a condensable gas component is calculated as,
\begin{gather}
H_{g, i} = H_{g, i} (T) = H_{l,i} + H_{v,i}.
\end{gather}
And the enthalpy of a non-condensable gas component can be computed using the gas-based method.

\noindent \textbf{Simple Hvap method.} The enthalpy of a liquid component uses,
\begin{gather}
H_{l,i} = H_{l,i}(T) = \int_{T_{ref}}^T {\left(cpl1_i + cpl2_i \cdot t + cpl3_i \cdot t^2 + cpl4_i \cdot
    t^3 + cpl5_i \cdot t^4\right)}dt,
\end{gather}
while the enthalpy of a condensable component uses,
\begin{gather}
H_{g,i} = H_{g,i}(T) = hvapr + \int_{T_{ref}}^T {\left(cpg1_i + cpg2_i \cdot t + cpg3_i \cdot t^2 + cpg4_i \cdot
    t^3 + cpg5_i \cdot t^4\right)}dt,
\end{gather}
where $hvapr$ is user input parameter.
The enthalpy of non-condensable gas component uses gas-based method.

If gas-based enthalpy is applied, $cpg1$, $cpg2$,  $cpg3$, $cpg4$, and $cpg5$ for water from CMG STARS are 34.49885 J/gmol-C,
-0.01426 J/gmol-C$^2$, 4.7356e-5 J/gmol-C$^3$, -3.56759e-8 J/gmol-C$^4$, and 9.35531e-12J/gmol-C$^5$.
$hvr$ and $ev$ are 4820 J/gmol-C$^{0.38}$, and 0.38. Other correlations also work. Another option is to use
steam enthalpy table for water component.

If enthalpy parameters for oil and gas components aren't provided, liquid-based method for oil components are
applied. Condensible components: $cpl1$ is 0.5 Btu/lb-F, $hvr$ is 0.25 Btu/lb-F, and $ev$ is 1. Heavy oil
components: $cpl1$ is 0.5 Btu/lb-F. Non-condensible gas components: $cpl1$ is 0.25 Btu/lb-F.

The internal energy for oil, gas, and water phases (\cite{cmg2015starguide}) are calculated as:
\begin{gather}
U_w = U_w (T) = H_w - p/\rho_w,\\
U_o = U_o (p, T, x_i) = H_o - p/\rho_o,\\
U_g = U_g (p, T, S_w, S_g, x_i, y_i) = H_g - p/\rho_g.
\end{gather}
For rock, a similar formula is used:
\begin{gather}
U_r = U_r (T) = cp1_r (T-T_{ref}) + \frac{cp2_r}{2}(T^2 - T_{ref}^2).
\end{gather}
One thing to notice is that the internal energy for rock has a unit of energy per unit volume, while others
have energy per unit amount of material. As mentioned above, there are two ways to calculate the volume of rock.
The following equation defines relationship among bulk volume $V_b$, rock volume $V_r$ and pore volume $V_p$,
\begin{gather}
    V_b = V_r + V_p,
\end{gather}
where $V_p$ is calculated by porosity correlations and $V_r$ is used when calculating the internal energy of
rock.
The first one
assumes the volume of rock (non-null) doesn't change, which is noted as constant rock in CMG STARS
(\verb|*VOLCONST *ROCK|). It assumes $V_r$ is constant, which preserves rock mass and heat, and $V_b$ changes
as $V_r$ changes.  The second one assumes
the volume of the grid block doesn't change, which is noted as constant bulk in CMG STARS
(\verb|*VOLCONST *BULK|). It assumes $V_b$ is constant and $V_r$ ($V_r = V_b - V_p$) changes as $V_p$ changes.
The default method is rock constant.

\subsection{Well Modeling}

A Peaceman's model is adopted for well modeling in this paper. A well may have many perforations, and
each perforation at a grid cell, its well rate for phase $\alpha$, $Q_{\alpha} = Vq_{\alpha}$, is 
calculated by the following formula (\cite{peaceman1978interpretation}):
\begin{gather}
    \label{well-model}
    Q_{\alpha, well} = WI\frac{\rho_{\alpha}k_{r\alpha}}{\mu_{\alpha}}
    \left(p_b - p_\alpha -\gamma_{\alpha}g(z_{bh} - z)\right),
\end{gather}
where WI is the well index and mobility is explicit or implicit. In CMG STARS, well rate can also be
calculated using a third method, unweighted method,
\begin{gather}
Q_{\alpha, well} = WI \left(p_b - p_\alpha -\gamma_{\alpha}g(z_{bh} - z)\right),
\end{gather}
where $WI$ is user input value.
A well index defines the relationship among a well bottom hole pressure, a flow rate and a grid block
pressure. $p_b$ is the bottom hole pressure defined at the reference depth z, $z_bh$ is the depth of the
perforation in grid cell, and $p_\alpha$ is the phase pressure in grid block $m$.
Well index can be read from modelling file and it can also be calculated using analytical method.
For a vertical well, it can be defined as:
\begin{gather}
WI= \frac{2\pi h_3 \sqrt{k_{11}k_{22}}}{\ln(\frac{r_e}{r_w})+s},
\end{gather}
where $r_e$ is equivalent radius. The calculation of well index can be controlled by several parameters, such
as \verb|geo|, \verb|geoa|, \verb|kh|, \verb|kha| and \verb|geofac|. Figure \ref{well-wi} shows a few commonly
used well index methods.
\begin{figure}[!htb]
\centering
\begin{evb}
# user input: 1e4
perf:
    # perf i,j,k wi
    1 1 1:10  1e4
    1 1 13:20 1e4
/

# user input: 1e4
perf: wi
    # perf i,j,k wi
    1 1 1:10  1e4
    1 1 13:20 1e4
/

# internal calculation: geo
perf: geo
    # perf i,j,k ff
    1 1 1:10  1
    1 1 13:20 0.81
/

# internal calculation: geoa
perf: geoa
    # perf i,j,k ff
    1 1 1:10  1
    1 1 13:20 0.81
/
\end{evb}
\caption{Well index types}
\label{well-wi}
\end{figure}
More details can be read 
from CMG STARS manual. Horizontal can be defined similarly. We should mention that well modeling is the most
complicated part in reservoir simulations and various operation constraints can be defined, such as fixed
bottom hole pressure, fix liquid and gas rate constraints and thermal constraints.

The bottom hole pressure update is handled differently in CMG STARS and in our simulator. In CMG STARS, the
bottom hole pressure is updated by change or in the beginning of each time step. For the change option, CMG
STARS updates bottom hole pressure if the change is large enough. In our simulator, the bottom hole pressure
is updated in each Newton iteration.

\subsubsection{Flash Calculation}

In equation (\ref{well-model}), $\rho_{\alpha}$, $k_{r\alpha}$ and $\mu_{\alpha}$ need to be calculated. For
production wells, they are from the grid block that contains the perforation, which are straightforward.
However, for injection wells, the mobility is the total mobility,
\begin{equation}
    \frac{k_{r}}{\mu} = \frac{k_{ro}}{\mu_{o}} + \frac{k_{rw}}{\mu_{w}} + \frac{k_{rg}}{\mu_{g}}.
\end{equation}
In CMG STARS, injection wells have three options, unweighted mobility, implicit mobility and explicit
mobility. The implicit mobility is
updated in each Newton iteration and the explicit mobility is updated at the beginning of each time step.
Production wells also have two options, implicit mobility and explicit mobility.
Iso-enthalpy flash calculation is required in each perforation to determine the status of injected fluids,
such as pressure, temperature, distributions in three phases, and density. For example, the injected water
(steam) can stay in liquid (pure water), steam (pure gas) and mixture of water and steam states depending on
the wellbore pressure and temperature of a perforation.

The choice of mobility models affects the calculations of well rates, Jacobian matrix, and numerical treatment
of each well. An injection well has three
options: unweighted mobility (including well index), implicit mobility and explicit
mobility. When unweighted mobility is applied, user input value is required for this well. The calculations
for explicit mobility is easier than implicit mobility, and many partial derivates are ignored when
assembling Jacobian matrix. The explicit mobility is less accurate then implicit mobility, but it can be
faster and more stable. When the implicit mobility is applied, many properties have to be updated in each
Newton iteration. A production well has two options: implicit mobility and explicit mobility.
In our simulator, each well can be assigned to any allowed mobility models. For example, well 1 is an
injection well and it applies unweighted mobility, well 2 is also an injection well and it applies explicit
mobility, and well 3 is a production well and it applies implicit mobility.

\subsubsection{Fixed Bottom Hole Pressure}

When the fixed bottom hole pressure condition is applied to a well, the well equation is written as,
\begin{gather}
    p_b = c,
\end{gather}
where $c$ is pressure and is a constant. The bottom hole pressure is defined at a reference depth or a grid
block. If neither is provided, the grid block contains the first perforation is served as reference grid block.

\subsubsection{Fixed Rate}

Fixed rate constraints are commonly used, including fixed oil rate, fixed water rate, fixed gas rate, and
fixed liquid rate (oil and water). The rate can be reservoir rate or surface rate. The volume of a fluid in
reservoir condition can be obtained easily. However, the volume of a fluid in surface condition requires flash
calculation to determine the distribution in oil, water and gas phases.
There are two ways to separate phase: segregated method and PT-flash method. The segregated method is easy but
the PT-flash is tricky. In CMG STARS, the segregated method is the default.
For phase $\alpha$, its fixed rate constraint is described by the following equation:
\begin{gather}
\sum_{m} \left( Q_{\alpha, well} \right)_{m} = c,
\end{gather}
where $c$ is a constant rate and known.
The fixed liquid rate is written as,
\begin{gather}
\sum_{m} \left(Q_{w, well} \right)_{m} + \sum_{m} \left(Q_{o, well} \right)_{m} = c,
\end{gather}
The fixed total fluid rate is written as,
\begin{gather}
\sum_{m} \left(Q_{w, well} \right)_{m} + \sum_{m} \left(Q_{o, well} \right)_{m} + \sum_{m} \left(Q_{g, well} \right)_{m} = c,
\end{gather}

\subsubsection{Constant Heat Transfer Model}

CMG STARS is the most popular thermal simulator, and it has many heater models, such as constant heat transfer
model (\verb|heatr| in CMG STARS), convective heat transfer model and heat well, which are applied to model heating stage. 
The first two types can be defined in any grid block. However, the heat well (\verb|HTWELL| in CMG STARS) can only
be defined in a real well, such as injection well and production well.
The constant heat transfer model means in some grid blocks, there exist heat transfter at certain rate, such
as 1,000 Btu/day. The energy exchange can occur in any grid block. The heat transfer can be turned on or off
using schedule.

\subsubsection{Convective Heat Transfer Model}
Constant heat transfer model simulates constant heat exchange while convective heat transfer model defines
dynamic heat transfer, which is controlled by two parameters: $\texttt{uhtr}$ (proportional heat transfer coefficient, 
Btu/day-F) and $\texttt{tmpset}$ (temperature
setpoint, F) 
(\verb|UHTR| and \verb|TEMSET| in CMG STARS).
If $\texttt{uhtr}$ is positive, it means to gain heat from source, the heat rate in a grid block 
is calculated as,
\begin{equation}
q = \left\{
 \begin{aligned}
     & \texttt{uhtr} * (\texttt{tmpset} - T), & &if \ \texttt{tmpset} > T; \\
     & 0,                   & &if \ \texttt{tmpset} <= T;
\end{aligned}
 \right.
\end{equation}
If $\texttt{uhtr}$ is negative, it means the reservoir loses heat, the heat rate in a grid block 
is calculated as,
\begin{equation}
q = \left\{
 \begin{aligned}
     & \texttt{uhtr} * (T - \texttt{tmpset}), & &if \ \texttt{tmpset} < T; \\
     & 0,                   & &if \ \texttt{tmpset} >= T;
\end{aligned}
 \right.
\end{equation}
where $T$ is the reservoir temperature.

\subsubsection{Heater Well}
As mentioned above, the constant and convective heat transfer models can be defined in any grid block. Another
heat model is also developed in CMG STARS and our simulator, which is noted as \verb|HTWELL| as in CMG STARS.
This type of heater is defined in a production or injection well, which has the same perforations as the well
contains the heater well. This heater well is more complicated than constant and convective heater transfer
models, which has more controls, such as heat rate model (\verb|HTWRATE| or \verb|HTWRATEPL| in CMG STARS),
temperature model (\verb|HTWTEMP| in CMG STARS), heat index model (\verb|HTWI| in CMG STARS), and dual rate/temperature
model. The dual rate/temperature model has two direction controls: uni-directed (\verb|UNIDIRECT| in CMG STARS) and
bi-directed (\verb|BIDIRECT| in CMG STARS).

For heat rate control (model), the heat rate in a perforation $m$ is calculated as,
\begin{equation}
    q = q_{hspec} = Q_{hspec} L_m / L_w,
\end{equation}
where $q$ is the heat rate, $Q_{hspec}$ is total heat rate defined by \verb|HTWRATE|, $L_m$ is the length of the layer
well completion, and $L_w$ is the total well length (sum of $L_m$).

For temperature model, the heat rate in a perforation is calculated as,
\begin{equation}
    q = q_{wspec} = I_{m} * (T_{wspec} - T_m),
\end{equation}
where $I_m$ is the heat conduct index (or heat index), $T_{wspec}$ is specify wellbore temperature, $T_m$ is grid block
temperature. We should mention that there are two method for calculating heat conduct index: 1) use thermal
conductivity formula introduced in mathematical model section; 2) use heat index introduced here (by turning
\verb|HTWI| on in CMG STARS).
For heat index model, the user input well index or internal index can be converted to heat index.

When dual rate/temperature model is enabled, the rate model and temperature model are switched automatically.
For heating ($Q_{hspec} > 0$), the heat rate in a layer is defined as,
\begin{equation}
    q = min\{I_m * \Delta T_m, q_{hspec}\},
\end{equation}
where $\Delta T_m$ is defined as,
\begin{equation}
\Delta T_m = \left\{
 \begin{aligned}
     & max\{T_{wspec} - T_k, 0\}, & & for \ \texttt{UNIDIRECT} \\
     &  T_{wspec} - T_k, & & for \ \texttt{BIDIRECT},
\end{aligned}
 \right.
\end{equation}
The $T_k$ is reservoir temperature in a grid block.
The \verb|UNIDIRECT| option shuts down heater when temperature difference is zero; while \verb|BIDIRECT|
allows heating and cooling (heat loss).

\begin{figure}[!htb]
    \centering
    \includegraphics[width=0.84\linewidth]{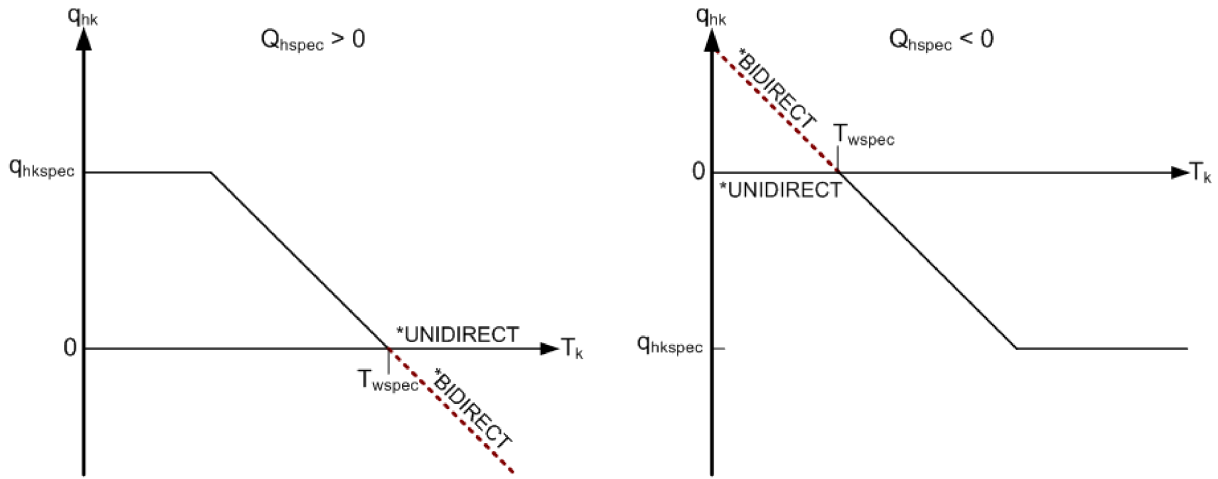}
    \caption{HTWELL: dual rate/temperature model \cite{cmg2015starguide}}
    \label{htwell-dual}
\end{figure}

For cooling ($Q_{hspec} < 0$), the heat rate in a layer is defined as,
\begin{equation}
    q = max\{I_m * \Delta T_m, q_{hspec}\},
\end{equation}
where $\Delta T_m$ is defined as,
\begin{equation}
\Delta T_m = \left\{
 \begin{aligned}
     & min\{T_{wspec} - T_k, 0\}, & & for \ \texttt{UNIDIRECT} \\
     &  T_{wspec} - T_k, & & for \ \texttt{BIDIRECT},
\end{aligned}
 \right.
\end{equation}
The \verb|UNIDIRECT| option shuts down cooling well, and the \verb|BIDIRECT| option allows bidirectional
heat transfer.
In both cases, the \verb|BIDIRECT| can simulate autoheater and autocooler.
Their meanings are shown by Figure \ref{htwell-dual} \cite{cmg2015starguide}.

\subsubsection{Subcool Control}
Subcool control is also known as steamtrap, which is used to prevent the production of live steam. It does
this by keeping the well's flowing bottomhole pressure (and hence the pressure in the grid block
containing the well) high enough that live steam does not appear in the well block \cite{cmg2015starguide}.

The well constraint equation solved is written as,
\begin{equation}
    T_{sat}(p_{wb}) - T_k = c,
\end{equation}
where $c$ is a pre-defined temperature difference, $T_{sat}$ is the steam saturation temperature corresponding
to wellbore pressure $p_{wb}$ defined in the perforation, and $T_k$ is the temperature defined in the grid
block that contains the perforation.

\subsubsection{Well Operations}

The thermal simulator supports the following well operations:
\begin{itemize}
    \item \textbf{BHP}: bottom hole pressure, reservoir condition. The bottom hole pressure is defined
        at a reference depth or a grid block. If neither is set, the first perforation is used as reference
        grid block.

    \item \textbf{STW}: water rate, surface condition. Flash calculation is required to calculate water
        volume at a given surface condition. For production wells, the water may from water phase or gas
        phase. For injection wells, the water is the total injected water component, which is measured as cold
        water at a given surface condition.

    \item \textbf{STO}: oil rate, surface condition. Flash calculation is required to calculate water,
        gas ad oil distribution, oil phase density at surface condition.
        The oil phase at surface condition may be from oil phase (reservoir condition) and gas phase
        (reservoir condition).

    \item \textbf{STG}: gas rate at surface condition, in which flash calculation is required.

    \item \textbf{STL}: liquid rate, surface condition. The rate is sum of oil rate and water rate.

    \item \textbf{STF}: total fluid rate, surface condition. The rate is sum of oil rate, water rate and
        gas rate.

    \item \textbf{BHW}: water rate, reservoir condition. No flash calculation is required.

    \item \textbf{BHO}: oil rate, reservoir condition. No flash calculation is required.

    \item \textbf{BHG}: gas rate, reservoir condition. No flash calculation is required.

    \item \textbf{BHL}: liquid rate, reservoir condition. No flash calculation is required.

    \item \textbf{BHF}: total fluid rate, reservoir condition. No flash calculation is required.

    \item \textbf{STEAM}: live steam rate, reservoir condition, cold water equivalent. No flash calculation is required.

    \item \textbf{STEAMTRAP}: increase bottom hole pressure such that no live steam is produced.

\end{itemize}

Each operation has a specifier (min and max). When multiple constraints are applied to a
well, if a constraint is picked as current operation, the equality relation is applied, such as fixed water
rate. Other constraints are applied as inequality, which is controlled by min or max specifier. For example,
the following operations can be applied to a well simultaneously,
\begin{figure}[!htb]
\centering
\begin{evb}
{
    operate: min bhp 17
    operate: max stw 100
    operate: max steam 10
    operate: max stl 1000
    operate: max stf 10000
    operate: max stg 800

    # steam trap at all perforations
    operate: steamtrap 10

    # steam trap at grid block (1, 1, 1), override 10
    operate: steamtrap 15 1 1 1

    # htwell: off, rate, temp (temperature), dual
    htwell: off
    htwrate: 3.4e6
    htwtemp: 611

    # bidirect or unidirect
    htwell_direction: bidirect

    # optional: on or off
    htwi: on

    # constant heater
    heatr: 9 1 1:4  1e6
    heatr: 5 5 1:4  1e6 

    # convective heater
    uhtr: 9 1 1:4  4e4 
    tmpset: 9 1 1:4  500

    # convective heater
    uhtr: 5 5 1:4  4e4 
    tmpset: 5 5 1:4  500 
}
\end{evb}
\caption{Well operations}
\end{figure}

\subsection{Boundary Conditions}

No flow boundary condition is applied to fluid, which is coupled with each mass conservation equation. For
energy conservation equation, heat loss to underburden and overburden is considered, which is modeled by
a semi-analytical method \cite{hloss-vin}. Each well may have multiple constraints, which is user input. They
are determined and switched dynamically during the simulations.

\subsection{Initial Conditions}

A few initial methods are supported. The easiest one is to use explicit initial conditions, such as pressure,
temperature, mole fraction and saturations. Another one is to use gravity average, in which the pressure is
calculated by depth difference to reference depth (grid block). The saturations, mole fractions and
temperature can be computed or be given by user input.

\section{Numerical Methods}
\label{sec-num}

In our previous previous work, a few reservoir simulators and their numerical methods have been reported
\cite{bos-pc,bos-3p,bos-3p-dpdp,poly-wang,jialuo}. The simulators share similar methods, such as time
discretization scheme, spatial discretization scheme, decoupling method, linear solver and preconditioners
\cite{bos-pc}. For the sake of completeness, the numerical methods are introduced in this section. There are
two main sets of unknowns: natural variables and overall variables. The natural variables use pressure,
temperature, saturations, and mole fractions. The overall variables use pressure, temperature and overall mole
fractions. Phase changes have to be checked in each nonlinear iteration and time step if we use natural
variables. When liquid and gas
phases co-exist, temperature and pressure are not independent, and only one variable is required, such as
pressure or temperature. Some researchers applied the variable substitution trick to switch unknowns and to
save computation. To overall variables, phase status is determined after obtaining solutions. In this paper,
fully implicit method is employed, which is friendly to large time step and to accuracy. However, it's
possible to apply some techniques to speed simulation, such as adaptive methods.

In this paper, one additional equation is adopted, when enables us to treat pressure and temperature as
independent variables through the entire simulation, which only introduces a little more computation but
simplifies the numerical treatment and linear systems.

\subsection{Time Discretization}

Let $u$ be a vector function, $u^n$ be the solution of $u$ at a given 
time step $n$, and $F$ be non-linear mathematical system of thermal reservoir model.
The backward Euler method is applied to discretize a time derivative,
\begin{equation}
    (\frac{\partial u}{\partial t})^{n+1} = \frac{u^{n+1} - u^{n}}{\Delta t} = F(u^{n + 1}, t^{n+1}),
\end{equation}
where $\Delta t$ is a time step. An implicit non-linear system is obtained,
which is solved at each time step using Newton method.

\subsection{Spatial Discretization}

The natural variables are applied as knowns, which are also called Type A variables, including pressure,
temperature, saturations, and mole fractions (oil components in oil phase and non-condensable gas in gas phase). 
The variables do not change during the simulation. However, depending on the gas phase status, one
constrainted equation is switched. If gas phase exist, the following equation is applied,
\begin{equation}
    \sum_i y_i = 1.
\end{equation}
If gas phase does not exist, the following equation is switched,
\begin{equation}
    S_g = 0.
\end{equation}
The status of gas phase has to be checked block by block in each Newton iteration.

When fluids move in a reservoir, there may be fluid exchange in two neighboring grid blocks,
which is described by transmissibility. 
Assuming $d  ~ (d = x, y, z)$ is a space direction and $A$ be the area of a face in the $d$ direction,
the transmissibility $K_{\alpha,d}$ for phase $\alpha ~(\alpha = o, w, g)$ is defined as
\begin{equation}
    T_{\alpha,d} = \frac{K A}{\Delta d} \times  \frac{K_{r\alpha}}{\mu_{\alpha}} \rho_{\alpha},
\end{equation}
where $\Delta d$ is the grid block length in the $d$ direction, $K$ is the absolute permeability,
$K_{r\alpha}$ is the relative permeability of phase $\alpha$, $\mu_\alpha$ is the viscosity of phase
$\alpha$ and $\rho_\alpha$ is the mole density of phase $\alpha$.
The transmissibility is defined on each face of a grid block. If a face is internal face shared shared by two
grid blocks, its value is the same for these two blocks. If the face is a boundary face, the transmissibility
is zero, as the no-flow boundary condition is applied.
Different weighting schemes must be applied to average different properties at an interface.
The left part, $\frac{K A}{\Delta d}$, is geometric properties,
and the harmonic averaging method is applied.
The right part, $\frac{K_{r\alpha}}{\mu_{\alpha}} \rho_{\alpha}$, relies on fluid properties,
and the upstream averaging method is applied \cite{Chen2007}.
The upsteam finite difference method is employed to descretize the model.

\subsection{Linear Solver}

The Jacobian matrix from Newton method is highly ill-conditioned, and the Krylov subspace
solvers are applied to solve the linear system $A x = b$.
The key to an effective solution method is to choose a proper preconditioner $M$,
which should be easy to setup and effective.
In our previous work, a family of scalable CPR-type
methods \cite{bos-pc} have been developed for reservoir simulations, which have been applied to black oil
model, compositional, in-situ combustion and the general thermal model in this paper.
The unknowns are numbered grid block by grid block and the resulted matrix in each iteration is block-wise,
\begin{equation}
    \label{mat-ja}
    A = \left(
    \begin{array}{llll}
        A_{11}         & \cdots       & \cdots & A_{1n}  \\
        A_{21}         & A_{22}       & \cdots & A_{2n}  \\
        \cdots         & \cdots       & \cdots & \cdots  \\
        A_{n1}         & A_{n2}       & \cdots & A_{nn}  \\
    \end{array}
    \right),
\end{equation}
where each sub-matrix $A_{ij}$ is a square matrix. In-house distributed-memory matrix, vector and their
operations have been developed, such as adding entries, assembling, getting sub-matrix (for CPR-type
preconditioners), factorization, sparse BLAS, and point-wise and block-wise matrices. Base on these
operations, internal parallel solvers, such as GMRES, LGMRES, CG, and BICGSTAB, and preconditioners, such as
RAS, AMG, CPR-FP, CPR-PF, CPR-FPF, ILU(k), and ILUT, have been implemented.

\subsection{Decoupling Methods}

A proper decoupling method is critical to the success of the CPR-type preconditioners.
In general, the decoupling method is applied before applying the CPR-type preconditioners, which
converts the original linear system to an equivalent linear system,
\begin{equation}\label{decoupled-eq}
    (D^{-1}A)x = D^{-1} b.
\end{equation}
Several decoupling methods have been proposed, such as Quasi-IMPES, True-IMPES \cite{Lacroix2001},
Alternate Block Factorization (ABF) \cite{Bank1989}, full row sum (FRS) and 
dynamic row sum (DRS) \cite{Sgries} methods.
The idea of ABF method is simple, which is defined as,
\begin{equation}
    \label{decoupled-abf}
    D_{abf} = diag(A_{11}, A_{22}, \cdots, A_{nn}).
\end{equation}
It converts the block diagonal part to identity matrix. This method requires to
calculate the inverse of each diagonal part, and the matrix-matrix multiplications are performed 
for each sub-matrix.
The FRS decoupling method is described as,
\begin{equation}
    \label{decoupled-frs}
    D^{-1}_{frs} = diag(D_{1}, D_{2}, \cdots, D_{n}),
\end{equation}
where,
\begin{equation}
    D_i = \left(
    \begin{array}{cccc}
        1  & 1  & \cdots & 1    \\
        0  & 1  & \cdots & 0    \\
        \cdots  & \cdots  & \cdots & \cdots   \\
        0  & \cdots  & 1 & 0   \\
        0  & 0       & 0 & 1 \\
    \end{array}
    \right).
\end{equation}
The diagonal part and the first row are 1 and all other locations are 0,
which means to add the all rows to the first row. The DFS decoupling method is a simplified version of the FRS
method, and details can be read in \cite{Sgries}.

The Guass-Jordan elimination (Gauss elimination, GJE) method has been used
to solve linear systems. 
Its idea is to convert $\left[ D | A | b \right]$ to an equivalent linear system $\left[ I | \tilde{A} |
\tilde{b} \right]$ by Gauss-Jordan elimination method, and the $\tilde{b}$ is final solution.
In this paper, it is adopted as a decoupling method and is applied grid block by grid block to turn the
diagonal matrices to identity matrix. Pivoting technique is used and only row reodering is involved.
Since the decoupling is processed block by block, no communication is required, which is friendly to parallel
computing. The GJE decoupling is more efficient than the ABF method.

When the CPR-type preconditioners are applied to reservoir simulations, it is important to keep the
pressure matrix positive definite. FRS method helps to enhance this property, from which the
CPR-type preconditioners can benefit.  In the  first stage, FRS or DRS methods are applied;
then ABF or GJE methods are used as the second stage.
In this case, two-stage decoupling methods are developed, which are noted as FRS+ABF, FRS+GJE, DRS+ABF and
DRS+GJE.

\subsection{Preconditioners}

Several scalable CPR-type preconditioner have been proposed \cite{bos-pc}, such as CPR-FP, CPR-PF, CPR-FPF,
and CPR-FFPF methods. According to our practices, the CPF-FPF method, which is 
a three-stage preconditioner, is effective for black oil model and thermal model. It is described by Algorithm
\ref{pc-fpf}, where the first
step is to solve an approximate solution using restricted additive Schwarz (RAS) method, the third step is to
solve the subproblem by algebraic multi-grid method (AMG), the fifth step is to get an approximate
solution again using restricted additive Schwarz method, and the second step and the forth step are to
calculate residual.

It is well-known that the RAS method is scalable for parallel computing. Parallel AMG method is also scalable.
However, multiple layers are applied inside AMG and each layer gets coarser and coarser, which
introduces complicated communication patterns. Also, since the layers get coarser and denser, more
communication could be introduced and scalability will be reduced if AMG has too many layers (levels).
Different coarsening algorithms and interpolation methods also affect the scalability and convergence of AMG
solver.
The CPR-FPF method is a combination of RAS method and AMG method, which is also scalable. Here we should also
mention that the setup phase of the parallel AMG
method is computationally intense. For small model or easy model, the RAS method should work well too.
In-house solvers and preconditioners have been developed, and the only external library that the thermal
simulator requires is the parallel AMG solver, Hypre. Figure \ref{fig-ds-amg-par} presents default parameters
for AMG preconditioner.
The default overlap of the RAS method is 1. If it's 0, then it's equivalent to block Jacobi method.
When there are too many MPI processes, we may increase the overlap to maintain convergence of the
preconditioner, such as 2 and 3. However, more communications are introduced in the setup phase of the RAS
method, which needs to construct a local sub-problem by requesting more entries of the distributed matrix from
other MPI processes.
The sub-problem in each MPI process (CPU core) from RAS method is solved by ILUT by default, which can also be solved by ILU(k)
or block ILU(k) \cite{bos-pc}. The size of the lower triangular matrix and the upper triangular matrix can be
reduced by dropping small entries or using smaller $p$ for ILUT and $k$ for ILUK, and the convergence of ILU
methods should be well balanced. The recommended level ($k$) for ILUK is 1.
The AMG method (solver) is more complicated, whose default parameters are listed as following,
\begin{figure}[!htb]
\centering
\begin{evb}
{
    /* maxit */               1,
    /* num_funcs */           -1,
    /* max_levels */          -1,
    /* strength */            0.5,
    /* max_row_sum */         0.9,
    /* trunc error */         1e-2,

    /* coarsen_type */        Falgout,
    /* cycle_type */          v-cycle,
    /* relax_type */          hybrid Gauss-Seidel-forward,
    /* coarsest_relax_type */ hybrid symmetric Gauss-Seidel,
    /* interp type */         cmi,
    /* itr relax */           2,
};
\end{evb}
\caption{Default parameters of AMG method}
\label{fig-ds-amg-par}
\end{figure}

\begin{algorithm}[!htb]
    \caption{The CPR-FPF Method}
    \label{pc-fpf}
    \begin{algorithmic}[1]
        \STATE $y = \textit{RAS}(A)^{-1} f$
        \STATE $y = y + \varPi_p \textit{AMG}(A_{PP})^{-1} \varPi_r  r $
        \STATE $y = \textit{RAS}(A)^{-1} f$
    \end{algorithmic}
\end{algorithm}

The design and implementation details of the linear solver and preconditioners can be found in \cite{sca-sol}.
The thermal simulator and some other reservoir simulators base on the in-house platform, PRSI \cite{sca-plat},
which provides gridding, DOF (degrees of freedom), mappling, solver and preconditioner, well modeling, keyword
parsing, option parsing, visualization, parallel input and output through MPI-IO, memory management, and
communication management. The platform is implemented by C and utilizes MPI for communications. It is highly
scalable and previous studies have shown that the platform and in-house simulators have ideal scalability\cite{sca-plat}.

\section{Implementations}
The simulator bases on our in-house platform, which provides
gridding, well modelling, option parsing, keyword parsing, visualization, linear solver, preconditioner,
communication and memory management.

\subsection{Option}
Many applicatios have command line options, which can change the behavoir of an application such as \verb|ls|,
which is shown by Figure \ref{fig-cl-par}, where 
\begin{itemize}
    \item \verb|-a| means to show hidden files;
    \item \verb|-l| means to use a long listing format;
    \item \verb|--color=tty| means to use color;
    \item \verb|-S| means to sort files by size (largest first);
    \item \verb|-r| means to reverse order while sorting.
\end{itemize}
Options provide a way to
change its behavoirs without re-compiling codes. Also, by using options, testing, benchmarking, debugging and
simulations can be arranged automatically with script programming, such as Python and Bash. Figure
\ref{fig-cl-replace} is an example, which replaces all \verb|abc| to \verb|xyz| for all \verb|c| files in the
current directory. If there are dozens of files, it could take a developer for some time to do the work. By
using Bash script, the work is done automatically and efficiently. Figure \ref{fig-cl-jobs} is a Bash script,
which runs all reservoir models start with "cmg" using 8 MPIs, GMRES solver, RAS preconditioner, ILUT
solver for local linear system from RAS method, and level 2 overlap. It is obvious that option is a powerful
tool for simulations.

\begin{figure}[!htb]
\centering
\begin{evb}
    ls -a -l --color=tty -S -r
\end{evb}
\caption{Command line options}
\label{fig-cl-par}
\end{figure}

\begin{figure}[!htb]
\centering
\begin{evb}
    for f in $(ls *.c); do sed -i 's/abc/xyz/g' $f; done
\end{evb}
\caption{Bash scripts: text replacement}
\label{fig-cl-replace}
\end{figure}

\begin{figure}[!htb]
\centering
\begin{evb}
    for f in $(ls cmg*.dat);
        do mpirun -np 8 ./simulator -model $f -solver gmres -pc ras -ras_solver ilut -overlap 2;
    done
\end{evb}
\caption{Batch simulation running}
\label{fig-cl-jobs}
\end{figure}

A option parsing module is implemented, which has three main parts: 1) registration; 2) parsing; 3) cleanup.
Figure \ref{fig-option} shows the usage and Figure \ref{fig-option-run} shows to how to use options.
\verb|opt_register| registers one option, which has four components: keyword, help info, type and address of a
variable. \verb|opt_parse| parses all command line parameters and maintains internal options and internal
variables: 1) parse options; 2) check if an option is legal or not; 3) call proper subroutines to set correct
values to registered variables, such as string to integer and string to floating point number; 4) manage
internal memory, status and data structures. It also checks if parameters for one option are legal or not. For
example, 3.4 isn't legal for integer, and 1.z isn't legal for floating point number. \verb|opt_parse| may have
some internal options, such as \verb|-h| and \verb|-help|, which are used to show registered options and help
info.

\begin{figure}[!htb]
\centering
\begin{evb}
int main(int argc, char **argv)
{
    INT m;
    FLOAT distance;
    VEC_INT vm;
    VEC_FLOAT vd;

    /* step 1 */
    opt_register("-m", "restart", OPT_T_INT, &m);
    opt_register("-d", "distance from place A to B", OPT_T_FLOAT, &distance);
    opt_register("-vec_m", "vector of integer: vm", OPT_T_VEC_INT, &vm);
    opt_register("-vec_d", "vector of floating point number: vd", OPT_T_VEC_FLOAT, &vd);

    /* step 2 */
    opt_parse(&argc, &argv);

    /* step 3 */
    opt_cleanup();
}
\end{evb}
\caption{Options calling sequence}
\label{fig-option}
\end{figure}

\begin{figure}[!htb]
\centering
\begin{evb}
    ./app -help
    ./app -m 3 -d 2.2 -vec_m "1 3 55 4" -vec_d "1.2 1e-8 3.1415926 2.718281"
\end{evb}
\caption{Command line options}
\label{fig-option-run}
\end{figure}

Depending on the internal implementation and compiler, the option parsing module support \verb|INT_MAX|
option keywords, where \verb|INT_MAX| is 2147483647 for 32-bit system. Some other functions are also designed
to work with option, such as \verb|opt_preset|, which is called before \verb|opt_parse| to serve as input
options, but it can be overridden by user option input. In Figure \ref{fig-option-preset}, if no option is
provided when running, \verb|m| will be parsed as 2 and \verb|distance| will be parsed as 2.33. If user
options are provided, such as \verb|"./app -m 8 -d 3.54"|, then \verb|m| will be parsed as 8 and \verb|distance| will
be parsed as 3.54.

\begin{figure}[!htb]
\centering
\begin{evb}
int main(int argc, char **argv)
{
    INT m;
    FLOAT distance;

    /* preset options */
    opt_preset("-m 2");
    opt_preset("-d 2.33");

    /* step 1 */
    opt_register("-m", "restart", OPT_T_INT, &m);
    opt_register("-d", "distance from place A to B", OPT_T_FLOAT, &distance);

    /* step 2 */
    opt_parse(&argc, &argv);

    /* step 3 */
    opt_cleanup();
}
\end{evb}
\caption{How to use preset}
\label{fig-option-preset}
\end{figure}

Some auxiliary functions should be implemented to support option parsing, such as 1) string to integer, where
integer could be integer, long integer and long long integer, 2) string to floating point number, and 3) vector
management. These functions have to check if input is legal. Now the option parsing module supports the
following types of option shown in Figure \ref{fig-option-type}.
\begin{figure}[!htb]
\centering
\begin{evb}
{
    OPT_T_BOOLEAN,    /* boolean */
    OPT_T_TRUE,       /* set variable to true if see the option */
    OPT_T_FALSE,      /* set variable to false if see the option */

    OPT_T_INT,        /* integer */
    OPT_T_FLOAT,      /* floating point number */

    OPT_T_VEC_INT,    /* vector of integer */
    OPT_T_VEC_FLOAT,  /* vector of floating point number */
    OPT_T_USER1,      /* user defined type 1, conversion function is required */
    .....
    OPT_T_USER16,     /* user defined type 16, conversion function is required */
}
\end{evb}
\caption{Option types}
\label{fig-option-type}
\end{figure}

\subsection{Keywords Parsing}
A model file provides various parameters for reservoir properties, chemical properties, well definition and
schedule changes, which contains many types of inputs, such as string, integer, floating
point number, vector of string, vector of integer, vector floating point number, table, well and schedule, as
shown in Figure \ref{fig-keyword-sample}, from which we can observe string, floating point number, vector of
floating point number, vector of string, modifier, table, well and schedule.
A reservoir model may have hundreds or thousands of lines of input parameters. 
A complicated model may have millions of lines of parameters. 

\begin{figure}[!htb]
\centering
\begin{evb}
unit: field

grid: 90 50 40
dx: 90*29.17
dy: 50*29.17
dz: 40*10

por: 0.3
cpor: 5e-5
porform: linear
volconst: rock
thconr: 24
thconw: 24
thcono: 24
thcong: 24

mod: permx
1 1 1:25         1000
8 8 4            5
9:12 8:18 4:20   5
2:22 7 5         2000 /

swt
#   sw       krw         krow        pcw
    0.45     0.0         0.4
    0.47     0.000056    0.361
    0.50     0.000552    0.30625
    0.55     0.00312     0.225
    0.60     0.00861     0.15625
    0.70     0.03088     0.05625
    0.75     0.04871     0.025
    0.77     0.05724     0.016
    0.80     0.07162     0.00625
    0.82     0.08229     0.00225
    0.85     0.1         0.0 /

solver: bicgstab
pc: cpr

sw: zvar 0.18 20*0.2 19*0.25

# well
well: Inj-No_1
type: injector
tinjw: 450
qual: 0.6
skin: 0
weight: mobweight implicit

operate: max bhp 3000
operate: max stw 100

perf: geo
    8 8 1:40 1
/


# schedule
run
time 30
time 100

time 365
stop
\end{evb}
\caption{Model sample}
\label{fig-keyword-sample}
\end{figure}

A trivial way is to write some codes for a specific keyword, such as \verb|cpor| for compressibility of
porosity, which is a floating point numer. The problem is that the design isn't flexible and dozens of
thousands of lines of codes may have to be written. Also, a keyword may have different usages. For example,
Figure \ref{fig-keyword-usage} defines several ways to define porosity. Porosity can be defined as constant,
layer by layer, grid block by grid block, or by an input file. The input file defines porosity block by block. 
By using a file for heterogeneous input, the model file is easy to read and maintain if the model is large,
such as one million. It isn't a good idea to write of millions of parameters in one file, such as
permeability, coordinates, saturations, and porosity.  Here MPI-IO is required to
read the file and to distribute the values to the right MPIs according to grid partition.

\begin{figure}[!htb]
\centering
\begin{evb}
# constant
por: 0.3
por: const 0.3
por: con 0.3

# by layer (xyz and/or ijk)
por: ivar 0.25 9*0.22 8*0.23 0.35
por: jvar 0.25 9*0.22 8*0.23 0.35
por: kvar 0.25 9*0.22 8*0.23 0.35
por: xvar 0.25 9*0.22 8*0.23 0.35
por: yvar 0.25 9*0.22 8*0.23 0.35
por: zvar 0.25 9*0.22 8*0.23 0.35

# by grid block (total size equals to grid block size)
# ordered by grid block global index
por: all 0.25 9*0.22 800*0.23 111*0.35

# by file
# porosity.dat has porosity for each grid block
# entry size >= grid block size
# one million grid blocks requires at least one million file entries
# assuming blocks and file entries have the same global order
por: file porosity.dat
\end{evb}
\caption{Keyword different usages}
\label{fig-keyword-usage}
\end{figure}

\begin{figure}[!htb]
\centering
\begin{evb}
typedef struct KEYWORD_INFO {
    char *key;
    void *addr;
    KEYWORD_TYPE type;

} KEYWORD_INFO;
\end{evb}
\caption{Keyword info data structure}
\label{fig-keyword-item}
\end{figure}

In our simulator, a much simpler method is designed, which also adopts three steps: 1) register; 2) parse; and 3)
cleanup. The first step is to register a keyword with one of the following data types shown in Figure
\ref{fig-keyword-item}, a key name, and its memory address. Its info is represented by \verb|KEYWORD_INFO|.
The key name is used to decide which keyword is parsing, data type is used to decide which subroutine will be
called to convert its parameters, and the address is where to store the parse information. Only one line of
code is required to register a keyword.
Figure \ref{fig-keyword-type} represents data types our simulator supports. Here are their brief explanations.
\begin{itemize}
    \item \verb|KEYWORD_T_TRUE|, a variable is set to \verb|TRUE| if parsed.

    \item \verb|KEYWORD_T_FALSE|, a variable is set to \verb|FALSE| if parsed.

    \item \verb|KEYWORD_T_BOOLEAN|, boolean variable. Its value could be \verb|TRUE| or \verb|FALSE|,
        depending on the parameter.

    \item \verb|KEYWORD_T_STRING|, string variable.

    \item \verb|KEYWORD_T_INT|,  integer variable.

    \item \verb|KEYWORD_T_FLOAT|, floating point number variable.

    \item \verb|KEYWORD_T_TABLE|, table variable. A model may have several tables, such as relative
        permeability tables and viscosity tables. Each table may have different columns. The relative tables
        may have four columns and the number of columns of viscosity tables depends on how many chemicals
        considered in the model. Each column may have different types, such as boolean, string, integer and
        floating point number.

    \item \verb|KEYWORD_T_MOD_FLOAT|, sub-domain modifier type (floating point number). Values are only
        defined on some sub-domains, which are used to modify certain properties, such as saturation,
        pressure, and permeabilities. Figure \ref{fig-keyword-sample} defines a modifier to modify x direction
        permeability.

    \item \verb|KEYWORD_T_MOD_INT|, sub-domain modifier type (integer).

    \item \verb|KEYWORD_T_VEC_STRING|, vector of string.

    \item \verb|KEYWORD_T_VEC_INT|, vector of integer.

    \item \verb|KEYWORD_T_VEC_FLOAT|, vector of floating point number.

    \item \verb|KEYWORD_T_VEC_TABLE|, vector of table. Some model has different relative permeability tables
        for different rock types. In thermal simulations, relative permeability tables may depend on
        temperature, in which one set of relative permeability tables are required for each temperature.
        Viscosity may depend on pressure, and one viscosity table is required for each given pressure. In
        these cases, multiple tables are needed, which are handled by vector of table.

    \item \verb|KEYWORD_T_VEC_MOD_FLOAT|, vector of sub-domain modifier (floating point number).

    \item \verb|KEYWORD_T_WELL|, well section. In this section, a well is defined, including skin, factor,
        well type, well name, direction, perforation and well index type.

    \item \verb|KEYWORD_T_SCHL|, schedule section. Well changes and report time are defined.

    \item \verb|KEYWORD_T_REACT|, chemical reaction section.

\end{itemize}

\begin{figure}[!htb]
\centering
\begin{evb}
{
    KEYWORD_T_TRUE,            /* true */
    KEYWORD_T_FALSE,           /* false */
    KEYWORD_T_BOOLEAN,         /* boolean */

    KEYWORD_T_STRING,          /* string */
    KEYWORD_T_INT,             /* int */
    KEYWORD_T_FLOAT,           /* float */

    KEYWORD_T_TABLE,           /* table */
    KEYWORD_T_MOD_FLOAT,       /* sub-domain modifier */
    KEYWORD_T_MOD_INT,         /* sub-domain modifier */

    KEYWORD_T_VEC_STRING,      /* string vector */
    KEYWORD_T_VEC_INT,         /* int vector */
    KEYWORD_T_VEC_FLOAT,       /* float vector */
    KEYWORD_T_VEC_TABLE,       /* table vector */
    KEYWORD_T_VEC_MOD_FLOAT,   /* vector of sub-domain modifier */

    KEYWORD_T_WELL,            /* well section */
    KEYWORD_T_SCHL,            /* schedule section */
    KEYWORD_T_REACT,           /* chemical reaction */
}
\end{evb}
\caption{Keyword types}
\label{fig-keyword-type}
\end{figure}

The second step is to sort all registered keywords by key name, to read model file and to parse the user
input. In this step, the model file is read line by line. Here "$\backslash$" at the end of one line means to
continue the current line. In this case, one line is allowed to split into multiple lines. Figure
\ref{fig-keyword-equiv} shows a few equivalent inputs. If a key is found, proper conversion subroutine is
called to parse values and to store them at provided address. Otherwise meaningful error is output as key
isn't found. A few hundreds of lines of codes are enough, which can handle arbitrary keywords, such as
1,000,000 keywords.

The third step is to clean internal variables and memory, which loops all registered keywords and releases
memory if needed. One hundred lines of codes are enough for this step.

\begin{figure}[!htb]
\centering
\begin{evb}
    # x (i) direction permeability, assigned by z (k) layer
    permx: kvar 1 333 1e-3 1 200 1000 3000 20 666 90 88 88 88 88 88
    permx: zvar 1 333 1e-3 1 200 1000 3000 20 666 90 88 88 88 88 88

    permx: \
        kvar \
        1 333 1e-3 1 200 1000  \
        3000 20 666 90 88 88 88 88 88

    permx: \
        zvar \
        1 333 1e-3 1 200 1000  \
        3000 20 666 90 \
        88 88 \
        88 88 88

    permx: kvar 1 333 1e-3 1 \
        200 1000 3000 20 666 90 5*88

    permx: kvar 1 \
        333 1e-3 \
        1 200 \
        1000 \
    # comment
    # comment
    # comment
    # comment
    # comment
    # comment
    # comment
        3000 \
        20 \
        666 \
    # comment \
    # comment \
    # comment \
        90 88 4*88

    permx: kvar 1 \
    # comment
        333 1e-3 \
        1 200 \
    # comment again \
        1000 \
        3000 \
        20 \
    # comment again 
        666 \
        90 88 2*88 2*88
\end{evb}
\caption{Equivalent inputs}
\label{fig-keyword-equiv}
\end{figure}

\subsection{Load Balancing}

\begin{figure}[!htb]
\begin{center}
\includegraphics[width=0.5 \linewidth,angle=270]{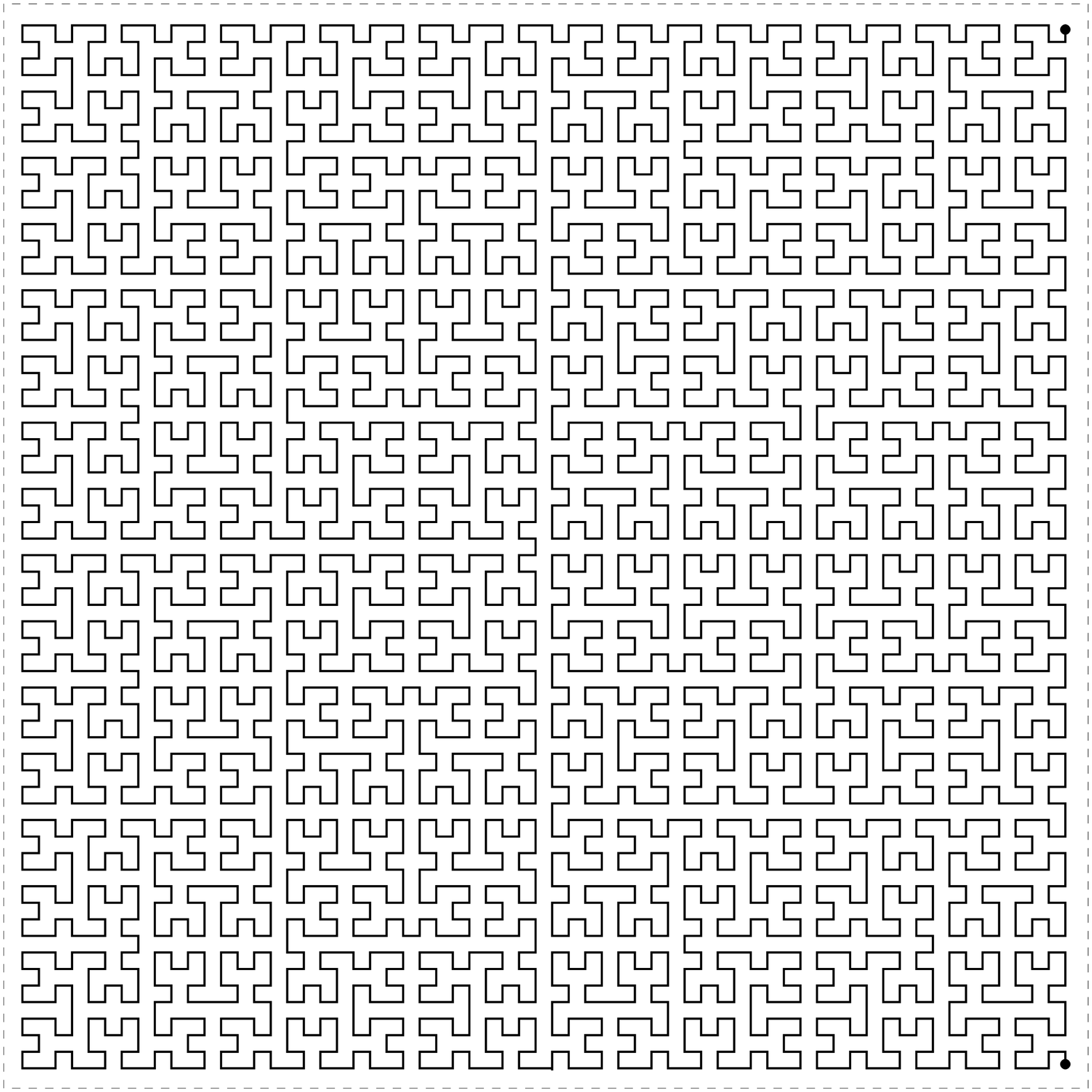}
\caption{Hilbert space-filling curve, level 6}
\label{fig-sfc-h6}
\end{center}
\end{figure}

The default load balancing (grid partitioning) method is Hilbert space-filling curve method.
Space-filling curves are those curves that fill an entire $n$-dimensional unit hypercube,
which have many types, such as Hilbert space-filling curves, Morton space-filling curves, and Sierpi\'{n}ski
space-filling curves.
Among those curves, Hilbert space-filling curves show better locality, which have been
widely applied in many areas.
Figure \ref{fig-sfc-h6} (\cite{hsfc}) is a level 6 Hilbert space-filling curve. 
Denser curves can be obtained if using higher levels.

A space-filling curve can fill an entire $n$-dimensional unit hypercube, which introduces mappings between
one dimensional space and $n$-dimensional space. For grid partitioning, the three-dimensional domain is
converted to one dimensional space. The partitioning method only needs to handle 1D partitioning.
Algorithm \ref{alg-hsfc} \cite{plat-liu} shows the three steps required to partition a given grid.
The communication volume of this algorithm is low. In the thermal simulator, other methods are also available,
such as ParMETIS and Zoltan.

\begin{algorithm}[!htb]
\caption{Space-filling curve method}
\label{alg-hsfc}
\begin{algorithmic}[1]
\STATE Map a computational domain $\Omega$ to a subset of $(0, 1)^3$.
\STATE For any cell, calculate its mapping that belongs to $(0, 1)$.
\STATE Partition the interval $(0, 1)$ into $N_p$ sub-intervals and each sub-interval has the same number of cells.
\end{algorithmic}
\end{algorithm}

If there are not too many inactive grid blocks, one round Algorithm \ref{alg-hsfc} will partition a given
grid. However, if there are lots of inactive grid blocks, Algorithm \ref{alg-hsfc} can be applied twice: one
round for active blocks and one round for inactive blocks. By doing this, the computations are well balanced.

Each grid block and well may have different weight depending on if a grid block is active or not and how many
perforations a well has. The weight can be set automatically or by reservoir models, in which a weight file is
allowed to apply.

\subsection{Gridding}
After a grid is partitioned, each sub-grid belongs to a MPI process. The sub-grid has to be re-constructed,
including coordinates, topology, and neighbour information. Each grid block has a unique global index and a
local index. Each block stores neighbour information. If one neighbour belongs to the same grid, then a
pointer points to the local neighbour. If one neighbour belongs to another MPI, then the remote MPI rank, its
global index and its local index in remote MPI are stored in a special data structure. The reason is that each
MPI knows which MPI to contact when information exchange is required.

Each block constructs boundary conditions of each face, such as interior face, remote face and boundary face.
Here interior face means the face is shared with another grid block, remote face means neighbour is in another
MPI, and boundary face means there is no neighbour. Each face may have more than one face types.

\subsection{Communications}
In parallel thermal simulations, grid, well perforations, matrix and vectors are distributed in many MPIs.
Communications are always required for one MPI to get information from other MPIs. A grid block may need
information from neighbour blocks. Well information from each perforation should be gathered to model a well.
When solving a linear system, sparse matrix-vector multipliation is a key part. Each MPI has to obtain vector
values from other MPIs. Since the grid distribution
and well distribution are not known to users, the communication patterns can be complicated. In reservoir
simulations, grid and well communication patterns can be computed and stored after the grid is partitioned. However,
communication pattern for a matrix has to be re-computed once a new matrix is constructed.

\subsection{Parallel IO}
File read and write are involved in the following situation.
The first one is initial stage. In a heterogeneous reservoir model, each block may have different properties, 
such as permeabilities,
porosity, pressure, and saturation. They are required to read and to distribute to different MPI processes. A
trivial way is each MPI reads once and this is easy to write codes. The problem is that a file is read many
times, such as 200 times, if we have 200 MPI processes. This method isn't efficient and slow. A proper method
is each MPI reads a portion of the file and distribut the input values to proper MPIs that requires the input.
The parallel read is tricky, since an input value, such as 2.354, should be parsed correctly. It cannot be
parsed as two values, 2.3 in one MPI process and 54 in another MPI process. In our simulator, two data files
are supported: column-based formatted style and free style. The formatted style is shown in Figure
\ref{fig-keyword-column}, which has three columns and is separated by space. Each column can be integer,
string and floating point number. This type data file may have any lines, such as one billion lines.

\begin{figure}[!htb]
\centering
\begin{evb}
# this is comment
1.       2       2.3
0.99     8       3e-4
0.29     3       3e-5
0.19     8       3.333333334567

........

# empty line following

# empty line following

# empty line following

# another comment
1e-6     22      8.776644446699999999777555

# another comment
# another comment
# another comment
# another comment
0.39     28     -9.23e-6
\end{evb}
\caption{Formatted inputs, 3 columns}
\label{fig-keyword-column}
\end{figure}

\begin{figure}[!htb]
\centering
\begin{evb}
1. 2 2.3 0.99 8 3e-4 1 111 2 2 2 2 2 2 2 2 2 2 2 2 2 2 2 2 2 2 2 2 2 2 2 2 2 2 2 2 2 2 2.55 2.55 2.55 3333 4.556666
2.55
2.55
2.55
2.55
2.55 2
2
2
2
2 2 2 2 2 2 2 2 2 2
2
2 2 2
3 3.31 2.25 3 3.31 2.25 3 3.31 2.25 3 3.31 2.25 3 3.31 2.25
3 3.31 2.25
3 3.31 2.25
3 3.31 2.25 3 3.31 2.25
3
3.31 2.25

........

0.39 28 -9.23e-6 0.39 28 -9.23e-6 0.39 28 -9.23e-6
\end{evb}
\caption{Free style inputs}
\label{fig-keyword-free}
\end{figure}

When parsing a formatted data file, the parser should identify each line.
It also has to recognize empty line and comment line. A subroutine may require all columns or some column. For
example, we may have five columns: pressure, temperature, water saturation, oil saturation and gas saturation.
In this case, it is obvious that:
\begin{itemize}
    \item (5,1) (line five and column one) is pressure for 5-th grid block;
    \item (8,2) (line eight and column two) is temperature for 8-th grid block;
    \item (39,5) (line thirty nine and column five) is gas saturation for 39-th grid block.
\end{itemize}

We should mention that each line may have different length, and in parallel computing, the parallel read
algorithm should be well designed 
such that each MPI read some lines, and all MPIs read the entire data file. Each line is owned by only one MPI.
A free style data file is shown by Figure \ref{fig-keyword-free}. A free style data file may have any lines of
inputs depending on the model, such as 1,000,000 lines, and each line may have any values. For example, one
line may have 1, 2, 100 or 1,000 values. The difficult part is that each input should be parsed as one value.
For example, \verb|-9.23e-6| should be parsed as \verb|-9.23e-6| by one and only one MPI, not \verb|-9.2|
by one MPI and \verb|3e-6| by
another MPI. Again, each value is only owned and parsed by one MPI.

The second situation that IO is involved is to write computed results. Visualization is an important part of
reservoir simulation, from which we can study chemical distribution, temperature change and pressure change. 
During reservoir simulations, the computed results are saved for visualization and other post-processing
purposes. Depending on the model, various different properties are output, such as saturations, temperature,
pressure, relative permeabilities, and mole fraction. Since a grid is distributed in many MPIs, if data is
written to one file, MPI output is an efficient way. Binary write is relative simple. Here the binary may be
in little endian or big endian. One issue is that binary file isn't friendly to user as ASCII file. In our
simulator, two formats are supported: binary format and ASCII format. Each ASCII output may have different
lengths, such as:
\begin{itemize}
    \item "3": one letter;
    \item "1048576": seven letters;
    \item "3.14": four letters;
    \item "2.71828133333e-6": sixteen letters.
\end{itemize}
The parallel write subroutines should take care of this when using collective MPI-IO.

The third situation is restart. Sometimes simulation restart is needed. The restart can speed testing
and debugging, and it can save our work from emergency.
The status has to be written to disk,
and a user may run the simulation from the restart point. A well-designed simulator can save restart files for
certain MPIs, such as 100, and restart the simulations for different MPIs, such as 256 MPIs. The restart
requires parallel read and write.  

Also, results are output for testing, debugging and validation purposes. When unexpected results are observed,
a good trick is to dump output and to compare with commercial simulators.

\subsection{Data Management}

It is natural to use grid blocks to represent a reservoir. Block-centered data is applied, which is integer or
floating point number. Some well data is defined on each well or eac perforation, which is also integer or
floating point number. A few commonly used data set (DOF, degrees of freedom) types are defined as shown by
Figure \ref{fig-dof-type}. Each DOF is defined as vertex, cell, sub-domain, well perforation and well. Here
are some brief explanations of these types.
\begin{itemize}
    \item \verb|DOF_POINT|, \verb|DOF_POINT_INT| and \verb|DOF_POINT_BOOLEAN| are defined on each vertex.
        Floating point number, integer and boolean variables are supported. Since each vertex is shared by
        several grid blocks, one vertex may be distributed in several MPIs. Only one MPI owns the vertex.
        Sychronization may be required when values are changed. A point type DOF has the same elements as total
        vertices.

    \item \verb|DOF_CELL|, \verb|DOF_CELL_INT| and \verb|DOF_CELL_BOOLEAN| are defined on each grid
        block(cell). \verb|DOF_CELL| is for floating point data, which is used to represent chemical and
        reservoir properties, such as pressure, temperature, saturations and porosity.
        \verb|DOF_CELL_INT| is for integer cell data. A cell type DOF has the same elements as grid blocks.
        \verb|DOF_CELL_BOOLEAN| is for boolean cell data, such as gas phase status.
        In thermal simulation, cell data is applied. 

    \item \verb|DOF_DOMAIN|, \verb|DOF_DOMAIN_INT| and \verb|DOF_DOMAIN_BOOLEAN| are defined for each
        sub-domains. For example, each sub-grid in one MPI has the same MPI rank. Another situation is that a
        reservoir may have several rock types, and grid blocks have the same rock type form one sub-domain. A
        sub-domain type DOF has the same elements as the size of all defined sub-domains.

    \item \verb|DOF_WELL|, \verb|DOF_WELL_INT|, and \verb|DOF_WELL_BOOLEAN| are defined for each well, such as
        well bottom hole pressure, well rate and well status (open and shut-in). A well type DOF has the same
        elements as the size of wells in a reservoir model. For example, if a reservoir model has 10 wells,
        then a well type DOF has 10 elements.

    \item \verb|DOF_PERF|, \verb|DOF_PERF_INT| and \verb|DOF_PERF_BOOLEAN| are defined for each perforation. A
        well may have many perforations. For example, a model has 12 wells and each well has 10 perforations,
        then a perforation type DOF has 120 elements. A perforation type DOF is designed to store perforation
        related information, such as pressure difference between grid block pressure and bottom hole pressure,
        mobility, density at perforation, and well perforation rates (oil, water and gas rates). A perforation
        may be open or closed depending on the model settings, \verb|DOF_PERF_INT| and \verb|DOF_PERF_BOOLEAN|
        are applied to store the status and integer values.

    \item \verb|DOF_CONST|, \verb|DOF_CONST_INT| and \verb|DOF_CONST_BOOLEAN| are designed for floating point,
        integer and boolean constants.

\end{itemize}

\begin{figure}[!htb]
\centering
\begin{evb}
    DOF_POINT,           /* defined on each vertex, floating point number */
    DOF_POINT_INT,       /* defined on each vertex, integer */
    DOF_POINT_BOOLEAN,   /* defined on each vertex, boolean */

    DOF_CELL,            /* defined on each cell, floating point number */
    DOF_CELL_INT,        /* defined on each cell, integer */
    DOF_CELL_BOOLEAN,    /* defined on each cell, boolean */

    DOF_DOMAIN,          /* defined on each sub-domain, such as rock properties, floating point number */
    DOF_DOMAIN_INT,      /* defined on each sub-domain, such as rock type, integer */
    DOF_DOMAIN_BOOLEAN,  /* defined on each sub-domain, boolean */

    DOF_WELL,            /* defined on each well, floating point number */
    DOF_WELL_INT,        /* defined on each well, integer */
    DOF_WELL_BOOLEAN,    /* defined on each well, boolean */

    DOF_PERF,            /* defined on each perforation, floating point number */
    DOF_PERF_INT,        /* defined on each perforation, integer */
    DOF_PERF_BOOLEAN,    /* defined on each perforation, boolean */
    
    DOF_CONST,           /* constant, floatint point number */
    DOF_CONST_INT,       /* constant, integer */
    DOF_CONST_BOOLEAN,   /* constant, boolean */
\end{evb}
\caption{DOF types}
\label{fig-dof-type}
\end{figure}

\begin{figure}[!htb]
\centering
\begin{evb}
    X = dof_create(type, dim, name);           /* create a DOF */
    dof_destroy(&X);                           /* destroy a DOF */

    Y = dof_copy(X);                           /* copy, including memory allocation */
    dof_data_copy(D, S);                       /* data copy only, D = S */

    d = dof_get_value_by_block(X, block);      /* get value for a grid block */
    d = dof_get_value_by_index(X, idx);        /* get value from a block with local index idx */

    dof_set_value_by_block(X, block) = a;      /* set value for a grid block */
    dof_set_value_by_index(X, idx) = a;        /* set value for a grid block with local index idx */
    dof_set_value_by_file(X, file);            /* set values by file, parallel read and distrubution */
    dof_set_value_by_file(X, file, n);         /* set values by n-th column of a file, parallel read and distrubution */

    dof_dump_data_to_file(file, X);                   /* dump a DOF to file, parallel write */
    dof_dump_data_to_file_var(file, X, Y, Z, NULL);   /* dump DOFs to file, parallel write */
\end{evb}
\caption{DOF management functions}
\label{fig-dof-func}
\end{figure}

When a DOF is created, its dimension is set, such as 1 for pressure and 3 for
permeabilities. Its dimensional is allowed to be any positive integer. The advantage of using one DOF for all
unknowns is that the Jacobian matrix is defined as block-wise, and the linear solver has better convergence
than point-wise Jacobian matrix.
Figure \ref{fig-dof-func} shows some basic management functions, such as creating, destroying, setting values,
getting values and dumping DOF to file.
Neighbour data can be obtained when required, which is owned by other MPIs.
Figure \ref{fig-dof-op} gives a few algebraic operations, such as dot product,
which are required during simulation, such as update solution from Newton method.

\begin{figure}[!htb]
\centering
\begin{evb}
    Z = a * X + b * Y;
    Y = a * X + b * Y;
    Y = a * X + Y;
    X = a * Y;
    d = <X, Y>;
    X = Y;
\end{evb}
\caption{DOF algebraic operations}
\label{fig-dof-op}
\end{figure}

\subsection{Linear System}
A set of equations, including mass conservation laws, energy conservation laws and constraint equations,
are defined on each grid block and each well. Each grid block will form some rows of the Jacobian matrix and
corresponding components of vectors, such as unknowns and right-hand side.

Two matrix types are supported: point-wise matrix and block-wise matrix. The point-wise matrix is the usual
matrix we see, in which each entry is an scalar value. The block-wise matrix is consisted of smaller
sub-matrices. Depending on how many unknowns are defined on each grid block, these sub-matrices could be
$ 3\times 3$, $4 \times 4$ and $64 \times 64$.
In the thermal simulator, each MPI owns certain rows of the Jacobian matrix, and the matrix from well
equations is placed in the last MPI.
Depending the model, each grid block has several unknowns, such as pressure, temperature, water saturation,
gas saturation, mole fraction and well bottom hole pressure. A mapping is constructed between DOFs and linear
systems. This mapping defines global unknown indices and local indices of each DOF and Jacobian matrix.

Distributed-memory matrix and vector modules are developed to handle the
linear system and to implement in-house linear solvers and preconditioners. Communication pattern for sparse
matrix-vector multiplication is complicated and changing all the time, which is constructed for each matrix.
Each preconditioner also has its own communication pattern, such as domain decomposition preconditioner, which
requires information from overlap domains in other MPIs.

\begin{figure}[!htb]
\centering
\begin{evb}
    x = a * x;
    x = a * x + b * y;
    z = a * x + b * y;
    d = <x, y>;
    
    y = A * x;
    y = a * A * x;
    y = a * A * x + b * y;
    z = a * A * x + b * y;
\end{evb}
\caption{Matrix and vector operations}
\label{fig-mv}
\end{figure}

Figure \ref{fig-mv} gives some matrix and vector algebraic operations. Vector-vector operations are naturally
parallel without communication except dot product, which needs one round reduction. 
Sparse matrix-vector multiplication (SpMV) requires one round collective communication to gather off-process
components of vector $x$. The communication is completed by \verb|MPI_Alltoallv|, and if the grid partition is
good enough, each MPI only communicates with a few other MPIs, such as 10 MPI. In this case, the scalability
is good. However, if the grid patition isn't good enough, each MPI communicates with many MPIs, a bad
scalability may be observed. Also, the communication is either synchronous or asynchronous.
The implementation of SpMV for synchronization communication, such as \verb|MPI_Alltoallv|, is easier than
using asynchronization communication.
The asynchronization communication requires developer to check if communication is completed or not.
A simple trick is to separate matrix to two parts, where one part needs local vector and one part
needs remote vector. Through this method, communication and computation can be overlapped, and a better scalability
can be achieved. 

After these BLAS operations are implemented, parallel linear solvers can be developed straightforwardly.
Linear solvers and preconditioners mentioned above, and decoupling methods are implemented.
The ILU methods are serial, and their scalability is poor for large scale simulations. The Restricted Additive
Schwarz (RAS) method is adopted, which treats matrix as graph and defines sub-graph (matrix) on each MPI. If
direct neighbour vertices are included, each sub-graph becomes larger and has overlaps with some other
sub-graphs. Then the process is repeated, more neighbours can be included. Usually one level and two level are
used. Each MPI forms a sub-graph and a sub-domain problem, which is a local linear system. The ILU(k),
ILUT(p,tol) and other methods are employed to solve the local linear system. General preconditioners, such as
RAS method, and special preconditioners, such as CPR (Constraint Pressure Residual) method, are implemented.

\begin{figure}[!htb]
\centering
\begin{evb}
# vtk DataFile Version 2.0
Hex grid created by Hui
ASCII
DATASET UNSTRUCTURED_GRID

POINTS 8 double
 0.    0.    0.  
 1.    0.    0.  
 0.    1.    0.  
 1.    1.    0.  
 0.    0.    1.  
 1.    0.    1.  
 0.    1.    1.  
 1.    1.    1.  

CELLS 1 9
8   0   1   3   2   4   5   7   6  

CELL_TYPES 1
12  

CELL_DATA 1

SCALARS submesh int 1
LOOKUP_TABLE default
0  

SCALARS u double 1
LOOKUP_TABLE default
-1.  
\end{evb}
\caption{Legacy VTK format for unstructured grids}
\label{fig-vis-leg}
\end{figure}

\subsection{Visualization}

Three visualization formats are supported: 1) legacy VTK using binary; 2) legacy VTK using ASCII; and 3) XML
(VTU and PVD). The VTK visualization support many types, such as structured grid, unstructured grid, image
data, poly data and rectilinear grid. Many grid block types are supported, such as point, linear, triangle,
and hexahedron. In this simulator, the unstructured grid type is applied for hexahedron.

Figure \ref{fig-vis-leg} shows an ASCII legacy VTK format, which has standard format. VTK released a file
format document, which has more information. The first line is head 
(\verb|# vtk DataFile Version 2.0|), following by comment (second line),
text format (\verb|BINARY| or \verb|ASCII|), dataset type(\verb|UNSTRUCTURED_GRID|), 
points, cells, cell types, cell data and point data.
Dataset has many types, such as structured grid, unstructured grid, image
data, poly data, and each grid type is an integer, such as 1 for vertex, 2 for poly vertex, 3 for line, 4 for
poly line, 5 for triangle, and 12 for hexahedron.
Each data section has name of data set, data type, and size of components. The binary format has the same
elements as ASCII except the data is in binary format. Parallel output will be required when writing VTK file.
There are a few limits for legacy VTK format: 1) integer has 32 bits, which cannot be used for Exascale
computing; 2) no time step information is available when time series data is stored.

\begin{figure}[!htb]
\centering
\begin{evb}

<?xml version="1.0"?>
<VTKFile type="UnstructuredGrid" version="0.1" byte_order="LittleEndian">
    <UnstructuredGrid>
        <Piece NumberOfPoints="8" NumberOfCells="1">
            <Points>
                <DataArray type="Float64" Name="vertices" NumberOfComponents="3" format="ascii">
                 0. 0. 0. 1. 0. 0. 0. 1. 0. 1. 1. 0. 0. 0. 1. 1. 0. 1. 0. 1. 1. 1. 1. 1.   
                </DataArray>
            </Points>

            <Cells>
                <DataArray type="Int64" Name="connectivity" format="ascii">
                    0 1 3 2 4 5 7 6   
                </DataArray>
                <DataArray type="Int64" Name="offsets" format="ascii">
                    8   
                </DataArray>
                <DataArray type="Int64" Name="types" format="ascii">
                    12  
                </DataArray>
            </Cells>

            <CellData>
                <DataArray type="Int64" Name="submesh" format="ascii">
                    0  
                </DataArray>
                <DataArray type="Float64" Name="u_h" NumberOfComponents="1" format="ascii">
                    -1.   
                </DataArray>
            </CellData>

            <PointData>
                <DataArray type="Float64" Name="u_p" NumberOfComponents="1" format="ascii">
                    -1. 1 2 3 -3 2 3 4
                </DataArray>
            </PointData>
        </Piece>
    </UnstructuredGrid>
</VTKFile>

\end{evb}
\caption{XML format (VTU) for unstructured grids}
\label{fig-vis-xml}
\end{figure}

\begin{figure}[!htb]
\centering
\begin{evb}

<VTKFile type="Collection" version="0.1" byte_order="LittleEndian">
    <Collection>
        <DataSet timestep="1.0000000" file="keyword_4_0001.vtu"/>
        <DataSet timestep="2.6670000" file="keyword_4_0002.vtu"/>
        <DataSet timestep="5.4458890" file="keyword_4_0003.vtu"/>
        <DataSet timestep="10.078297" file="keyword_4_0004.vtu"/>
        <DataSet timestep="17.315484" file="keyword_4_0005.vtu"/>
        <DataSet timestep="18.657742" file="keyword_4_0006.vtu"/>
        <DataSet timestep="19.731548" file="keyword_4_0007.vtu"/>
        <DataSet timestep="20.000000" file="keyword_4_0008.vtu"/>
        <DataSet timestep="20.268452" file="keyword_4_0009.vtu"/>
        <DataSet timestep="20.715961" file="keyword_4_0010.vtu"/>
        <DataSet timestep="21.456624" file="keyword_4_0011.vtu"/>
        <DataSet timestep="22.317238" file="keyword_4_0012.vtu"/>
        <DataSet timestep="23.036065" file="keyword_4_0013.vtu"/>
        <DataSet timestep="32.000000" file="keyword_4_0014.vtu"/>
        <DataSet timestep="41.500000" file="keyword_4_0015.vtu"/>
    </Collection>
</VTKFile>

\end{evb}
\caption{PVD collection format for time series data}
\label{fig-vis-pvd}
\end{figure}

Figure \ref{fig-vis-xml} is a sample XML for VTU (unstructured VTK file), which has similar sections as legacy
VTK files, such as grid definition and data set. It has point data (\verb|Points|), cells (\verb|Cells|),
including connectivity data (\verb|connectivity|), offset data (\verb|offsets|) and cell
type data (\verb|types|).  The XML VTK file supports more data types, such as Int32, Int64, Float32 and Float64.
When writing XML file, collective parallel write subroutines mentioned above are employed.
The VTU format can be written as binary or ASCII files, little endian or big endian.
By default, ASCII output is applied.
Figure \ref{fig-vis-pvd} is a PVD collection
data file, which is also XML format. It has time step (\verb|timestep|), group info (\verb|group|), part info
(\verb|part|) and file info (\verb|file|). Since we write all
data to one file, group and part are optional. Visualization software, such as ParaView, can read the VTU and
PVD files and display the correct properties and time step.

\subsection{Schedule}
In real reservoir model, operations of a well may be changed during time, such as CSS, which has three stages,
and each stage has different well operations. A flexible schedule has to be designed. The thermal simulator
supports two types of schedule. One is shown by Figure \ref{fig-schedule-1}, where the keyword is \verb|time|,
whose unit is day.
Changes are defined following a time point. The figure shows well operation changes for well \verb|Prod_No-2|.
If no change follows a time point, then it means the simulator must compute results for this time point and
report the results. The second format we support is shown by Figure \ref{fig-schedule-2}. The keyword is
marked by \verb|date| (not case sensitive). Several date formats are supported, such as
\begin{itemize}
    \item Date 2016 3   1.3: it is equivalent to 7:12 AM March 1, 2016;

    \item Date 2016 6   18 12:24 PM: date and time are 12:24 PM, June 18, 2016;

    \item Date 2016 8   1  8:33 AM: date and time are 8:33AM, August 1, 2016;

    \item Date 2017 8   1  23:33:42: date and time are 23:33:42, August 1,2017.
\end{itemize}

After each time or date, well operation changes are defined. Other settings, such as restart and numerical
setting can also be changed.

\begin{figure}[!htb]
\centering
\begin{evb}
# schedule, format 1
run

# ----------------------------- day 1
time 1

# ----------------------------- day 10
time 10

# restart point
restart

# numerical change
numerical:

solver: bicgstab
pc: cpr-fpf
tolsol: 1e-3
tolnew: 1e-4
dtmax: 100
/

# ----------------------------- day 200
time 200

# well change
well: Prod_No-2
operate: min bhp   600
operate: max stl   600

htwell: temp
....
/

# numerical change
numerical:
solver: bicgstab
pc: cpr
tolnew: 1e-3
dtmax: 200
/

# ------------------------------ day 1065
time 1065

# restart point
restart

# well change
well: Prod_No-2
operate: min bhp   300
operate: max stl   1600
operate: max stf   10000
operate: steamtrap 10

htwell: off
....
/

# end simulation
time 3650
stop
\end{evb}
\caption{Reservoir simulation schedule, format one}
\label{fig-schedule-1}
\end{figure}

\begin{figure}[!htb]
\centering
\begin{evb}
# schedule, format 2
run

Date 2016 1   1

Date 2016 2   1
well: Prod_No-2
operate: min bhp   300
operate: max steam 10
operate: max stl   300
/

Date 2016 3   1.3

Date 2016 6   1  12:24 PM
Date 2016 7   1

Date 2016 8   1  8:33 AM

well: Prod_No-2
operate: min bhp   600
operate: max stl   600
/

Date 2016 12  23
Date 2017 5   25.76

Date 2017 8   1  23:33:42
Date 2018 1   1
stop
\end{evb}
\caption{Reservoir simulation schedule, format two}
\label{fig-schedule-2}
\end{figure}

\subsection{Restart}
Restart is an important feature to software developers and reservoir engineers. A reservoir model may need to
predict well production for 20 years, and one simulation run could take hours or days. When a design fault is
found, it may take a long time to reach the time step that has the bug. A developer may need to run dozens of
time to understand the bug and to fix it, which is a waste of time. For example, if we know the issue is found
at 2240 day, a restart mark can be set at a time point earlier, such as 2235.47 day. At this point, restart
files are written to disk. Then the simulator may restart from this point, and there is no need to from the
beginning. Figure \ref{fig-restart} shows a restart point. Lots of time can be saved. The restart function
requires parallel write and parallel read. This feature is also useful to reservoir engineers. The
\verb|restart| keyword can be place after any time point. Multiple restart files are written to disk using
parallel write, and restart can be initiated at any given time point.

\begin{figure}[!htb]
\centering
\begin{evb}
# schedule
run

time 1
time 10

time 200

# first restart point at day 200
restart

time 2235.47

# second restart point at day 2235.47
restart

time 2240

time 2340

# third restart point at day 2340
restart

time 3350

# forth restart point at day 3350
restart

time 3650
stop
\end{evb}
\caption{Dump restart files at given time points}
\label{fig-restart}
\end{figure}

\begin{figure}[!htb]
\centering
\begin{evb}
    ./simulator -restart -which_restart 3
\end{evb}
\caption{Restart simulation from the third restart point}
\label{fig-restart}
\end{figure}
\section{Numerical Studies}
\label{sec-exp}

\section{Numerical Studies}
\label{sec-exp}

Numerical experiments are presented here, which includes a few sections. The first section validates our
results against CMG STARS, which is the most widely applied thermal simulator. The purpose is to prove the
correctness of our numerical methods, models and implementation. 
The second section validates well control methods against CMG STARS.
The third section studies numerical performance of our methods. 
The forth section tests the scalability of our thermal simulator using some giant models.

We should mention that the models are randomly generated. The only purpose is to validate our results and CMG
STARS. If the model is the same, the results from our simulator and CMG STARS should be very close.
By comparing results from CMG STARS, the implementation and accuracy of various properties and well controls
can be verified.
In the numerical section, the injection wells and the production wells are placed to be close to each other
such that the models are hard to solve.

\subsection{Validation}
This section covers a few commonly used models, pure heavy oil, heavy oil and light oil, oil and
non-condensable gas (NCG).

\subsubsection{Heavy Oil}

\begin{figure}[H]
    \centering
    \includegraphics[width=0.53\linewidth, angle=270]{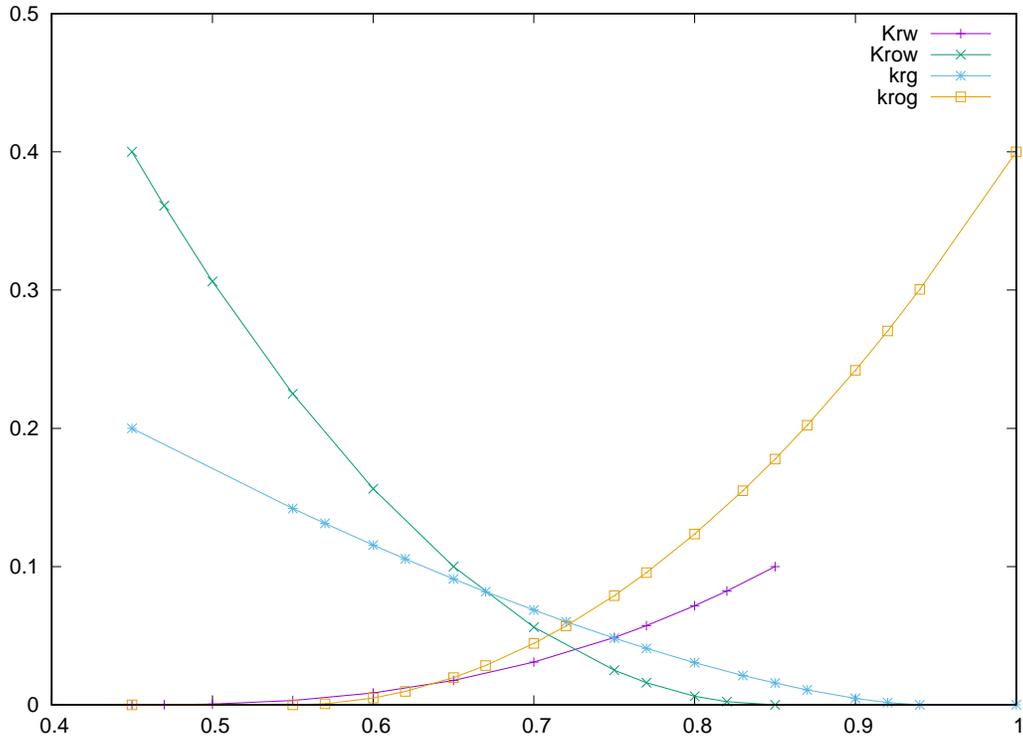}
    \caption{Example \ref{val-heavy}, heavy oil: relative permeability of the water-oil table and liquid-gas table}
    \label{fig-val-heavy-kr}
\end{figure}

\begin{figure}[H]
    \centering
    \includegraphics[width=0.53\linewidth, angle=270]{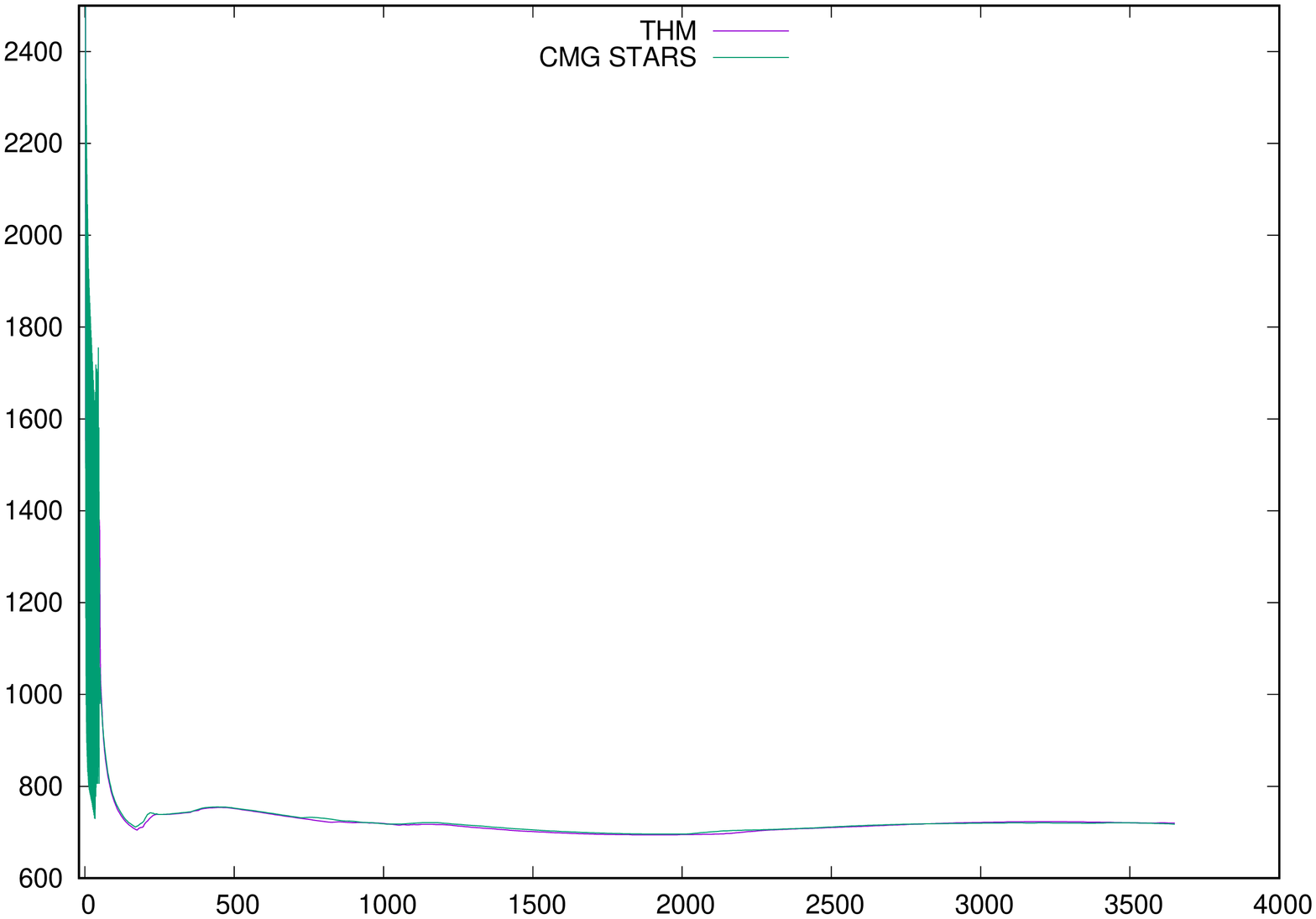}
    \caption{Example \ref{val-heavy}, heavy oil: injection well, bottom hole pressure (psi)}
    \label{fig-val-heavy-inj-bhp}
\end{figure}

\begin{figure}[H]
    \centering
    \includegraphics[width=0.53\linewidth, angle=270]{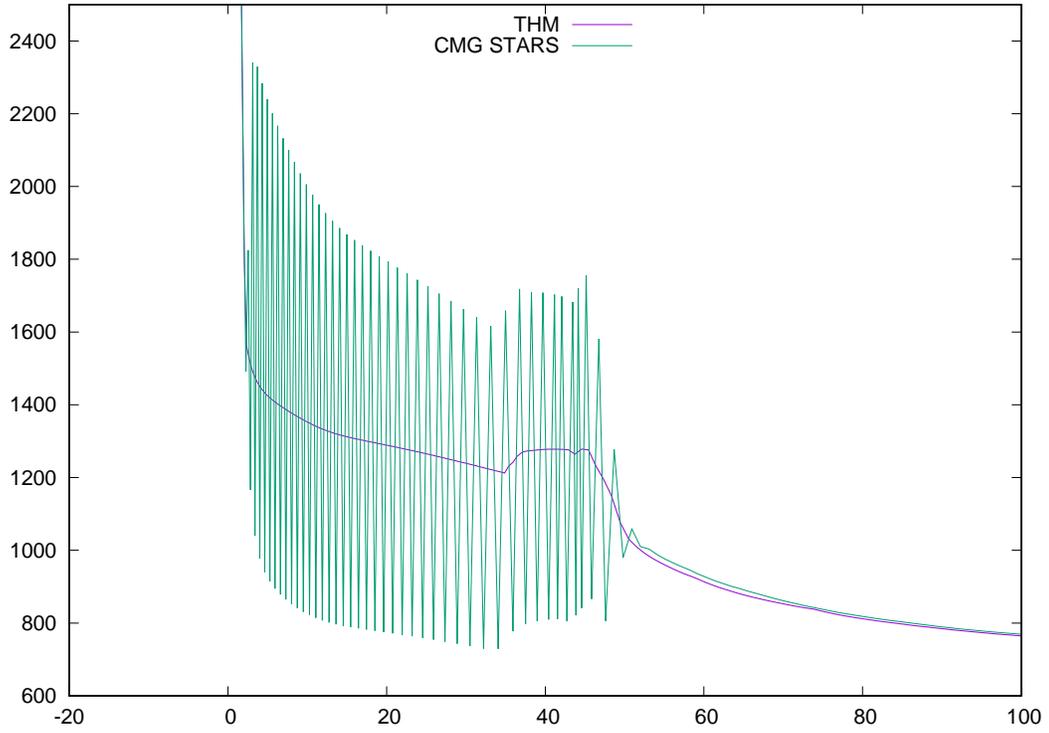}
    \caption{Example \ref{val-heavy}, heavy oil: injection well, bottom hole pressure (psi), first 100 days}
    \label{fig-val-heavy-inj-bhp-jump}
\end{figure}

\begin{figure}[H]
    \centering
    \includegraphics[width=0.53\linewidth, angle=270]{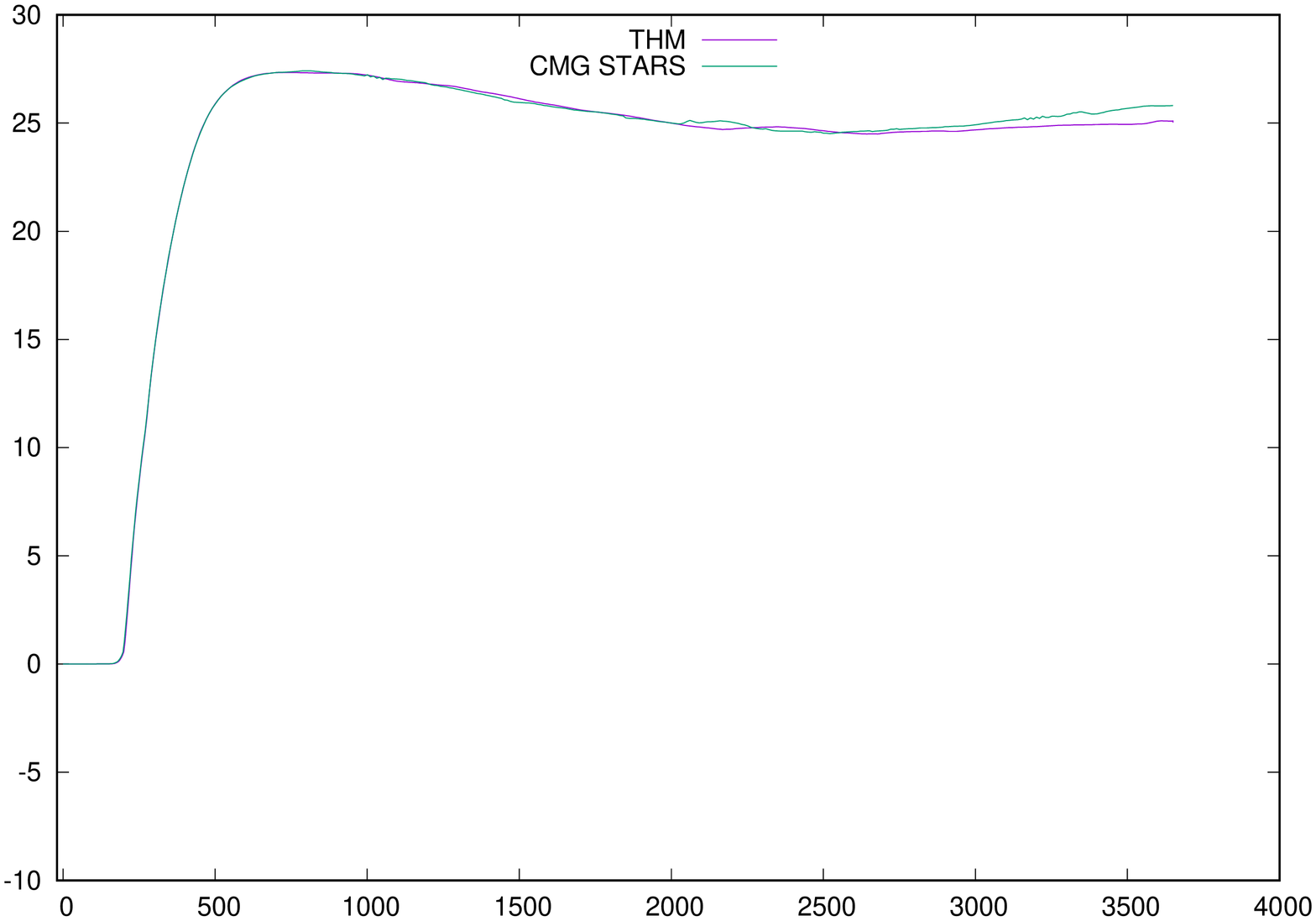}
    \caption{Example \ref{val-heavy}, heavy oil: water production rate (bbl/day), first production well}
    \label{fig-val-heavy-p1-pwr}
\end{figure}

\begin{figure}[H]
    \centering
    \includegraphics[width=0.53\linewidth, angle=270]{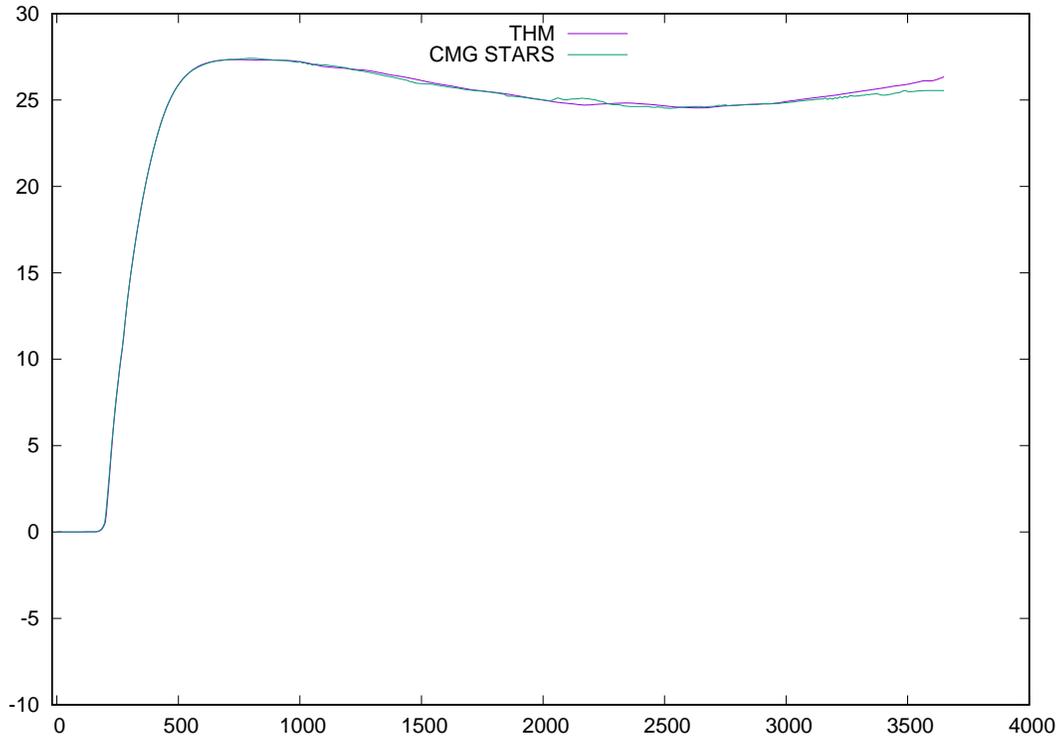}
    \caption{Example \ref{val-heavy}, heavy oil: water production rate (bbl/day), second production well}
    \label{fig-val-heavy-p2-pwr}
\end{figure}

\begin{figure}[H]
    \centering
    \includegraphics[width=0.53\linewidth, angle=270]{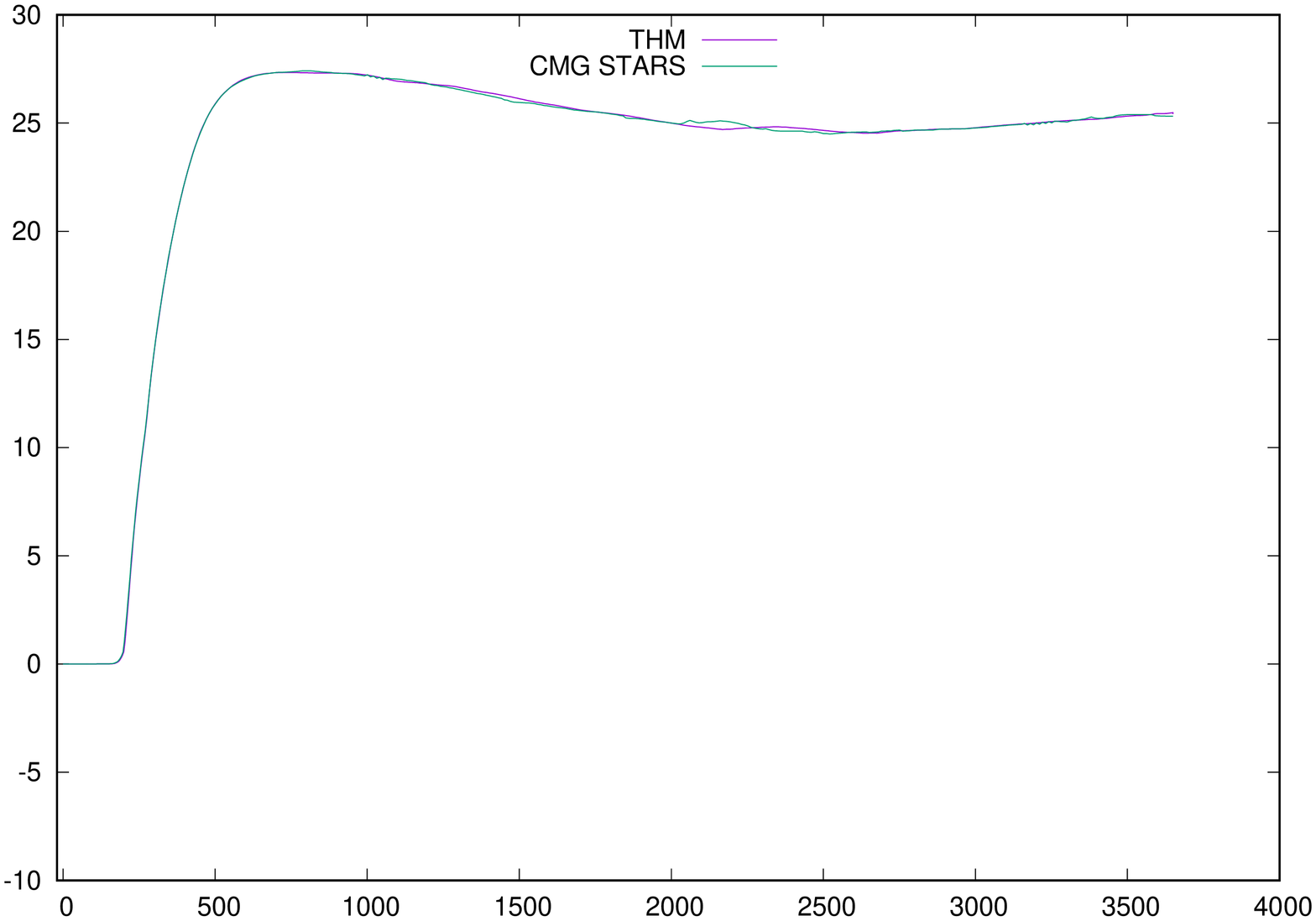}
    \caption{Example \ref{val-heavy}, heavy oil: water production rate (bbl/day), third production well}
    \label{fig-val-heavy-p3-pwr}
\end{figure}

\begin{figure}[H]
    \centering
    \includegraphics[width=0.53\linewidth, angle=270]{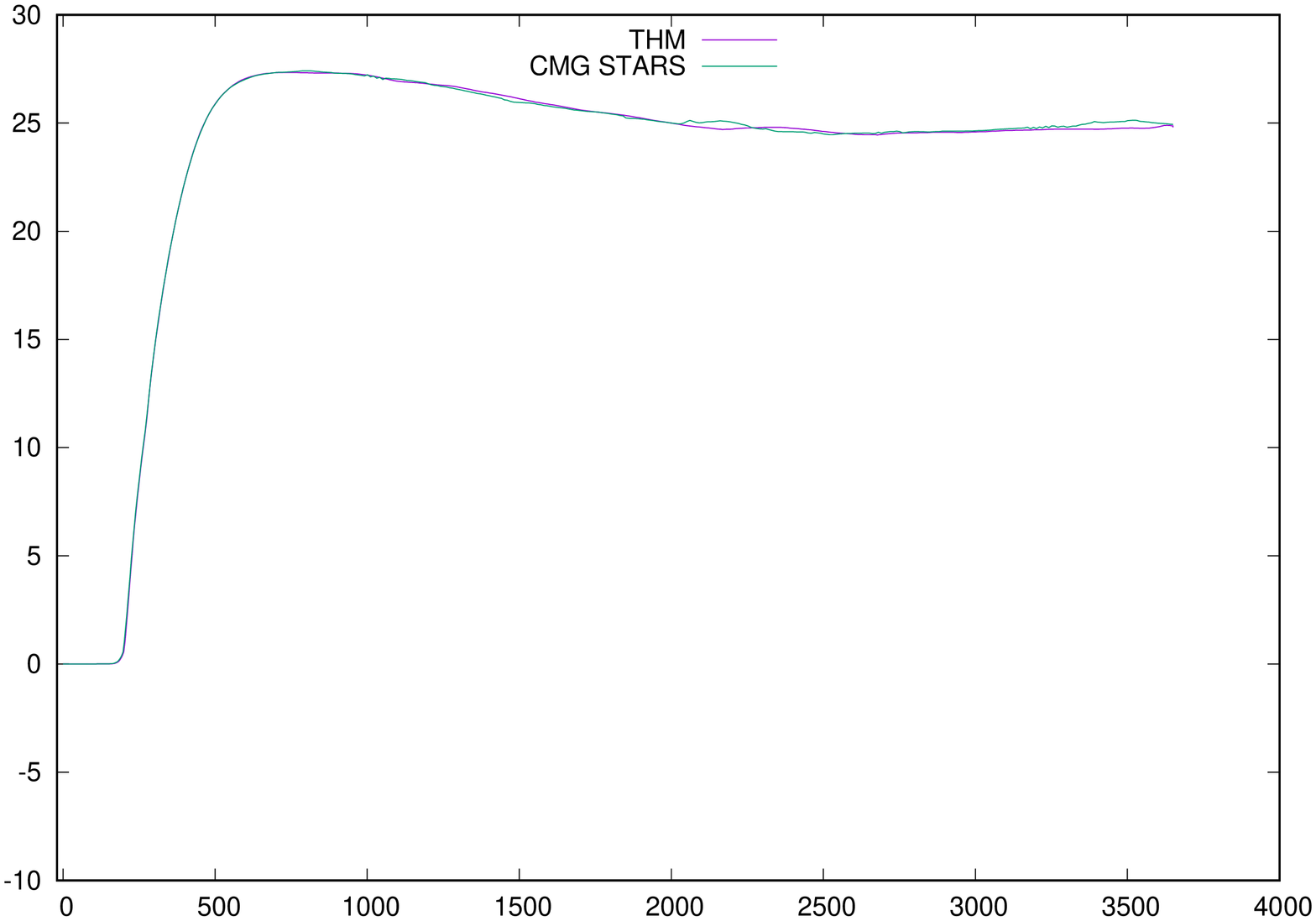}
    \caption{Example \ref{val-heavy}, heavy oil: water production rate (bbl/day), forth production well}
    \label{fig-val-heavy-p4-pwr}
\end{figure}

\begin{figure}[H]
    \centering
    \includegraphics[width=0.53\linewidth, angle=270]{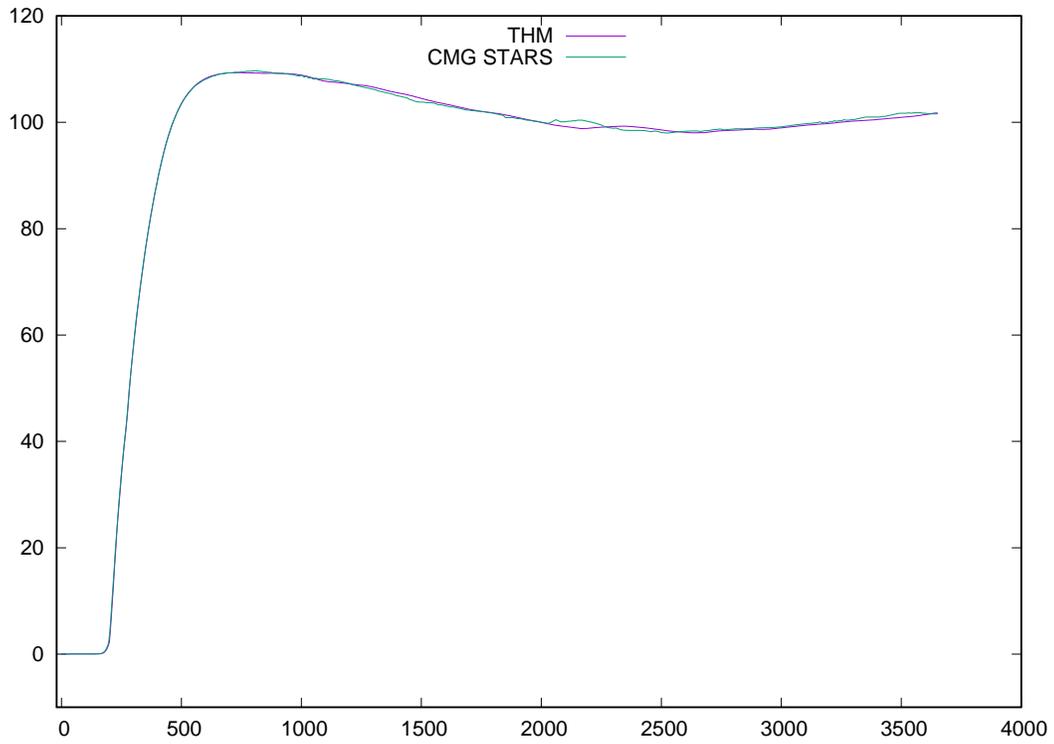}
    \caption{Example \ref{val-heavy}, heavy oil: total water production rate (bbl/day)}
    \label{fig-val-heavy-pwr}
\end{figure}

\begin{figure}[H]
    \centering
    \includegraphics[width=0.53\linewidth, angle=270]{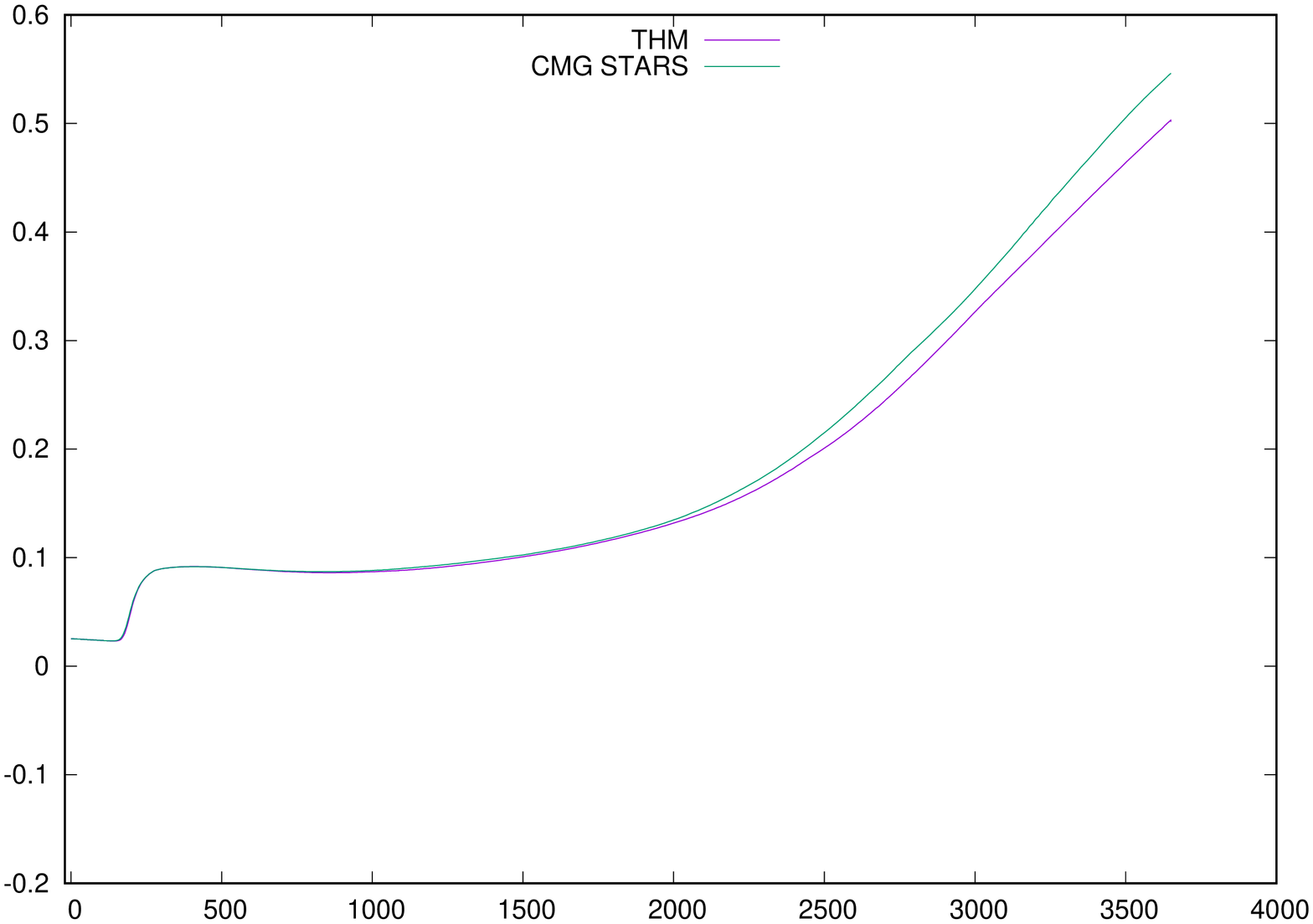}
    \caption{Example \ref{val-heavy}, heavy oil: oil production rate (bbl/day), first production well}
    \label{fig-val-heavy-p1-por}
\end{figure}

\begin{figure}[H]
    \centering
    \includegraphics[width=0.53\linewidth, angle=270]{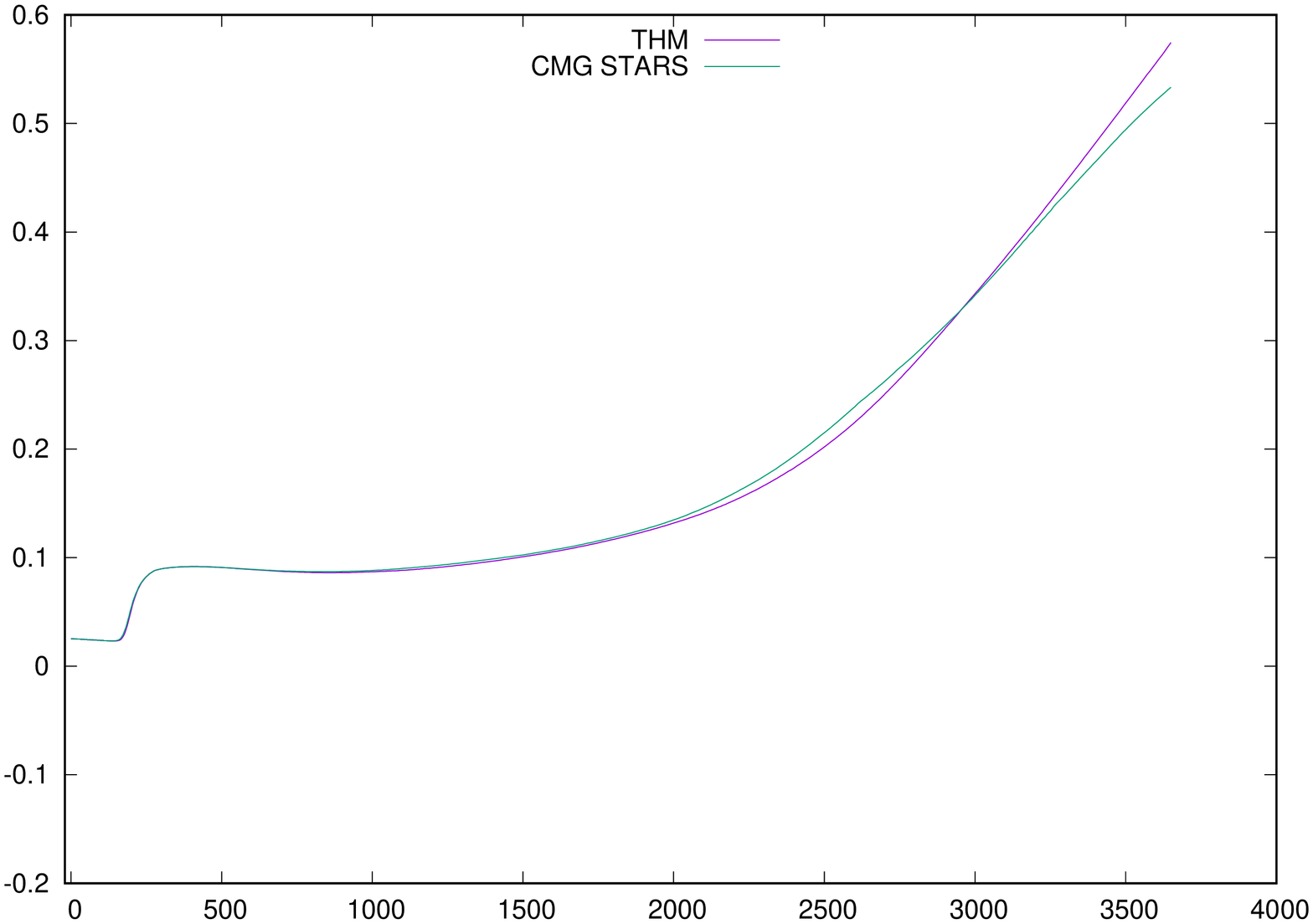}
    \caption{Example \ref{val-heavy}, heavy oil: oil production rate (bbl/day), second production well}
    \label{fig-val-heavy-p2-por}
\end{figure}

\begin{figure}[H]
    \centering
    \includegraphics[width=0.53\linewidth, angle=270]{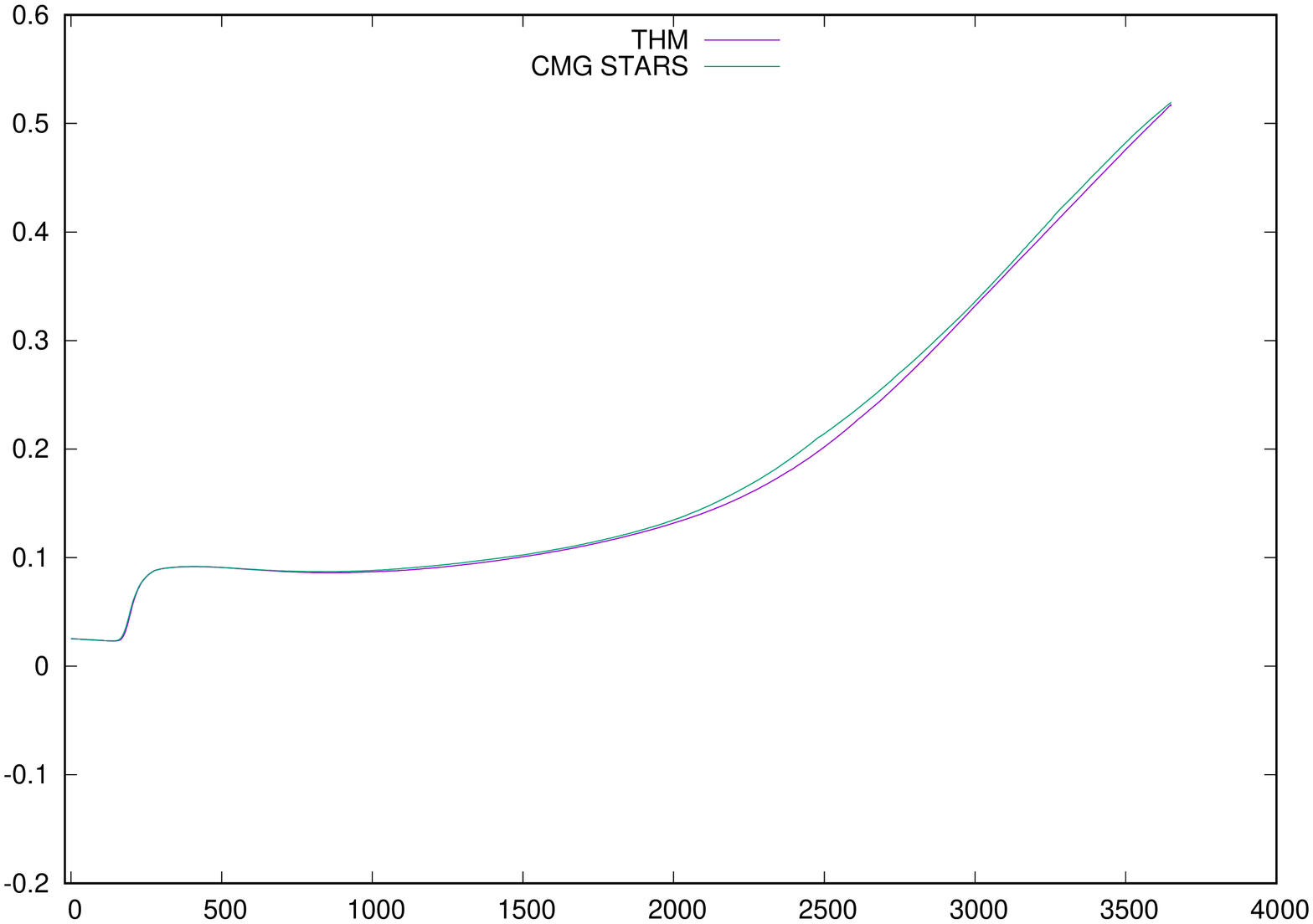}
    \caption{Example \ref{val-heavy}, heavy oil: oil production rate (bbl/day), third production well}
    \label{fig-val-heavy-p3-por}
\end{figure}

\begin{figure}[H]
    \centering
    \includegraphics[width=0.53\linewidth, angle=270]{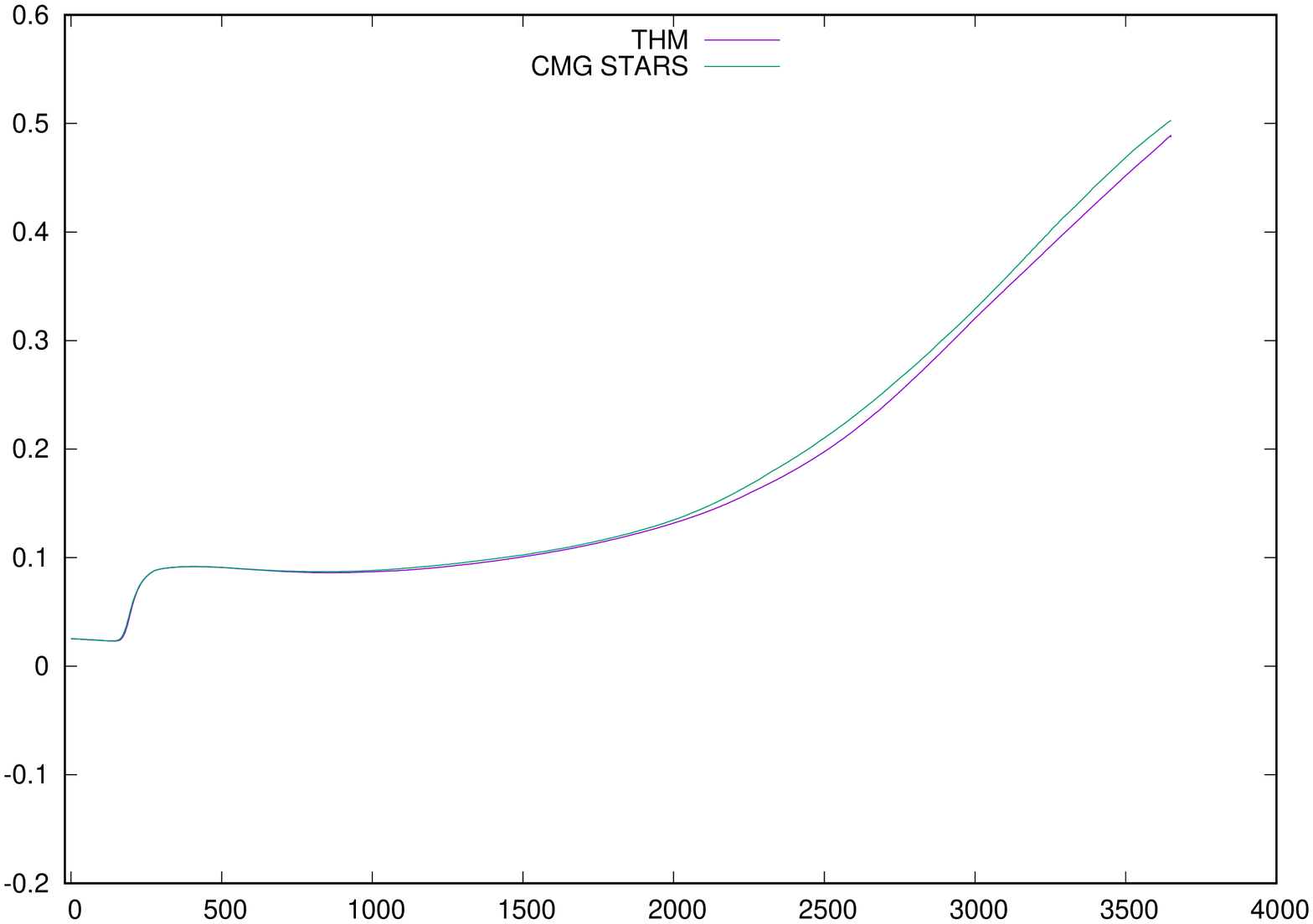}
    \caption{Example \ref{val-heavy}, heavy oil: oil production rate (bbl/day), forth production well}
    \label{fig-val-heavy-p4-por}
\end{figure}

\begin{figure}[H]
    \centering
    \includegraphics[width=0.53\linewidth, angle=270]{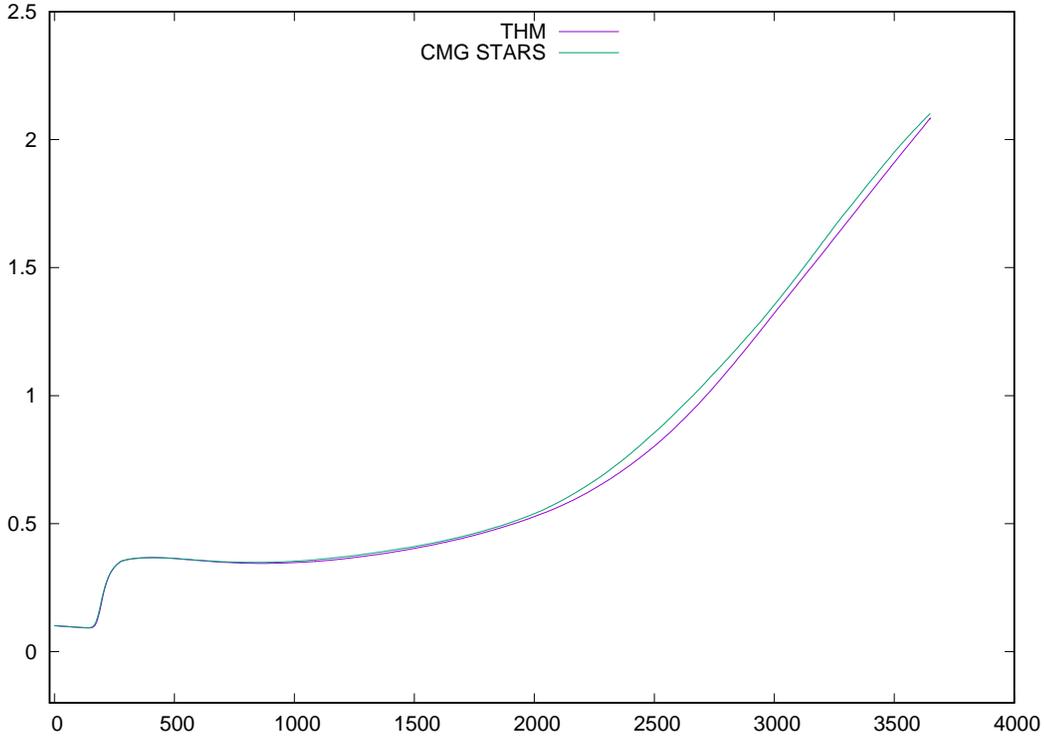}
    \caption{Example \ref{val-heavy}, heavy oil: total oil production rate (bbl/day)}
    \label{fig-val-heavy-por}
\end{figure}

\begin{example}
    \normalfont
    \label{val-heavy} The grid dimension of the model is $9 \times 9 \times 4$, with sizes of 29.17 ft, 29.17
    ft and 10 ft in $x$, $y$ and $z$ direction. Details of the model are presented in the following tables.
    Water component and one heavy oil
    component are simulated. As shown by Figure \ref{fig-val-heavy-kr}, the water-oil relative permeability
    and the liquid-gas relative permeability have sharp change.
    It has five vertical wells: one injection
    well in the center (5, 5), and four production wells in four corners, (1, 1), (1, 9), (9, 1) and (9, 9).
    The bottom hole pressure of the injection well, water rate and oil rate of each well are shown from Figure
    \ref{fig-val-heavy-inj-bhp} to Figure \ref{fig-val-heavy-por}. All results are compared with CMG STARS.
\end{example}

\begin{table}[!htb]
    \centering
    \begin{tabular}{c c c}
        \hline
        {$S_w$} & $k_{rw}$ & $k_{row}$ \\\hline
        0.45  &   0.0       &   0.4     \\
        0.47  &   0.000056  &  0.361    \\
        0.50  &   0.000552  &  0.30625  \\
        0.55  &   0.00312   &  0.225    \\
        0.60  &   0.00861   &  0.15625  \\
        0.65  &   0.01768   &  0.1      \\
        0.70  &   0.03088   &  0.05625  \\
        0.75  &   0.04871   &  0.025    \\
        0.77  &   0.05724   &  0.016    \\
        0.80  &   0.07162   &  0.00625  \\
        0.82  &   0.08229   &  0.00225  \\
        0.85  &   0.1       &  0.0      \\
        \hline
    \end{tabular}
    \caption{Input data for Example \ref{val-heavy} (cont'd).}
    \label{val-heavy-tab-input3}
\end{table}

\begin{table}[!htb]
    \centering
    \begin{tabular}{c c c}
        \hline
        {$S_l$} & $k_{rg}$ & $k_{rog}$ \\\hline
        0.45  &   0.2      &   0.0        \\
        0.55  &   0.14202  &   0.0        \\
        0.57  &   0.13123  &   0.00079    \\
        0.60  &   0.11560  &   0.00494    \\
        0.62  &   0.10555  &   0.00968    \\
        0.65  &   0.09106  &   0.01975    \\
        0.67  &   0.08181  &   0.02844    \\
        0.70  &   0.06856  &   0.04444    \\
        0.72  &   0.06017  &   0.05709    \\
        0.75  &   0.04829  &   0.07901    \\
        0.77  &   0.04087  &   0.09560    \\
        0.80  &   0.03054  &   0.12346    \\
        0.83  &   0.02127  &   0.15486    \\
        0.85  &   0.01574  &   0.17778    \\
        0.87  &   0.01080  &   0.20227    \\
        0.90  &   0.00467  &   0.24198    \\
        0.92  &   0.00165  &   0.27042    \\
        0.94  &   0.0      &   0.30044    \\
        1.    &   0.0      &   0.4        \\
        \hline
    \end{tabular}
    \caption{Input data for Example \ref{val-heavy} (cont'd).}
    \label{val-heavy-tab-input4}
\end{table}

\begin{table}[!htb]
    \centering
    \begin{tabular}{l c}
        \hline
        \textbf{Initial condition} &  \\
        \hline
        $k_{x,y,z}\ (md)$  & 313, 424, 535 \\
        $\phi$ & 0.3 \\
        $\phi_c$ & 5e-4 \\
        $p\ (psi)$ & 75 \\
        $T\ (^\circ F)$ & 125 \\
        $S_{w, o, g}$ & 0.45, 0.55, 0. \\
        \hline
    \end{tabular}
    \caption{Input data for Example \ref{val-heavy}}
    \label{val-heavy-tab-input1}
\end{table}

\begin{table}[!htb]
    \centering
    \begin{tabular}{c c c}
        \hline
        \textbf{Well conditions} & & \\\hline
        Injector & water $(bbl/day)$ & 100  \\
        & wi $(ft \cdot md)$ & 1e4 \\
        & tinjw $(^\circ F)$ & 450 \\
        & steam quality & 0.4 \\
        \hline

        Producer 1 & bhp $(psi)$ & 17 \\
        & wi $(ft \cdot md)$ & 1e4 \\
        & steamtrap $(^\circ F)$ & 10 \\
        \hline

        Producer 2 & bhp $(psi)$ & 17 \\
        & wi $(ft \cdot md)$ & 1e4 \\
        & steamtrap $(^\circ F)$ & 20 \\
        \hline

        Producer 3 & bhp $(psi)$ & 17 \\
        & wi $(ft \cdot md)$ & 1e4 \\
        & steamtrap $(^\circ F)$ & 30 \\
        \hline

        Producer 4 & bhp $(psi)$ & 17 \\
        & wi $(ft \cdot md)$ & 1e4 \\
        & steamtrap $(^\circ F)$ & 40 \\
        \hline
    \end{tabular}
    \caption{Input data for Example \ref{val-heavy} (cont'd).}
    \label{val-heavy-tab-input5}
\end{table}

\clearpage

Figure \ref{fig-val-heavy-inj-bhp} is the bottom hole pressure of the injection well.
Figure \ref{fig-val-heavy-p1-pwr} is the water production rate of the first production well.
Figure \ref{fig-val-heavy-p2-pwr} is the water production rate of the second production well.
Figure \ref{fig-val-heavy-p3-pwr} is the water production rate of the third production well.
Figure \ref{fig-val-heavy-p4-pwr} is the water production rate of the forth production well.
Figure \ref{fig-val-heavy-pwr} is the total water production rate of all production wells.
Figure \ref{fig-val-heavy-p1-por} is the oil production rate of the first production well.
Figure \ref{fig-val-heavy-p2-por} is the oil production rate of the second production well.
Figure \ref{fig-val-heavy-p3-por} is the oil production rate of the third production well.
Figure \ref{fig-val-heavy-p4-por} is the oil production rate of the forth production well.
Figure \ref{fig-val-heavy-por} is the total oil production rate of all production wells.
All figures show that our results match CMG STARS very well, which confirms our methods and implementation are
correct.

%%%%%%%%%%%%%%%%%%%%%%%%%%%%%%%%%%%%%%%%%%%%%%%%%%%%%%%%%%%%%%%%%%%%%%%%%%%%%%%%%%%%%%%%%%%%%%%%%%%%%%%%%%%%%%
\subsubsection{Heavy Oil and Light Oil}

\begin{figure}[H]
    \centering
    \includegraphics[width=0.53\linewidth, angle=270]{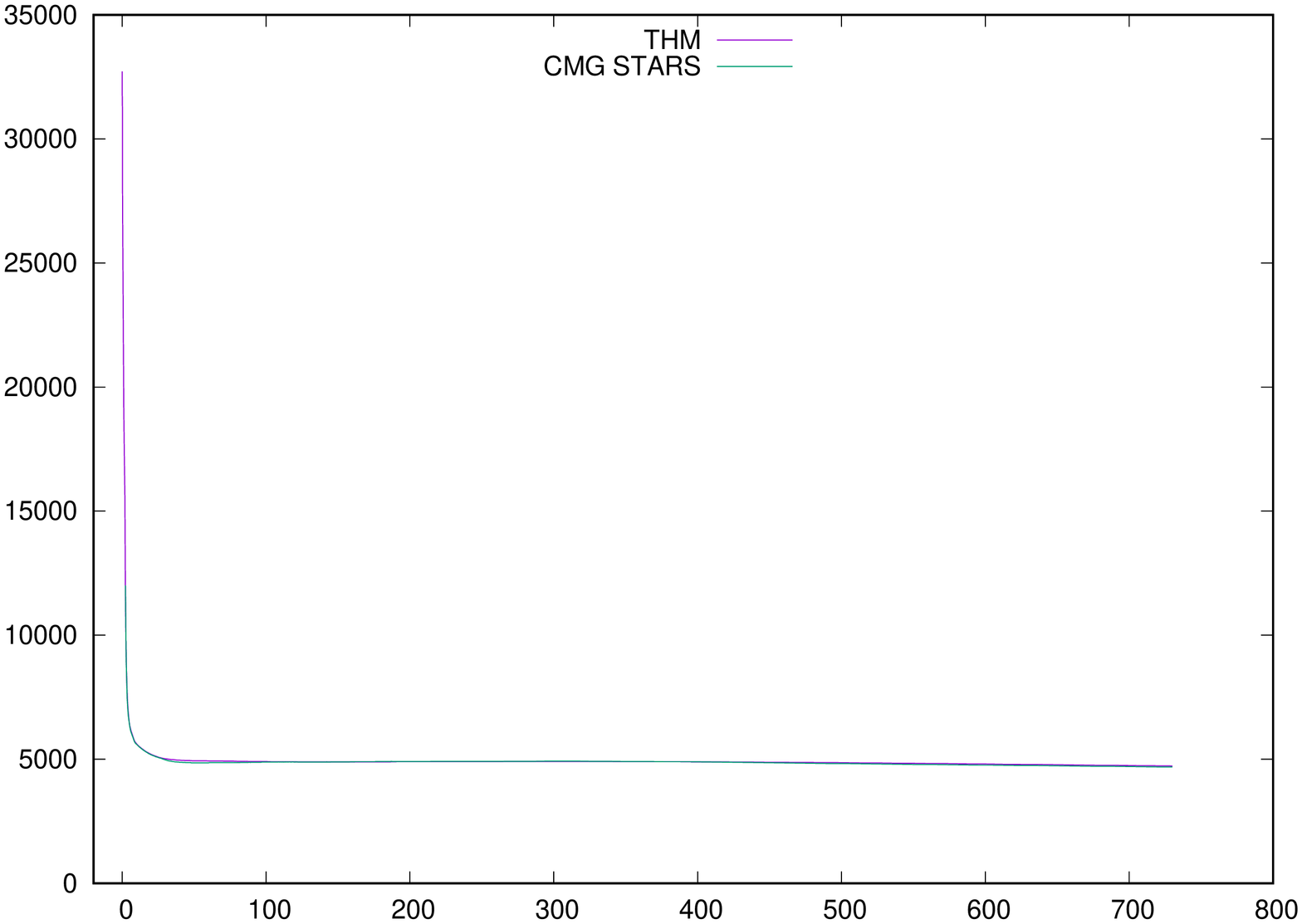}
    \caption{Example \ref{val-light}, light oil: injection well, bottom hole pressure (psi)}
    \label{fig-val-light-inj-bhp}
\end{figure}

\begin{figure}[H]
    \centering
    \includegraphics[width=0.53\linewidth, angle=270]{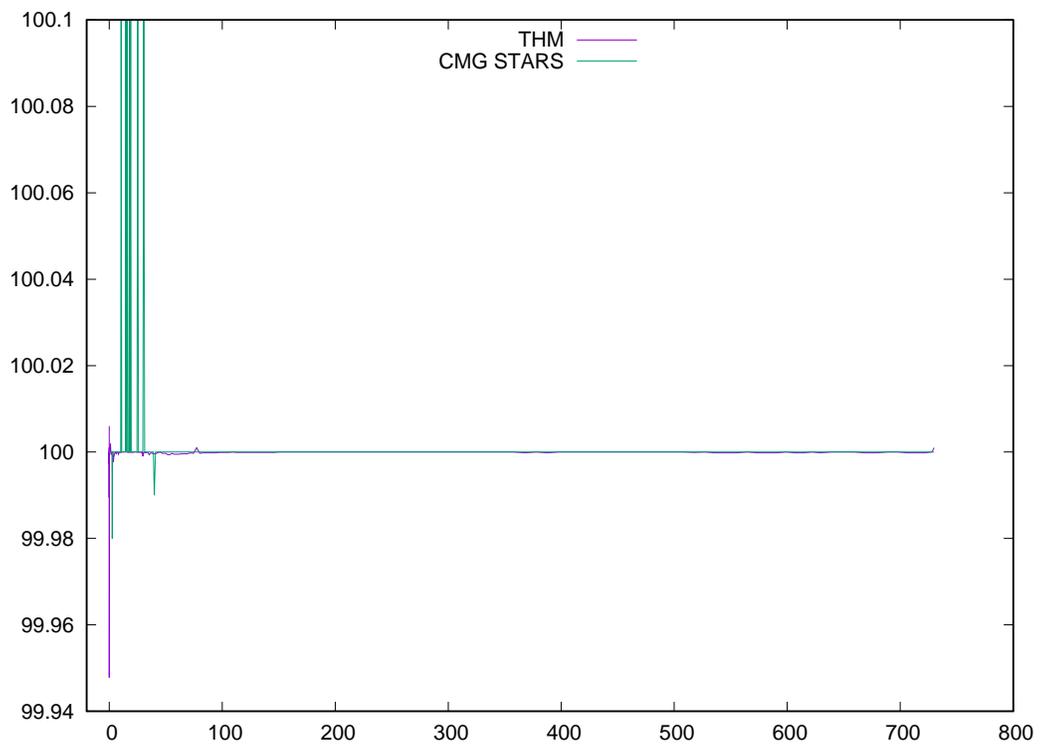}
    \caption{Example \ref{val-light}, light oil: injection well, water injection rate (bbl/day)}
    \label{fig-val-light-ir}
\end{figure}

\begin{figure}[H]
    \centering
    \includegraphics[width=0.53\linewidth, angle=270]{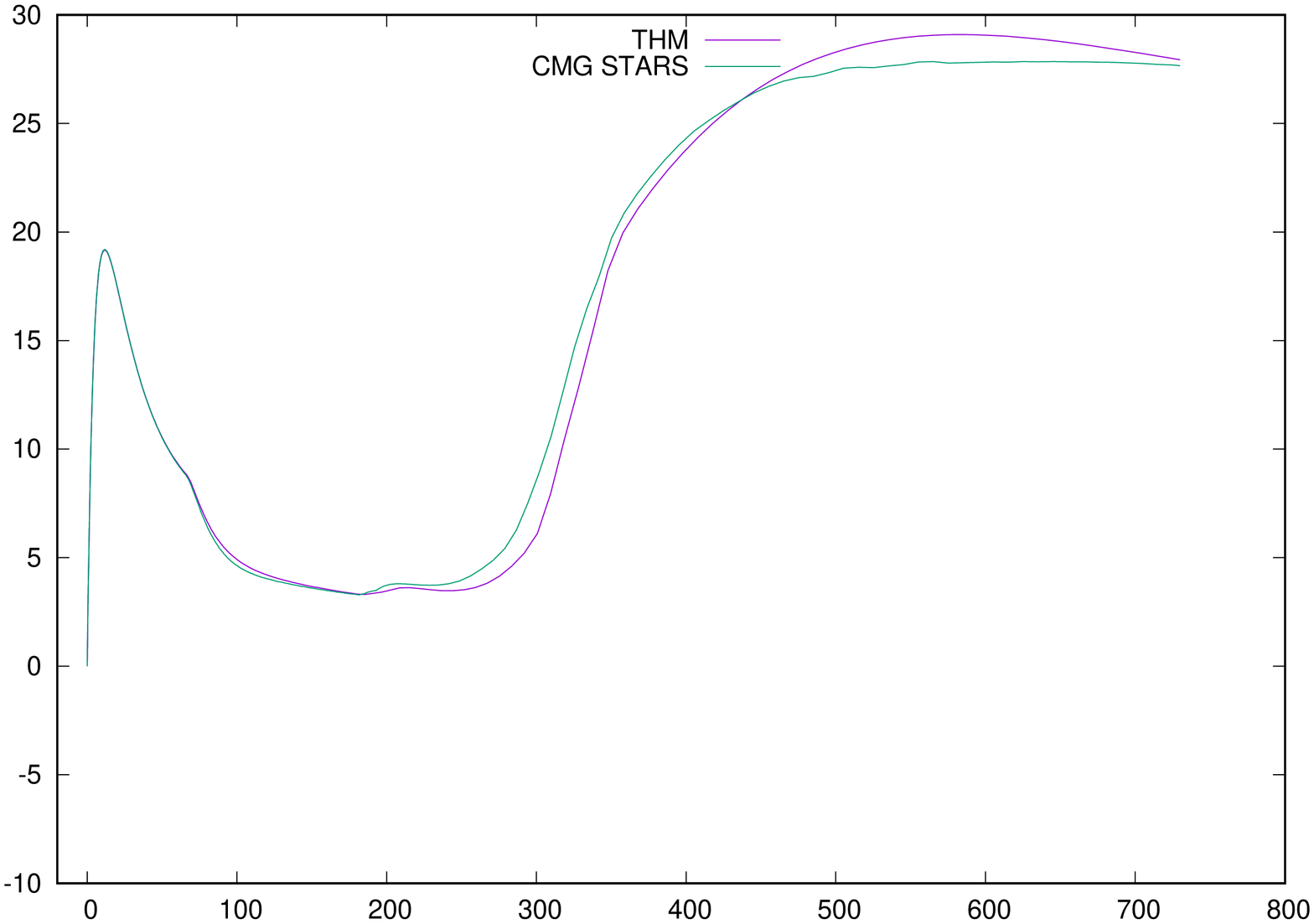}
    \caption{Example \ref{val-light}, light oil: water production rate (bbl/day), first production well}
    \label{fig-val-light-p1-pwr}
\end{figure}

\begin{figure}[H]
    \centering
    \includegraphics[width=0.53\linewidth, angle=270]{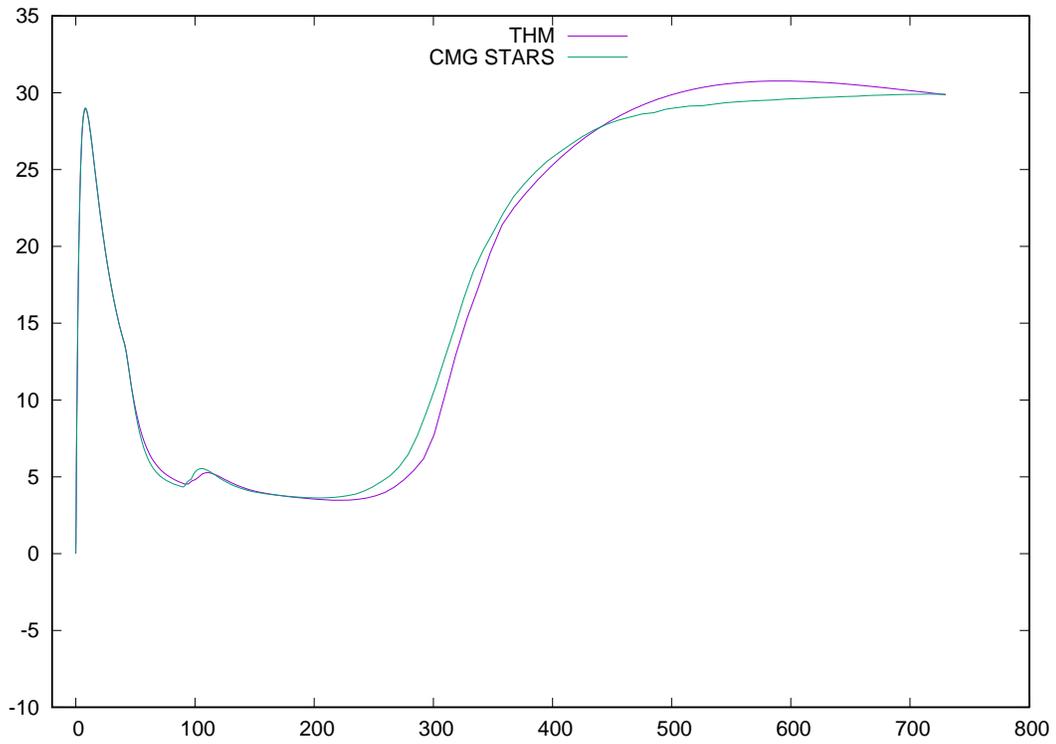}
    \caption{Example \ref{val-light}, light oil: water production rate (bbl/day), second production well}
    \label{fig-val-light-p2-pwr}
\end{figure}

\begin{figure}[H]
    \centering
    \includegraphics[width=0.53\linewidth, angle=270]{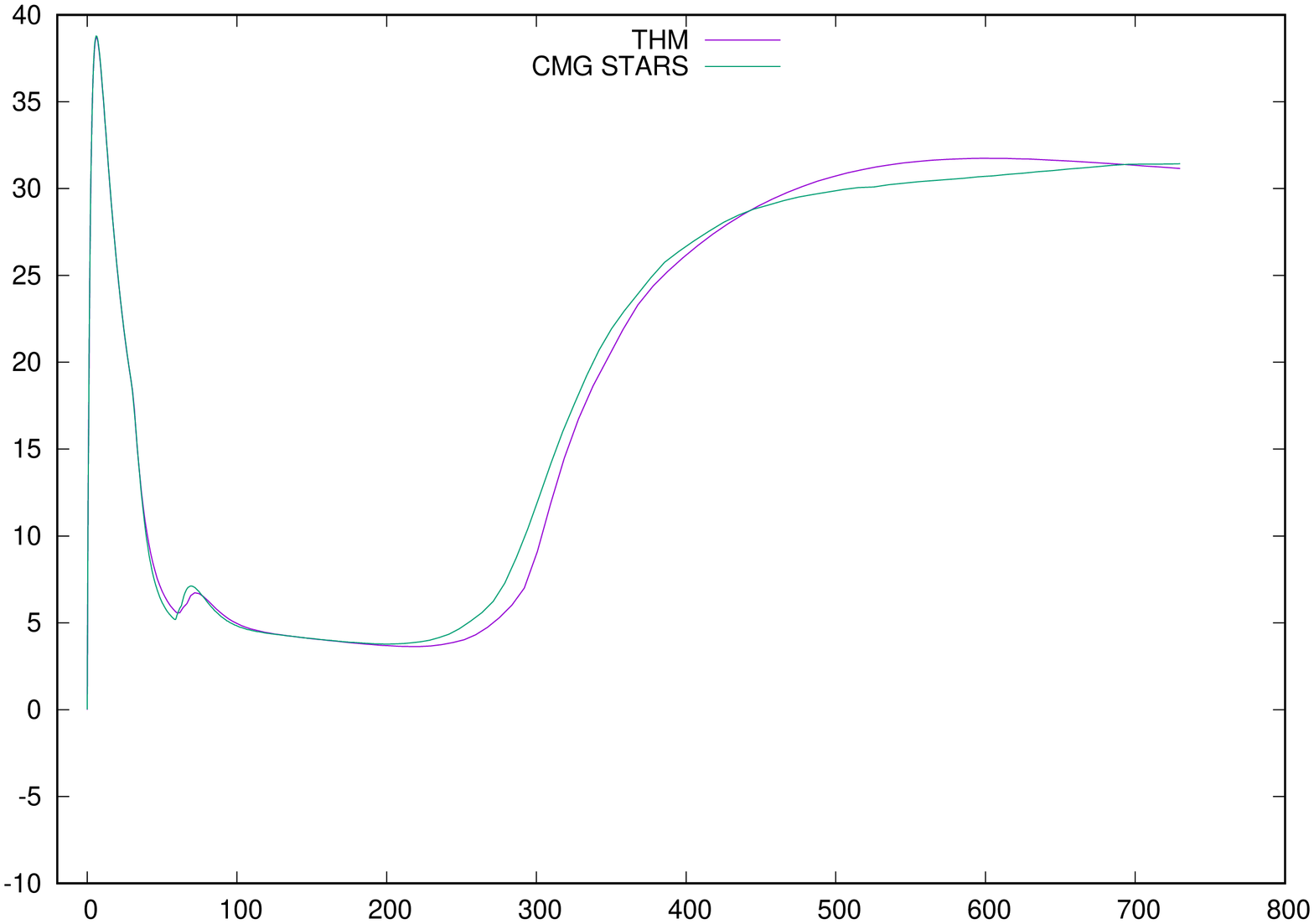}
    \caption{Example \ref{val-light}, light oil: water production rate (bbl/day), third production well}
    \label{fig-val-light-p3-pwr}
\end{figure}

\begin{figure}[H]
    \centering
    \includegraphics[width=0.53\linewidth, angle=270]{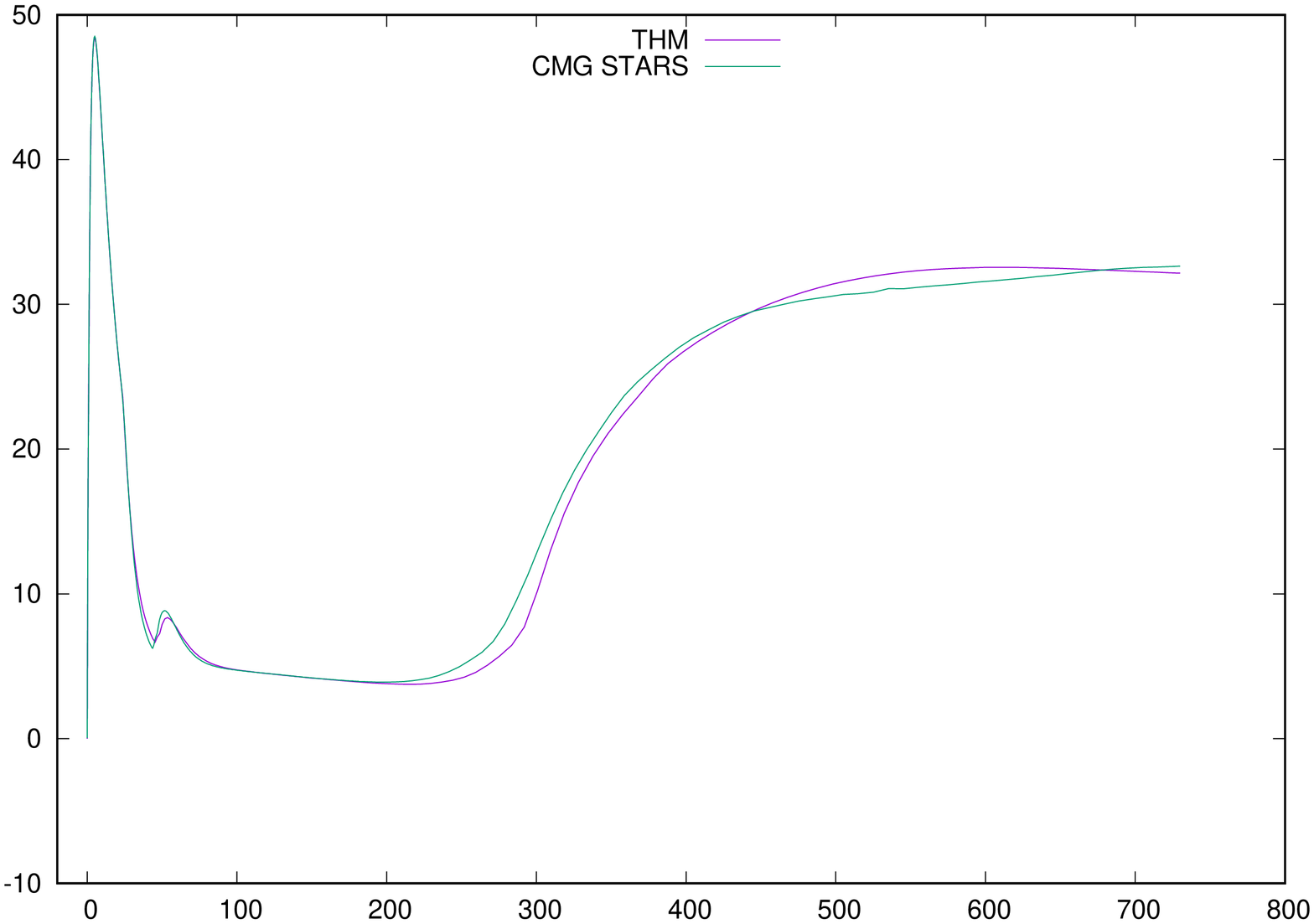}
    \caption{Example \ref{val-light}, light oil: water production rate (bbl/day), forth production well}
    \label{fig-val-light-p4-pwr}
\end{figure}

\begin{figure}[H]
    \centering
    \includegraphics[width=0.53\linewidth, angle=270]{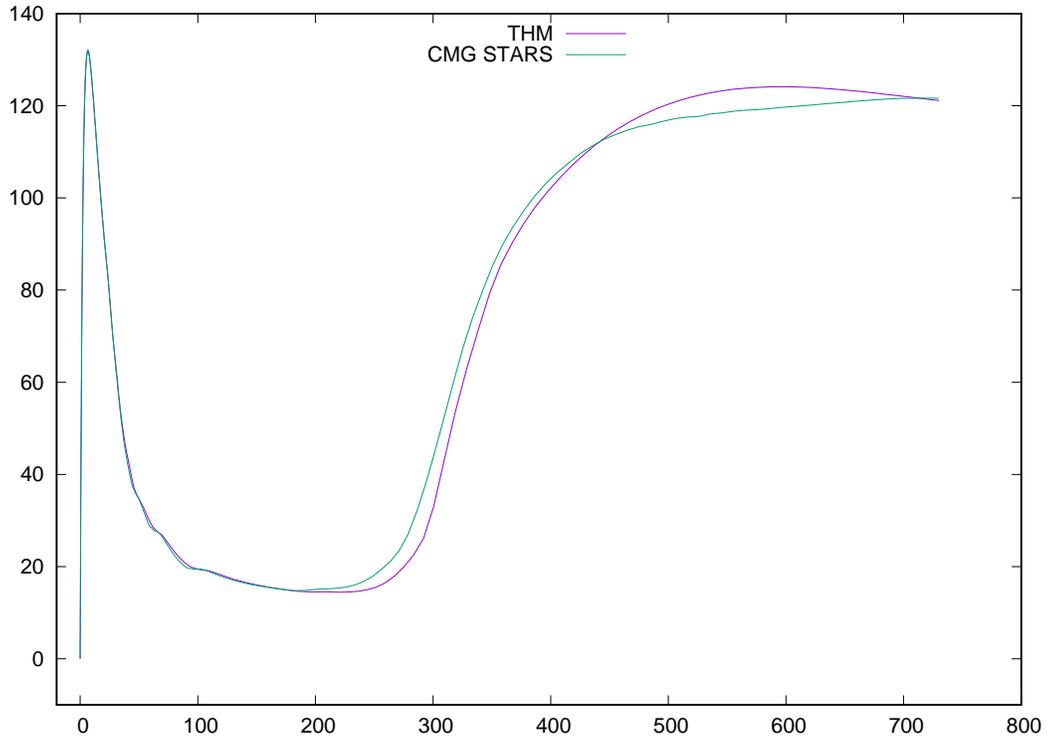}
    \caption{Example \ref{val-light}, light oil: total water production rate (bbl/day)}
    \label{fig-val-light-pwr}
\end{figure}

\begin{figure}[H]
    \centering
    \includegraphics[width=0.53\linewidth, angle=270]{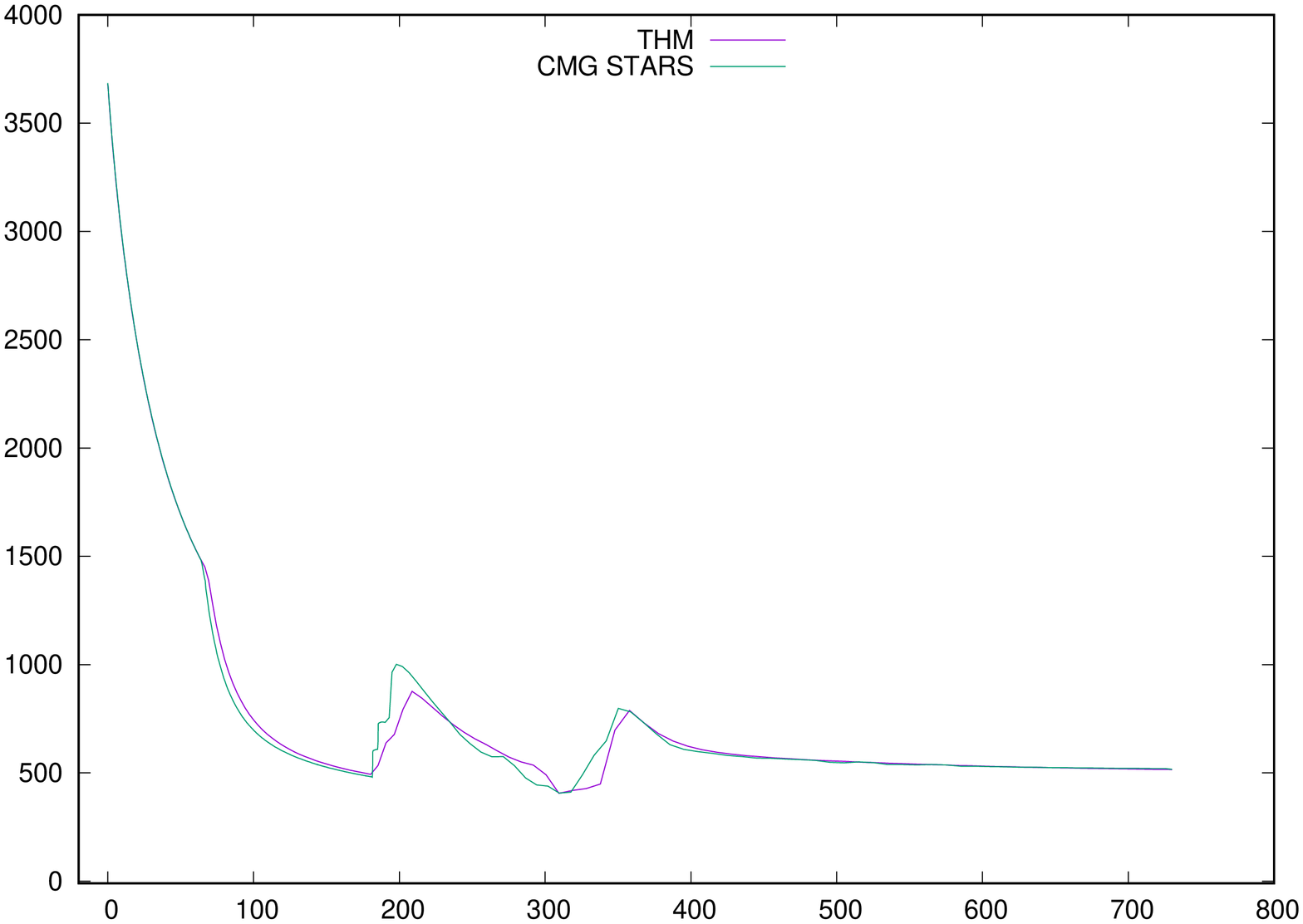}
    \caption{Example \ref{val-light}, light oil: gas production rate ($ft^3$/day), first production well}
    \label{fig-val-light-p1-pgr}
\end{figure}

\begin{figure}[H]
    \centering
    \includegraphics[width=0.53\linewidth, angle=270]{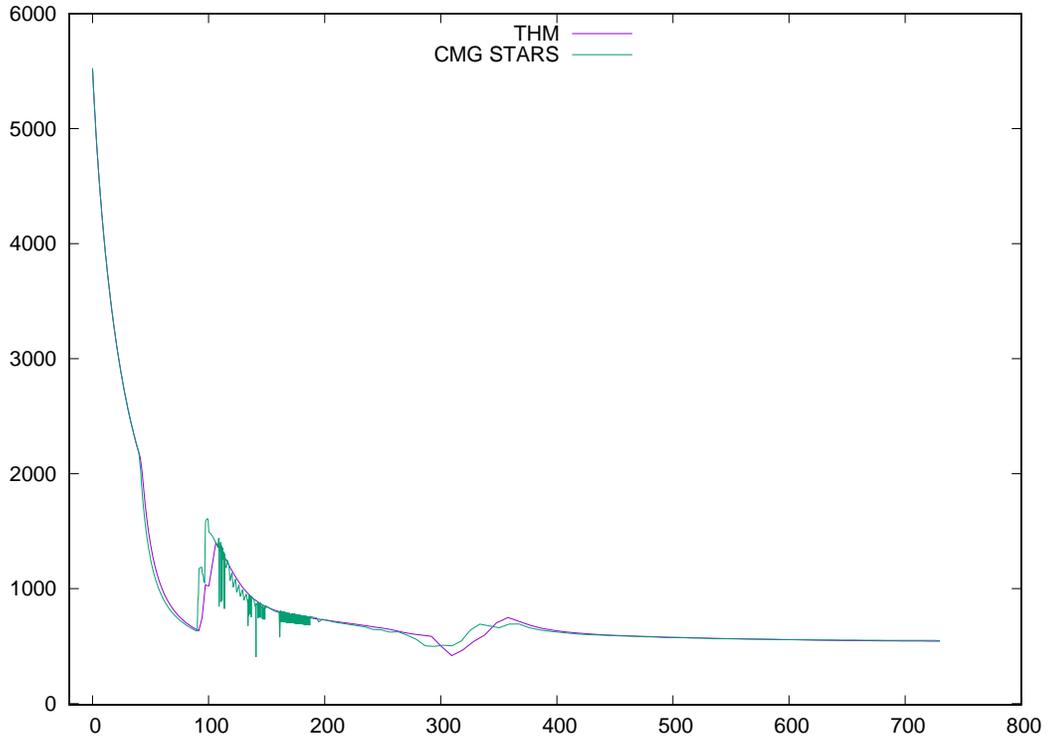}
    \caption{Example \ref{val-light}, light oil: gas production rate ($ft^3$/day), second production well}
    \label{fig-val-light-p2-pgr}
\end{figure}

\begin{figure}[H]
    \centering
    \includegraphics[width=0.53\linewidth, angle=270]{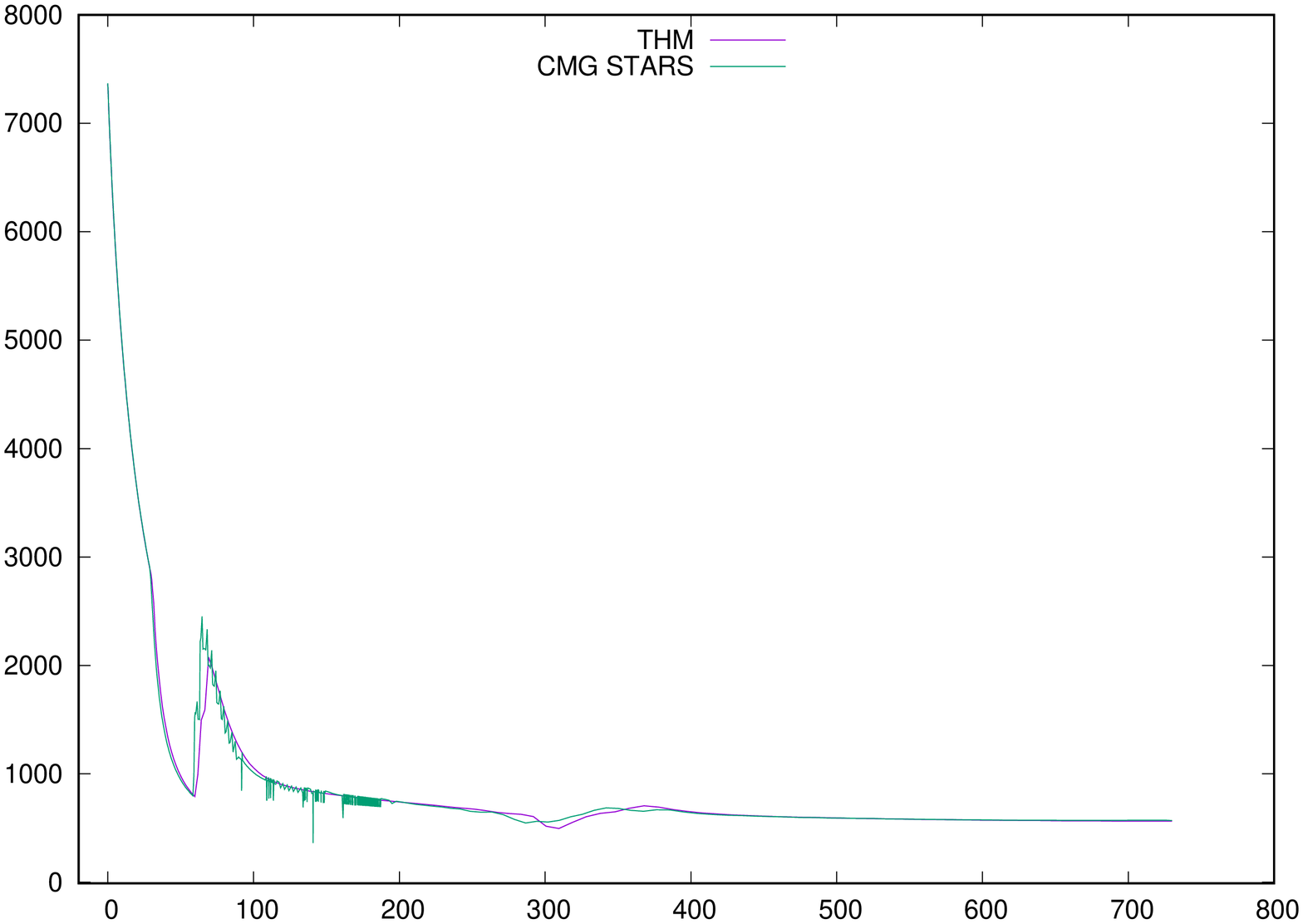}
    \caption{Example \ref{val-light}, light oil: gas production rate ($ft^3$/day), third production well}
    \label{fig-val-light-p3-pgr}
\end{figure}

\begin{figure}[H]
    \centering
    \includegraphics[width=0.53\linewidth, angle=270]{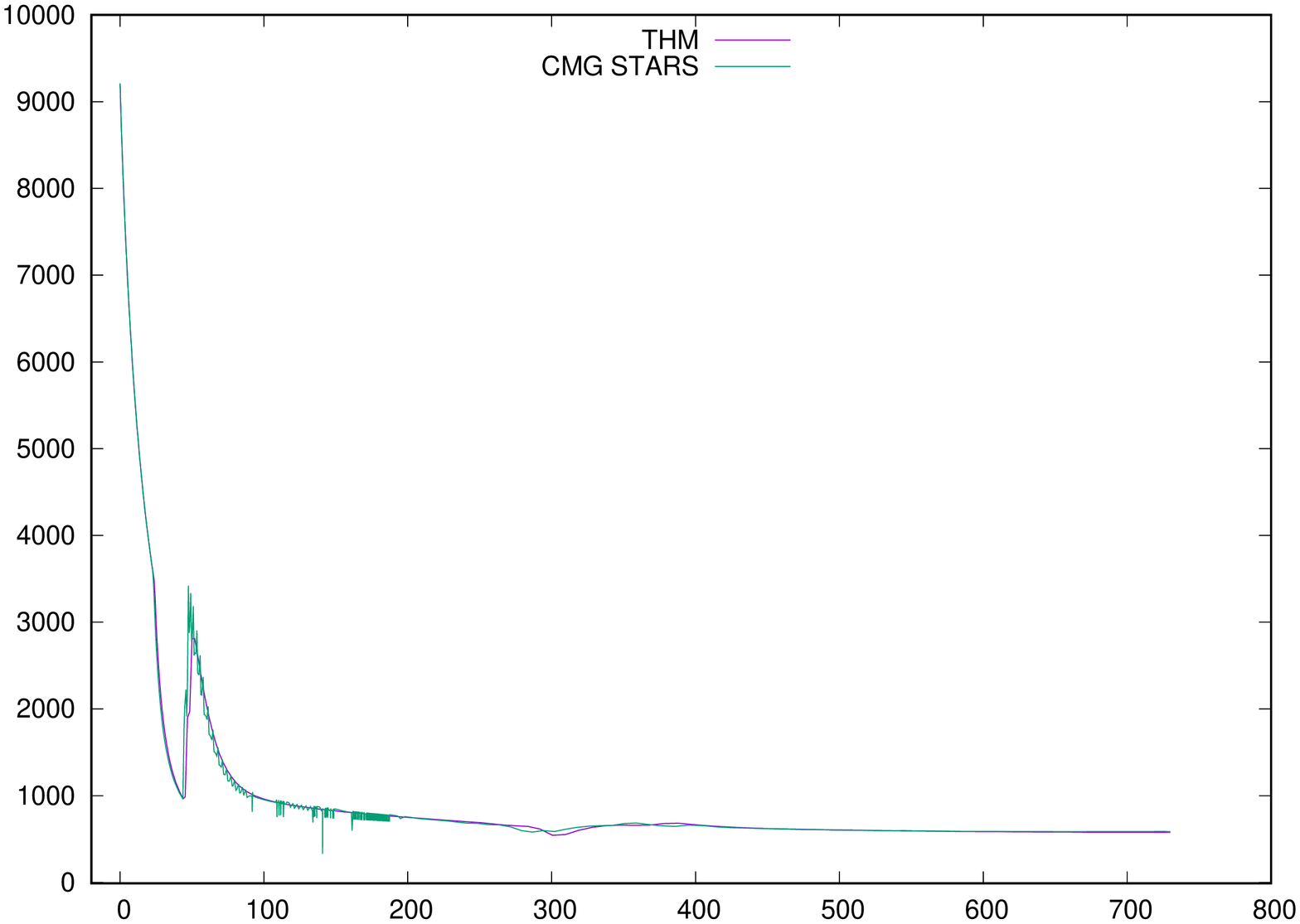}
    \caption{Example \ref{val-light}, light oil: gas production rate ($ft^3$/day), forth production well}
    \label{fig-val-light-p4-pgr}
\end{figure}

\begin{figure}[H]
    \centering
    \includegraphics[width=0.53\linewidth, angle=270]{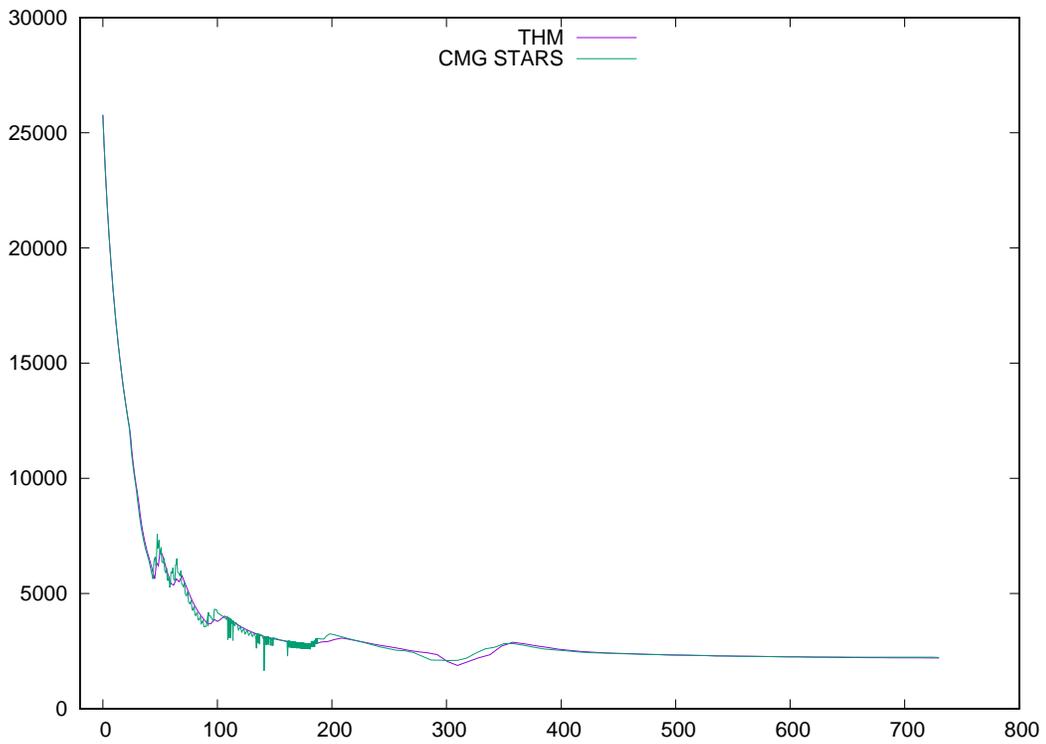}
    \caption{Example \ref{val-light}, light oil: total gas production rate ($ft^3$/day)}
    \label{fig-val-light-pgr}
\end{figure}

\begin{figure}[H]
    \centering
    \includegraphics[width=0.53\linewidth, angle=270]{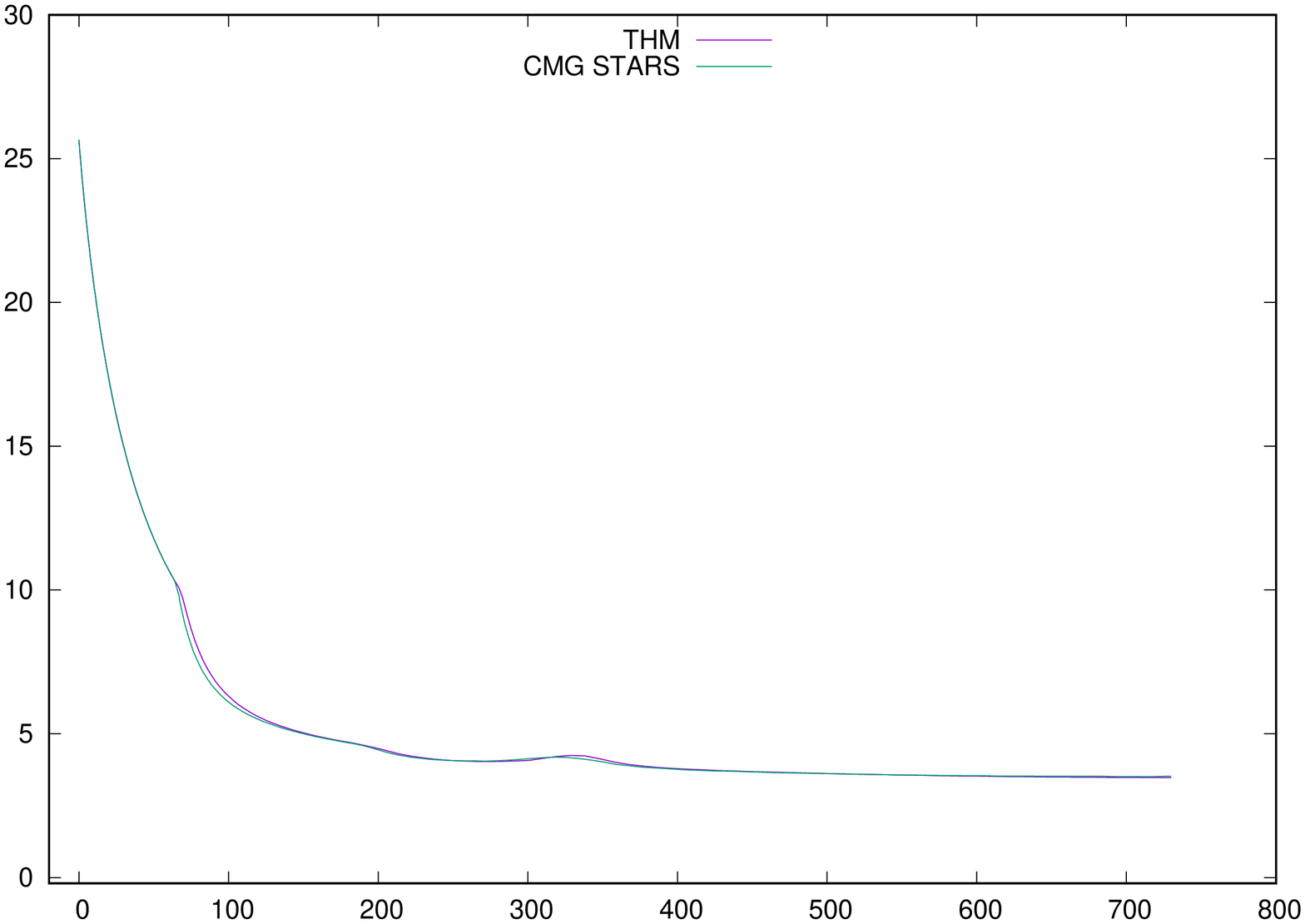}
    \caption{Example \ref{val-light}, light oil: oil production rate (bbl/day), first production well}
    \label{fig-val-light-p1-por}
\end{figure}

\begin{figure}[H]
    \centering
    \includegraphics[width=0.53\linewidth, angle=270]{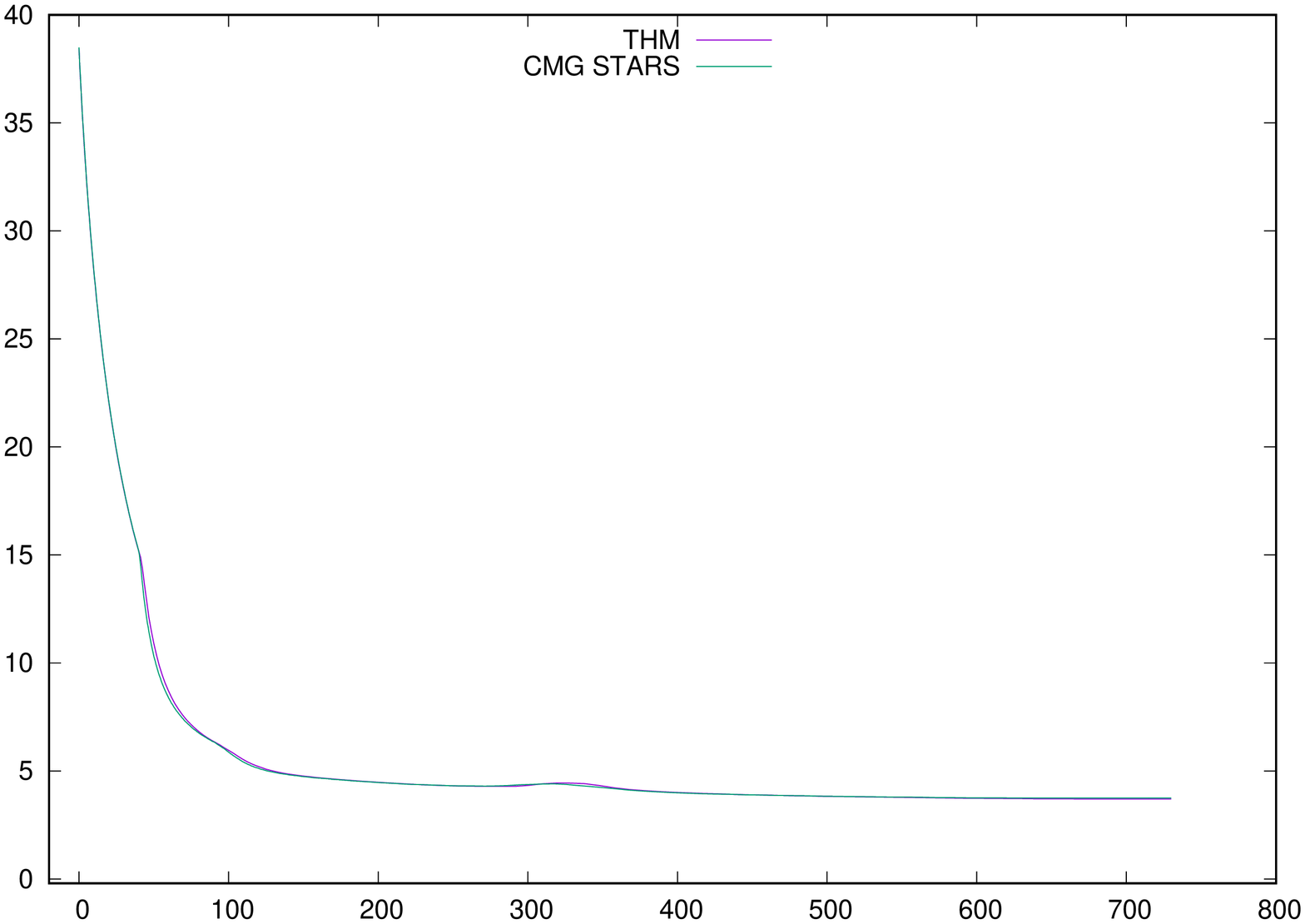}
    \caption{Example \ref{val-light}, light oil: oil production rate (bbl/day), second production well}
    \label{fig-val-light-p2-por}
\end{figure}

\begin{figure}[H]
    \centering
    \includegraphics[width=0.53\linewidth, angle=270]{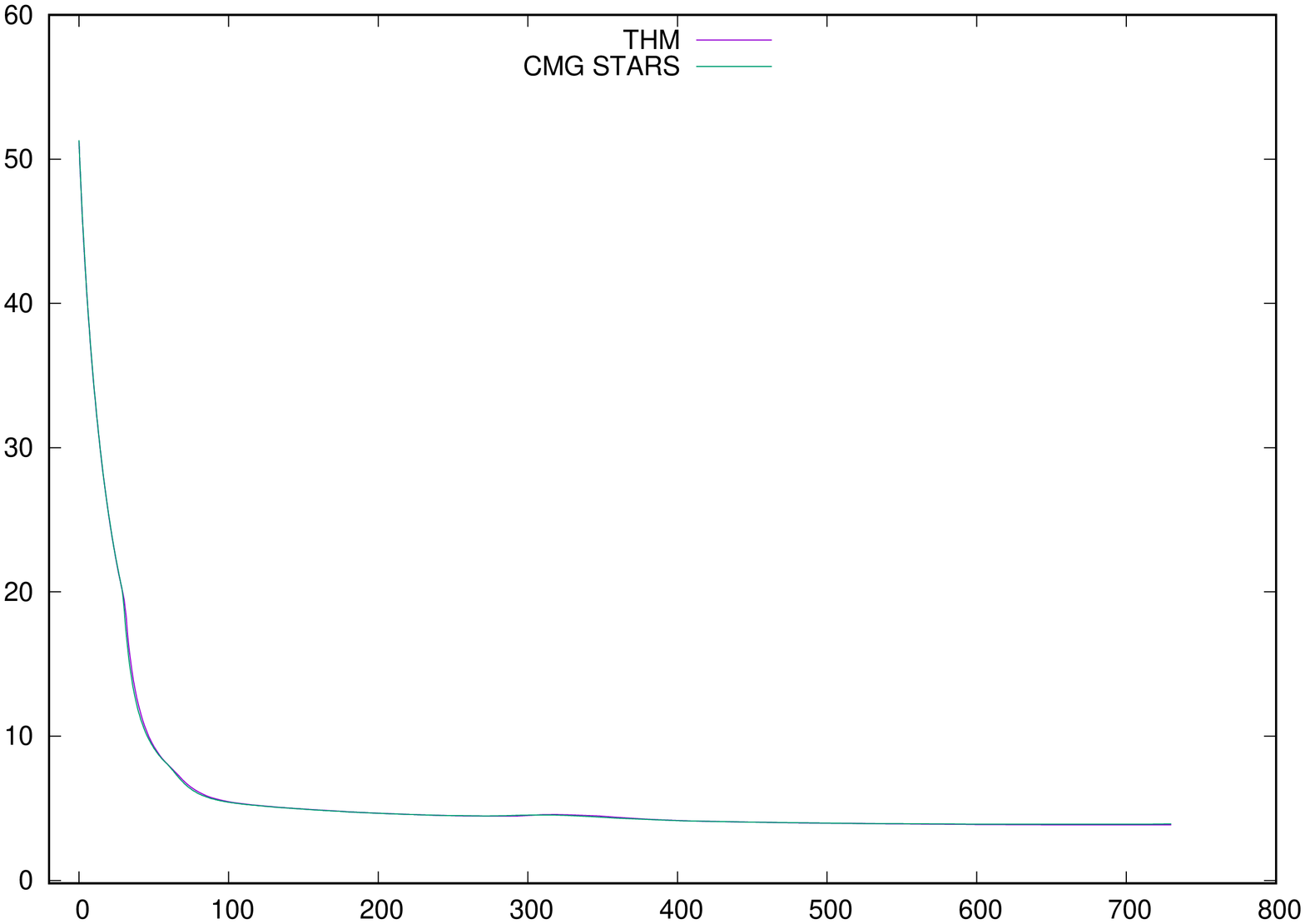}
    \caption{Example \ref{val-light}, light oil: oil production rate (bbl/day), third production well}
    \label{fig-val-light-p3-por}
\end{figure}

\begin{figure}[H]
    \centering
    \includegraphics[width=0.53\linewidth, angle=270]{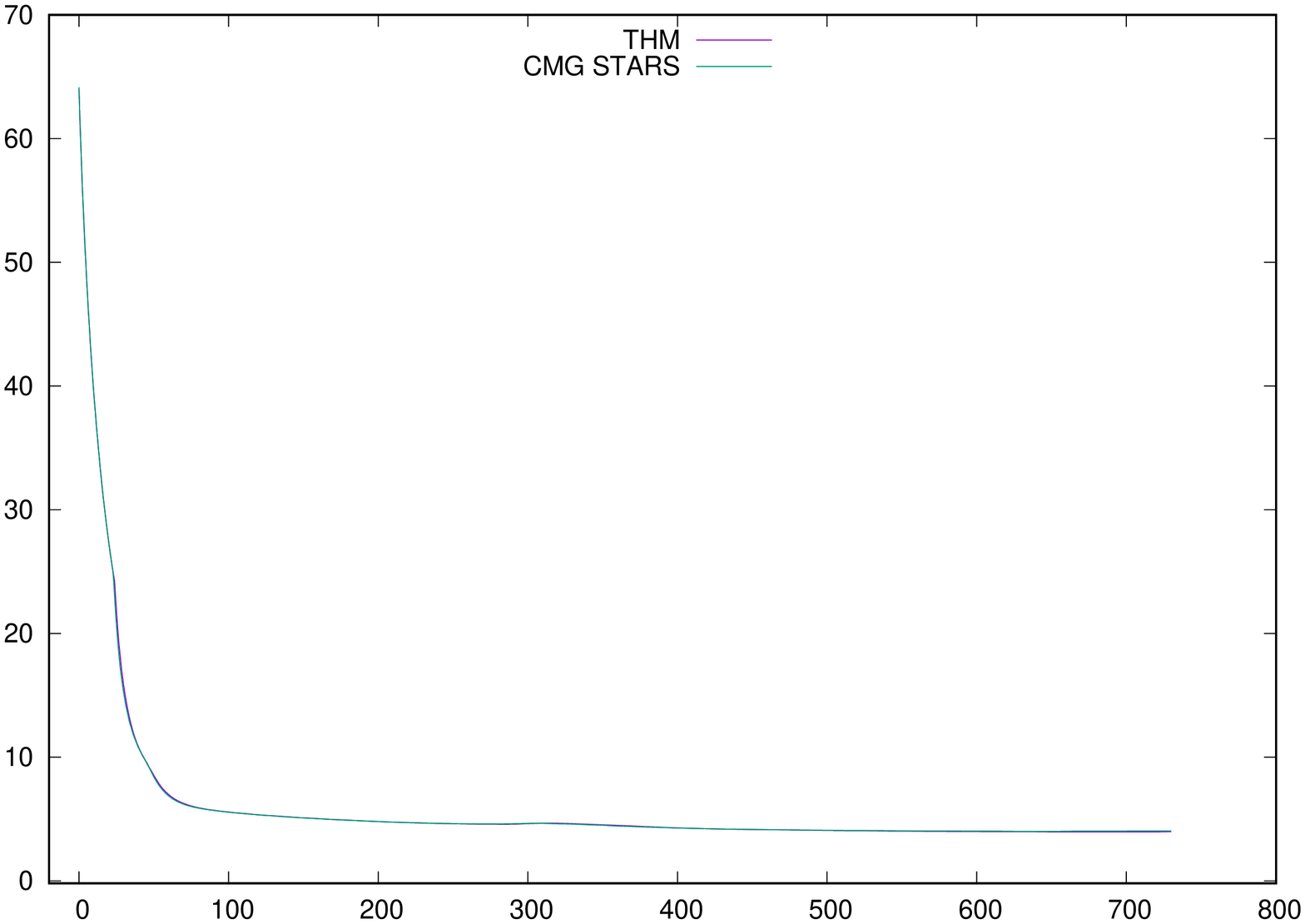}
    \caption{Example \ref{val-light}, light oil: oil production rate (bbl/day), forth production well}
    \label{fig-val-light-p4-por}
\end{figure}

\begin{figure}[H]
    \centering
    \includegraphics[width=0.53\linewidth, angle=270]{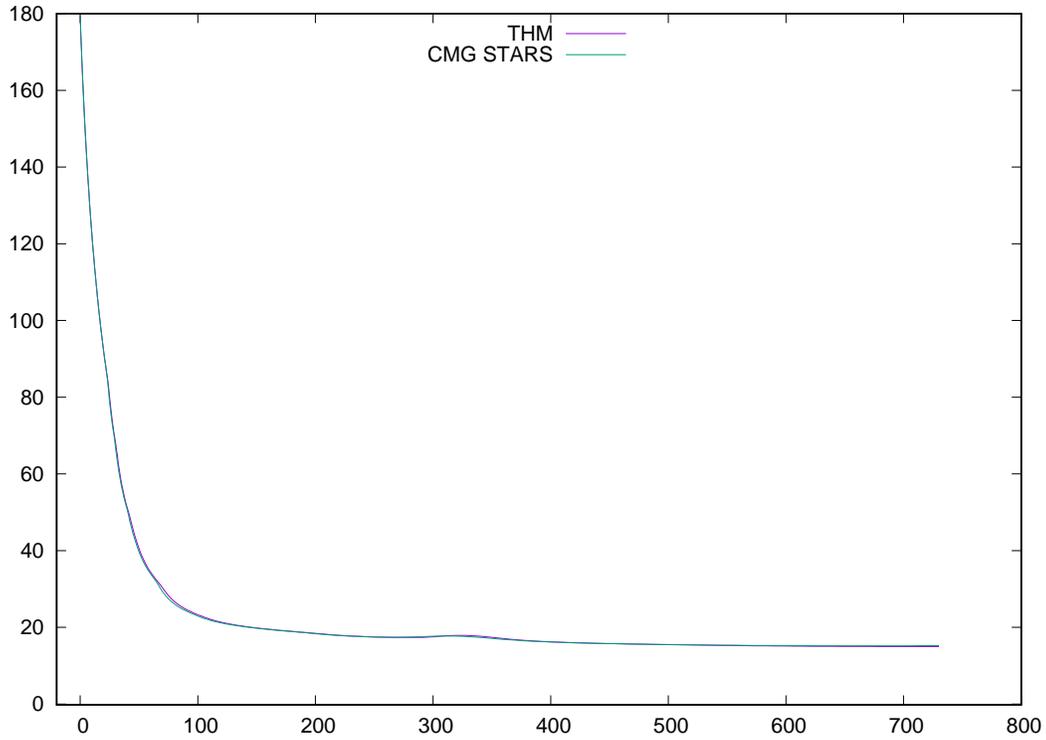}
    \caption{Example \ref{val-light}, light oil: total oil production rate (bbl/day)}
    \label{fig-val-light-por}
\end{figure}

\begin{example}
    \normalfont
    \label{val-light} This model is similar as Example \ref{val-heavy} except that a light oil component is
    added and the well  operations are changed.
    It has five vertical wells: one injection
    well in the center (5, 5), and four production wells in four corners, (1, 1), (1, 9), (9, 1) and (9, 9).
    The bottom hole pressure of the injection well, water rate and oil rate of each well are shown from Figure
    \ref{fig-val-light-inj-bhp} to Figure \ref{fig-val-light-por}. All results are compared with CMG STARS.
\end{example}

\begin{table}[!htb]
    \centering
    \begin{tabular}{l c}
        \hline
        \textbf{Initial condition} &  \\
        \hline
        $k_{x,y,z}\ (md)$  & 313, 424, 535 \\
        $\phi$ & 0.3 \\
        $\phi_c$ & 5e-4 \\
        $p\ (psi)$ & 4000 \\
        $T\ (^\circ F)$ & 125 \\
        $S_{w, o, g}$ & 0.45, 0.55, 0. \\
        $x$ & 0.6, 0.4 \\
        \hline
    \end{tabular}
    \caption{Input data for Example \ref{val-light}}
    \label{val-light-tab-input1}
\end{table}

\begin{table}[!htb]
    \centering
    \begin{tabular}{l c }
        \hline
        \textbf{Properties} & \textbf{LO} \\
        $M\ (lb/lbmole)$ & 250  \\
        $p_{crit}\ (psi) $ & 225   \\
        $T_{crit}\ ({}^\circ F)$ & 800  \\
        \hline

        $\rho_{ref} (lbmole/ft^3)$ & 0.2092 \\
        $cp\ (1/psi)$ & 5.e-6 \\
        $ct1\ (1/{^\circ F})$ & 3.8e-4\\
        \hline

        $cpg1\ (Btu/({^\circ F} \cdot lbmol))$ & 247.5  \\
        $hvr\ (Btu/({^\circ F}^{ev} \cdot lbmol))$ & 657   \\
        $ev$ & 0.38   \\
        \hline
        $avg\ (cp/{^\circ F})$ & 5.e-5  \\
        $bvg$ & 0.9 \\
        $avisc\ (cp)$ & 0.287352 \\
        $bvisc\ ({^\circ F})$ & 3728.2 \\
        \hline
        $kv1\ (psi)$ & 7.9114e4  \\
        $kv4\ ({^\circ F})$ & -1583.71  \\
        $kv5\ ({^\circ F})$ & -446.78 \\
        \hline
    \end{tabular}
    \caption{Input data for Example \ref{val-light}}
    \label{val-light-tab-input2}
\end{table}

\begin{table}[!htb]
    \centering
    \begin{tabular}{c c c}
        \hline
        \textbf{Well conditions} & & \\\hline
        Injector & water $(bbl/day)$ & 100  \\
        & wi $(ft \cdot md)$ & 1e4 \\
        & tinjw $(^\circ F)$ & 450 \\
        & steam quality & 0.3 \\
        \hline

        Producer 1 & bhp $(psi)$ & 17 \\
        & wi $(ft \cdot md)$ & 2e4 \\
        \hline

        Producer 2 & bhp $(psi)$ & 17 \\
        & wi $(ft \cdot md)$ & 3e4 \\
        \hline

        Producer 3 & bhp $(psi)$ & 17 \\
        & wi $(ft \cdot md)$ & 4e4 \\
        \hline

        Producer 4 & bhp $(psi)$ & 17 \\
        & wi $(ft \cdot md)$ & 5e4 \\
        \hline
    \end{tabular}
    \caption{Input data for Example \ref{val-light} (cont'd).}
    \label{val-light-tab-input3}
\end{table}

Figure \ref{fig-val-light-inj-bhp} is the bottom hole pressure of the injection well.
Figure \ref{fig-val-light-p1-pwr} is the water production rate of the first production well.
Figure \ref{fig-val-light-p2-pwr} is the water production rate of the second production well.
Figure \ref{fig-val-light-p3-pwr} is the water production rate of the third production well.
Figure \ref{fig-val-light-p4-pwr} is the water production rate of the forth production well.
Figure \ref{fig-val-light-pwr} is the total water production rate of all production wells.
Figure \ref{fig-val-light-p1-pgr} is the gas production rate of the first production well.
Figure \ref{fig-val-light-p2-pgr} is the gas production rate of the second production well.
Figure \ref{fig-val-light-p3-pgr} is the gas production rate of the third production well.
Figure \ref{fig-val-light-p4-pgr} is the gas production rate of the forth production well.
Figure \ref{fig-val-light-pgr} is the total gas production rate of all production wells.
Figure \ref{fig-val-light-p1-por} is the oil production rate of the first production well.
Figure \ref{fig-val-light-p2-por} is the oil production rate of the second production well.
Figure \ref{fig-val-light-p3-por} is the oil production rate of the third production well.
Figure \ref{fig-val-light-p4-por} is the oil production rate of the forth production well.
Figure \ref{fig-val-light-por} is the total oil production rate of all production wells.
All figures show that our results match CMG STARS very well, which confirms our methods and implementation are
correct.

%%%%%%%%%%%%%%%%%%%%%%%%%%%%%%%%%%%%%%%%%%%%%%%%%%%%%%%%%%%%%%%%%%%%%%%%%%%%%%%%%%%%%%%%%%%%%%%%%%%%%%%%%%%%%%
\subsubsection{Non-condensable Gas}

\begin{figure}[H]
    \centering
    \includegraphics[width=0.53\linewidth, angle=270]{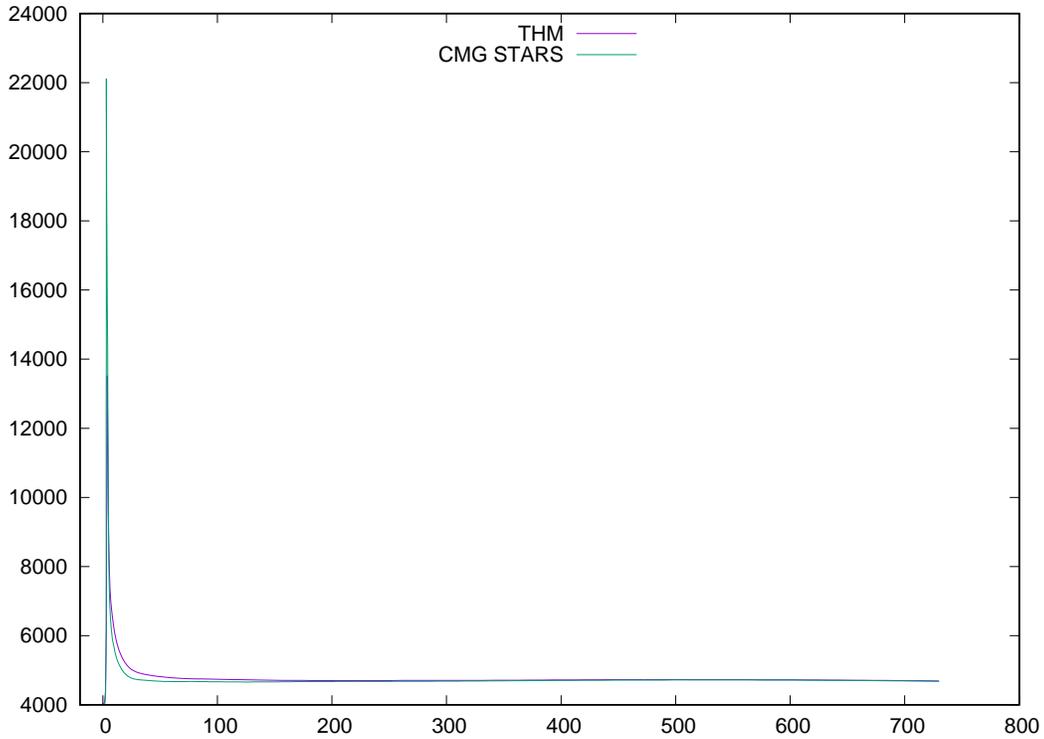}
    \caption{Example \ref{val-gas}, NCG: injection well, bottom hole pressure (psi)}
    \label{fig-val-gas-inj-bhp}
\end{figure}

\begin{figure}[H]
    \centering
    \includegraphics[width=0.53\linewidth, angle=270]{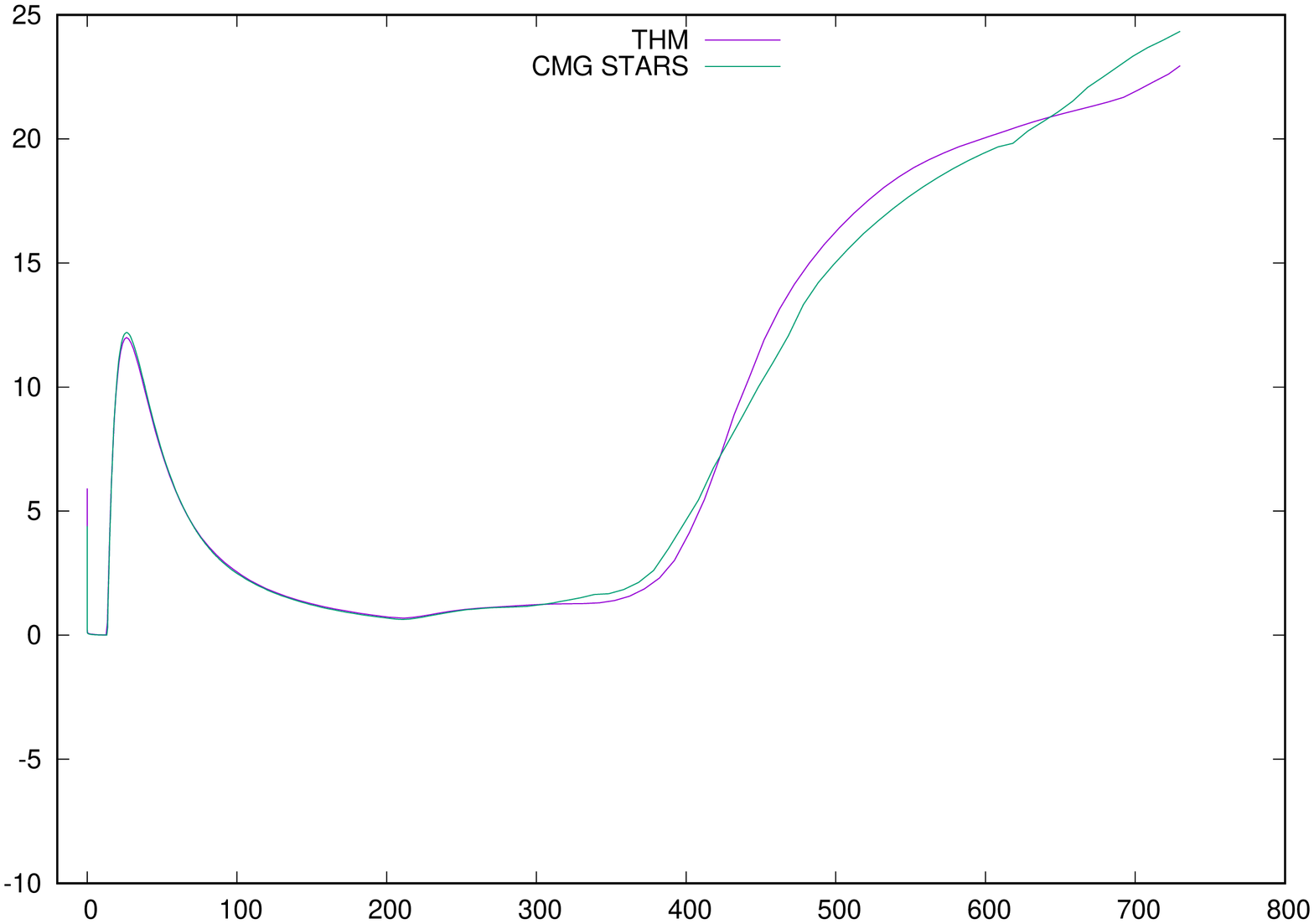}
    \caption{Example \ref{val-gas}, NCG: water production rate (bbl/day), first production well}
    \label{fig-val-gas-p1-pwr}
\end{figure}

\begin{figure}[H]
    \centering
    \includegraphics[width=0.53\linewidth, angle=270]{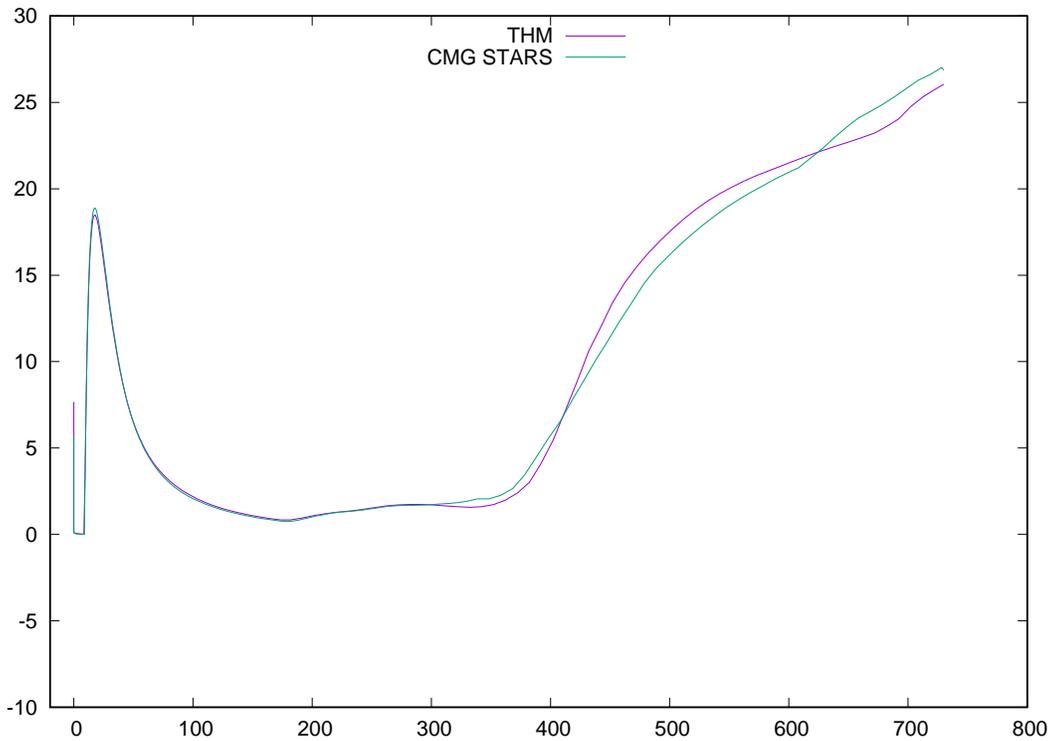}
    \caption{Example \ref{val-gas}, NCG: water production rate (bbl/day), second production well}
    \label{fig-val-gas-p2-pwr}
\end{figure}

\begin{figure}[H]
    \centering
    \includegraphics[width=0.53\linewidth, angle=270]{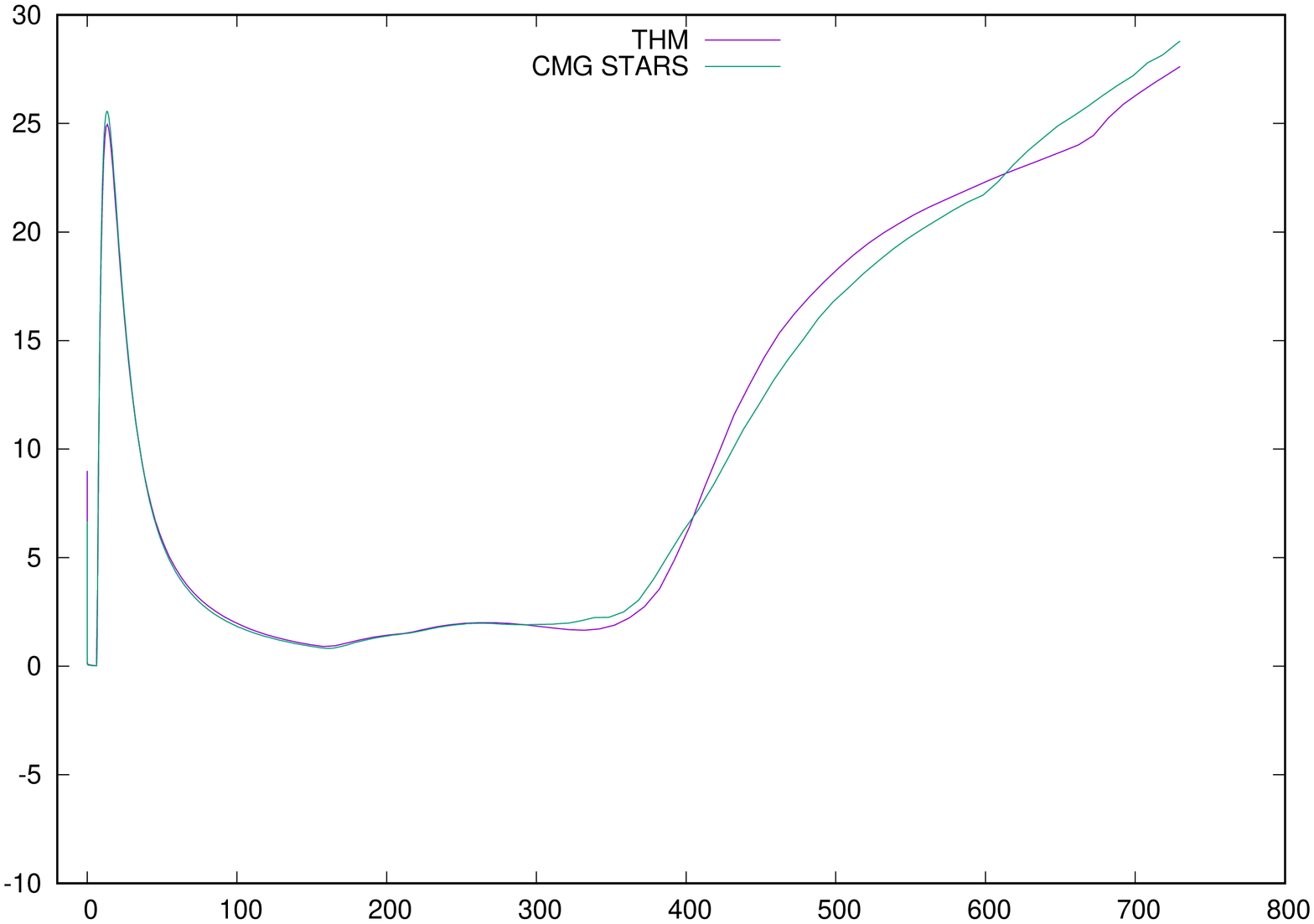}
    \caption{Example \ref{val-gas}, NCG: water production rate (bbl/day), third production well}
    \label{fig-val-gas-p3-pwr}
\end{figure}

\begin{figure}[H]
    \centering
    \includegraphics[width=0.53\linewidth, angle=270]{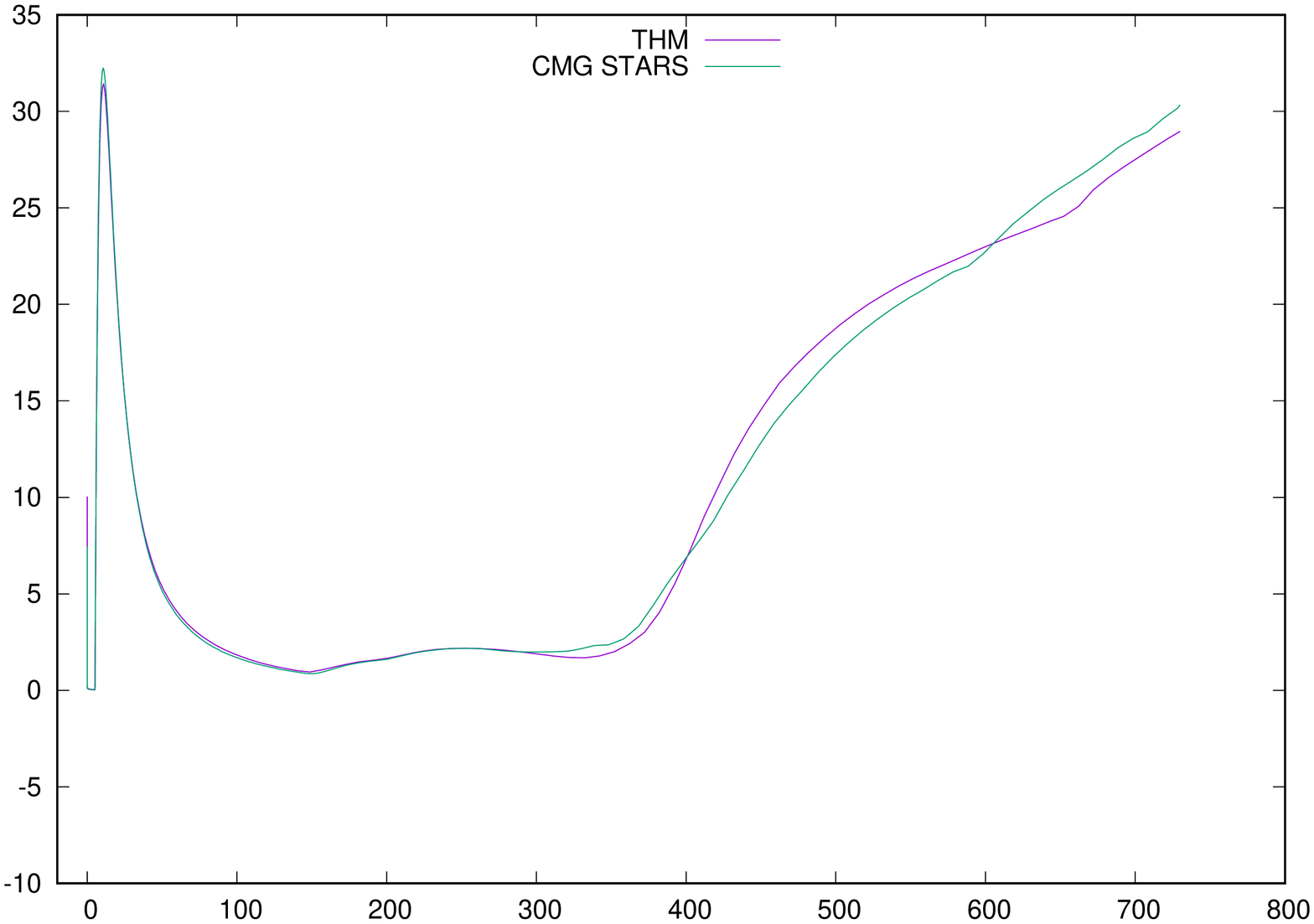}
    \caption{Example \ref{val-gas}, NCG: water production rate (bbl/day), forth production well}
    \label{fig-val-gas-p4-pwr}
\end{figure}

\begin{figure}[H]
    \centering
    \includegraphics[width=0.53\linewidth, angle=270]{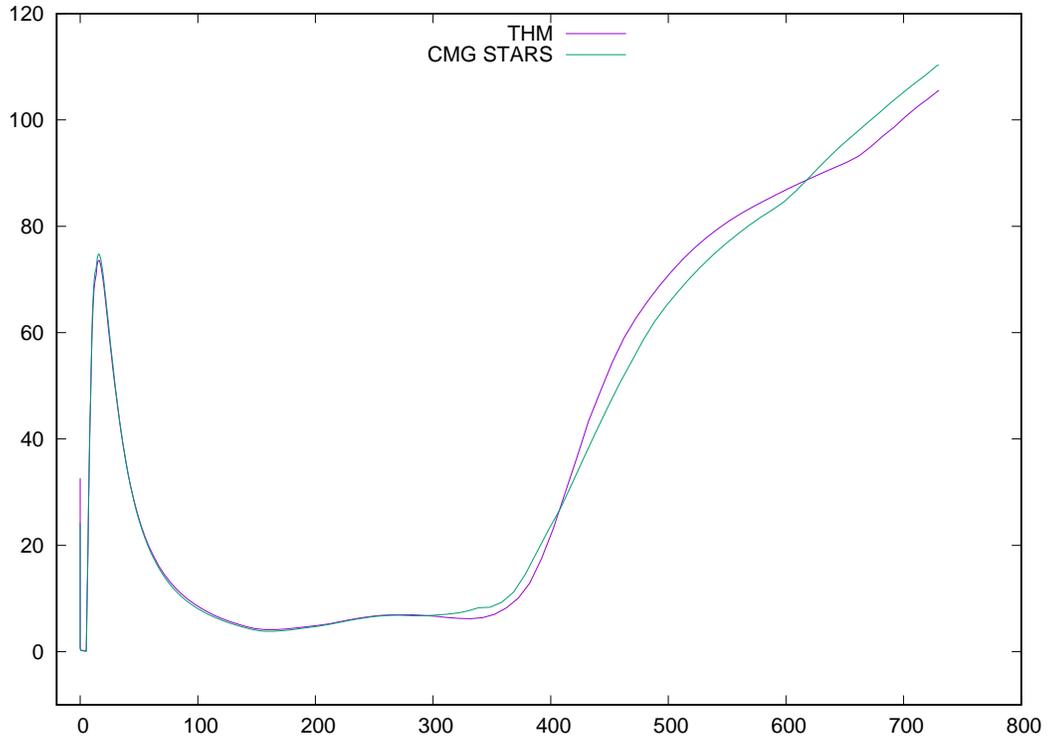}
    \caption{Example \ref{val-gas}, NCG: total water production rate (bbl/day)}
    \label{fig-val-gas-pwr}
\end{figure}

\begin{figure}[H]
    \centering
    \includegraphics[width=0.53\linewidth, angle=270]{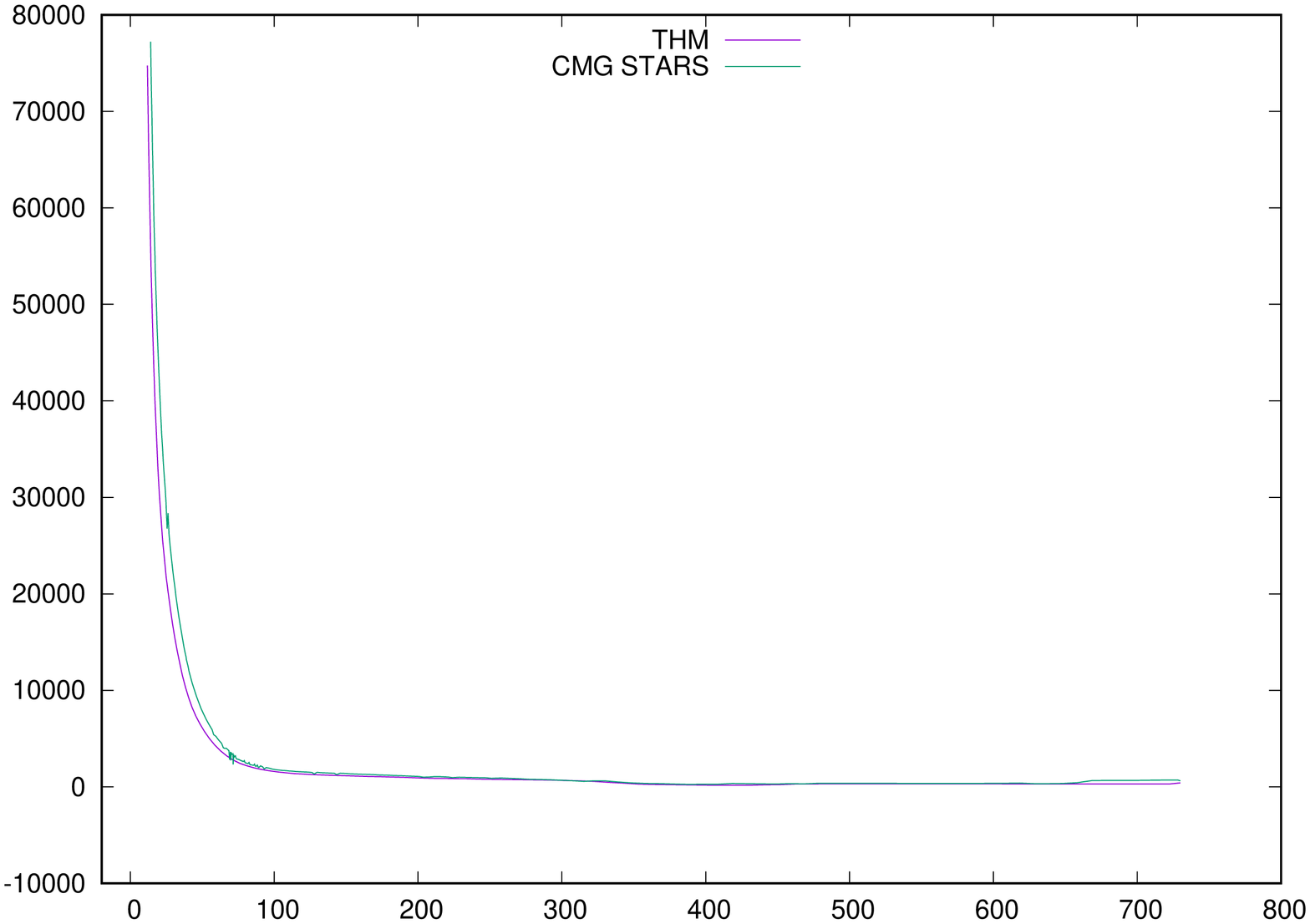}
    \caption{Example \ref{val-gas}, NCG: gas production rate ($ft^3$/day), first production well}
    \label{fig-val-gas-p1-pgr}
\end{figure}

\begin{figure}[H]
    \centering
    \includegraphics[width=0.53\linewidth, angle=270]{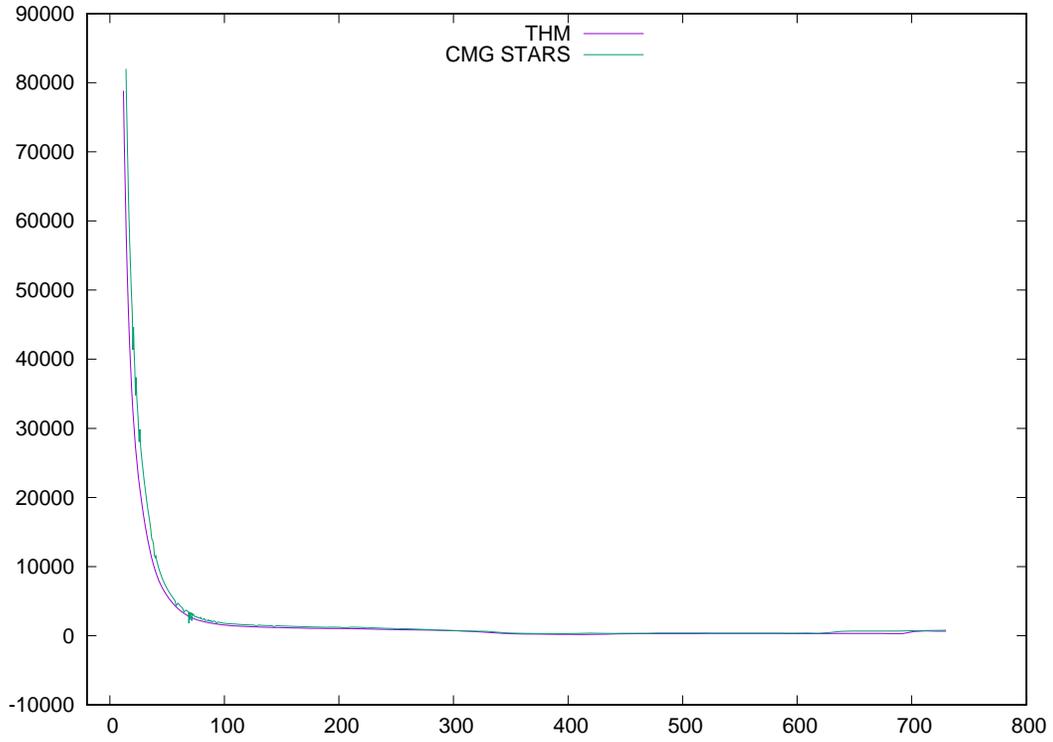}
    \caption{Example \ref{val-gas}, NCG: gas production rate ($ft^3$/day), second production well}
    \label{fig-val-gas-p2-pgr}
\end{figure}

\begin{figure}[H]
    \centering
    \includegraphics[width=0.53\linewidth, angle=270]{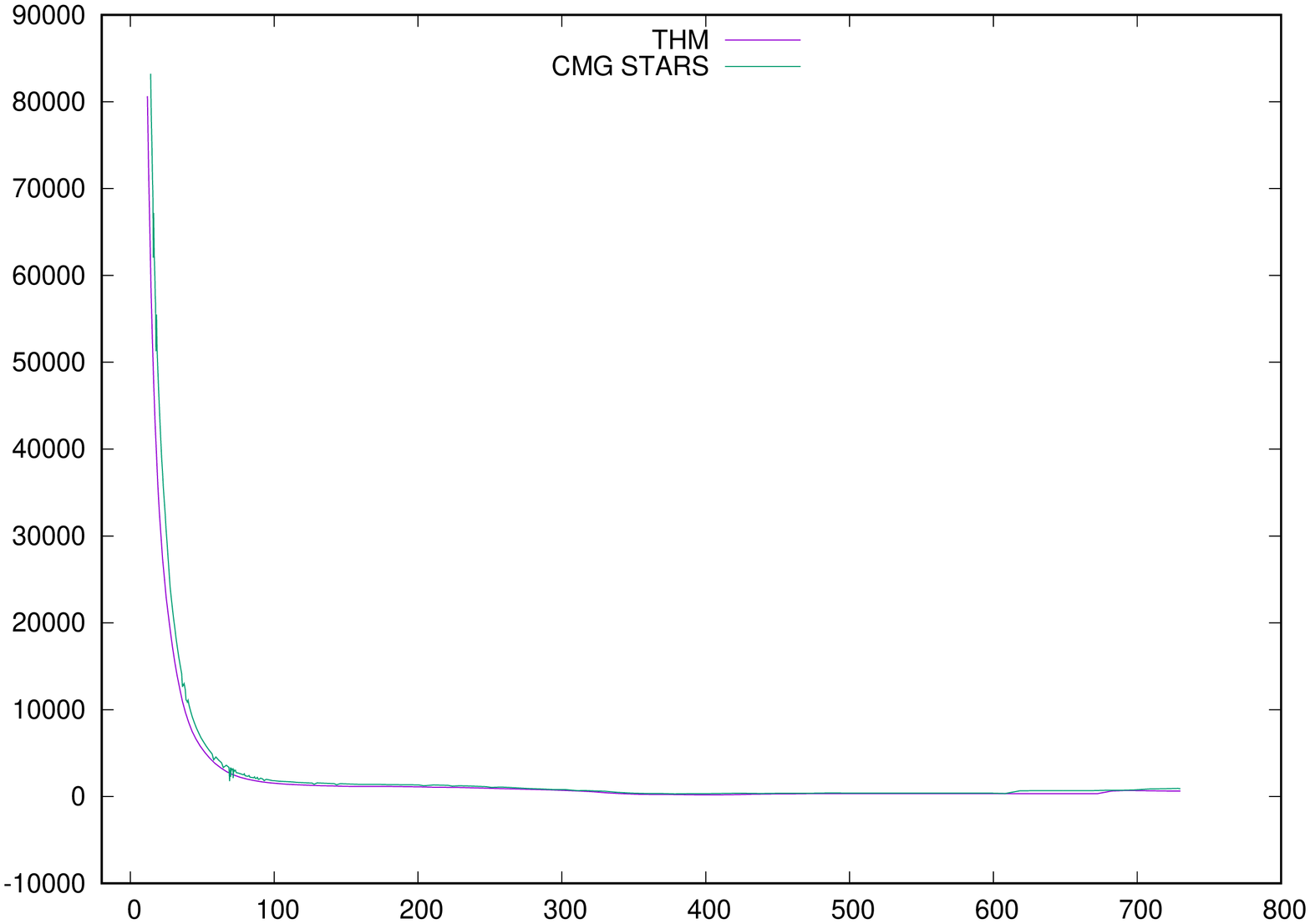}
    \caption{Example \ref{val-gas}, NCG: gas production rate ($ft^3$/day), third production well}
    \label{fig-val-gas-p3-pgr}
\end{figure}

\begin{figure}[H]
    \centering
    \includegraphics[width=0.53\linewidth, angle=270]{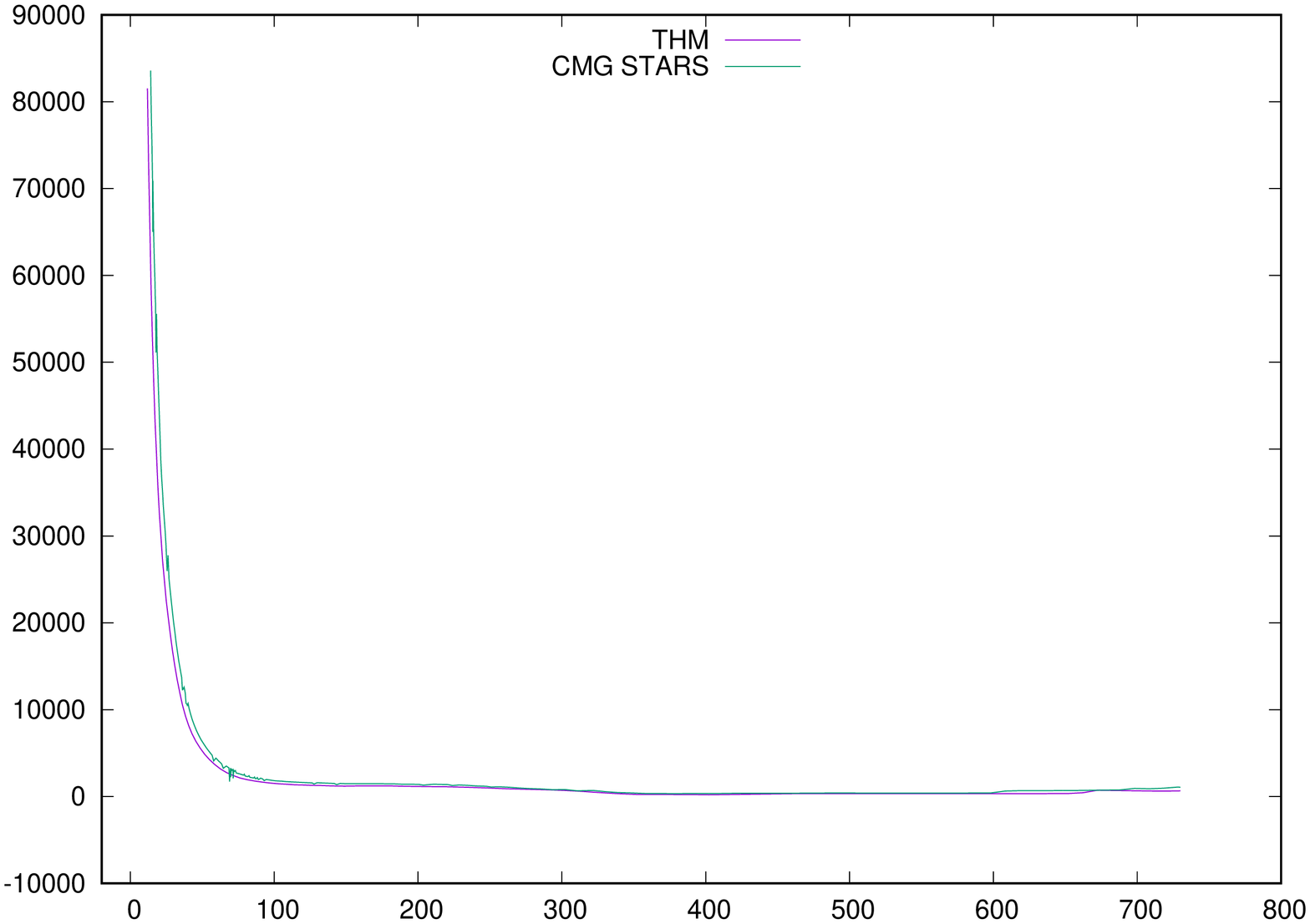}
    \caption{Example \ref{val-gas}, NCG: gas production rate ($ft^3$/day), forth production well}
    \label{fig-val-gas-p4-pgr}
\end{figure}

\begin{figure}[H]
    \centering
    \includegraphics[width=0.53\linewidth, angle=270]{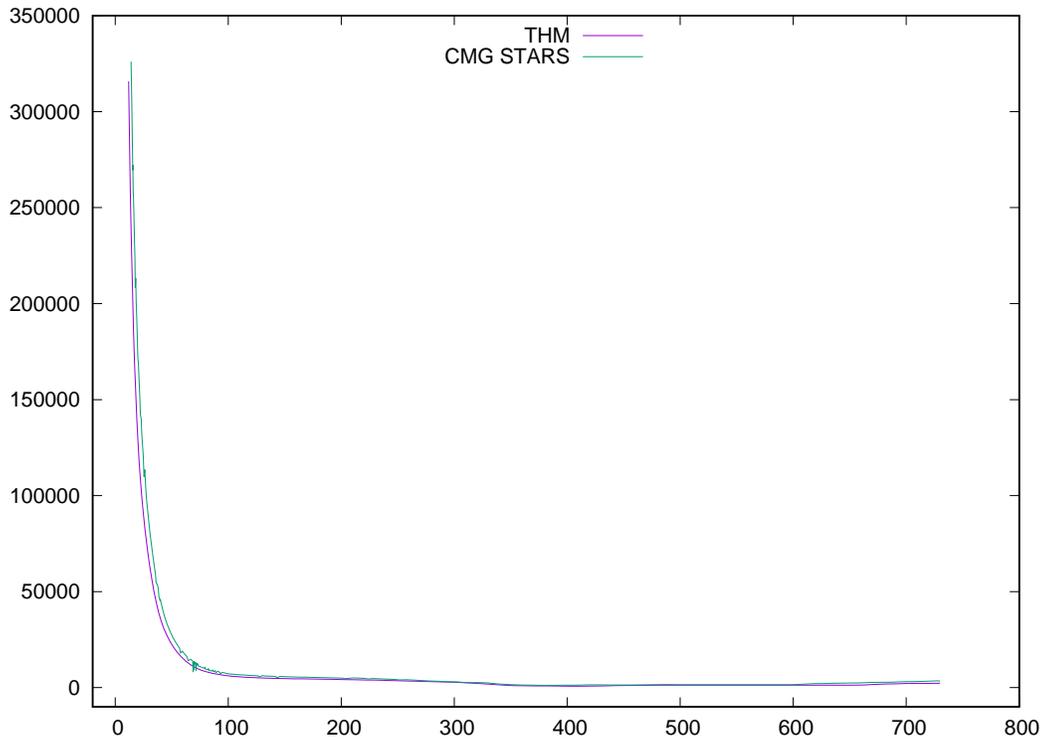}
    \caption{Example \ref{val-gas}, NCG: total gas production rate ($ft^3$/day)}
    \label{fig-val-gas-pgr}
\end{figure}

\begin{figure}[H]
    \centering
    \includegraphics[width=0.53\linewidth, angle=270]{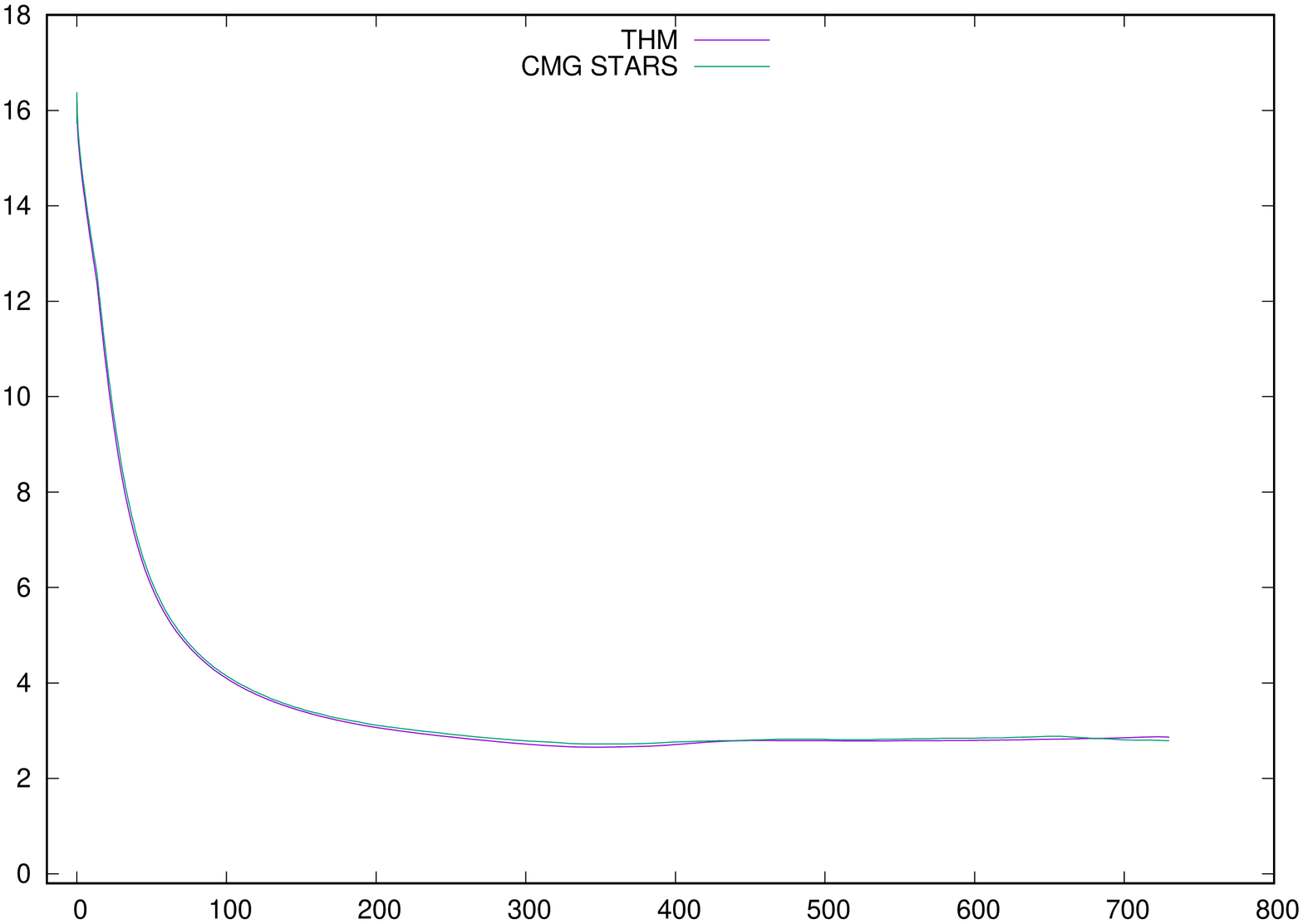}
    \caption{Example \ref{val-gas}, NCG: oil production rate (bbl/day), first production well}
    \label{fig-val-gas-p1-por}
\end{figure}

\begin{figure}[H]
    \centering
    \includegraphics[width=0.53\linewidth, angle=270]{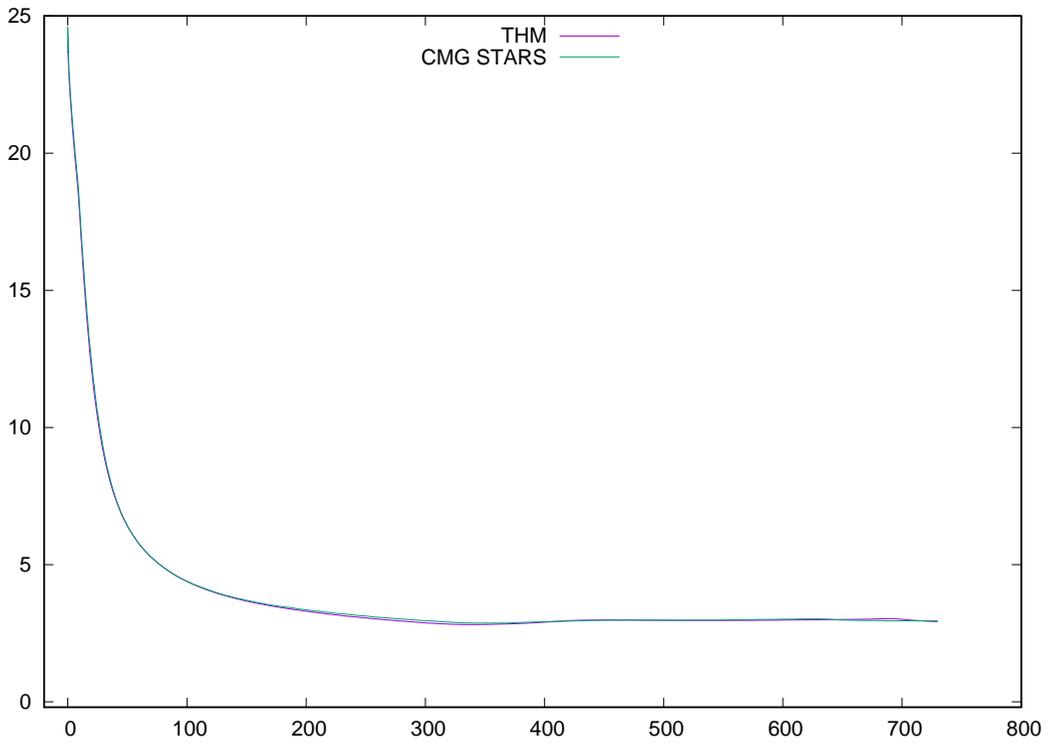}
    \caption{Example \ref{val-gas}, NCG: oil production rate (bbl/day), second production well}
    \label{fig-val-gas-p2-por}
\end{figure}

\begin{figure}[H]
    \centering
    \includegraphics[width=0.53\linewidth, angle=270]{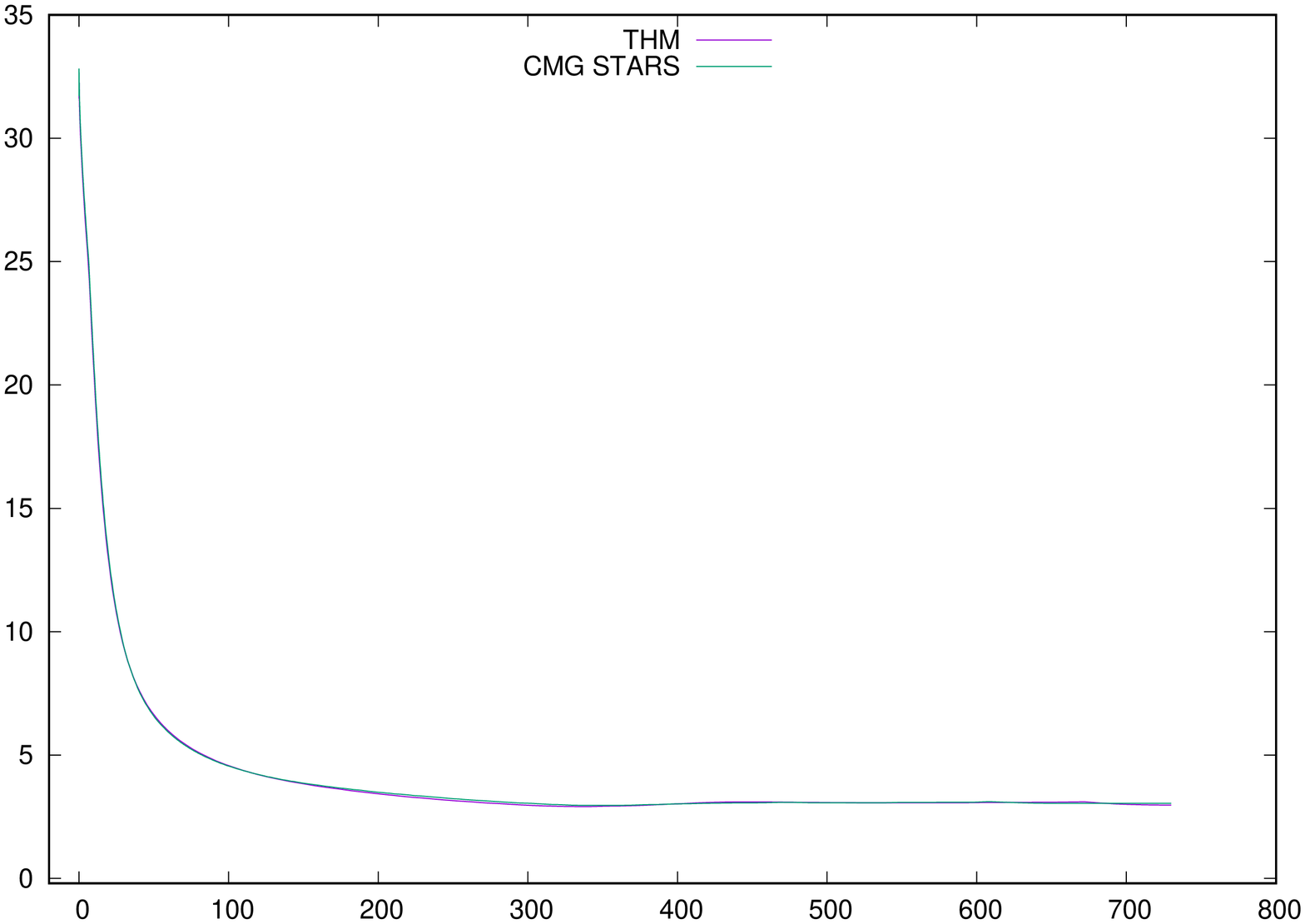}
    \caption{Example \ref{val-gas}, NCG: oil production rate (bbl/day), third production well}
    \label{fig-val-gas-p3-por}
\end{figure}

\begin{figure}[H]
    \centering
    \includegraphics[width=0.53\linewidth, angle=270]{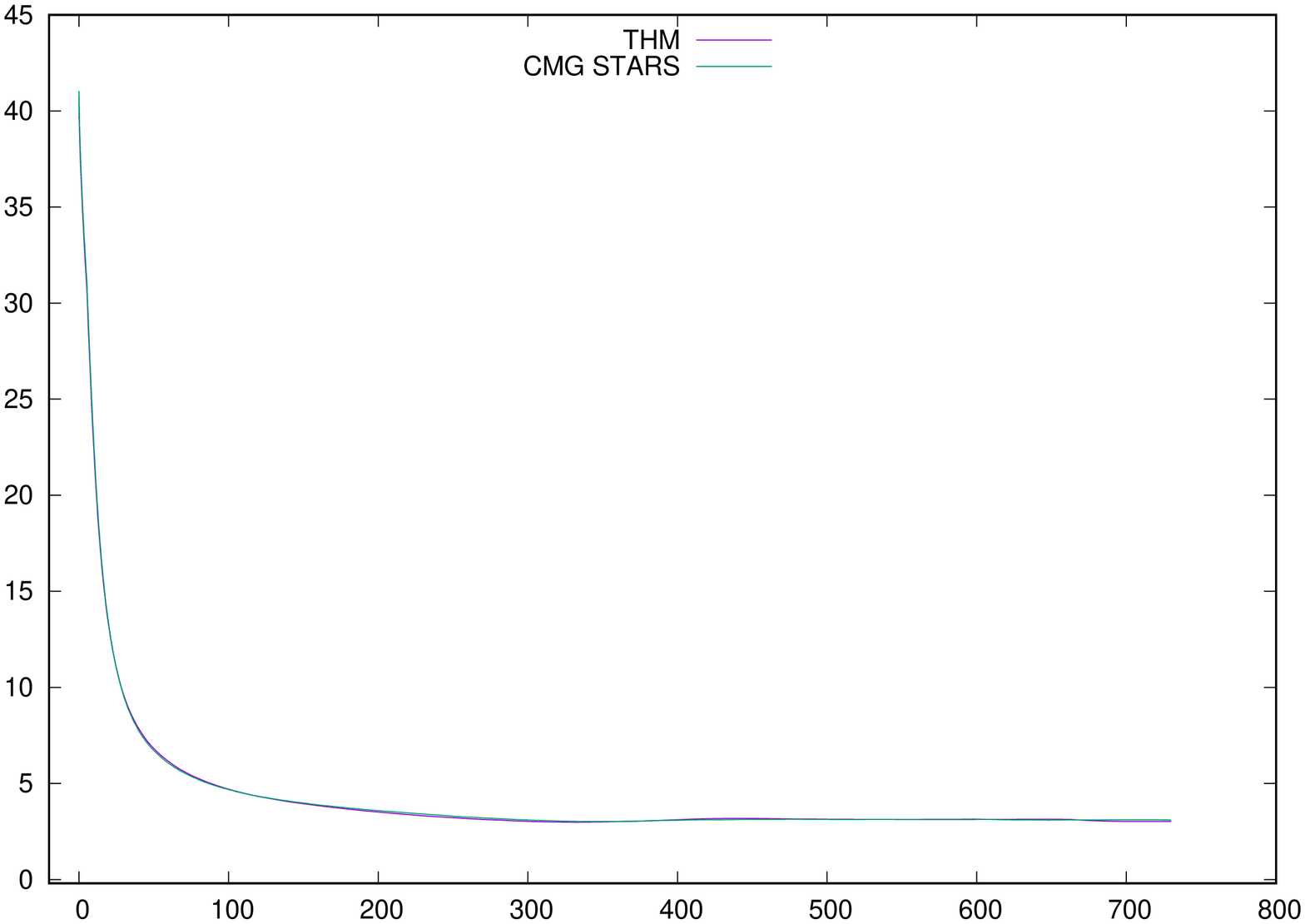}
    \caption{Example \ref{val-gas}, NCG: oil production rate (bbl/day), forth production well}
    \label{fig-val-gas-p4-por}
\end{figure}

\begin{figure}[H]
    \centering
    \includegraphics[width=0.53\linewidth, angle=270]{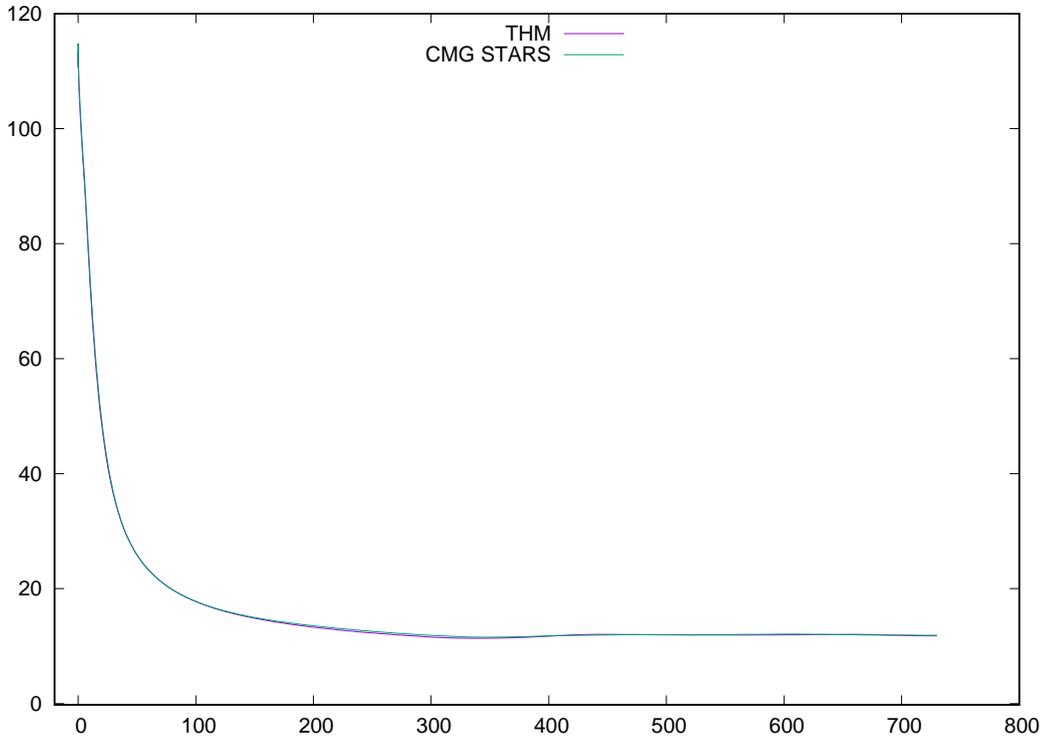}
    \caption{Example \ref{val-gas}, NCG: total oil production rate (bbl/day)}
    \label{fig-val-gas-por}
\end{figure}

\begin{example}
    \normalfont
    \label{val-gas} This model is similar as Example \ref{val-light} except that two non-condensable gas (NCG)
    components are added. Data is provided in Table \ref{val-gas-tab-input1} ad Table
    \ref{val-gas-tab-input1}.
    It also has five vertical wells: one injection
    well in the center (5, 5), and four production wells in four corners, (1, 1), (1, 9), (9, 1) and (9, 9).
    The bottom hole pressure of the injection well, water rate and oil rate of each well are shown from Figure
    \ref{fig-val-gas-inj-bhp} to Figure \ref{fig-val-gas-por}. All results are compared with CMG STARS.
\end{example}

\begin{table}[!htb]
    \centering
    \begin{tabular}{l c}
        \hline
        \textbf{Initial condition} &  \\
        \hline
        $k_{x,y,z}\ (md)$  & 313, 424, 535 \\
        $\phi$ & 0.3 \\
        $\phi_c$ & 5e-4 \\
        $p\ (psi)$ & 4000 \\
        $T\ (^\circ F)$ & 125 \\
        $S_{w, o, g}$ & 0.4, 0.5, 0.1 \\
        $x$ & 0.6, 0.4 \\
        $y$ & 4.73644e-4, 0, 0.486126, 0.2, 0.3134 \\
        \hline
    \end{tabular}
    \caption{Input data for Example \ref{val-gas}}
    \label{val-gas-tab-input1}
\end{table}

\begin{table}[!htb]
    \centering
    \begin{tabular}{l c c}
        \hline
        \textbf{Properties} & \textbf{N2} & \textbf{Isert}\\
        $M\ (lb/lbmole)$ & 28 & 40.8  \\
        $p_{crit}\ (psi) $ & 730 & 500   \\
        $T_{crit}\ ({}^\circ F)$ & -181 & -232  \\
        \hline

        $cpg1\ (Btu/({^\circ F} \cdot lbmol))$ & 6.713 & 7.44  \\
        $cpg2\ (Btu/({^\circ F} \cdot lbmol))$ & -4.883e-7 & -0.0018 \\
        $cpg3\ (Btu/({^\circ F} \cdot lbmol))$ & 1.287e-6 & 1.975e-6  \\
        $cpg4\ (Btu/({^\circ F} \cdot lbmol))$ & -4.36e-10 & -4.78e-10 \\
        \hline
        $avg\ (cp/{^\circ F})$ & 2.1960e-4 & 2.1267e-4  \\
        $bvg$ &  0.721 & 0.702 \\
        \hline
    \end{tabular}
    \caption{Input data for Example \ref{val-gas}}
    \label{val-gas-tab-input2}
\end{table}

Figure \ref{fig-val-gas-inj-bhp} is the bottom hole pressure of the injection well.
Figure \ref{fig-val-gas-p1-pwr} is the water production rate of the first production well.
Figure \ref{fig-val-gas-p2-pwr} is the water production rate of the second production well.
Figure \ref{fig-val-gas-p3-pwr} is the water production rate of the third production well.
Figure \ref{fig-val-gas-p4-pwr} is the water production rate of the forth production well.
Figure \ref{fig-val-gas-pwr} is the total water production rate of all production wells.
Figure \ref{fig-val-gas-p1-pgr} is the gas production rate of the first production well.
Figure \ref{fig-val-gas-p2-pgr} is the gas production rate of the second production well.
Figure \ref{fig-val-gas-p3-pgr} is the gas production rate of the third production well.
Figure \ref{fig-val-gas-p4-pgr} is the gas production rate of the forth production well.
Figure \ref{fig-val-gas-pgr} is the total gas production rate of all production wells.
Figure \ref{fig-val-gas-p1-por} is the oil production rate of the first production well.
Figure \ref{fig-val-gas-p2-por} is the oil production rate of the second production well.
Figure \ref{fig-val-gas-p3-por} is the oil production rate of the third production well.
Figure \ref{fig-val-gas-p4-por} is the oil production rate of the forth production well.
Figure \ref{fig-val-gas-por} is the total oil production rate of all production wells.
All figures show that our results match CMG STARS very well, which confirms our methods and implementation are
correct.

\subsection{Well Controls}

The water (steam) injection rate, water production and oil production rates, bottom hole pressure for each
well are reported. All rates are surface rates, and flash calculations are required to convert reservoir rates
to surface rates. The injection rate is measured as cold water equivalent. As we mentioned, the well modeling
is the most complicated, and we will change well operation constraints to test our simulator.

If there is no special statement, the non-linear method is the standard Newton method with a tolerance 1e-6
and maximal iterations of 10,
the linear solver is BICGSTAB with a tolerance 1e-4 and maximal iterations of 100, 
and the preconditioner is CPR-FPF method. All wells use implicit numerical methods, though the explicit method
has been implemented.

\subsubsection{Fixed Bottom Hole Pressure}

\begin{figure}[H]
    \centering
    \includegraphics[width=0.53\linewidth, angle=270]{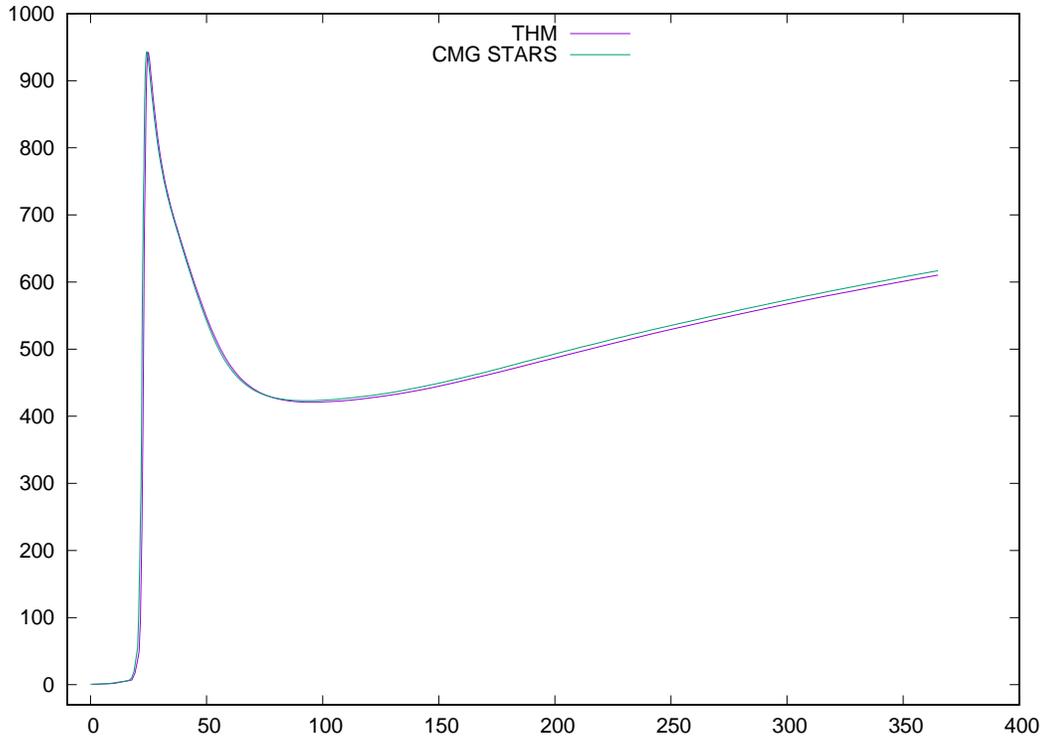}
    \caption{Example \ref{ex-fixed-bhp}, fixed bottom hole pressure: water injection rate (bbl/day)}
    \label{fig-ex-fixed-bhp-ir}
\end{figure}

\begin{figure}[H]
    \centering
    \includegraphics[width=0.53\linewidth, angle=270]{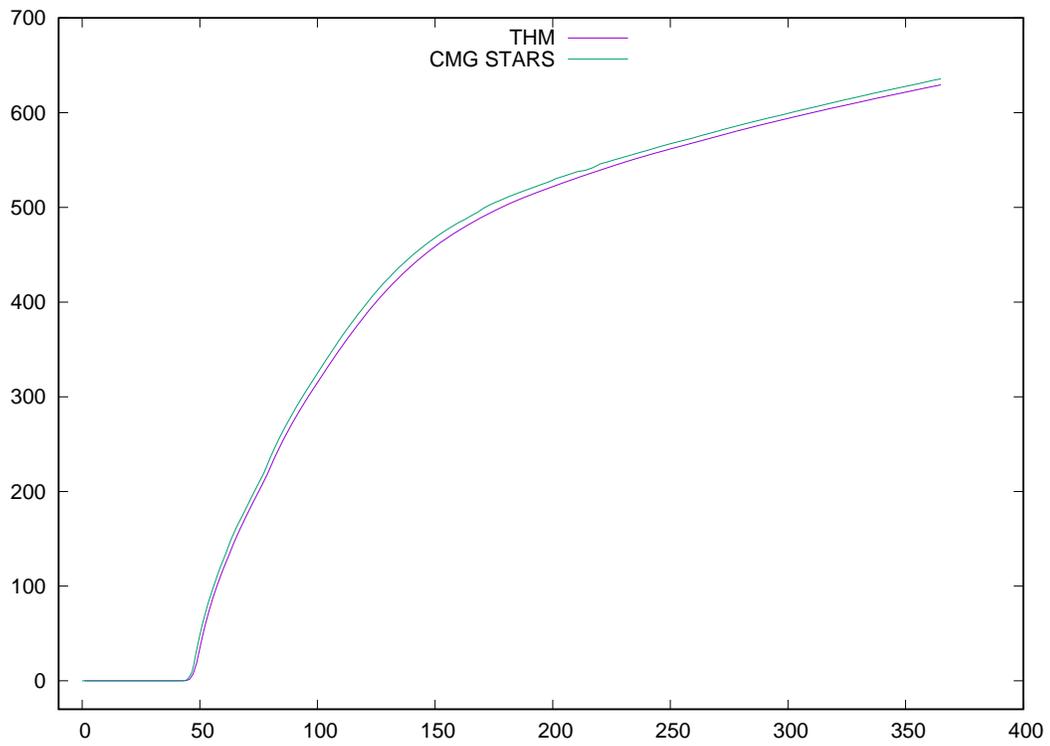}
    \caption{Example \ref{ex-fixed-bhp}, fixed bottom hole pressure: water production rate (bbl/day)}
    \label{fig-ex-fixed-bhp-pwr}
\end{figure}

\begin{figure}[H]
    \centering
    \includegraphics[width=0.53\linewidth, angle=270]{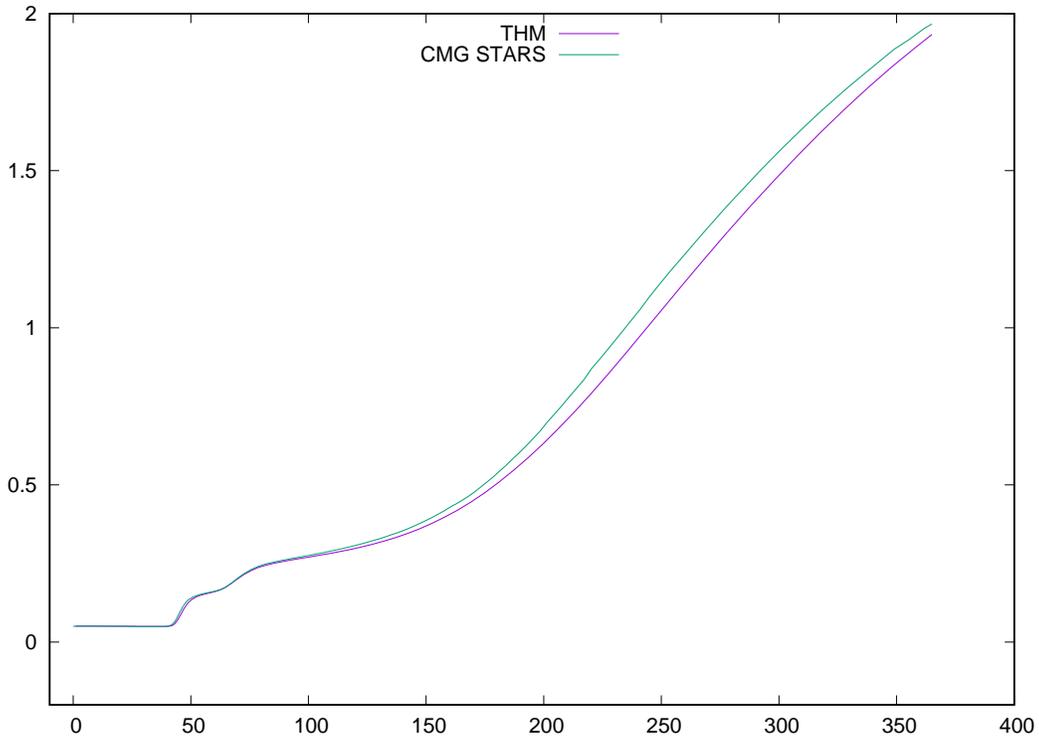}
    \caption{Example \ref{ex-fixed-bhp}, fixed bottom hole pressure: oil production rate (bbl/day)}
    \label{fig-ex-fixed-bhp-por}
\end{figure}

\begin{example}
    \normalfont
    \label{ex-fixed-bhp}
    The injection well operates at 1500 psi. The steam quality is 0, and its temperature is 450 F.
    Two production wells operates are 17 psi. The simulation period is 365 days. 
    Figure \ref{fig-ex-fixed-bhp-ir}, \ref{fig-ex-fixed-bhp-pwr},
    and \ref{fig-ex-fixed-bhp-por} show the water injection rate, total water production and total oil
    production. The rates are surface rate.  All results are compared with CMG STARS. From these figures, we
    can see that our results match CMG STARS.
\end{example}

\subsubsection{Fixed Rate}

\begin{figure}[H]
    \centering
    \includegraphics[width=0.53\linewidth, angle=270]{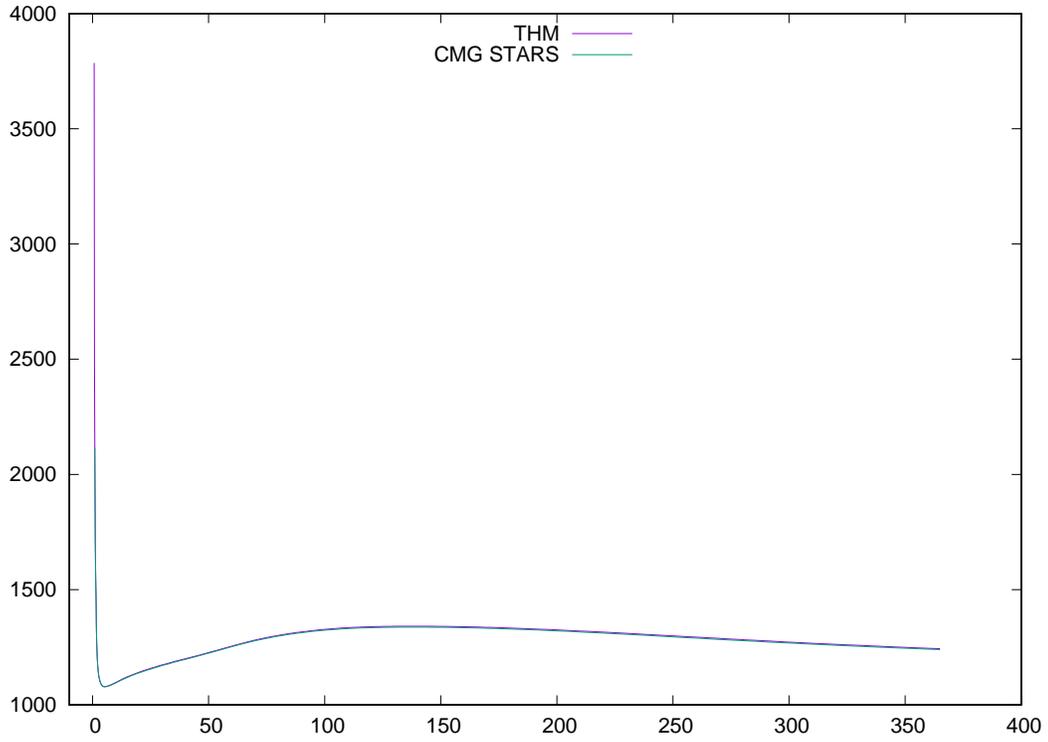}
    \caption{Example \ref{ex-fixed-rate}, fixed rate: injection well, bottom hole pressure (psi)}
    \label{fig-ex-fixed-rate-inj-bhp}
\end{figure}

\begin{figure}[H]
    \centering
    \includegraphics[width=0.53\linewidth, angle=270]{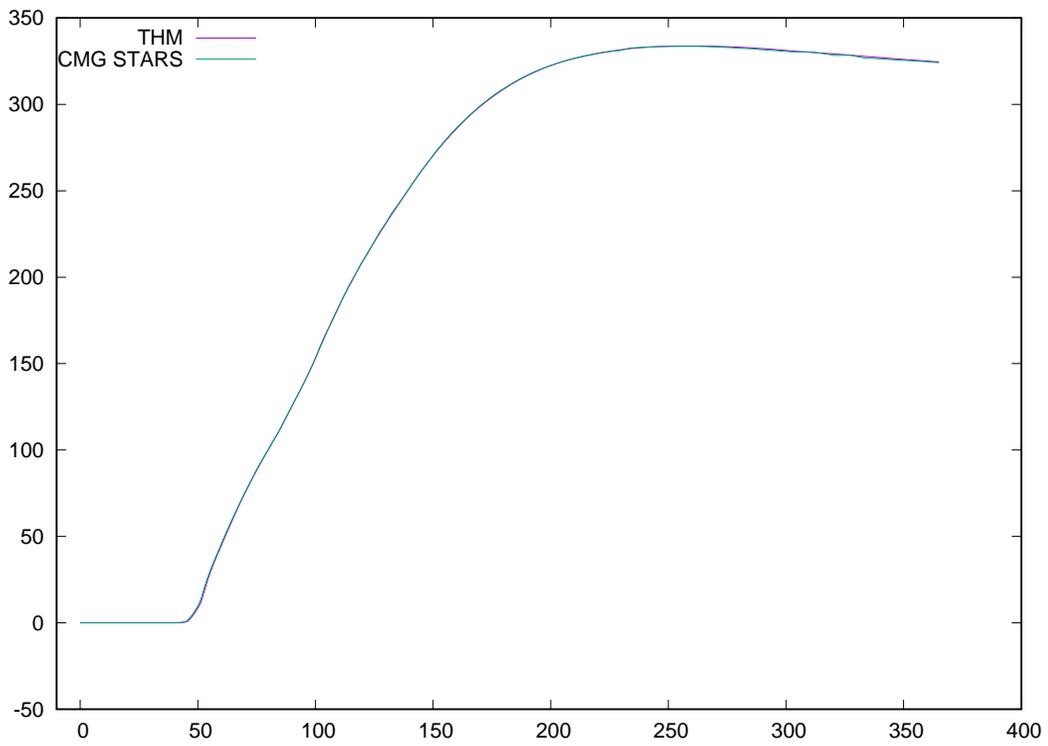}
    \caption{Example \ref{ex-fixed-rate}, fixed rate: water production rate (bbl/day)}
    \label{fig-ex-fixed-rate-pwr}
\end{figure}

\begin{figure}[H]
    \centering
    \includegraphics[width=0.53\linewidth, angle=270]{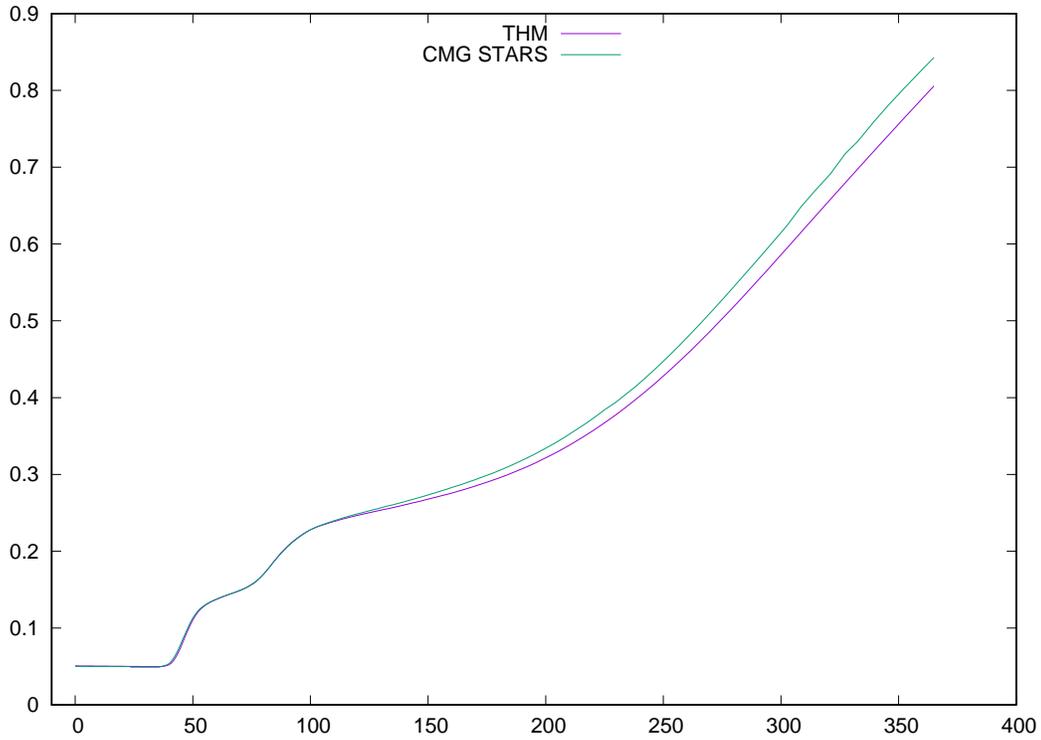}
    \caption{Example \ref{ex-fixed-rate}, fixed rate: oil production rate (bbl/day)}
    \label{fig-ex-fixed-rate-por}
\end{figure}

\begin{example}
    \normalfont
    \label{ex-fixed-rate} The injection well operates at 300 bbl/day, and the production wells operates at 17
    psi. The simulation period is 365 days.
    Figure \ref{fig-ex-fixed-rate-inj-bhp}, \ref{fig-ex-fixed-rate-pwr}, and \ref{fig-ex-fixed-rate-por}
    are bottom hole pressure for injection well, total water production and total oil production.
    All results are compared with CMG STARS.
\end{example}

When fixed rate constraint is applied to a well, its rate is known, but its bottom hole pressure is unknown,
which should be obtained by Newton methods. Figure \ref{fig-ex-fixed-rate-inj-bhp} represents the bottom hole
pressure of the injection well, from which we can see that our results match CMG STARS exactly. It means the
methods and the implementation are correct. Figure \ref{fig-ex-fixed-rate-pwr} is the total water production
rate, which also match CMG STARS exactly. Figure \ref{fig-ex-fixed-rate-por} is the total oil production rate
(bbl/day). The results match CMG STARS exactly in the first 100 days, and after that, there is slight
difference. The reason is that each simulator has its own numerical settings and automatical numerical
tunings. For example, the density, bottom hole pressure update and mobility for wells in CMG STARS have many
parameters to control, and CMG has automatical bottom hole pressure update algorithms depending on time step
and pressure changes, whose details are unknown to us.

\subsubsection{Rate and Pressure Constraints}

A well may have many operation constraints, such as maximal injection rate with maximal bottom hole pressure
for injection well, maximal oil production rate with minimal bottom hole pressure for production well, 
and maximal liquid rate with minimal bottom hole pressure for production well.

\begin{figure}[H]
    \centering
    \includegraphics[width=0.53\linewidth, angle=270]{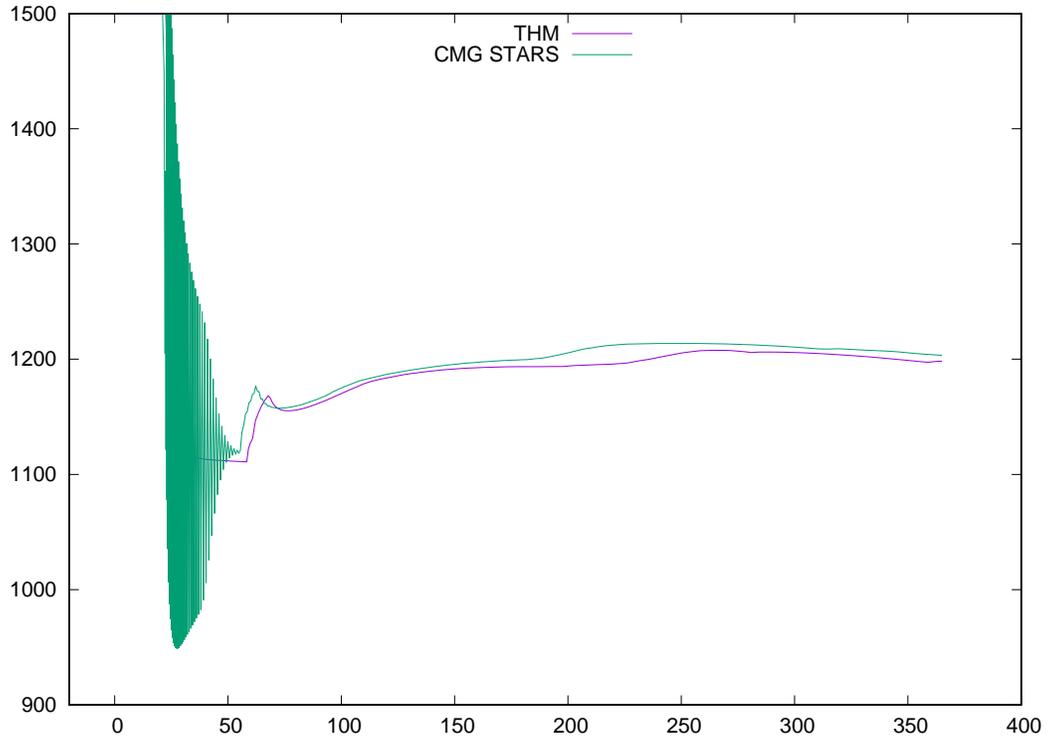}
    \caption{Example \ref{ex-max-rate}, rate and pressure control: injection well, bottom hole pressure (psi)}
    \label{fig-ex-max-rate-inj-bhp}
\end{figure}

\begin{figure}[H]
    \centering
    \includegraphics[width=0.53\linewidth, angle=270]{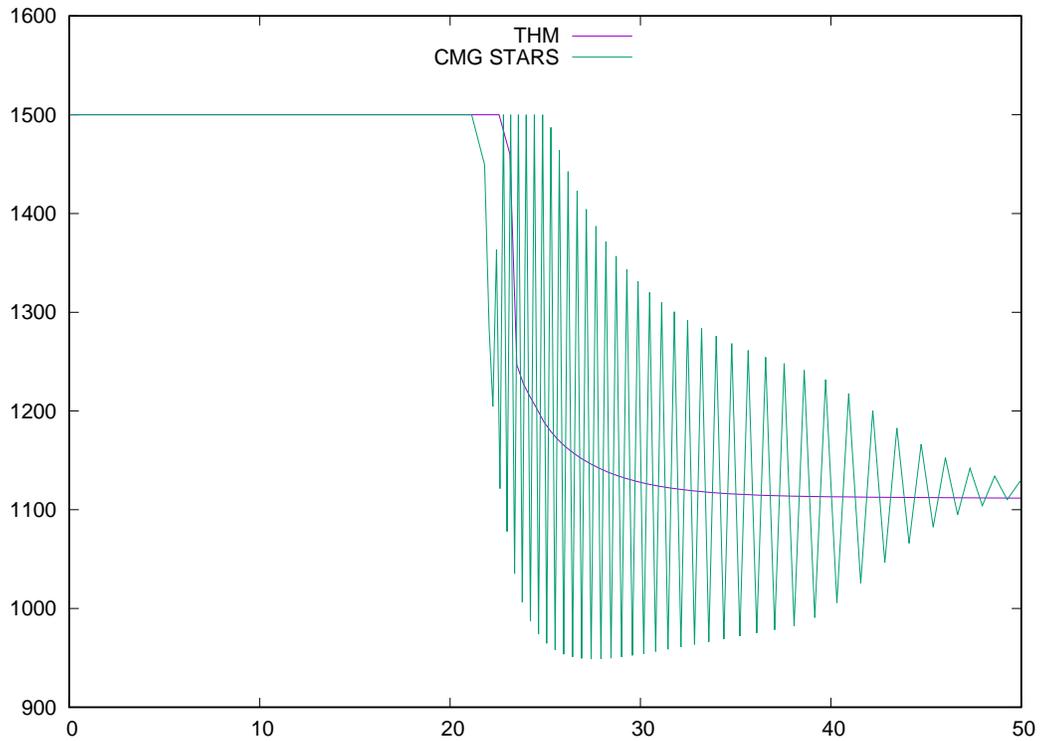}
    \caption{Example \ref{ex-max-rate}, rate and pressure control: injection well, bottom hole pressure (psi),
    first 50 days}
    \label{fig-ex-max-rate-inj-bhp-jump}
\end{figure}

\begin{figure}[H]
    \centering
    \includegraphics[width=0.53\linewidth, angle=270]{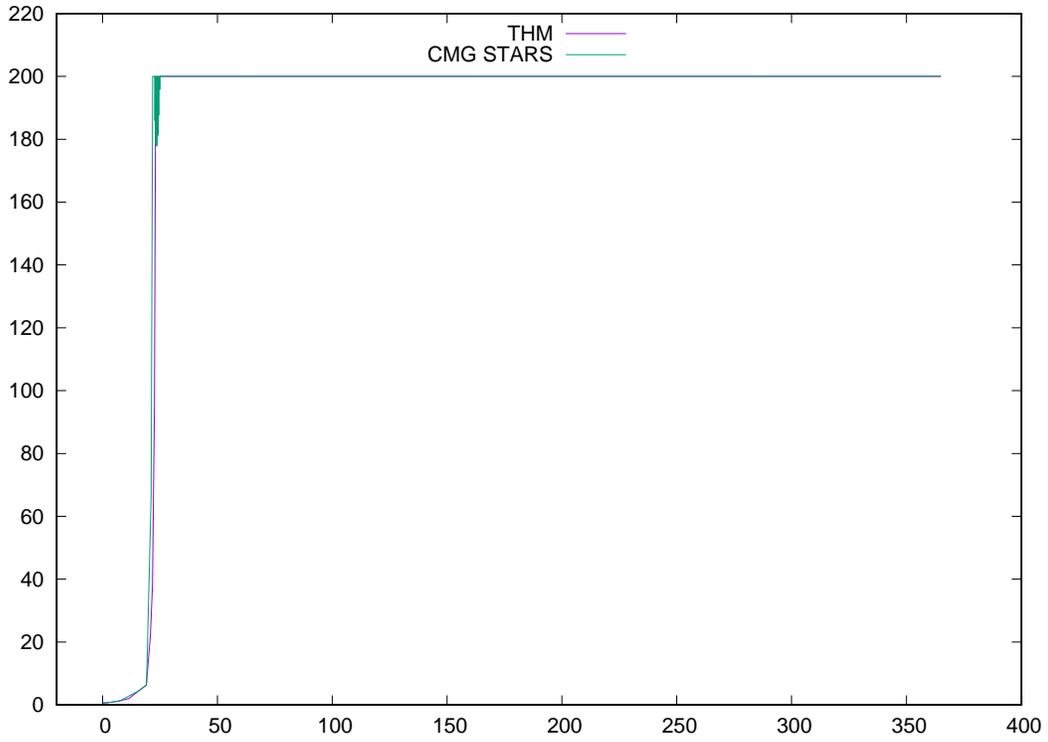}
    \caption{Example \ref{ex-max-rate}, rate and pressure control: water injection rate (bbl/day)}
    \label{fig-ex-max-rate-ir}
\end{figure}

\begin{figure}[H]
    \centering
    \includegraphics[width=0.53\linewidth, angle=270]{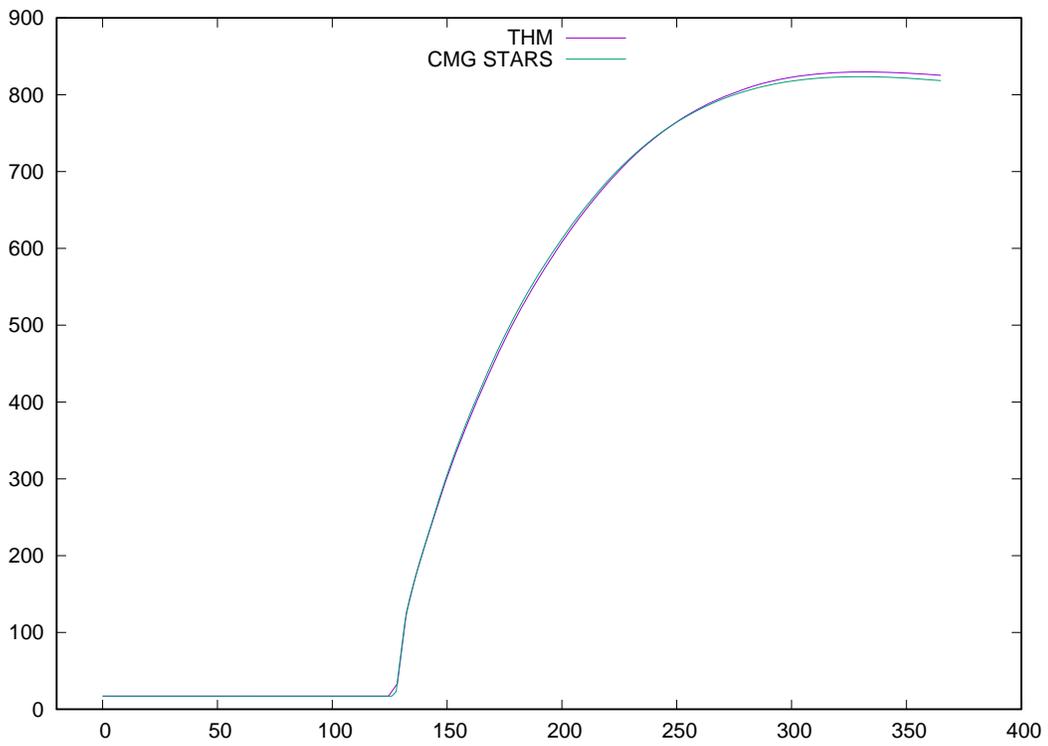}
    \caption{Example \ref{ex-max-rate}, rate and pressure control: first production well, bottom hole pressure (psi)}
    \label{fig-ex-max-rate-p2-bhp}
\end{figure}

\begin{figure}[H]
    \centering
    \includegraphics[width=0.53\linewidth, angle=270]{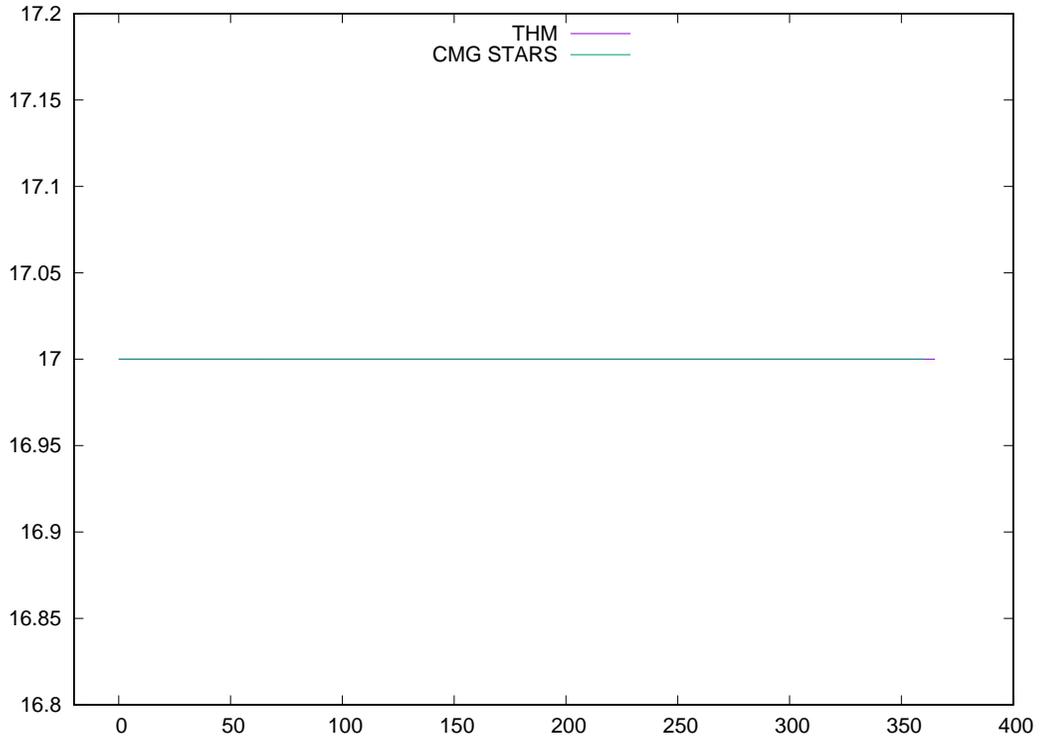}
    \caption{Example \ref{ex-max-rate}, rate and pressure control: second production well, bottom hole pressure (psi)}
    \label{fig-ex-max-rate-p3-bhp}
\end{figure}

\begin{figure}[H]
    \centering
    \includegraphics[width=0.53\linewidth, angle=270]{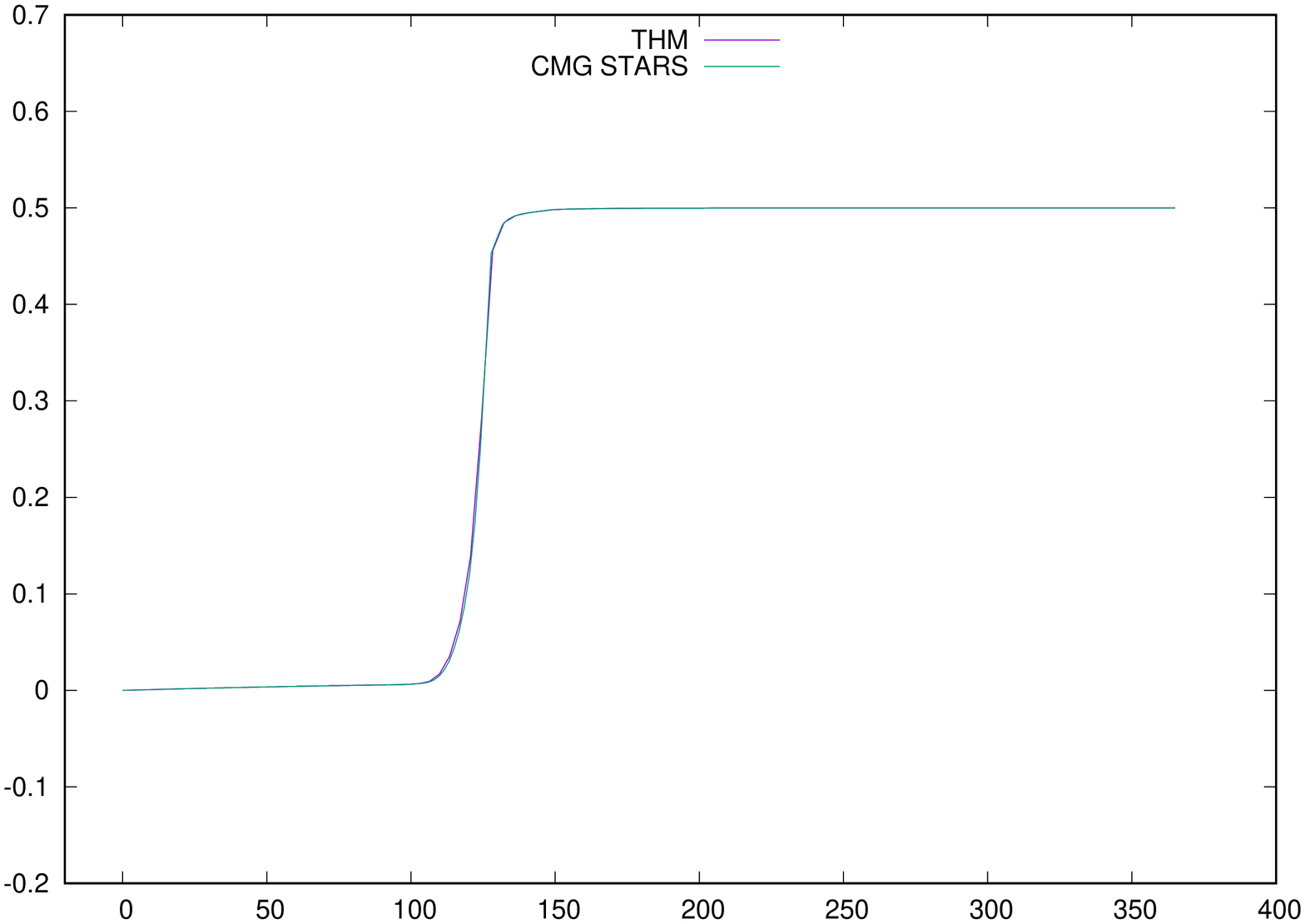}
    \caption{Example \ref{ex-max-rate}, rate and pressure control: water production rate (bbl/day), first production well}
    \label{fig-ex-max-rate-p2-pwr}
\end{figure}

\begin{figure}[H]
    \centering
    \includegraphics[width=0.53\linewidth, angle=270]{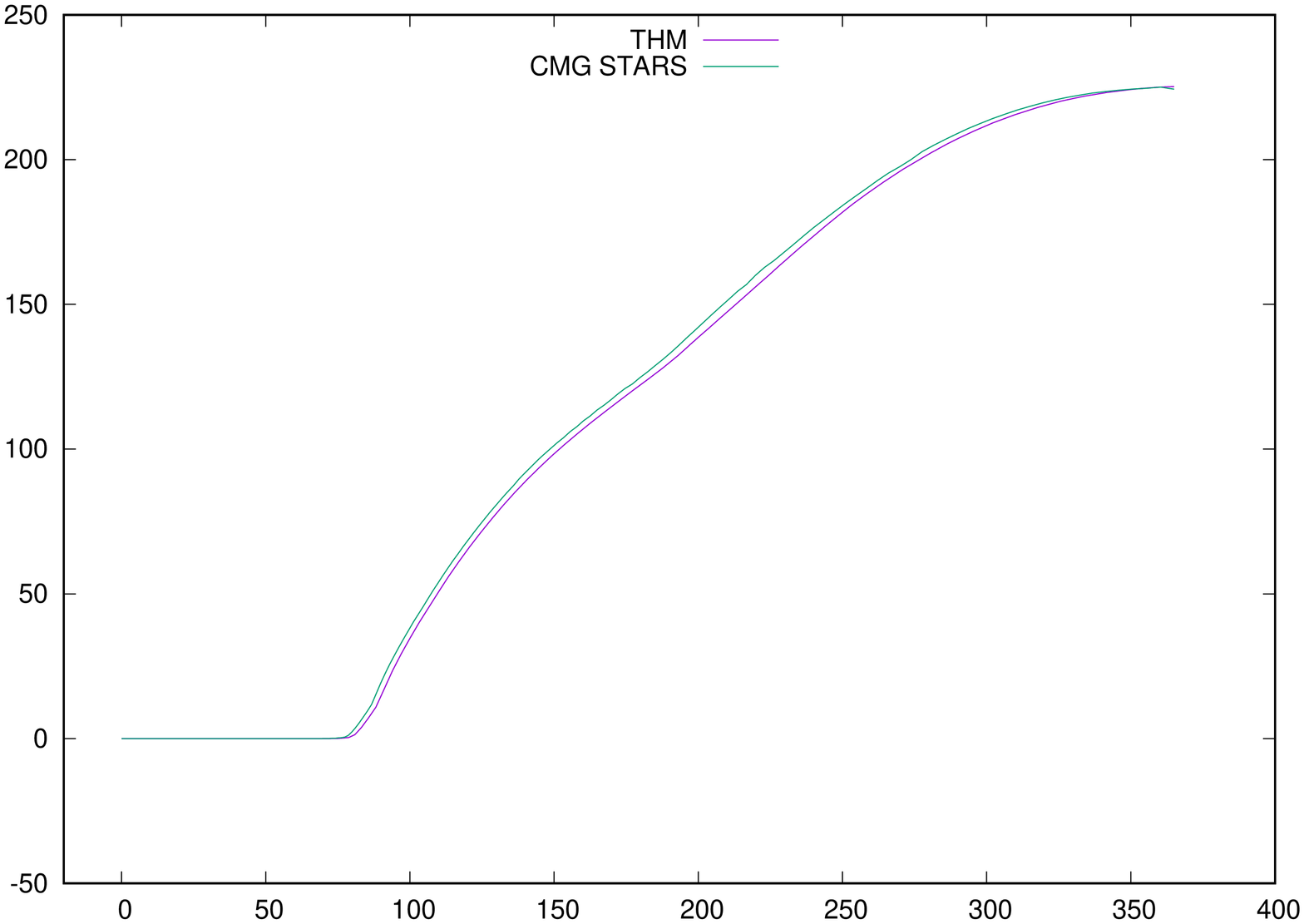}
    \caption{Example \ref{ex-max-rate}, rate and pressure control: water production rate (bbl/day), second production well}
    \label{fig-ex-max-rate-p3-pwr}
\end{figure}

\begin{figure}[H]
    \centering
    \includegraphics[width=0.53\linewidth, angle=270]{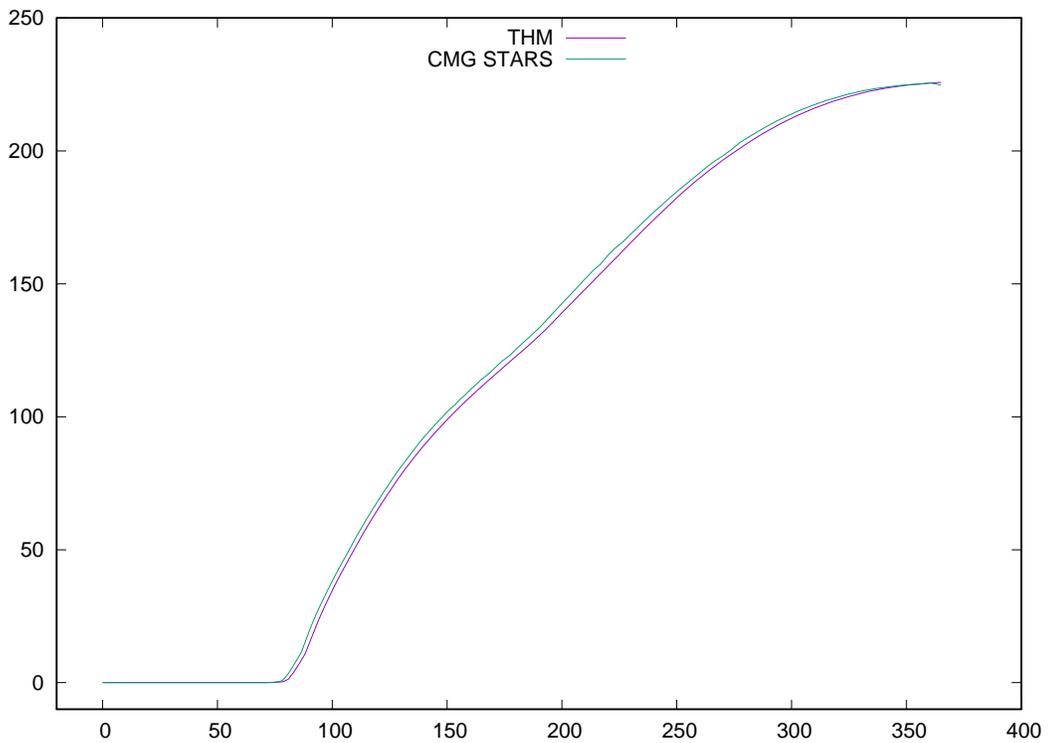}
    \caption{Example \ref{ex-max-rate}, rate and pressure control: total water production rate (bbl/day)}
    \label{fig-ex-max-rate-pwr}
\end{figure}

\begin{figure}[H]
    \centering
    \includegraphics[width=0.53\linewidth, angle=270]{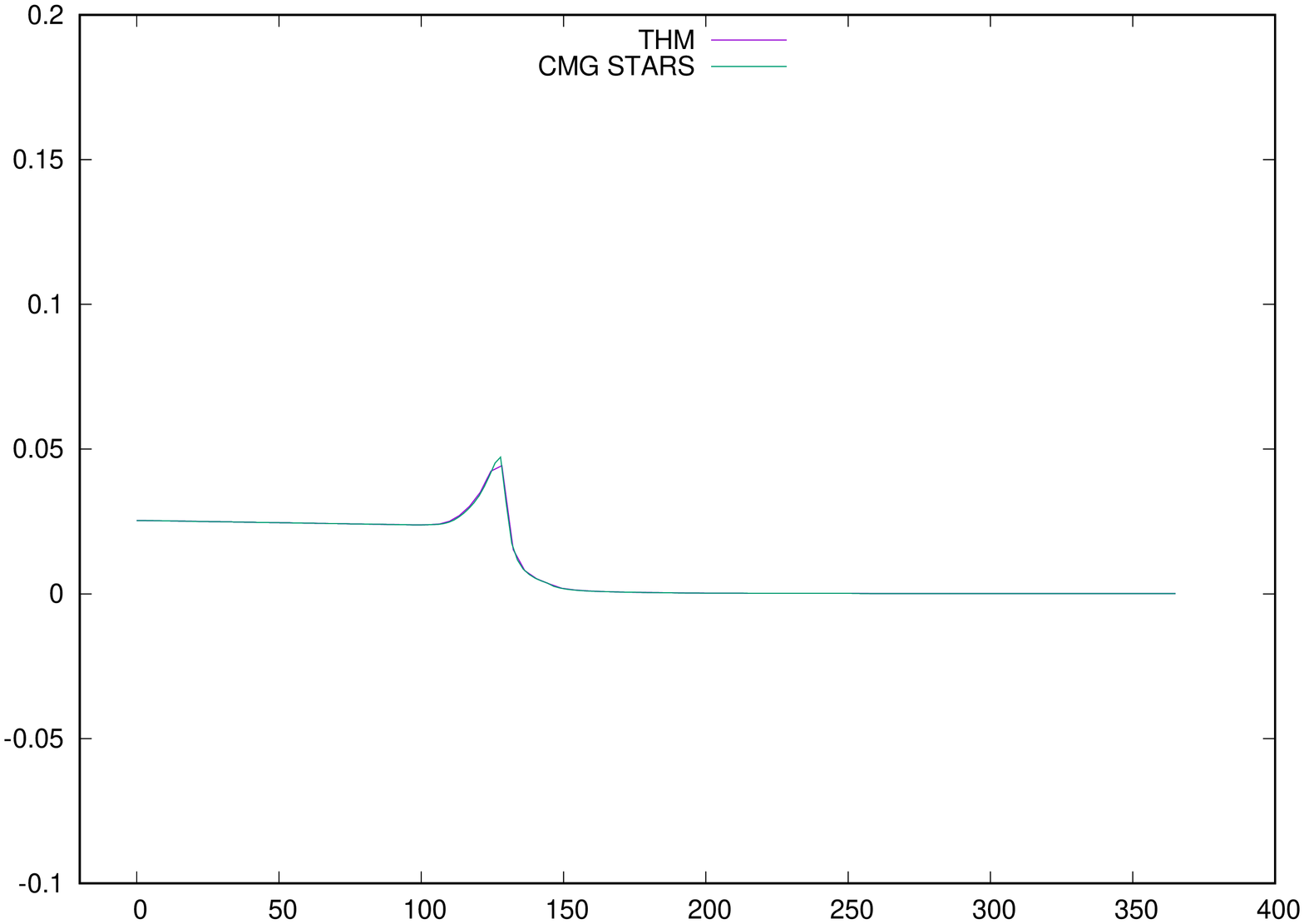}
    \caption{Example \ref{ex-max-rate}, rate and pressure control: oil production rate (bbl/day), first production well}
    \label{fig-ex-max-rate-p2-por}
\end{figure}

\begin{figure}[H]
    \centering
    \includegraphics[width=0.53\linewidth, angle=270]{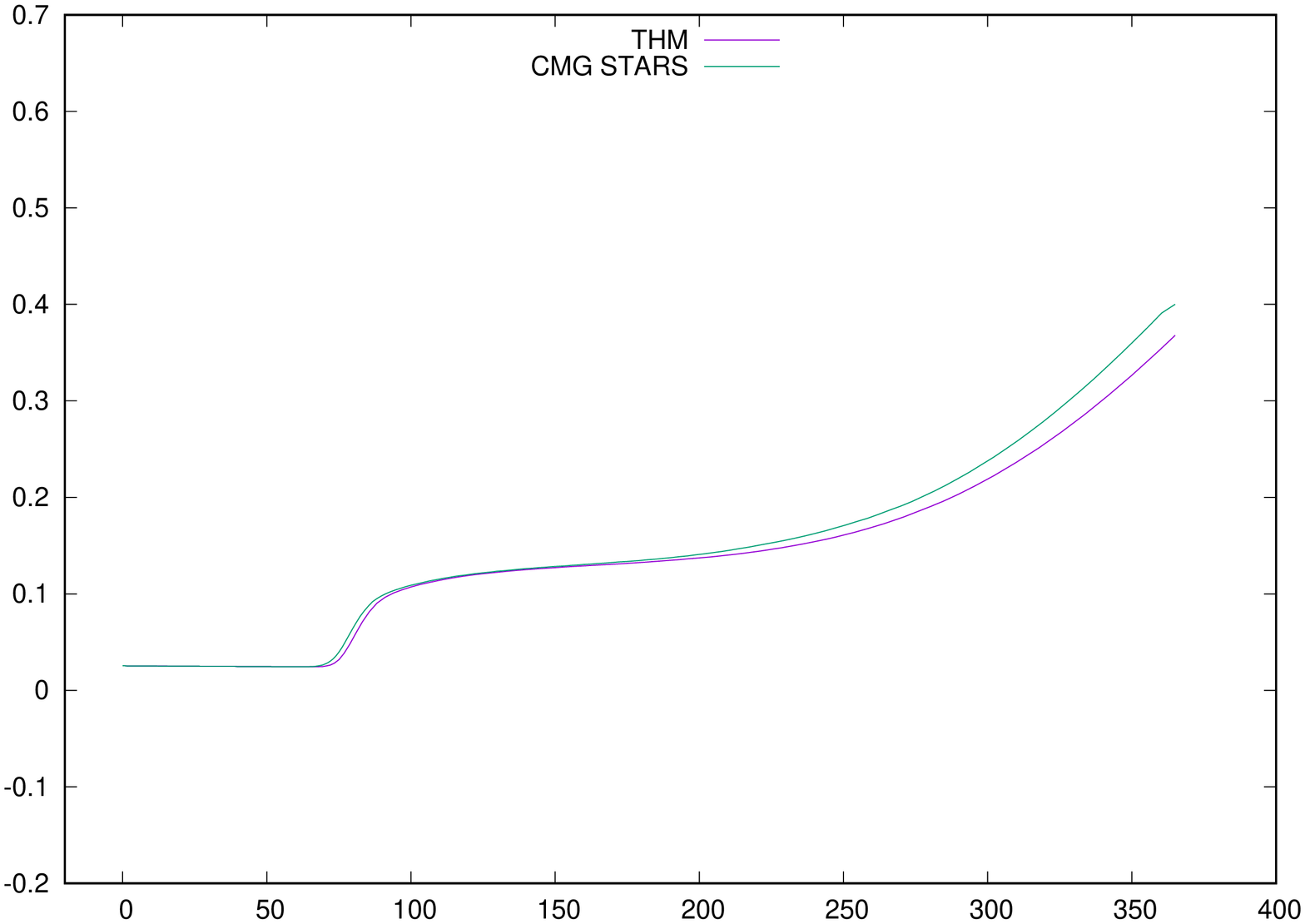}
    \caption{Example \ref{ex-max-rate}, rate and pressure control: oil production rate (bbl/day), second production well}
    \label{fig-ex-max-rate-p3-por}
\end{figure}

\begin{figure}[H]
    \centering
    \includegraphics[width=0.53\linewidth, angle=270]{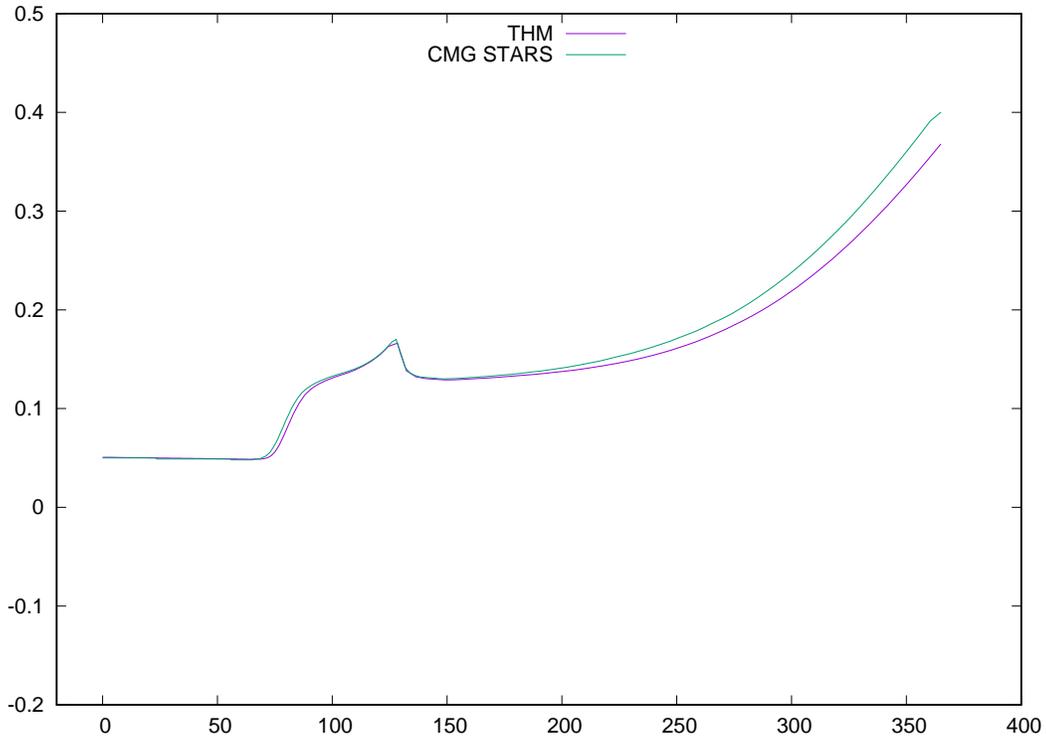}
    \caption{Example \ref{ex-max-rate}, rate and pressure control: total oil production rate (bbl/day)}
    \label{fig-ex-max-rate-por}
\end{figure}

\begin{example}
    \normalfont
    \label{ex-max-rate} The injection well has a maximal injection rate of 200 bbl/day and a maximal bottom
    hole pressure of 1500 psi. The steam has a steam quality of 0.3 and temperature of 450 F. The first
    production well has a miximal liquid rate of 0.5 bbl/day and a minimal bottom hole pressure of 17 psi. The
    second production well has a maximal oil rate of 0.4 bbl/day and a minimal bottom hole pressure of 17 psi.
    The simulation period is 365 days.
    The bottom hole pressures for each well, water rates and oil rates for each well are presented from Figure
    \ref{fig-ex-max-rate-inj-bhp} to Figure \ref{fig-ex-max-rate-por}. All results are compared with CMG
    STARS.
\end{example}

Figure \ref{fig-ex-max-rate-inj-bhp}, \ref{fig-ex-max-rate-p2-bhp} and \ref{fig-ex-max-rate-p3-bhp} show the
bottom hole pressure for injection well and production wells. The results for production wells match well.
Our Newton method shows good convergence but CMG STARS shows severe convergence issues.
From Figure \ref{fig-ex-max-rate-p2-pwr} to Figure \ref{fig-ex-max-rate-por}, we can see that the water 
rate and oil rate for each well, total water rate and total oil rate match CMG STARS well.

\subsubsection{Constant Heat Transfer Model}

\begin{figure}[H]
    \centering
    \includegraphics[width=0.53\linewidth, angle=270]{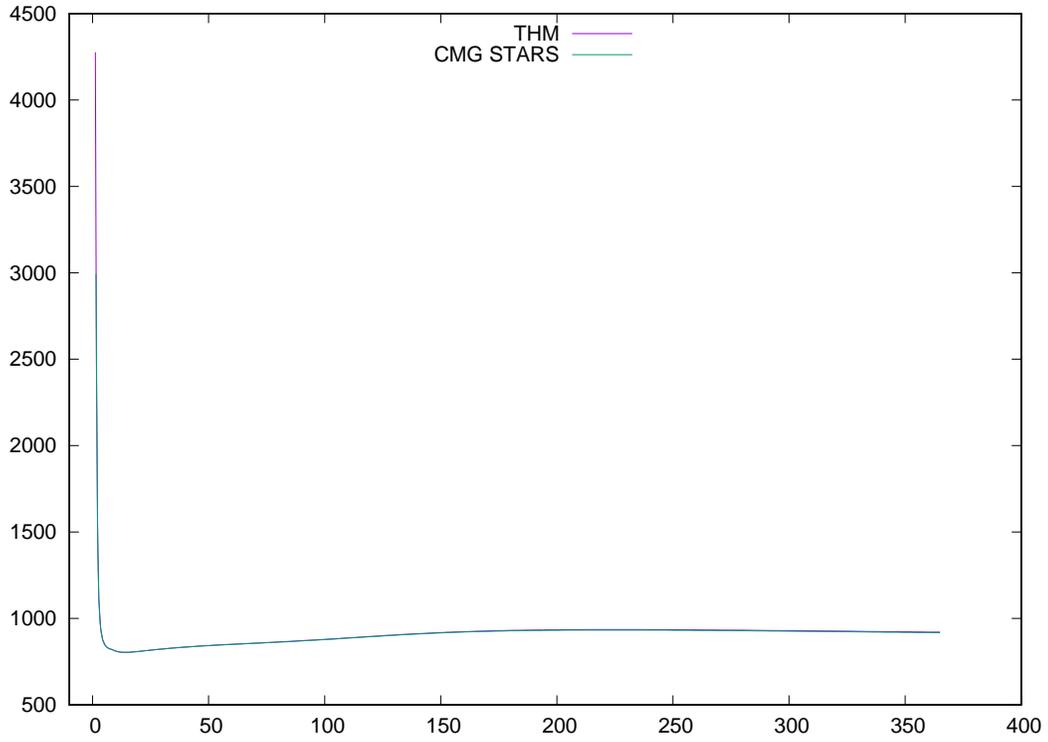}
    \caption{Example \ref{ex-cons-heat}, constant heat transfer model: injection well, bottom hole pressure (psi)}
    \label{fig-ex-cons-heat-inj-bhp}
\end{figure}

\begin{figure}[H]
    \centering
    \includegraphics[width=0.53\linewidth, angle=270]{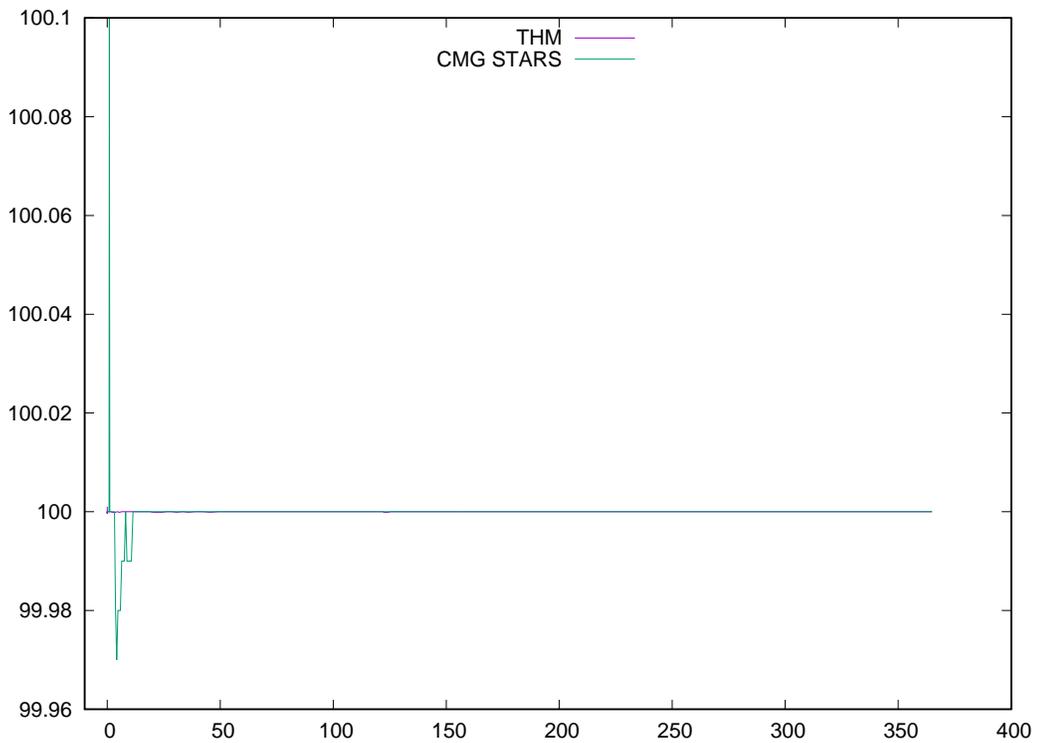}
    \caption{Example \ref{ex-cons-heat}, constant heat transfer model: water injection rate (bbl/day)}
    \label{fig-ex-cons-heat-ir}
\end{figure}

\begin{figure}[H]
    \centering
    \includegraphics[width=0.53\linewidth, angle=270]{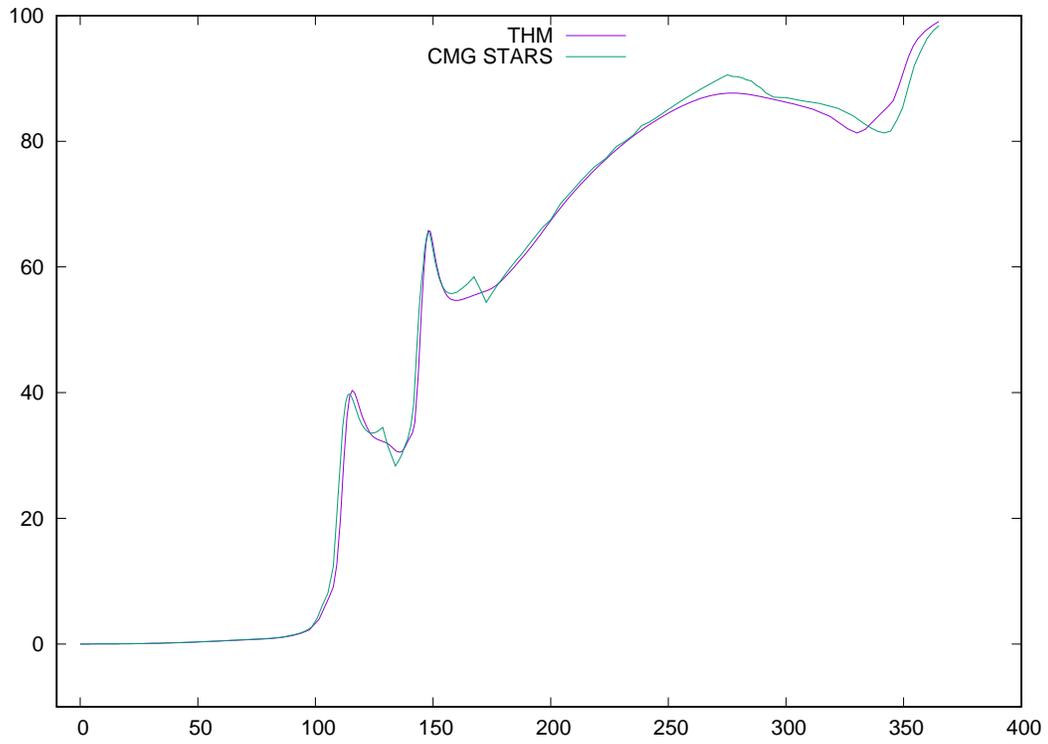}
    \caption{Example \ref{ex-cons-heat}, constant heat transfer model: water production rate (bbl/day)}
    \label{fig-ex-cons-heat-pwr}
\end{figure}

\begin{figure}[H]
    \centering
    \includegraphics[width=0.53\linewidth, angle=270]{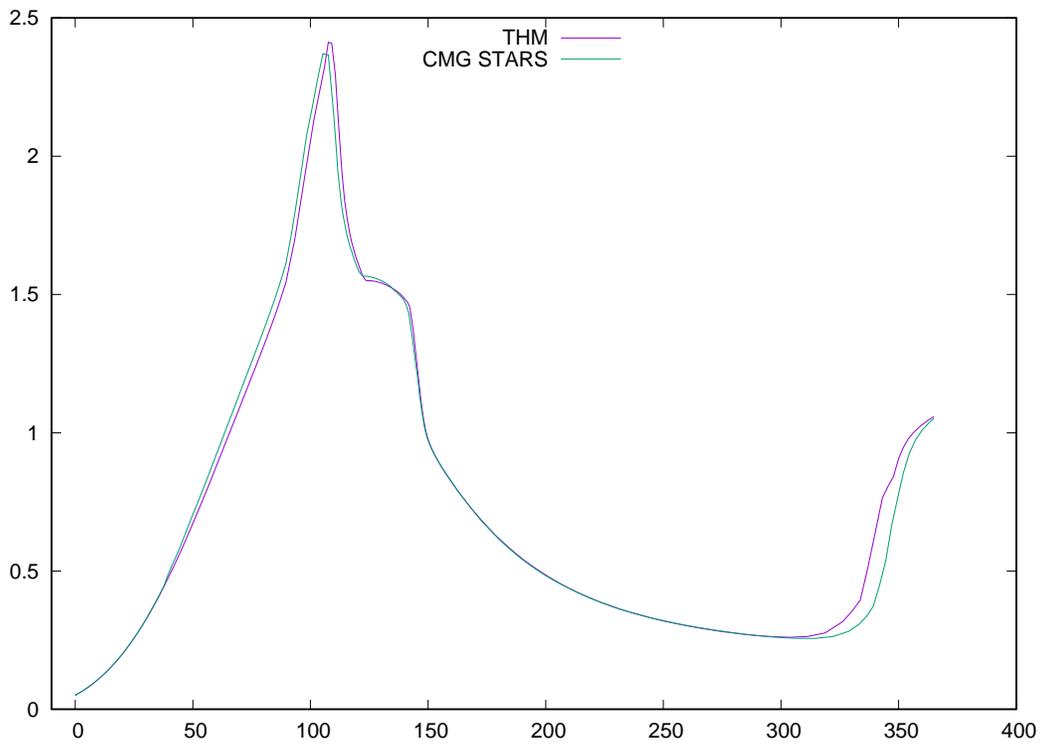}
    \caption{Example \ref{ex-cons-heat}, constant heat transfer model: oil production rate (bbl/day)}
    \label{fig-ex-cons-heat-por}
\end{figure}

\begin{example}
    \normalfont
    \label{ex-cons-heat} In this example, the injection well operates at a fixed steam injection rate of 100
    bbl/day, and the steam quality is 0 at a temperature of 450 F. Each production well operates at fixed
    bottom hole pressure of 17 psi. Constant heat transfers to each perforation at a rate of 1e6 Btu/day. The
    simulation period is 365 days. Figure \ref{fig-ex-cons-heat-inj-bhp} to Figure \ref{fig-ex-cons-heat-por}
    show simulated results and they are compared with CMG STARS.
\end{example}

Figure \ref{fig-ex-cons-heat-inj-bhp} is the bottom hole pressure and compared with CMG STARS. We can see that
the match is exact. For injection rate shown by Figure \ref{fig-ex-cons-heat-ir}, our convergence is smoother
than CMG STARS. The water and oil production rates match very well as shown in Figure
\ref{fig-ex-cons-heat-por} and Figure \ref{fig-ex-cons-heat-pwr}.

\subsubsection{Convective Heat Transfer Model}

\begin{figure}[H]
    \centering
    \includegraphics[width=0.53\linewidth, angle=270]{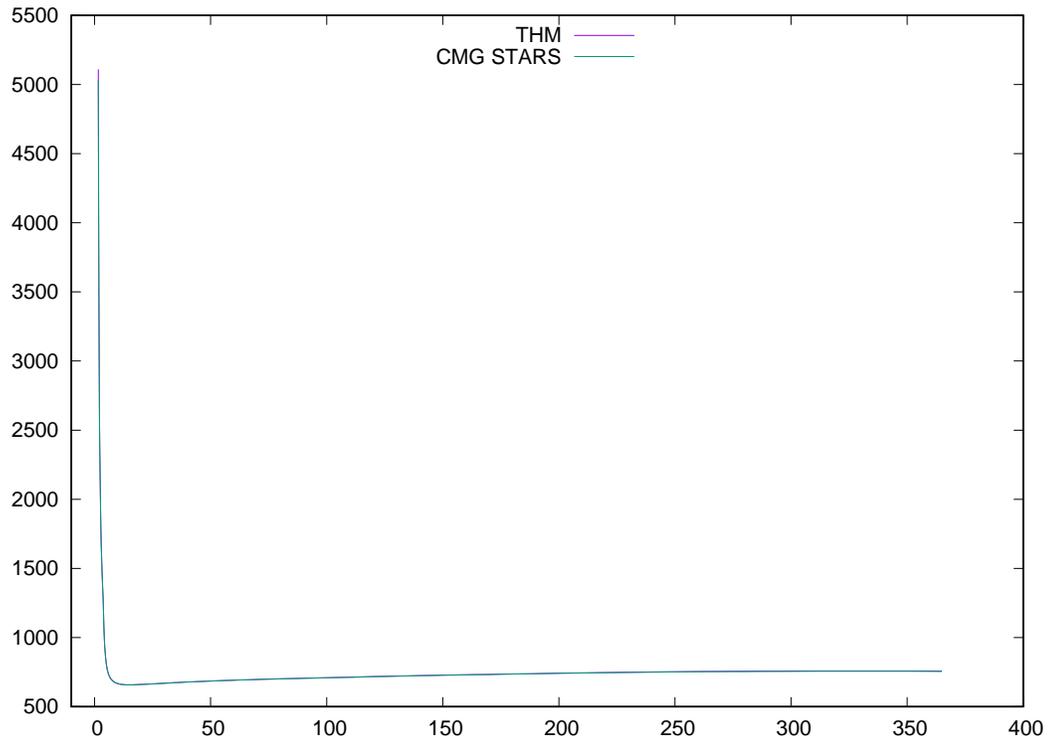}
    \caption{Example \ref{ex-conv-heat}, convective heat transfer model: injection well, bottom hole pressure (psi)}
    \label{fig-ex-conv-heat-inj-bhp}
\end{figure}

\begin{figure}[H]
    \centering
    \includegraphics[width=0.53\linewidth, angle=270]{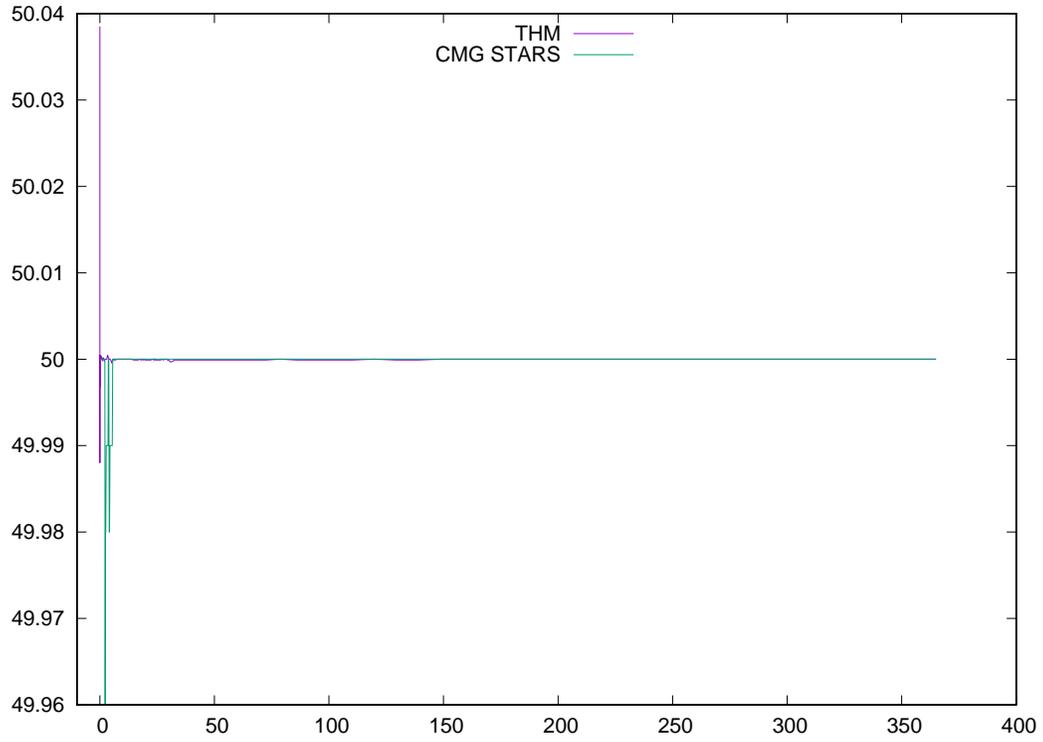}
    \caption{Example \ref{ex-conv-heat}, convective heat transfer model: water injection rate (bbl/day)}
    \label{fig-ex-conv-heat-ir}
\end{figure}

\begin{figure}[H]
    \centering
    \includegraphics[width=0.53\linewidth, angle=270]{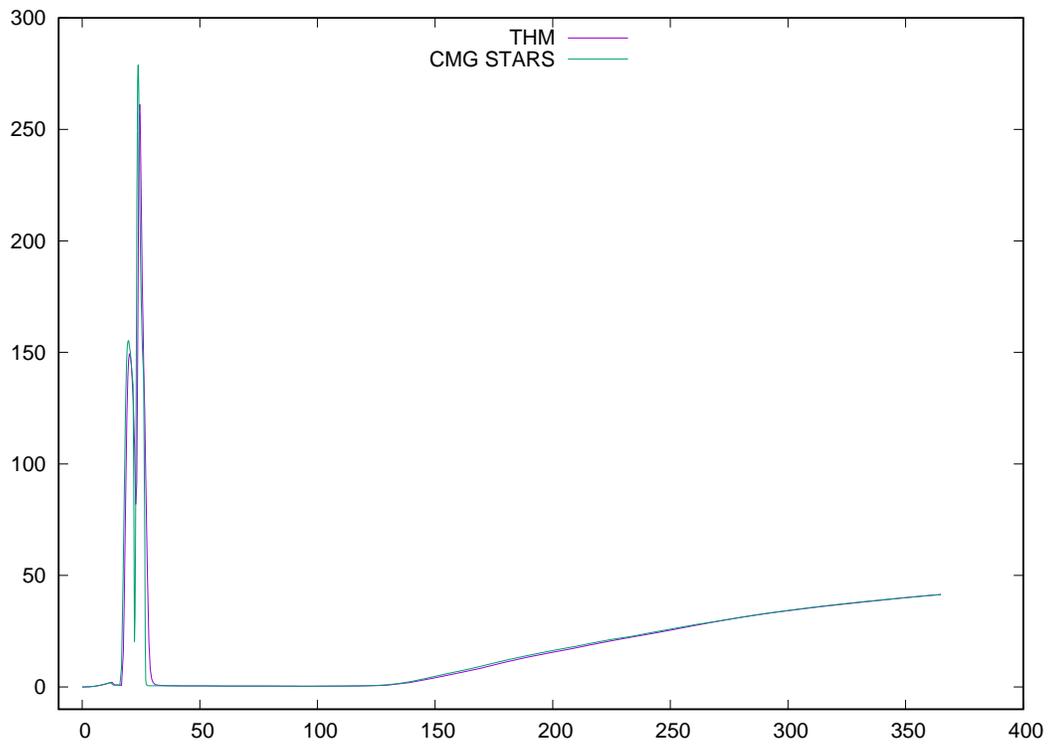}
    \caption{Example \ref{ex-conv-heat}, convective heat transfer model: water production rate (bbl/day)}
    \label{fig-ex-conv-heat-pwr}
\end{figure}

\begin{figure}[H]
    \centering
    \includegraphics[width=0.53\linewidth, angle=270]{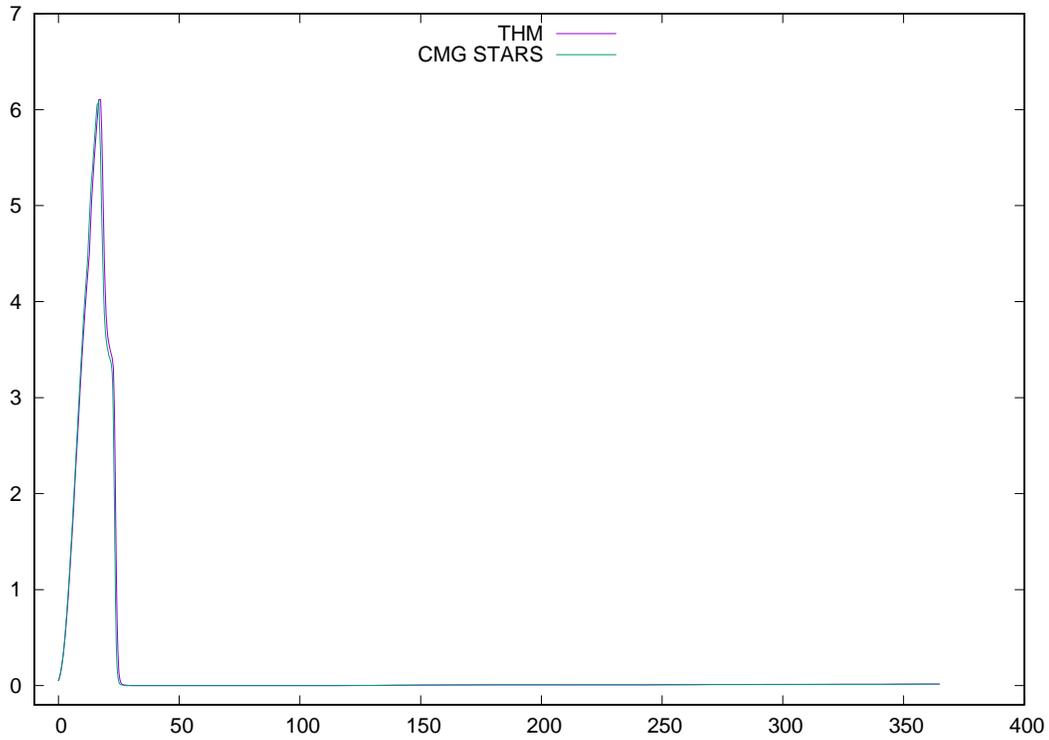}
    \caption{Example \ref{ex-conv-heat}, convective heat transfer model: oil production rate (bbl/day)}
    \label{fig-ex-conv-heat-por}
\end{figure}

\begin{example}
    \normalfont
    \label{ex-conv-heat} Here the injection rate is 50 bbl/day, and production wells have fixed bottom hole
    pressure. Each perforation of production wells have a $\texttt{uhtr}$ of 4e4 btu/day-F and a temperature setpoint
    ($\texttt{tmpset}$) of 500 F. Again the simulation period is 365 days. All results are compared with CMG STARS.
    Figure \ref{fig-ex-conv-heat-inj-bhp}, \ref{fig-ex-conv-heat-ir}, \ref{fig-ex-conv-heat-pwr} and
    \ref{fig-ex-conv-heat-por} are injection well bottom hole pressure, injection surface rate, water
    production surface rate and oil production surface rate. From these figures, we can see the match between
    our simulator and CMG STARS is excellent.
\end{example}

\subsubsection{Heater Well}

\begin{figure}[H]
    \centering
    \includegraphics[width=0.53\linewidth, angle=270]{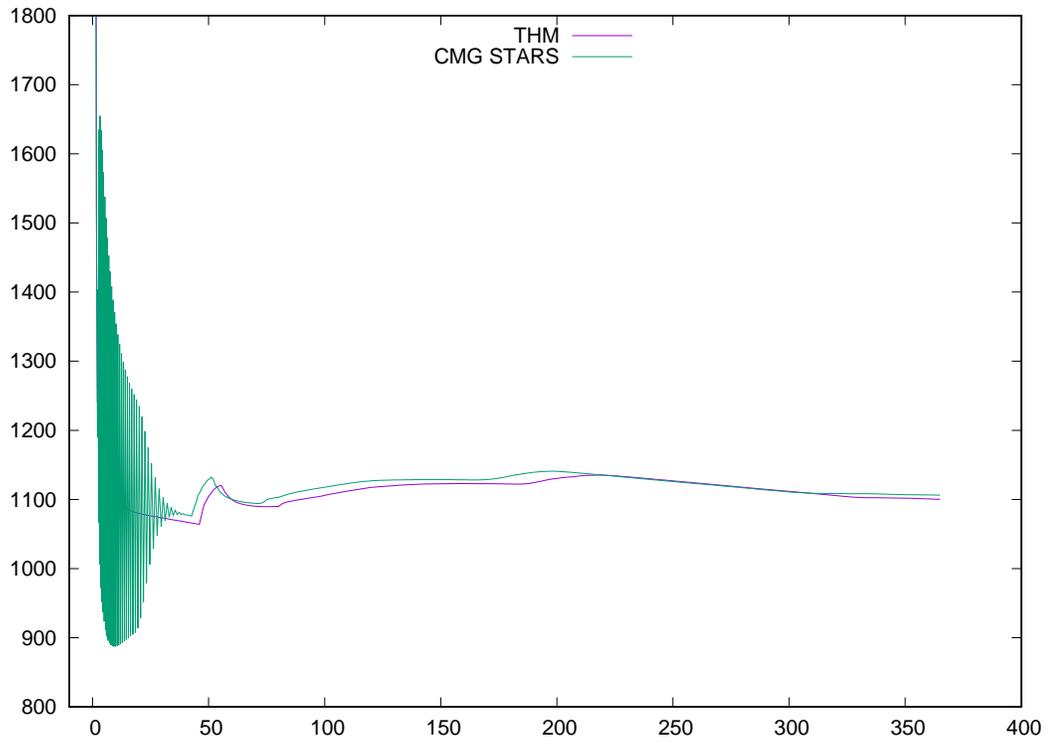}
    \caption{Example \ref{ex-htwell}, heater well: injection well, bottom hole pressure (psi)}
    \label{fig-ex-htwell-inj-bhp}
\end{figure}

\begin{figure}[H]
    \centering
    \includegraphics[width=0.53\linewidth, angle=270]{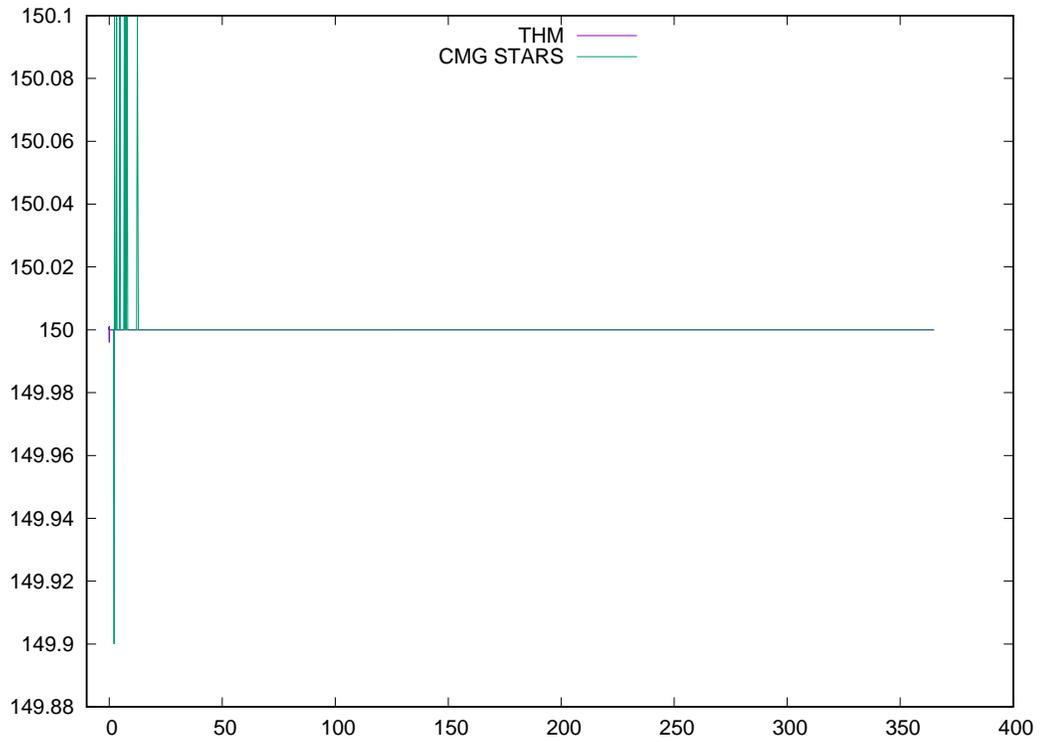}
    \caption{Example \ref{ex-htwell}, heater well: water injection rate (bbl/day)}
    \label{fig-ex-htwell-ir}
\end{figure}

\begin{figure}[H]
    \centering
    \includegraphics[width=0.53\linewidth, angle=270]{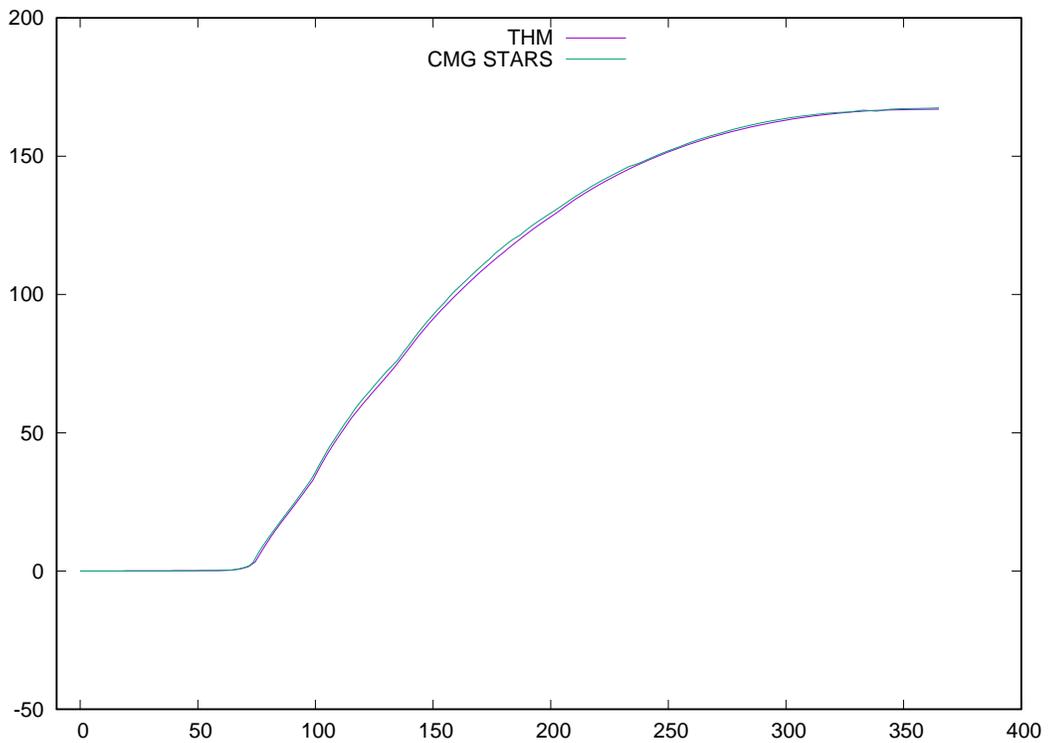}
    \caption{Example \ref{ex-htwell}, heater well: water production rate (bbl/day)}
    \label{fig-ex-htwell-pwr}
\end{figure}

\begin{figure}[H]
    \centering
    \includegraphics[width=0.53\linewidth, angle=270]{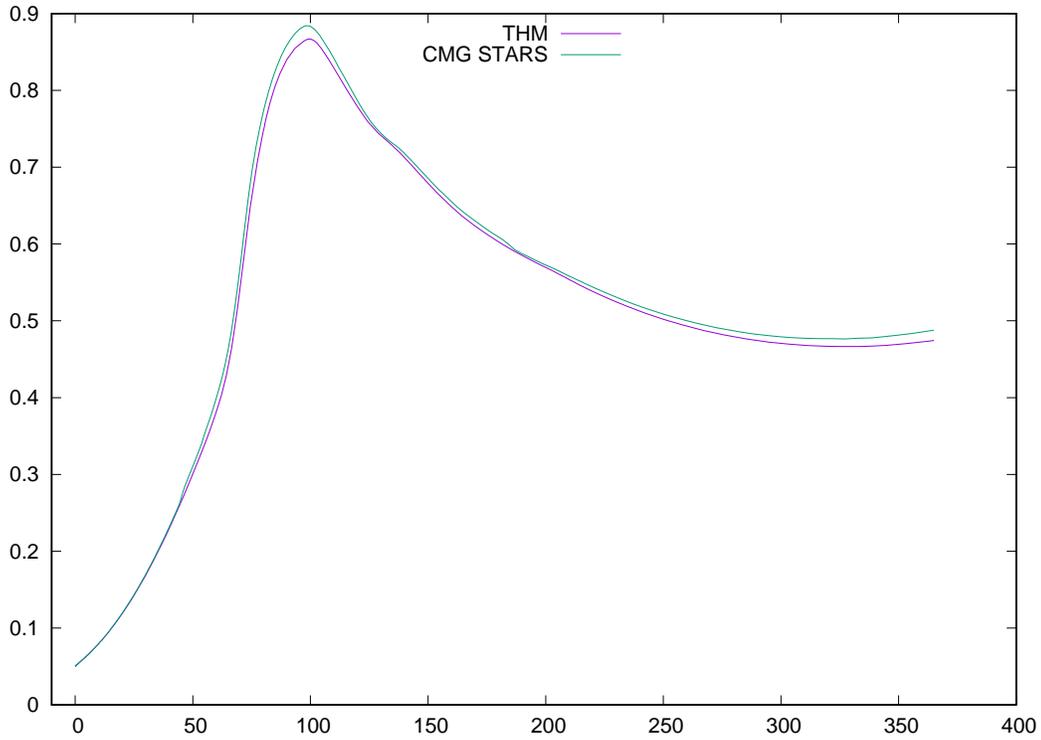}
    \caption{Example \ref{ex-htwell}, heater well: oil production rate (bbl/day)}
    \label{fig-ex-htwell-por}
\end{figure}

\begin{example}
    \normalfont
    \label{ex-htwell} The injection well operates at a fixed water rate 150 bbl/day. The injected water has a
    steam quality of 0.3 and temperature of 450 F. The production wells operate at fixed bottom hole pressure
    of 17  psi. The first production well uses temperature model at 600 F, and the second production well uses
    rate model with heat rate 3.4e6 Btu/day. The simulation period is 365 days. 
    Figure \ref{fig-ex-htwell-inj-bhp}, \ref{fig-ex-htwell-ir},
    \ref{fig-ex-htwell-pwr}, and \ref{fig-ex-htwell-por} present bottom hole pressure of injection well,
    injection rate, total water production rate and total oil production rate. All results are compared with
    CMG STARS.
\end{example}

Figure \ref{fig-ex-htwell-inj-bhp} and \ref{fig-ex-htwell-ir} show the bottom hole pressure and injection
rate, from which we can see the match is good except the first 20 days. CMG STARS shows convergence issue
while our simulator is more robust. The total water and oil production rates have good match, which are
demonstrated by Figure \ref{fig-ex-htwell-pwr} and Figure \ref{fig-ex-htwell-por}, respectively.

\begin{figure}[H]
    \centering
    \includegraphics[width=0.53\linewidth, angle=270]{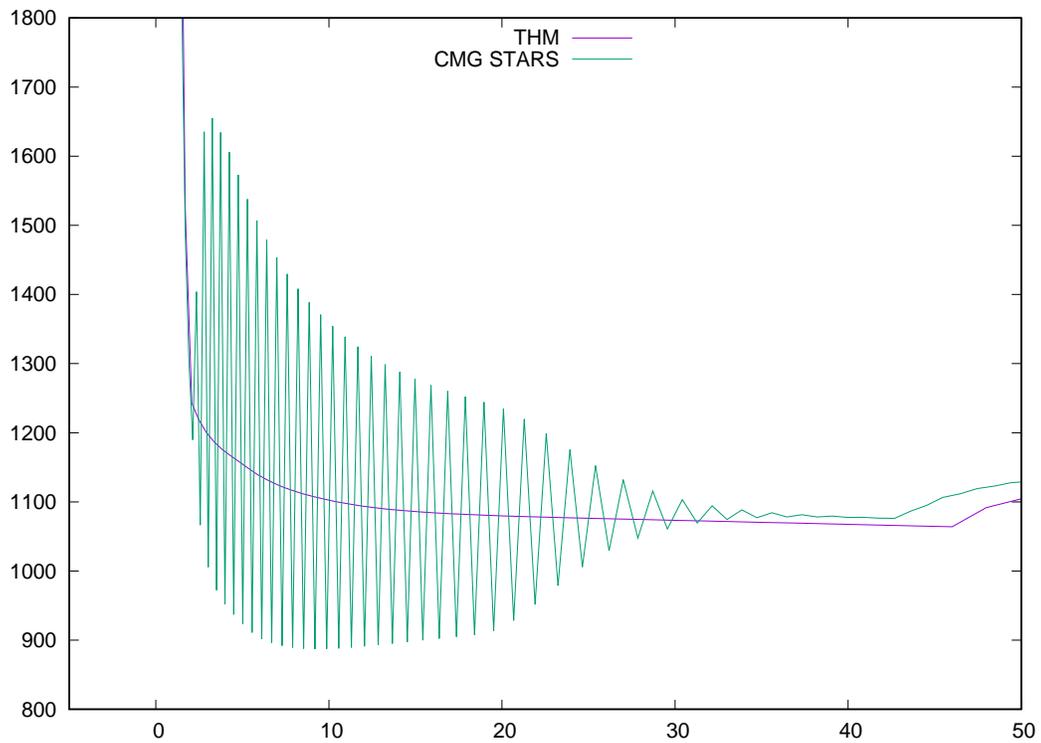}
    \caption{Example \ref{ex-htwell}, heater well: injection well, bottom hole pressure (psi), first 50 days}
    \label{fig-ex-htwell-inj-bhp-jump}
\end{figure}

\subsubsection{Heater Constraints}
The heater controls can be applied simultaneously in one thermal model. They can applied to different wells,
and for a well, it may use a combination of constant heater, convective heater, and heater well model.

\begin{figure}[H]
    \centering
    \includegraphics[width=0.53\linewidth, angle=270]{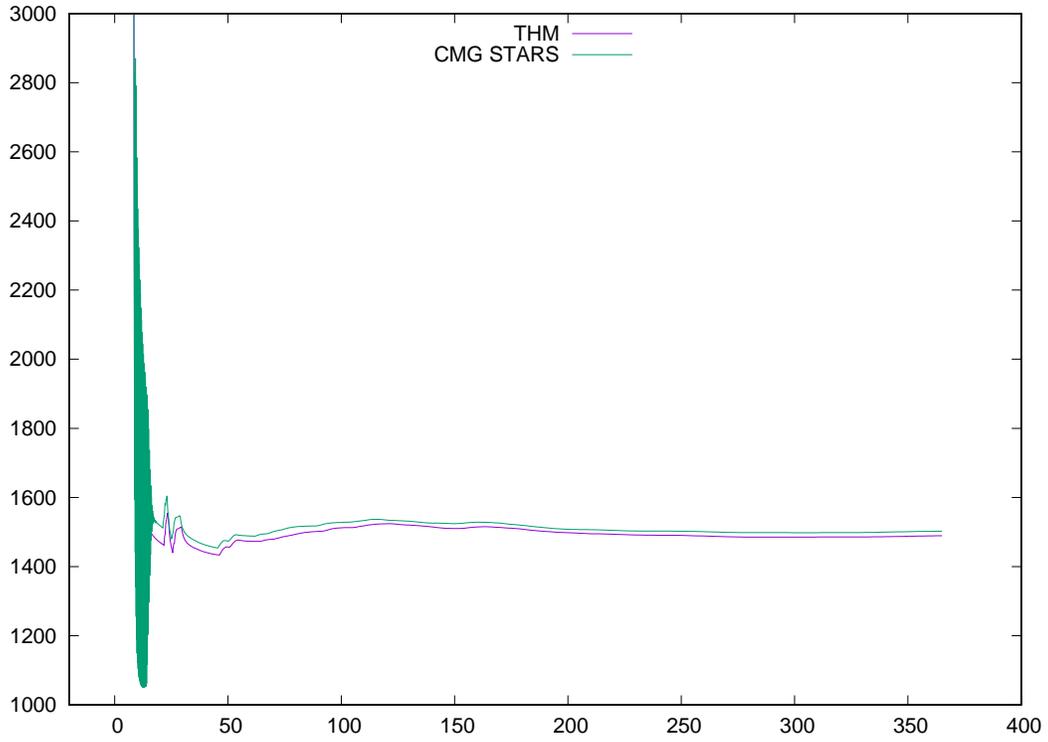}
    \caption{Example \ref{ex-heater}, combination of multiple heat models: injection well, bottom hole pressure (psi)}
    \label{fig-ex-heater-inj-bhp}
\end{figure}

\begin{figure}[H]
    \centering
    \includegraphics[width=0.53\linewidth, angle=270]{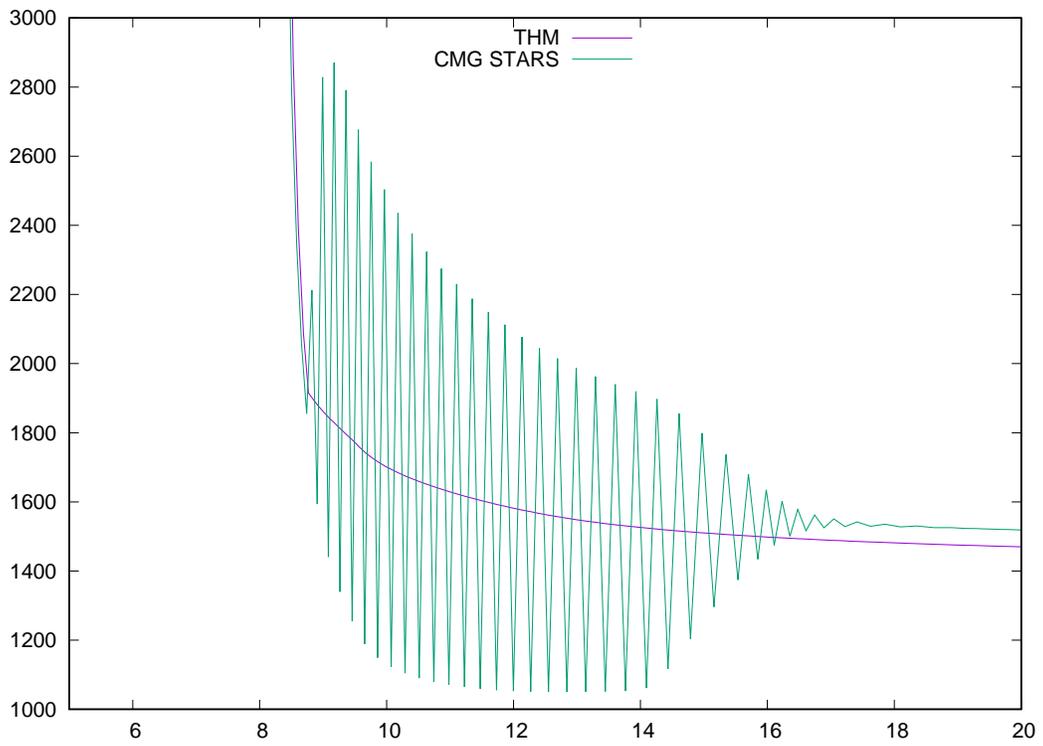}
    \caption{Example \ref{ex-heater}, combination of multiple heat models: injection well, bottom hole
    pressure (psi), first 20 days}
    \label{fig-ex-heater-inj-bhp-jump}
\end{figure}

\begin{figure}[H]
    \centering
    \includegraphics[width=0.53\linewidth, angle=270]{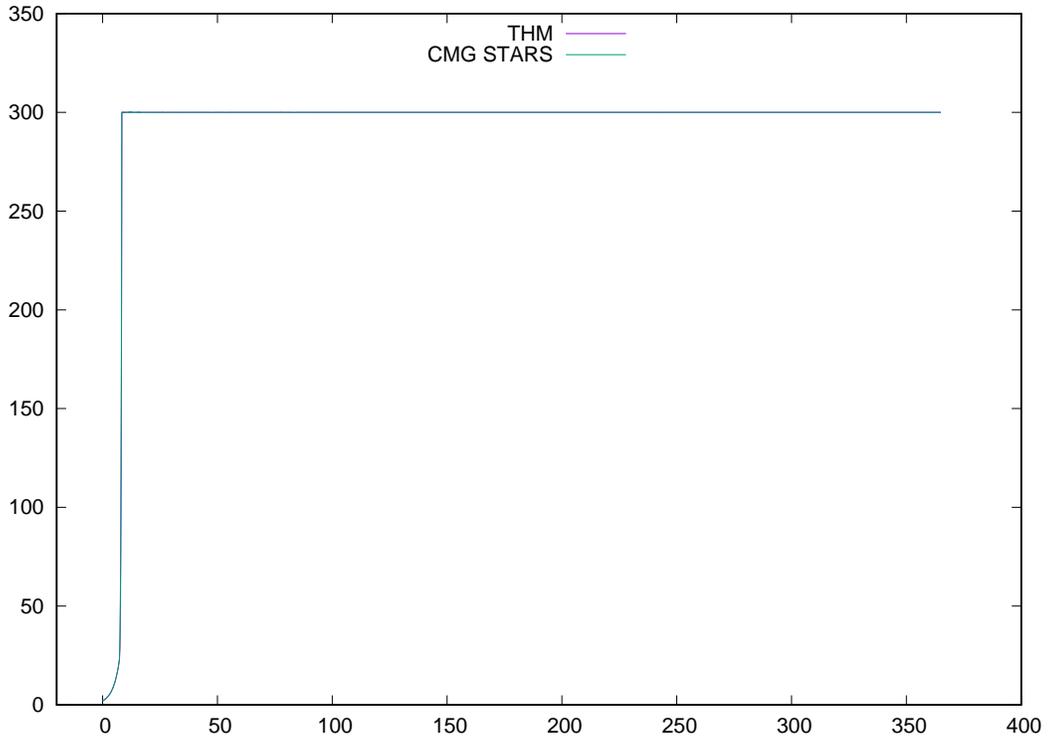}
    \caption{Example \ref{ex-heater}, combination of multiple heat models: total water injection rate (bbl/day)}
    \label{fig-ex-heater-ir}
\end{figure}

\begin{figure}[H]
    \centering
    \includegraphics[width=0.53\linewidth, angle=270]{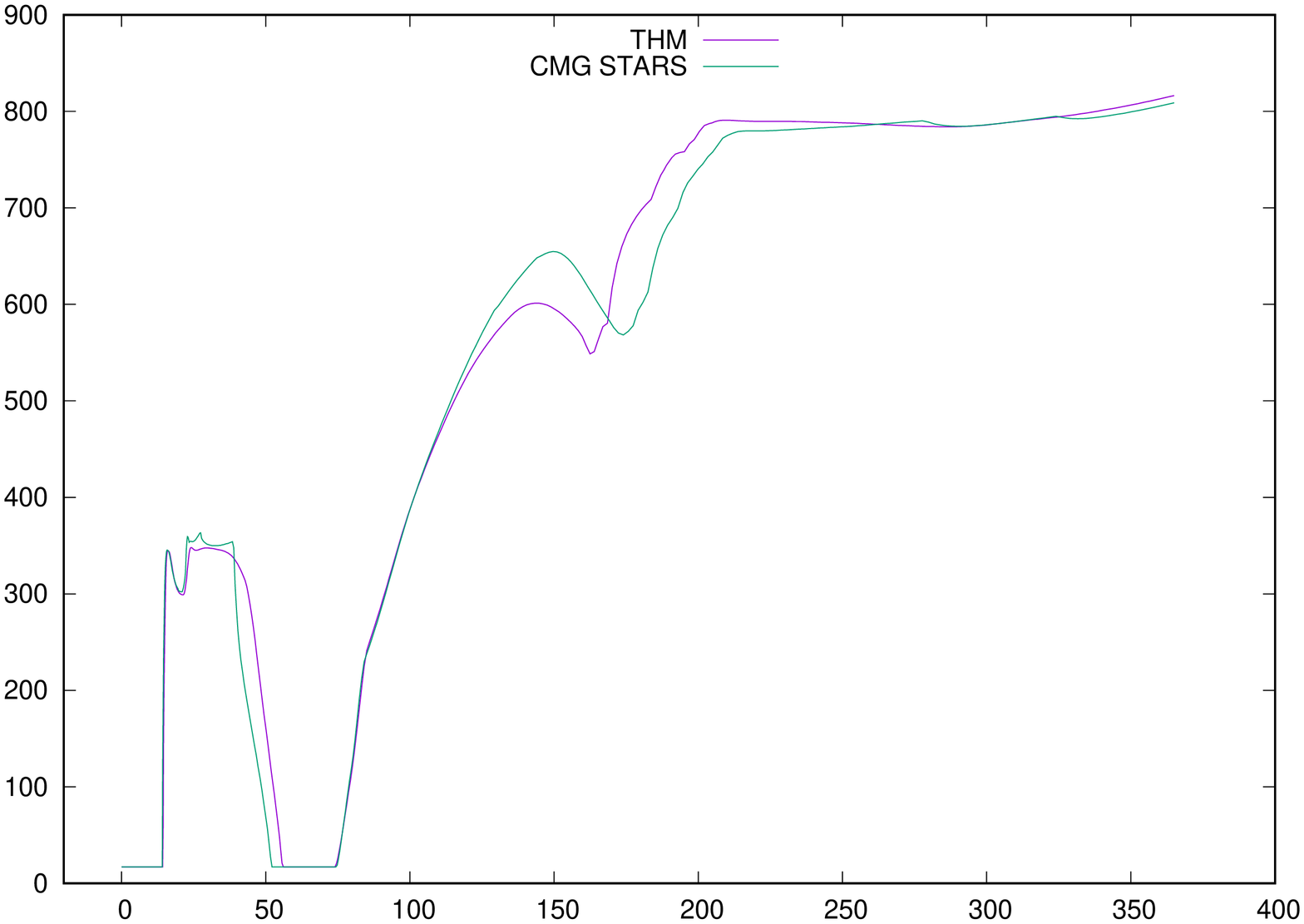}
    \caption{Example \ref{ex-heater}, combination of multiple heat models: first production well, bottom hole pressure (psi)}
    \label{fig-ex-heater-p2-bhp}
\end{figure}

\begin{figure}[H]
    \centering
    \includegraphics[width=0.53\linewidth, angle=270]{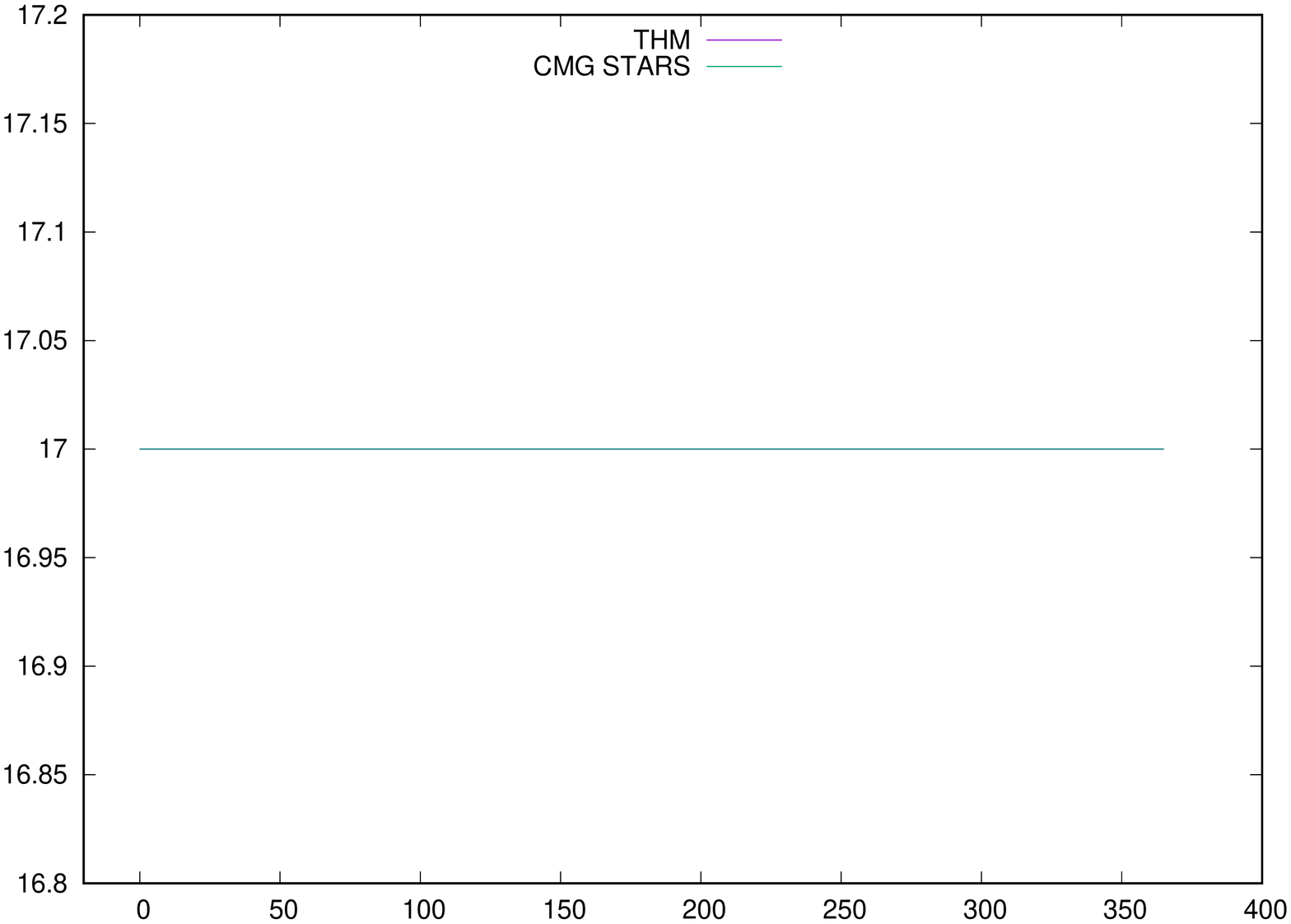}
    \caption{Example \ref{ex-heater}, combination of multiple heat models: second production well, bottom hole pressure (psi)}
    \label{fig-ex-heater-p3-bhp}
\end{figure}

\begin{figure}[H]
    \centering
    \includegraphics[width=0.53\linewidth, angle=270]{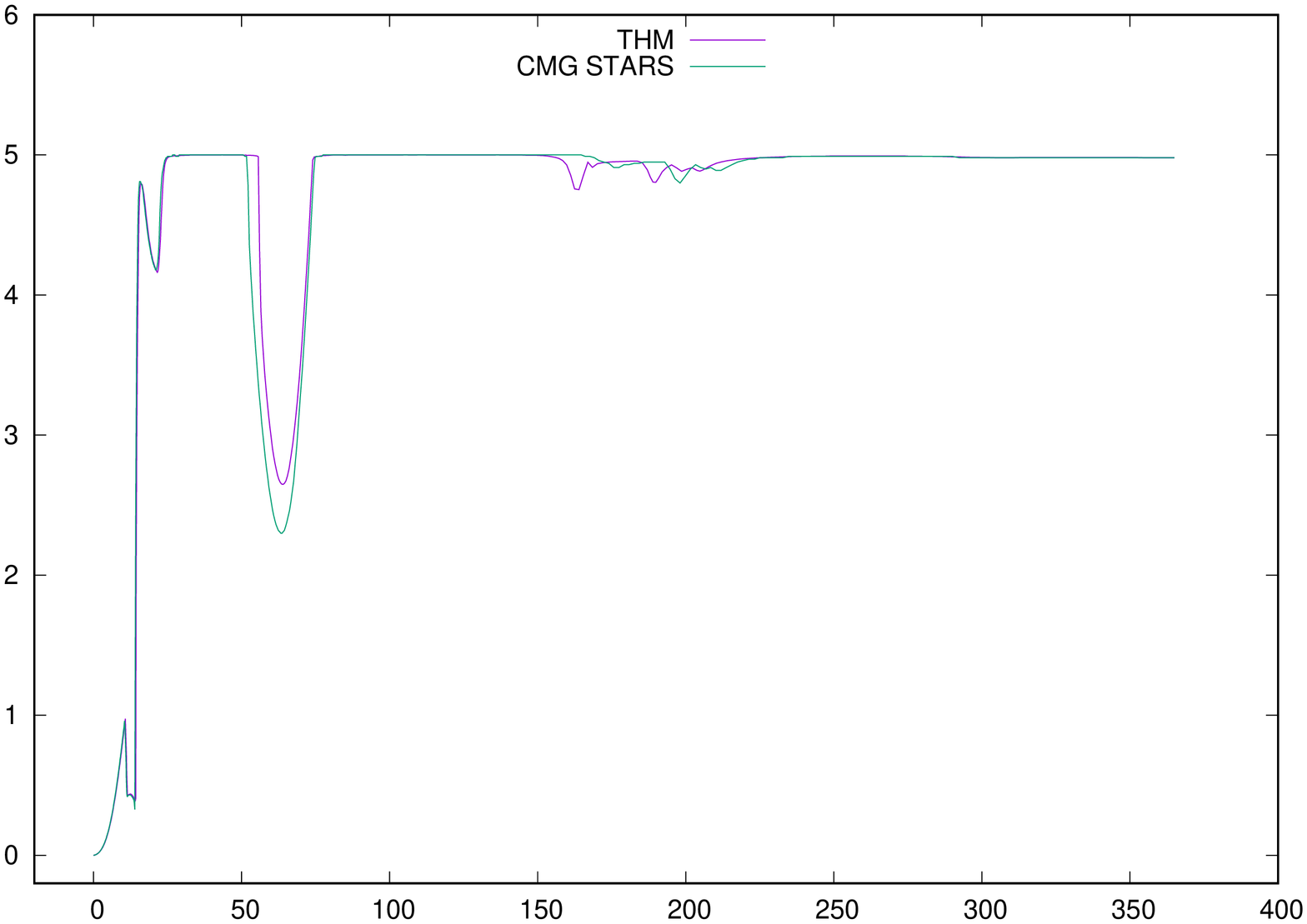}
    \caption{Example \ref{ex-heater}, combination of multiple heat models: water production rate (bbl/day), first production well}
    \label{fig-ex-heater-p2-pwr}
\end{figure}

\begin{figure}[H]
    \centering
    \includegraphics[width=0.53\linewidth, angle=270]{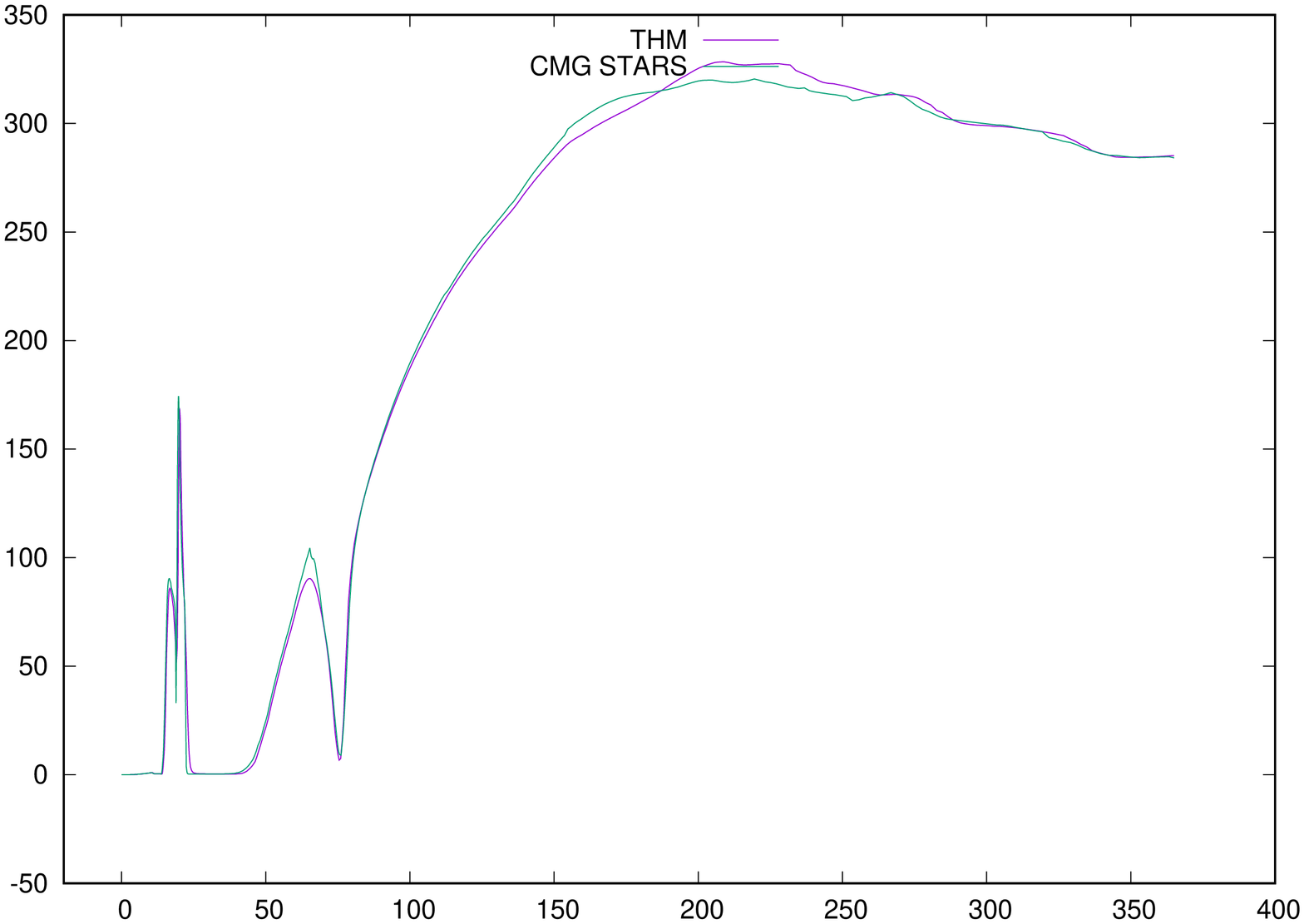}
    \caption{Example \ref{ex-heater}, combination of multiple heat models: water production rate (bbl/day), second production well}
    \label{fig-ex-heater-p3-pwr}
\end{figure}

\begin{figure}[H]
    \centering
    \includegraphics[width=0.53\linewidth, angle=270]{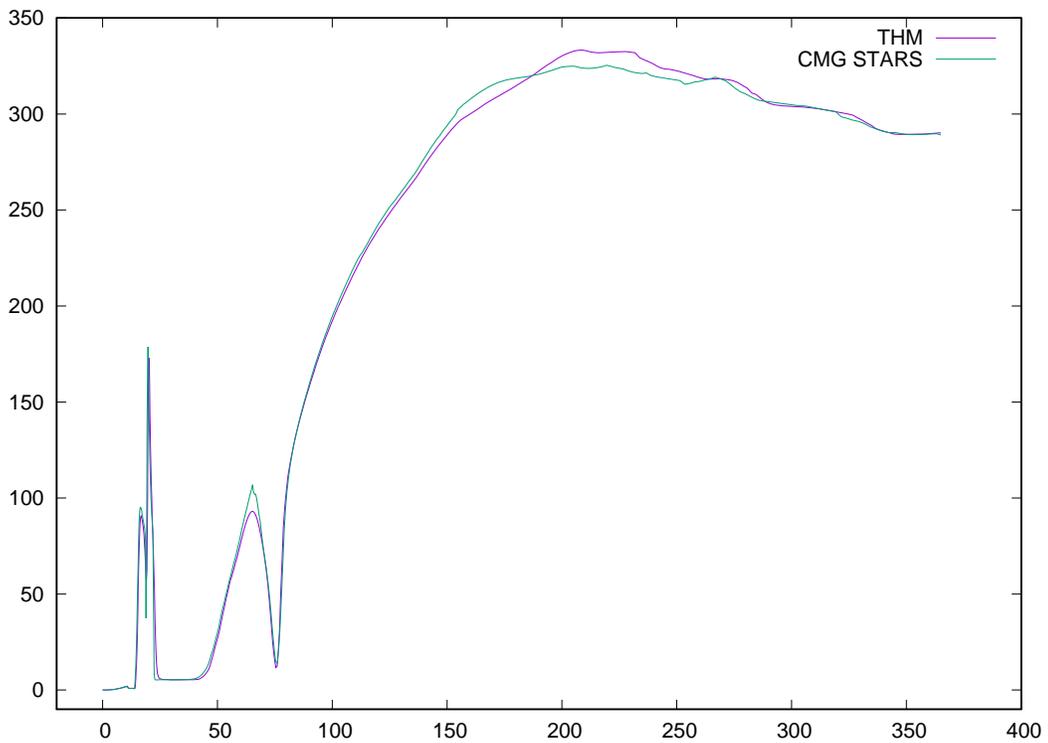}
    \caption{Example \ref{ex-heater}, combination of multiple heat models: total water production rate (bbl/day)}
    \label{fig-ex-heater-pwr}
\end{figure}

\begin{figure}[H]
    \centering
    \includegraphics[width=0.53\linewidth, angle=270]{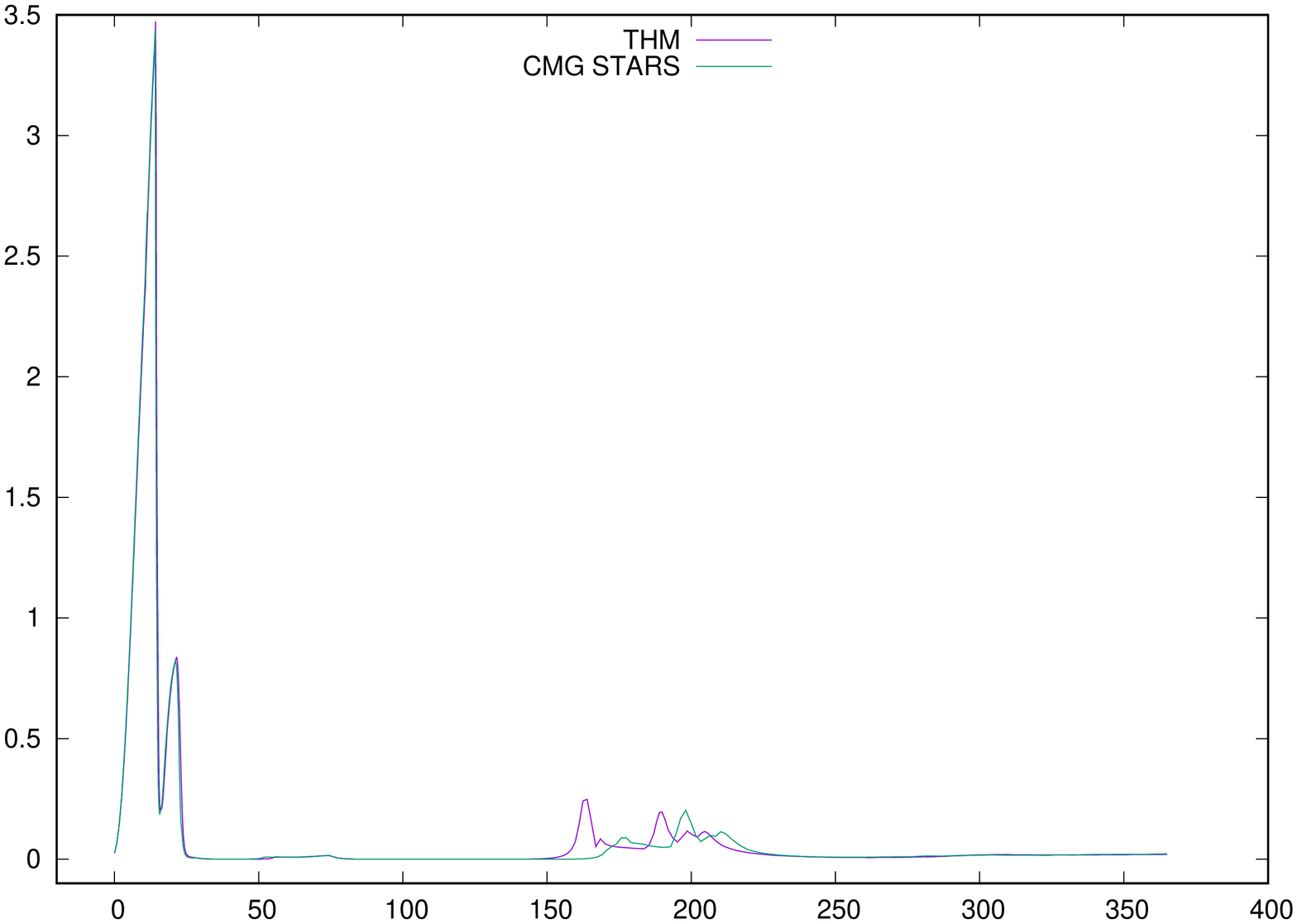}
    \caption{Example \ref{ex-heater}, combination of multiple heat models: oil production rate (bbl/day), first production well}
    \label{fig-ex-heater-p2-por}
\end{figure}

\begin{figure}[H]
    \centering
    \includegraphics[width=0.53\linewidth, angle=270]{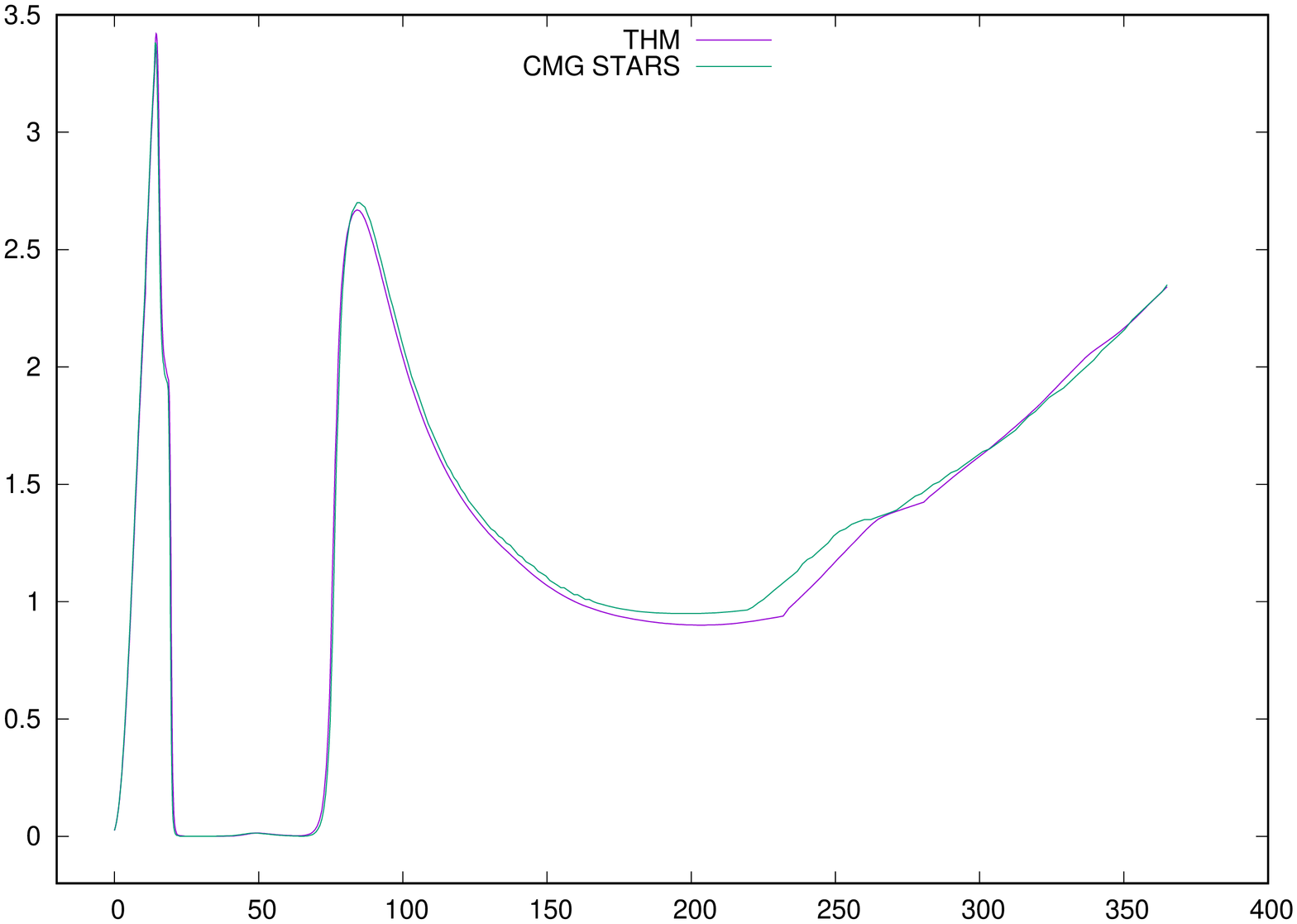}
    \caption{Example \ref{ex-heater}, combination of multiple heat models: oil production rate (bbl/day), second production well}
    \label{fig-ex-heater-p3-por}
\end{figure}

\begin{figure}[H]
    \centering
    \includegraphics[width=0.53\linewidth, angle=270]{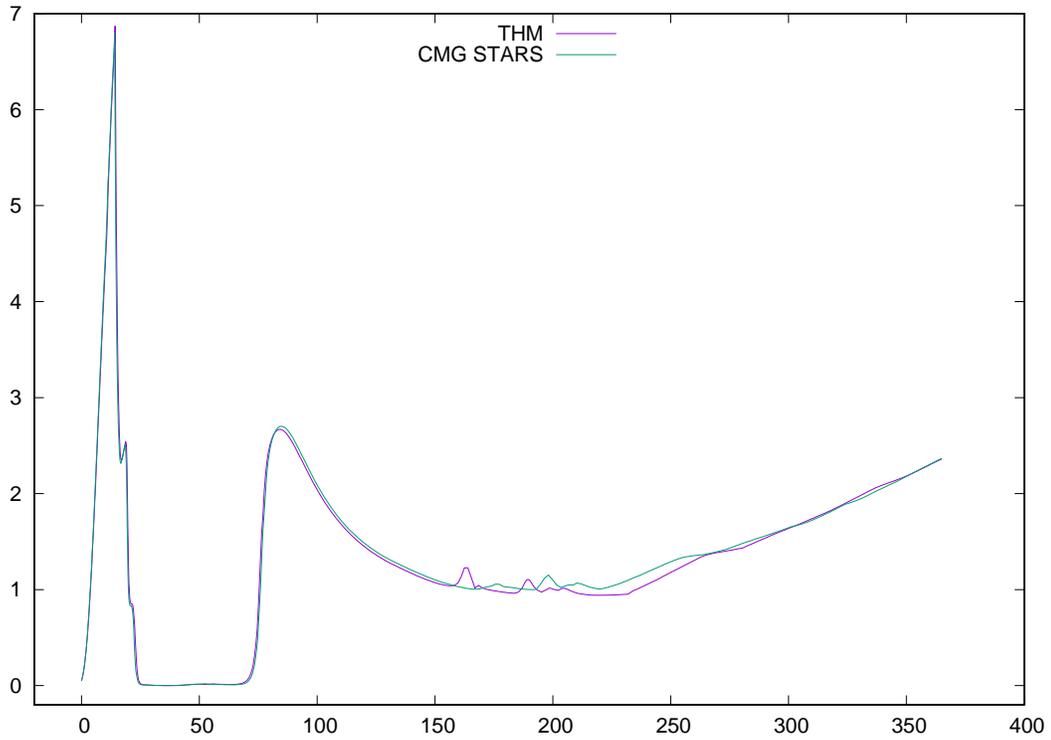}
    \caption{Example \ref{ex-heater}, combination of multiple heat models: total oil production rate (bbl/day)}
    \label{fig-ex-heater-por}
\end{figure}

\begin{example}
    \normalfont
    \label{ex-heater} The injection well operates at maximal injection rate of 300 bbl/day water and maximal
    bottom hole pressure of 5,000 psi. Its steam quality is 0.5. The first production well operates at minimal
    bottom hole pressure of 17 psi and maximal liquid rate of 5 bbl/day. The temperature heater model is
    applied with a specify temperature of 600 F. The second production well operates at minimal bottom hole
    pressure of 17 psi and maximal oil rate of 4 bbl/day. The dual rate/temperature model is applied with a
    specify heat rate of 3.4e6 Btu/day and a specify temperature of 611. Constant heat transfer model is
    applied to each perforation at a constant heat rate of 1e6 Btu/day. The convective heat transfer model is
    also applied to each perforation at 4e4 Btu/day-F and a temperature setpoint of 500. Results are shown
    from Figure \ref{fig-ex-heater-inj-bhp} to Figure \ref{fig-ex-heater-por}. Bottom hole pressure, water
    rate and oil rate of each well are compared with CMG STARS.
\end{example}

All results match CMG STARS well. For injection well, Figure \ref{fig-ex-heater-inj-bhp} shows CMG has
convergence issues while our simulator and numerical methods are more stable.

\subsubsection{Subcool Control}

\begin{figure}[H]
    \centering
    \includegraphics[width=0.53\linewidth, angle=270]{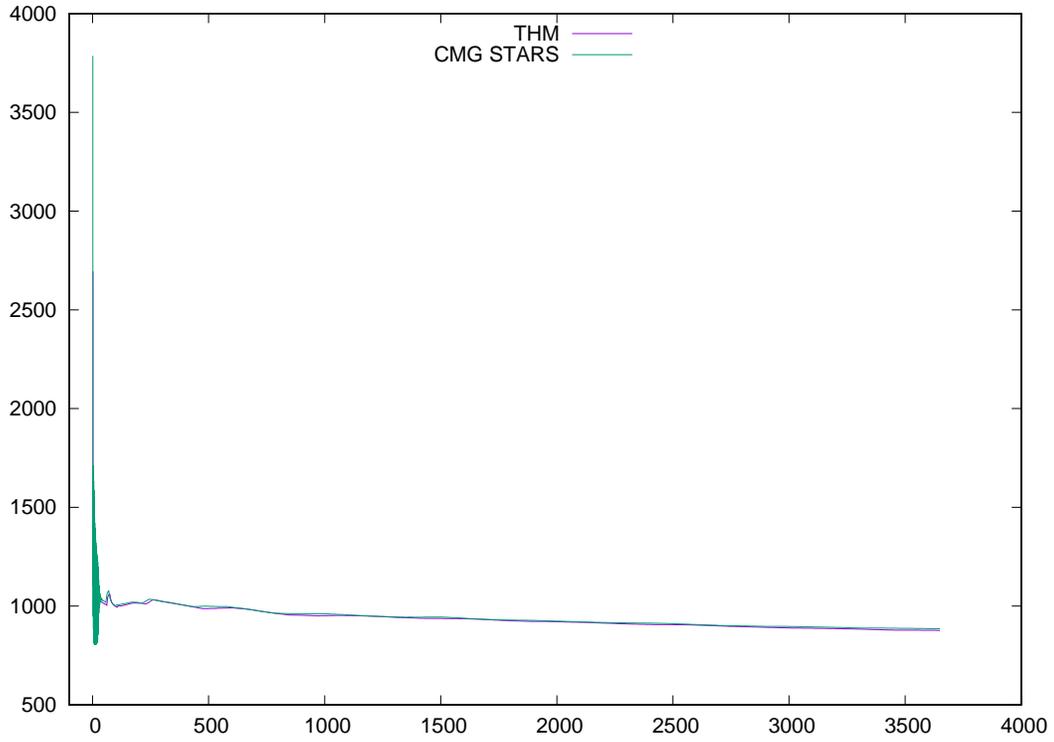}
    \caption{Example \ref{ex-subcool}, subcool (steam trap): injection well, bottom hole pressure (psi)}
    \label{fig-ex-subcool-inj-bhp}
\end{figure}

\begin{figure}[H]
    \centering
    \includegraphics[width=0.53\linewidth, angle=270]{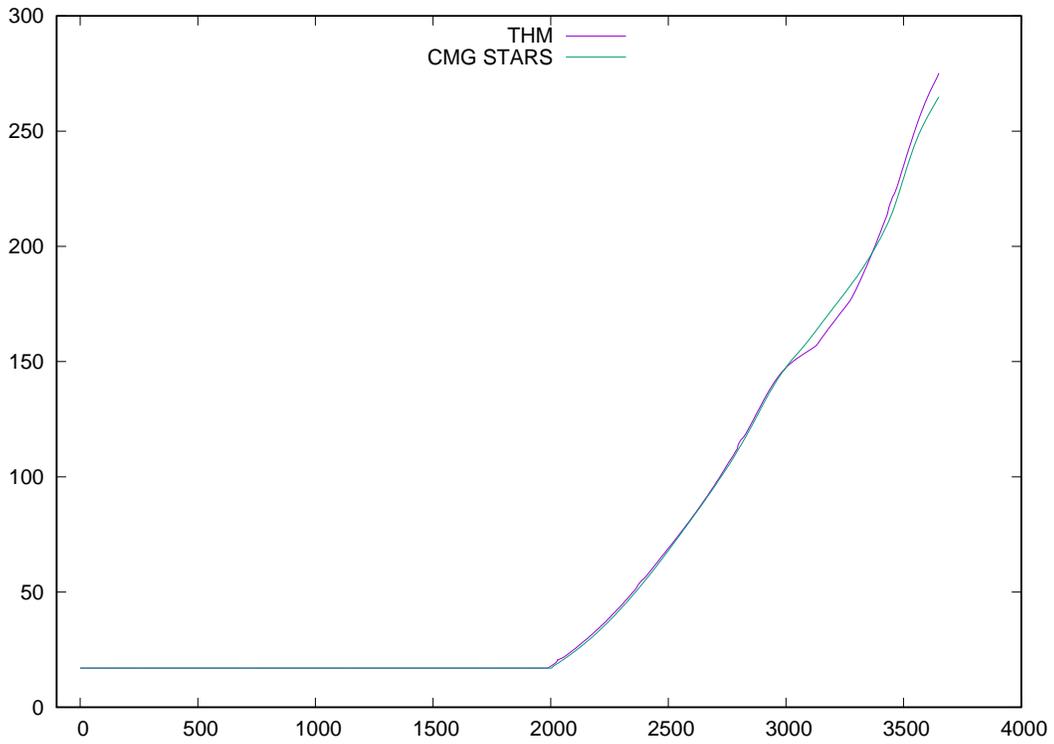}
    \caption{Example \ref{ex-subcool}, subcool (steam trap): first production well, bottom hole pressure (psi)}
    \label{fig-ex-subcool-p2-bhp}
\end{figure}

\begin{figure}[H]
    \centering
    \includegraphics[width=0.53\linewidth, angle=270]{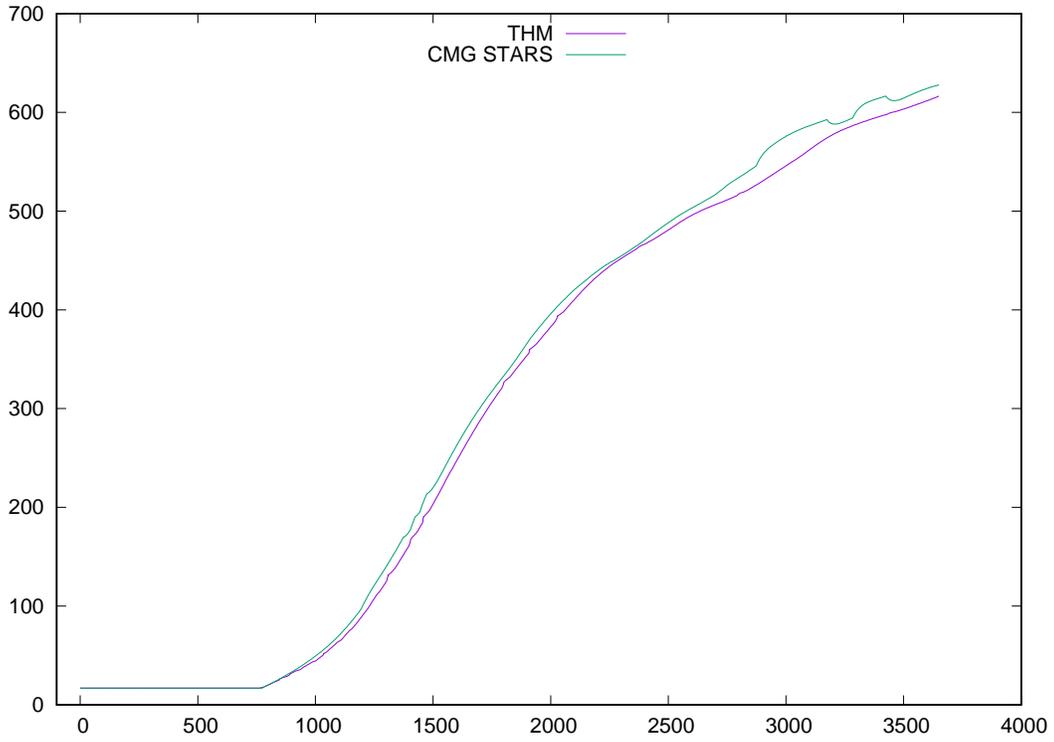}
    \caption{Example \ref{ex-subcool}, subcool (steam trap): second production well, bottom hole pressure (psi)}
    \label{fig-ex-subcool-p3-bhp}
\end{figure}

\begin{figure}[H]
    \centering
    \includegraphics[width=0.53\linewidth, angle=270]{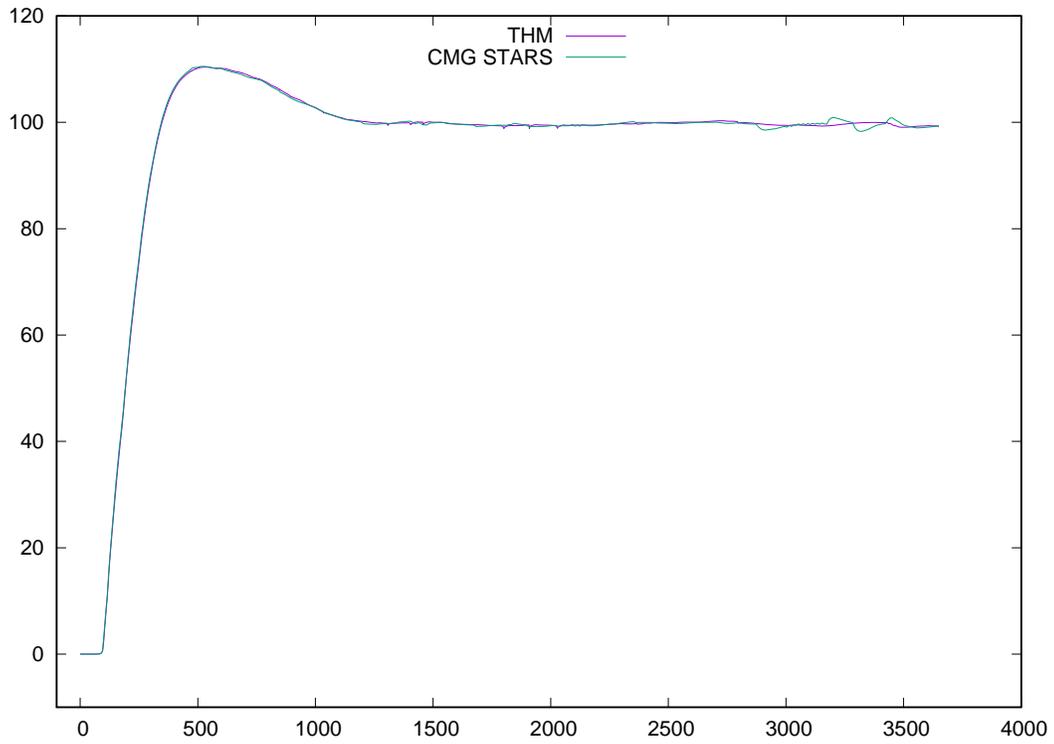}
    \caption{Example \ref{ex-subcool}, subcool (steam trap): water production rate (bbl/day)}
    \label{fig-ex-subcool-pwr}
\end{figure}

\begin{figure}[H]
    \centering
    \includegraphics[width=0.53\linewidth, angle=270]{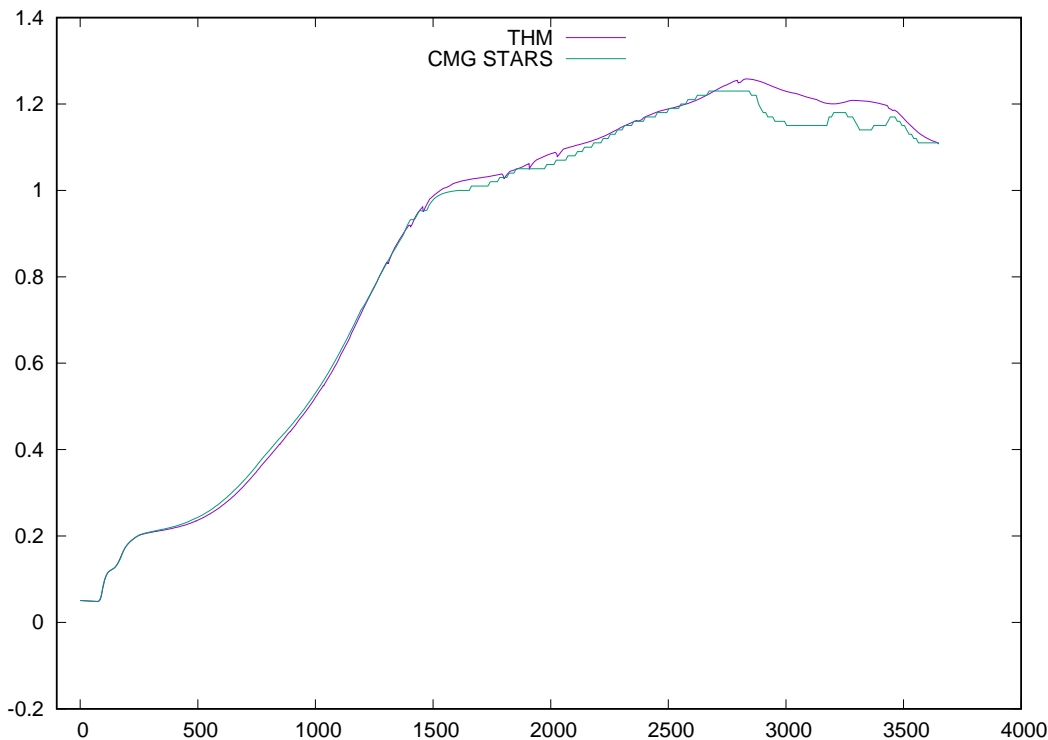}
    \caption{Example \ref{ex-subcool}, subcool (steam trap): oil production rate (bbl/day)}
    \label{fig-ex-subcool-por}
\end{figure}

\begin{example}
    \normalfont
    \label{ex-subcool} The injection well operates at fixed injection rate of 100 bbl/day.
    Both production wells operate at fixed bottom hole pressure of 17 psi. The steamtrap temperature
    differences are 20 F and 30 F respectively.
    Bottom hole pressure of each well is presented, and total water and oil production rates are also
    presented, from Figure \ref{fig-ex-subcool-inj-bhp} to Figure \ref{fig-ex-subcool-por}.
\end{example}

Figure \ref{fig-ex-subcool-inj-bhp} is bottom hole pressure of injection well, and our results match CMG STARS
exactly. Figure \ref{fig-ex-subcool-p2-bhp} and Figure \ref{fig-ex-subcool-p3-bhp} show that the steamtrap
works, as steam is injected into reservoir to heat reservoir and fluid, their temperature increases. The
steamtrap works by increasing the wellbore pressure to prevent live steam production. The water and oil
production also match CMG STARS well.

\subsection{Numerical Performance}

\begin{example}
    \normalfont
    \label {num-sagd} This example tests a SAGD model with 25 well pairs, which includes one water component, one heavy
    component, one light component and two 
    inert gase components, and their properties are the same as Example \ref{val-gas}. The grid dimension is $100
    \times 100 \times 6$ and gird size is $10 ft \times 10 ft \times 1 ft$. 
    The simulation time is 200 days and the maximal time step is 10 days. The Newton tolerance is 1e-3 and its maximal
    iterations are 15. The linear solver is BICGSTAB, its tolerance is also 1e-3 and its maximal iterations
    are 60. GJE is the decoupling method. Table \ref{table-num-sagd-well} provides the well info. All
    injectors operate at 3 bbl/day water injection with steam quality of 0.2 and temperature of 450 F. All
    producers operate at bottom hole pressure of 2000 psi and steam trap temperature difference of 20 F. The
    equivalent CMG STARS model is simulated. However, after 12 hours run, CMG STARS always has time steps
    around 1e-4 day, and it simulates 0.3269 days after 415 time steps, so we have to terminate it. Numerical
    summaries of our simulator are shown in Table \ref{table-num-sagd}. Since the model is small, only one
    computing node is employed. 
\end{example}

\begin{table}[!htb]
    \centering
    \begin{tabular}{|c|c|c|c|}
        \hline
        Injector perforation & Well index &  Producer perforation   &  Well index     \\
        \hline
    1:100  2 3 & 1e5 &    1:100  2 6 & 1e5 \\
    1:100  6 3 & 1e5 &    1:100  6 6 & 1e5 \\
    1:100 10 3 & 1e5 &    1:100 10 6 & 1e5 \\
    1:100 14 3 & 1e5 &    1:100 14 6 & 1e5 \\
    1:100 18 3 & 1e5 &    1:100 18 6 & 1e5 \\
    1:100 22 3 & 1e5 &    1:100 22 6 & 1e5 \\
    1:100 26 3 & 1e5 &    1:100 26 6 & 1e5 \\
    1:100 30 3 & 1e5 &    1:100 30 6 & 1e5 \\
    1:100 34 3 & 1e5 &    1:100 34 6 & 1e5 \\
    1:100 38 3 & 1e5 &    1:100 38 6 & 1e5 \\
    1:100 42 3 & 1e5 &    1:100 42 6 & 1e5 \\
    1:100 46 3 & 1e5 &    1:100 46 6 & 1e5 \\
    1:100 50 3 & 1e5 &    1:100 50 6 & 1e5 \\
    1:100 54 3 & 1e5 &    1:100 54 6 & 1e5 \\
    1:100 58 3 & 1e5 &    1:100 58 6 & 1e5 \\
    1:100 62 3 & 1e5 &    1:100 62 6 & 1e5 \\
    1:100 66 3 & 1e5 &    1:100 66 6 & 1e5 \\
    1:100 70 3 & 1e5 &    1:100 70 6 & 1e5 \\
    1:100 74 3 & 1e5 &    1:100 74 6 & 1e5 \\
    1:100 78 3 & 1e5 &    1:100 78 6 & 1e5 \\
    1:100 82 3 & 1e5 &    1:100 82 6 & 1e5 \\
    1:100 86 3 & 1e5 &    1:100 86 6 & 1e5 \\
    1:100 90 3 & 1e5 &    1:100 90 6 & 1e5 \\
    1:100 94 3 & 1e5 &    1:100 94 6 & 1e5 \\
    1:100 98 3 & 1e5 &    1:100 98 6 & 1e5 \\
        \hline
    \end{tabular}
    \caption{Well info of Example \ref{num-sagd}}
    \label{table-num-sagd-well}
\end{table}

\begin{table}[!htb]
    \centering
    \begin{tabular}{|c|c|c|c|c|c|c|}
        \hline
        CPU cores  &  \# Time steps & \# Newton & Avg. Newton & \# Linear solver & Avg. Linear & Time    \\
        \hline
         4         &   96(4)        &  314      & 3.27        & 5587             & 17.80       & 1420.63 \\
         8         &  100(6)        &  338      & 3.38        & 6157             & 18.22       & 816.69  \\
         16        &  101(4)        &  326      & 3.23        & 6215             & 19.06       & 558.22  \\
        \hline
    \end{tabular}
    \caption{Numerical summaries of Example \ref{num-sagd}}
    \label{table-num-sagd}
\end{table}

Table \ref{table-num-sagd} provides numerical summaries for time steps (and time cut), total Newton
iterations, total linear solver iterations, total simulation time, average Newton iterations per time step,
average linear iterations per Newton iteration. As expected, when more CPU cores (MPIs) are used, time steps
and linear iterations increase. Even through, the results show that our numerical methods are effective, which
can solve a time step in less than 4 Newton iterations and solve a linear system in less than 20 iterations.
When more CPU cores are used, the simulation time is cut, which shows that parallel computing is a powerful
tool for reservoir simulation.

\begin{example}
    \normalfont
    \label {num-light} This example tests one water component, one heavy component and one light component.
    SAGD process with 756 well pairs is simulated.
    The grid has a dimension of $60 \times 220 \times 85$ and size of $20 ft \times 10 ft
    \times 1 ft$. All wells are horizontal wells along $x$ direction, if the index of $y$ direction of a grid
    block equals to 3, 7, 11, 15, 19, 23, 27, 31, 35, 39, 43, 47, 51, 55, 59, 63, 67, 71, 75, 79, 83, 87, 91, 95,
    99, 103, 107, 111, 115, 119, 123, 127, 131, 135, 139, 143, 147, 151, 155, 159, 163, 167, 171, 175, 179, 183,
    187, 191, 195, 199, 203, 207, 211, or 215, and the index of $z$ direction equals to 4, 10, 16, 22, 28, 34, 40,
    46, 52, 58, 64, 70, 76, or 82, then an injection well is defined. For example, (1:60, 3, 10) defines an
    injection well at (3,10) of $yz$-plance, and its perforations are from 1 to 60. This defines 756 injection
    wells. A production well is defined two blocks under an injection well. Therefore, 756 well pairs and total
    1512 wells are defined in the model. All injection wells operate at 10 bbl/day water injection, with a steam
    quality of 0.2 and temperature of 450F. All production wells operate at fixed bottom hole pressure of 100 psi.
    Each perforation of an injection well is heated at rate of 1e5 btu/day. The model file has around 20,000
    lines.  Their properties are the same as Example \ref{val-light}. The simulation time is 100 days. 8 CPU
    cores (8 MPIs) are employed. The Newton tolerance is 1e-4 and its maximal iterations are 15. The linear
    solver is BICGSTAB, its tolerance is also 1e-4 and its maximal iterations are 100.
\end{example}

\begin{table}[!htb]
    \centering
    \begin{tabular}{|c|c|c|c|c|}
        \hline
        Preconditioner &  Decoupling   &  Time steps  & \# Newton & \# Linear solver    \\
        \hline
        CPR-FP         &  NONE         &  NA          & NA        & NA  \\
        CPR-FP         &  FRS          &  NA          & NA        & NA \\
        CPR-FP         &  DRS          &  NA          & NA        & NA \\
        CPR-FP         &  ABF          &  NA          & NA        & NA \\
        CPR-FP         &  GJE          &  101 (12)    & 598       & 8856 \\
        CPR-FP         &  DRS+ABF      &  NA          & NA        & NA  \\
        CPR-FP         &  DRS+GJE      &  95 (10)     & 559       & 8090 \\
        CPR-FP         &  FRS+ABF      &  NA          & NA        & NA \\
        CPR-FP         &  FRS+GJE      & 96 (10) & 552 & 7608 \\
        \hline
        CPR-PF         &  NONE         &   NA          & NA        & NA \\
        CPR-PF         &  FRS          &   NA          & NA        & NA \\
        CPR-PF         &  DRS          &   NA          & NA        & NA \\
        CPR-PF         &  ABF          & 125 (15)     & 738        & 18680 \\
        CPR-PF         &  GJE          & 105 (10)     & 585        & 11263 \\
        CPR-PF         &  DRS+ABF      & 103 (9)      & 563        & 12642  \\
        CPR-PF         &  DRS+GJE      & 109 (11)     & 636        & 12947  \\
        CPR-PF         &  FRS+ABF      & NA           & NA        & NA \\
        CPR-PF         &  FRS+GJE      & 109 (12)     & 640 & 13050 \\
        \hline
        CPR-FPF        &  NONE         &  NA          & NA        & NA \\
        CPR-FPF        &  FRS          &  NA          & NA        & NA \\
        CPR-FPF        &  DRS          &  NA          & NA        & NA  \\
        CPR-FPF        &  ABF          &  NA & NA & NA \\
        CPR-FPF        &  GJE          & 97 (10) & 566 &  7887 \\
        CPR-FPF        &  DRS+ABF      &  NA          & NA        & NA  \\
        CPR-FPF        &  DRS+GJE      &  97 (10)     & 566       & 7813 \\
        CPR-FPF        &  FRS+ABF      & NA & NA & NA \\
        CPR-FPF        &  FRS+GJE      & 97 (10)& 559 & 7666 \\
        \hline
    \end{tabular}
    \caption{Numerical summary of Example \ref{num-light}}
    \label{table-num-light}
\end{table}

Table \ref{table-num-light} presents numerical results, including preconditioners, decoupling methods, total
time steps and time cuts, total Newton iterations, and total linear iteration. Here \verb|NA| means the
combination fails to simulate the model. The results clearly show that a proper decoupling method is critical
to the success of linear solver and CPR-type precondtioners. The GJE decoupling and the FRS+GJE decoupling
work better than the ABF decoupling.

\begin{example}
    \label{num-gas} \normalfont This example tests one water component, one heavy component, one light component and two
    inert gase components.  Their properties are the same as Example \ref{val-gas}.  The grid and well
    configurations are the same as Example \ref{num-light}.  The simulation time is 100 days. 8 CPU cores (8
    MPIs) are employed. The Newton tolerance is 1e-4 and its maximal iterations are 15. The linear solver is
    BICGSTAB, its tolerance is also 1e-4 and its maximal iterations are 100.
\end{example}

\begin{table}[!htb]
    \centering
    \begin{tabular}{|c|c|c|c|c|}
        \hline
        Preconditioner &  Decoupling   &  Time steps  & \# Newton & \# Linear solver    \\
        \hline
        CPR-FP         &  NONE         &   NA          & NA        & NA \\
        CPR-FP         &  FRS          &   NA          & NA        & NA \\
        CPR-FP         &  DRS          &   NA          & NA        & NA \\
        CPR-FP         &  ABF          &  NA          & NA        & NA \\
        CPR-FP         &  GJE          &   133 (17)   & 588       & 8625 \\
        CPR-FP         &  DRS+ABF      &  NA          & NA        & NA  \\
        CPR-FP         &  DRS+GJE      &  141 (15)    & 771       & 17780 \\
        CPR-FP         &  FRS+ABF      &  NA & NA & NA \\
        CPR-FP         &  FRS+GJE      &  137 (15) & 750 & 17656 \\
        \hline
        CPR-PF         &  NONE         &   NA          & NA        & NA \\
        CPR-PF         &  FRS          &   NA          & NA        & NA \\
        CPR-PF         &  DRS          &   NA          & NA        & NA \\
        CPR-PF         &  ABF          &  NA & NA & NA \\
        CPR-PF         &  GJE          &  153 (10)  & 729 & 20324 \\
        CPR-PF         &  DRS+ABF      &  NA          & NA        & NA  \\
        CPR-PF         &  DRS+GJE      &  158 (14)    & 780       & 22515  \\
        CPR-PF         &  FRS+ABF      &  NA & NA & NA \\
        CPR-PF         &  FRS+GJE      &  144(10)     & 593       & 3380 \\
        \hline
        CPR-FPF         &  NONE         &   NA          & NA        & NA \\
        CPR-FPF         &  FRS          &   NA          & NA        & NA \\
        CPR-FPF         &  DRS          &   NA          & NA        & NA \\
        CPR-FPF         &  ABF          &  NA & NA & NA \\
        CPR-FPF         &  GJE          & 145 (17)  & 805 & 18441 \\
        CPR-FPF         &  DRS+ABF      & NA          & NA        & NA   \\
        CPR-FPF         &  DRS+GJE      & NA          & NA        & NA   \\
        CPR-FPF         &  FRS+ABF      & NA & NA & NA \\
        CPR-FPF         &  FRS+GJE      & 142 (15) & 803 & 18461 \\
        \hline
    \end{tabular}
    \caption{Numerical summary of Example \ref{num-gas}}
    \label{table-num-gas}
\end{table}

Table \ref{table-num-gas} presents numerical results, including preconditioners, decoupling methods, total
time steps and time cuts, total Newton iterations, and total linear iteration. Again, \verb|NA| means the
combination fails to simulate the model.
The linear systems from this example are much larger than those from previous example, and they are more
difficult to solve. The results show that all ABF decoupling and FRS+ABF methods fail.

\begin{example}
    \normalfont
    \label {num-11m} This example tests one water component, one heavy component and one light component.
    SAGD process with 7406 well pairs (14812 wells, 7406 injectors and 7406 producers) is simulated.
    The grid has a dimension of $60 \times 2200 \times 85$, 11 million grid blocks, and size of $20 ft \times 10 ft
    \times 2 ft$. All wells are horizontal wells along $x$ direction.
    All injection wells operate at 5 bbl/day water injection, with a steam
    quality of 0.2 and temperature of 450F. All production wells operate at fixed bottom hole pressure of 300 psi.
    All wells are modelled by implicit method. The model file has around 185,000 lines. 
    Their properties are the same as Example \ref{val-light}. The simulation time is 100 time steps due to
    system running time limit. The initial time step is 1e-6 days.
    10 nodes and 200 CPU cores (200 MPIs) are employed on Niagara, Compute Canada.
    The Newton tolerance is 1e-3 and its maximal iterations are 15. The linear
    solver is BICGSTAB, its tolerance is also 1e-4 and its maximal iterations are 100. The maximal changes in
    a time step for pressure, saturation, mole fraction and temperature are 500 psi, 0.1, 0.1 and 15 F.
\end{example}

\begin{table}[!htb]
    \centering
    \begin{tabular}{|c|c|c|c|c|}
        \hline
        Preconditioner &  Decoupling   &  Time steps  & \# Newton & \# Linear solver    \\
        \hline
        CPR-FP         &  GJE          & 100       & 284       & 1245 \\
        CPR-FP         &  FRS+GJE      & 100       & 267       & 1194 \\
        \hline
        CPR-PF         &  GJE          & 100       & 299       & 1540 \\
        CPR-PF         &  FRS+GJE      & 100       & 282       & 1183 \\
        \hline
        CPR-FPF        &  GJE          & 100       & 308       & 1801 \\
        CPR-FPF        &  FRS+GJE      & 100       & 298       & 1255 \\
        \hline
    \end{tabular}
    \caption{Numerical summary of Example \ref{num-11m}}
    \label{table-num-11m}
\end{table}

Table \ref{table-num-11m} shows that all tests pass. The Newton method converges in around three iterations,
while linear solver converges in four to five iterations in average. For a specific preconditioner, 
the FRS+GJE decoupling method is always better than the GJE method.

\begin{example}
    \normalfont
    \label {num-44m} This example tests one water component, one heavy component and one light component.
    SAGD process with 15106 well pairs (30212 wells, 15106 injectors and 15106 producers) is simulated.
    The grid has a dimension of $60 \times 4400 \times 85$ and size of $20 ft \times 10 ft
    \times 4 ft$. All wells are horizontal wells along $x$ direction.
    All injection wells operate at 5 bbl/day water injection, with a steam
    quality of 0.2 and temperature of 450F. All production wells operate at fixed bottom hole pressure of 300 psi.
    All wells are modelled by implicit method. The model file has around 377,000 lines. 
    Their properties are the same as Example \ref{val-light}. The simulation time is 100 time steps due to
    system running time limit. The initial time step is 1e-6 days.
    20 nodes and 400 CPU cores (400 MPIs) are employed on Niagara, Compute Canada.
    The Newton tolerance is 1e-3 and its maximal iterations are 15. The linear
    solver is BICGSTAB, its tolerance is also 1e-3 and its maximal iterations are 100. The maximal changes in
    a time step for pressure, saturation, mole fraction and temperature are 500 psi, 0.1, 0.1 and 15 F.
\end{example}

\begin{table}[!htb]
    \centering
    \begin{tabular}{|c|c|c|c|c|}
        \hline
        Preconditioner &  Decoupling   &  Time steps  & \# Newton & \# Linear solver    \\
        \hline
        CPR-FP         &  GJE          & 100       & 334       & 1954 \\
        CPR-FP         &  FRS+GJE      & 100       & 291       & 897 \\
        \hline
        CPR-PF         &  GJE          & 100       & 310       & 1389 \\
        CPR-PF         &  FRS+GJE      & 100       & 293       & 1195 \\
        \hline
        CPR-FPF        &  GJE          & 97 (failed) & 279       & 1439 \\
        CPR-FPF        &  FRS+GJE      & 100       & 269       & 571 \\
        \hline
    \end{tabular}
    \caption{Numerical summary of Example \ref{num-44m}}
    \label{table-num-44m}
\end{table}

Table \ref{table-num-44m} shows the numerical summary for SAGD simulation with 15106 well pairs. All tests
pass except one (CPR-FPF + GJE) due to an internal error. Again, the table shows that the Newton method and
linear solver are efficient. The FRS+GJE decoupling method works much better than the GJE decoupling method.

\subsection{Scalability}
The parallel computers from Compute Canada are employed. The Niagara supercomputer consists of 1500 nodes,
and each node has 40 Intel Skylake cores at 2.4GHz, for a total of 60,000 cores.
Each node has 202 GB (188 GiB) RAM, and EDR Infiniband network is used to communicate.
The Cedar supercomputer has a hybrid architecture, which uses Intel E5-2683 v4 "Broadwell" at 2.1Ghz, E5-2650
v4 at 2.2GHz, Intel E7-4809 v4 "Broadwell" at 2.1Ghz, and Intel Platinum 8160F "Skylake" at 2.1Ghz. It has a
total of 58,416 CPU cores for computation, and 584 GPU devices.

\begin{figure}[!htb]
    \centering
    \includegraphics[width=0.53\linewidth, angle=270]{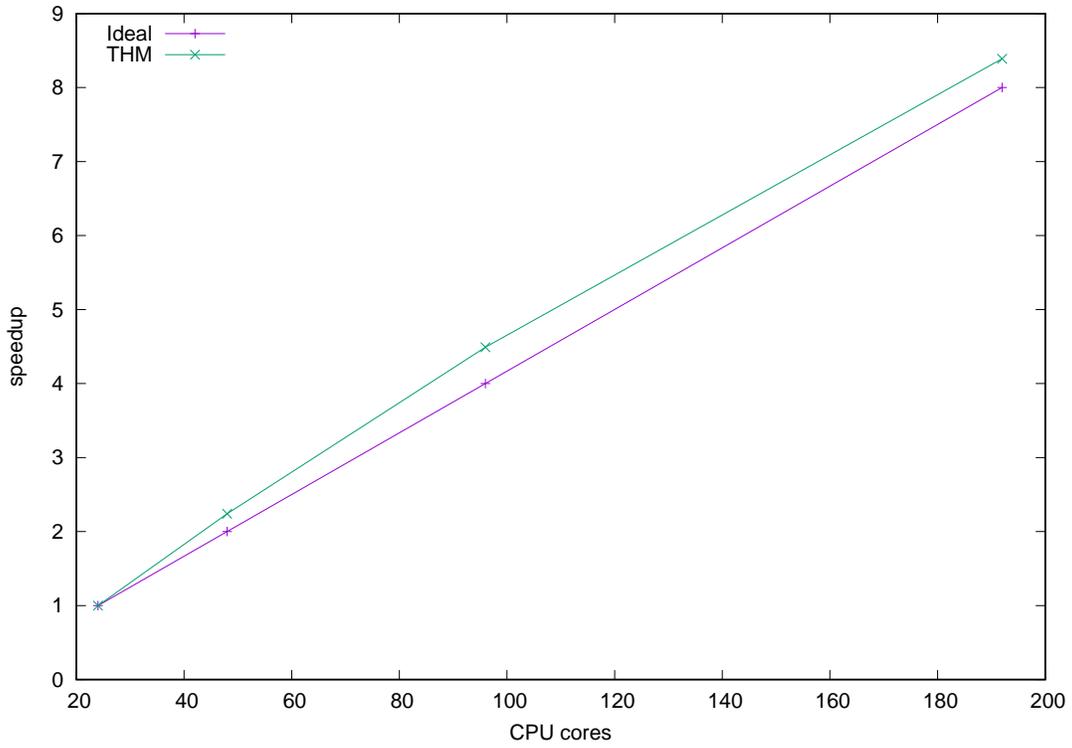}
    \caption{Example \ref{sca-230m}: scalability curve}
    \label{fig-sca-230m}
\end{figure}

\begin{example}
    \normalfont
    \label{sca-230m} This example studies a large thermal model with a grid dimension of $360 \times 400
    \times 1600$,
    230 million grid blocks. 12 nodes are empolyed using the Niagara supercomputer, and up to 192 CPU cores are
    used. The Newton method is applied with a tolerance of 1e-6 and maximal iterations of 10. The linear
    solver is BICGSTAB with a tolerance of 1e-5 and maximal iterations of 100. The preconditioner is the
    CPR-FPF method. Table \ref{table-sca-230m} presents running time and memory used. Figure
    \ref{fig-sca-230m} shows the scalability.
\end{example}

Table \ref{table-sca-230m} shows that huge amount of memory is required, which is not possible for desktop
computers. The running time and Figure \ref{fig-sca-230m} show the simulator, linear solver and preconditioner
have good scalability. The solver and preconditioner can solve linear systems with billions of unknowns.

\begin{table}[!htb]
    \centering
    \begin{tabular}{|c|c|c|c|c|}
        \hline
        CPU cores &  Total time (s)  &   Solver time (s) & Overall speedup & Memory (GB)   \\
        \hline
        24        & 2448.78          & 927.92            & 1.00 (100\%)    & 1,945.92       \\
        48        & 1094.55          & 380.40            & 2.24 (112\%)    & 1,959.28       \\
        96        & 545.20           & 194.81            & 4.49 (112\%)    & 1,970.83       \\
        192       & 291.88           & 107.32            & 8.38 (105\%)    & 1,994.25       \\
        \hline
    \end{tabular}
    \caption{Summary of Example \ref{sca-230m}}
    \label{table-sca-230m}
\end{table}

\begin{figure}[!htb]
    \centering
    \includegraphics[width=0.53\linewidth, angle=270]{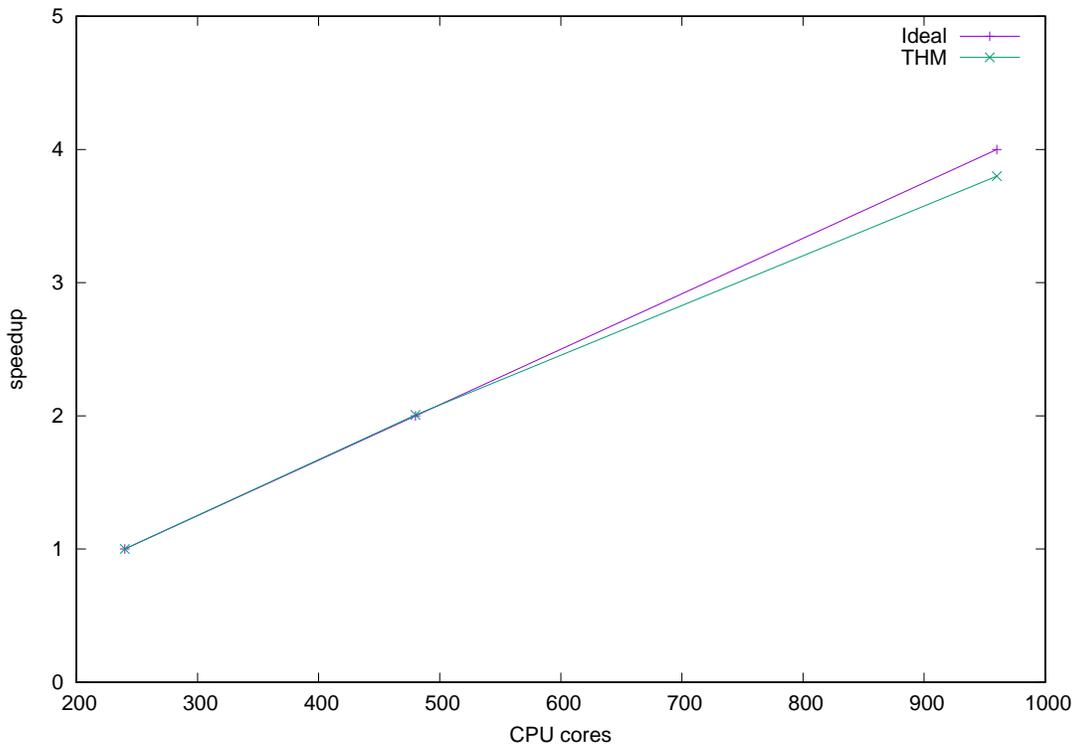}
    \caption{Example \ref{sca-1-2b}: scalability curve}
    \label{fig-sca-1-2b}
\end{figure}

\begin{example}
    \normalfont
    \label{sca-1-2b} This example studies a large thermal model with a grid dimension of $360 \times 2000
    \times 1600$,
    1.2 billion grid blocks. 120 nodes are empolyed using the Cedar supercomputer, and up to 960 CPU cores are
    used. The Newton method is applied with a tolerance of 1e-10 and maximal iterations of 10. The linear
    system has 6 billion unknowns, and the linear
    solver BICGSTAB is applied, which uses a tolerance of 1e-10 and maximal iterations of 100. The preconditioner is the
    CPR-FPF method. Table \ref{table-sca-1-2b} presents running time and memory used. Figure
    \ref{fig-sca-1-2b} shows the scalability.
\end{example}

Table \ref{table-sca-1-2b} and Figure \ref{fig-sca-1-2b} show the simulator, linear solver and preconditioner
have excellent scalability. The simulator can handle large-scale models, 
and the linear solver and preconditioner can solve linear systems with billions of unknowns.

\begin{table}[!htb]
    \centering
    \begin{tabular}{|c|c|c|c|c|}
        \hline
        CPU cores &  Total time (s)  &   Solver time (s) & Overall speedup & Memory (GB)\\
        \hline
        240       & 1802.24          & 934.92            & 1.00 (100\%)    & 9,839       \\
        480       & 897.69           & 455.47            & 2.00 (100\%)    & 9,906       \\
        960       & 474.29           & 227.89            & 3.80 (95.0\%)   & 9,996       \\
        \hline
    \end{tabular}
    \caption{Summary of Example \ref{sca-1-2b}}
    \label{table-sca-1-2b}
\end{table}

\begin{example}
    \normalfont
    \label{sca-3b} This example studies a large thermal model with a grid dimension of $1080 \times 2000
    \times 1600$, 3.46 billion grid blocks and the resulted linear systems have 17.3 billion unknowns. 360
    nodes are empolyed using the Cedar supercomputer. The Newton method is applied with a tolerance of 1e-10
    and maximal iterations of 10. The linear solver is BICGSTAB with a tolerance of 1e-10 and maximal
    iterations of 100. The preconditioner is the CPR-FPF method with GJE decoupling. Table \ref{table-sca-3b}
    presents running time and memory used.
\end{example}

Table \ref{table-sca-3b} show the simulator, linear solver and preconditioner have excellent scalability. This
example proves that our thermal simulator can handle extreme large-scale models. If more computing resource is
available, larger model can be simulated. Our linear solver and
preconditioner can solve linear systems with dozens of billions of unknowns. In ideal case, when the size of
MPIs doubles, the simulation time should be cut by half and the ideal speedup should be 2. This example shows
a speedup of 1.65 and an efficiency of 82.5\%, and we believe the reason is that when more CPU cores are used
in one node, these processors compete memory and computing, which reduces the effective memory communication
bandwidth. Therefore, the speedup is reduced.

\begin{table}[!htb]
    \centering
    \begin{tabular}{|c|c|c|c|c|}
        \hline
        CPU cores          &  Total time (s)  &   Solver time (s) & Overall speedup & Memory (GB)\\
        \hline
        2880  (360 X 8)    & 1247.74          & 996.76            & 1.00 (100\%)    & 30,101.46   \\
        5760  (360 X 16)   & 757.70           & 578.32            & 1.65 (82.5\%)   & 33,490.12   \\
        \hline
    \end{tabular}
    \caption{Summary of Example \ref{sca-3b}}
    \label{table-sca-3b}
\end{table}

\subsection{Scalability on Another Supercomputer}

A Cray XC30 supercomputer deployed is employed. Each computation node contains two 2.7 GHz, 12-core Intel E5-2697 v2
CPUs, and 64 GB of memory is shared between the two processors. The memory bandwidth is around 103 Gb/s.
The memory acces is non-uniform access (NUMA): each processor owns a single NUMA region
of 32 GB. Accessing its own region has a lower latency than accessing the other NUMA region.
Also, these 24 cores compete the memory channels.
The Aries interconnect connects all computation nodes in a Dragonfly topology.

\begin{figure}[!htb]
    \centering
    \includegraphics[width=0.53\linewidth, angle=270]{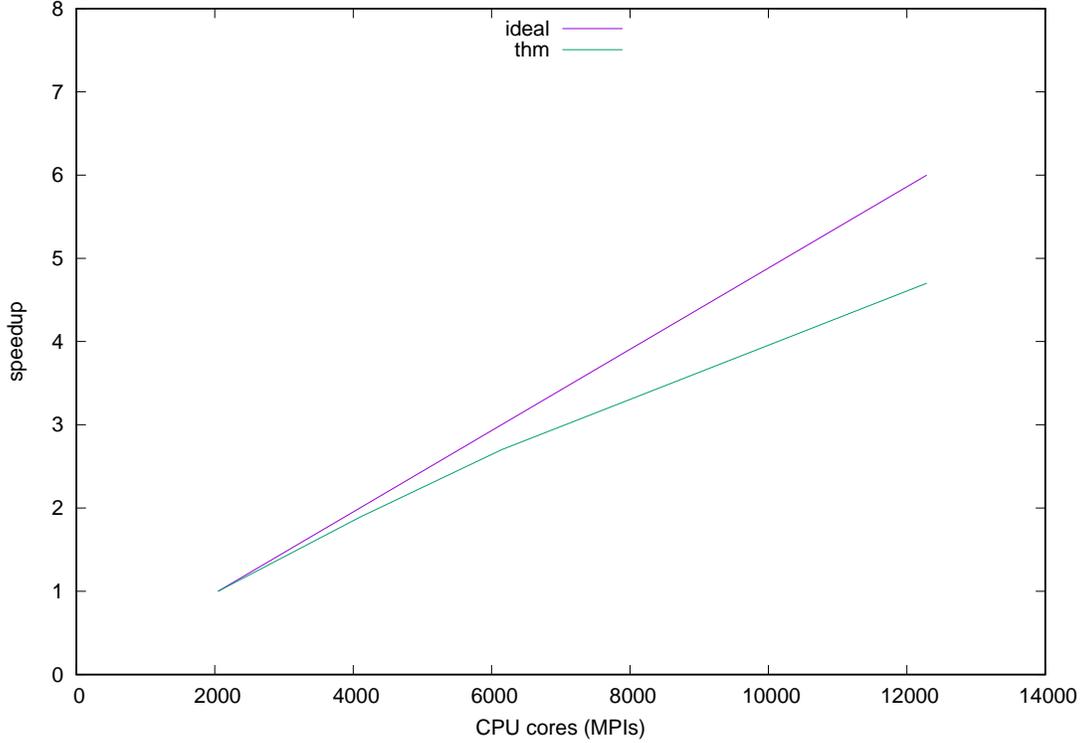}
    \caption{Example \ref{sca-4.6b}: scalability curve}
    \label{fig-sca-4.6b}
\end{figure}

\begin{example}
    \normalfont
    \label{sca-4.6b}
    This example studies a large thermal model with a grid dimension of $1440 \times 2000
    \times 1600$, 4.6 billion grid blocks and the resulted linear systems have 23 billion unknowns.
    1024 nodes are empolyed.
    The Newton method is applied with a tolerance of 1e-4
    and maximal iterations of 10. The linear solver is BICGSTAB with a tolerance of 1e-3 and maximal
    iterations of 100. The preconditioner is the CPR-FPF method with GJE decoupling. Table \ref{table-sca-4.6b}
    presents running time and memory used and Figure \ref{fig-sca-4.6b} shows the scalability.
\end{example}

    Table \ref{table-sca-4.6b} shows overall time, linear solver time, linear solver speedup, 
    overall speedup and total memory.
    When 4096 and 6144 cores are used, the scalalabilities are 1.89 and 2.7, respectively, while the best
    scalalabilities should be 2 and 3. In this case, the parallel efficiencies are 94\% and 90\%, which are
    good for parallel numerical simulations. However, this example shows linear solver has better speedup and
    parallel efficiency. 
    If special optimization techniques are applied, such as
    multi-level load balancing that consider the architecture of the system and multi-layer communications,
    the communication volume and latency will be reduced and scalability can be improved. When 12288 cores are
    employed, each node run 12 cores and 12 MPIs, and each processor uses its 6 cores. In this case, memory
    access may be an important issue, which may reduce the effective memory bandwidth of each MPI and increase
    computation time.

\begin{table}[!htb]
    \centering
    \begin{tabular}{|c|c|c|c|c|c|}
        \hline
        CPU cores &  Total time (s)  &   Solver time (s)& Solver speedup & Overall speedup & Memory (GB) \\
        \hline
        2048      & 793.68           &  594.70          & 1.00 (100\%)   & 1.00 (100\%)    & 43,090.70   \\
        4096      & 419.45           &  305.93          & 1.94 (97.0\%)  & 1.89 (94\%)     & 41,542.27   \\
        6144      & 293.85           &  213.48          & 2.78 (92.7\%)  & 2.70 (90\%)     & 45,118.68    \\
        12288     & 168.97           &  118.23          & 5.03 (83.0\%)  & 4.70 (78\%)     & 44,063.20    \\
        \hline
    \end{tabular}
    \caption{Summary of Example \ref{sca-4.6b}}
    \label{table-sca-4.6b}
\end{table}

\begin{figure}[!htb]
    \centering
    \includegraphics[width=0.53\linewidth, angle=270]{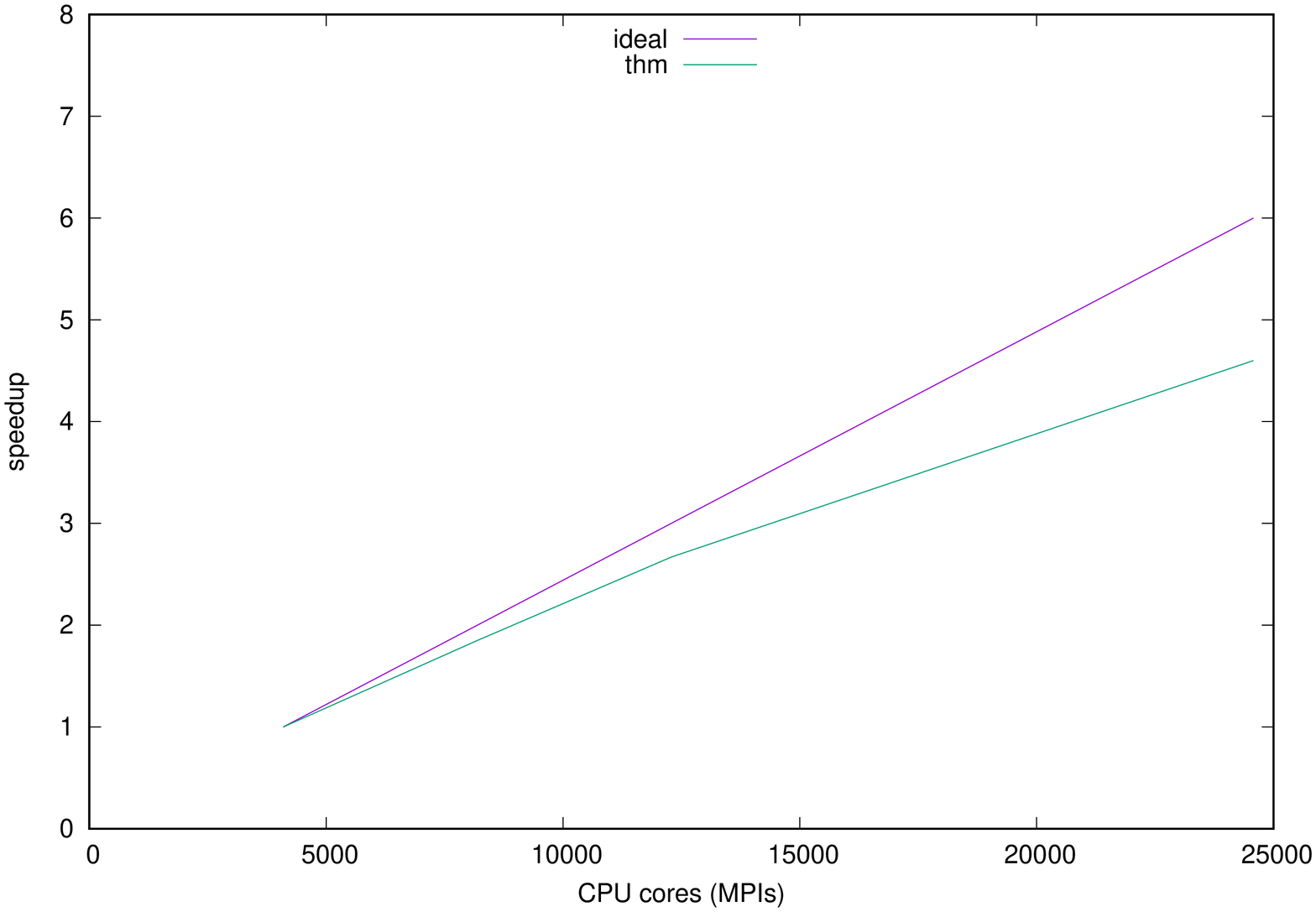}
    \caption{Example \ref{sca-9.2b}: scalability curve}
    \label{fig-sca-9.2b}
\end{figure}

\begin{example}
    \normalfont
    \label{sca-9.2b}
    This example studies a large thermal model with a grid dimension of 9.2 billion grid blocks and the
    resulted linear systems have 46 billion unknowns.  2048 nodes are empolyed.
    The Newton method is applied with a tolerance of 1e-4
    and maximal iterations of 10. The linear solver is BICGSTAB with a tolerance of 1e-3 and maximal
    iterations of 100. The preconditioner is the CPR-FPF method with GJE decoupling. Table \ref{table-sca-9.2b}
    presents running time and memory used, and Figure \ref{fig-sca-9.2b} shows the scalability (overall
    speedup) curve.
\end{example}

\begin{table}[!htb]
    \centering
    \begin{tabular}{|c|c|c|c|c|c|}
        \hline
        CPU cores &  Total time (s)  &   Solver time (s) & Solver speedup & Overall speedup    & Memory (GB) \\
        \hline
        4096      & 804.68           &  600.54           & 1.00 (100\%)    & 1.00 (100\%)      & 83,970.30   \\
        8192      & 434.8            &  314.37           & 1.91 (95.5\%)   & 1.85 (92.5\%)     & 86,522.03   \\
        12288     & 300.83           &  216.14           & 2.77 (92.0\%)   & 2.67 (89.0\%)     & 92,951.34    \\
        24576     & 174.59           &  118.70           & 5.05 (84.2\%)   & 4.60 (76.66\%)    & 90,880.31    \\
        \hline
    \end{tabular}
    \caption{Summary of Example \ref{sca-9.2b}}
    \label{table-sca-9.2b}
\end{table}

\begin{figure}[!htb]
    \centering
    \includegraphics[width=0.53\linewidth, angle=270]{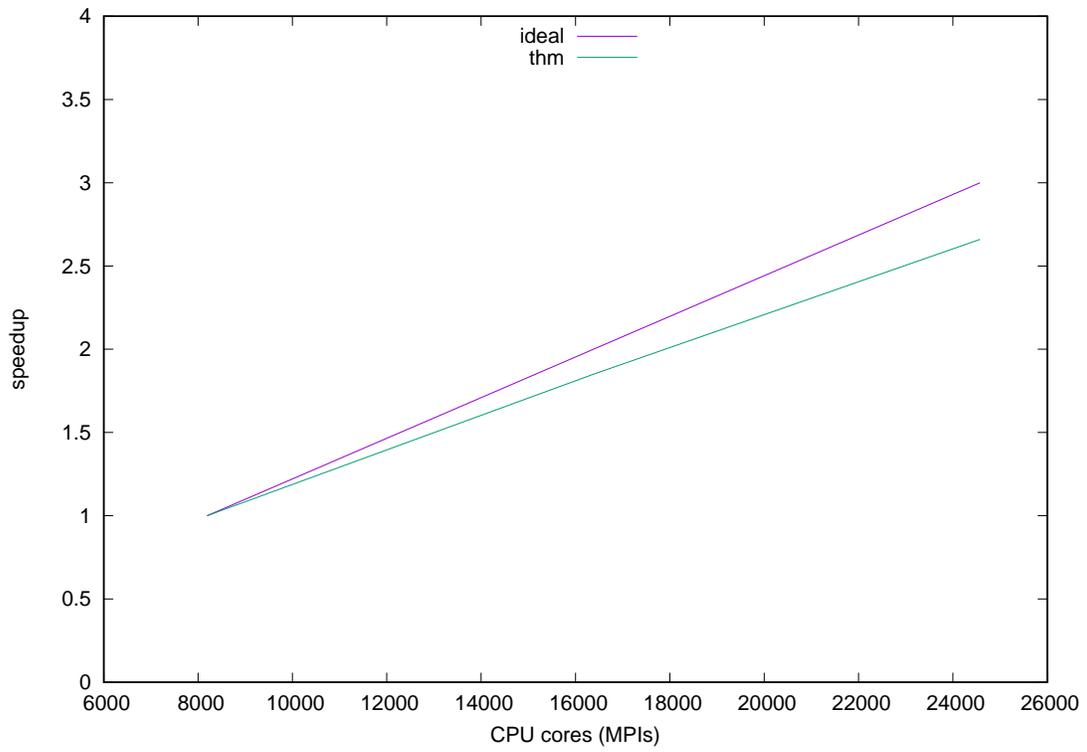}
    \caption{Example \ref{sca-20b}: scalability curve}
    \label{fig-sca-20b}
\end{figure}

\begin{example}
    \normalfont
    \label{sca-20b}
    This example studies a large thermal model with a grid dimension of 20 billion grid blocks using 4096
    nodes and the resulted linear systems have 100 billion unknowns.
    The Newton method is applied with a tolerance of 1e-4
    and maximal iterations of 10. The linear solver is BICGSTAB with a tolerance of 1e-3 and maximal
    iterations of 100. The preconditioner is the CPR-FPF method with GJE decoupling. Table \ref{table-sca-20b}
    presents running time and memory used, and Figure \ref{fig-sca-20b} is the scalability.
\end{example}

\begin{table}[!htb]
    \centering
    \begin{tabular}{|c|c|c|c|c|c|}
        \hline
        CPU cores &  Total time (s)  &   Solver time (s) & Solver speedup  & Overall speedup    & Memory (GB) \\
        \hline
        8192      & 886.10           &  656.64           & 1.00 (100\%)    & 1.00 (100\%)       & 182,530.13   \\
        16384     & 476.95           &  341.02           & 1.92 (96.0\%)   & 1.85 (92.5\%)      & 184,450.12   \\
        24576     & 332.46           &  236.71           & 2.77 (92.3\%)   & 2.66 (88.6\%)      & 206,267.72   \\
        \hline
    \end{tabular}
    \caption{Summary of Example \ref{sca-20b}}
    \label{table-sca-20b}
\end{table}

Example \ref{sca-9.2b} and Example \ref{sca-20b} show similar results. The overall scalability is good but the
linear solver has better scalability. The memory consumption is proportional to grid blocks, which means if
more computation resource is available, larger models can be studied. Since the thermal reservoir simulator
has good scalability, the same model can be run faster if using more computation nodes. These examples
indicate that our thermal simulator can handle giant thermal models. Meantime, we have obserbed that when more
cores of a processor are employed, the scalability tends to reduce, which could be caused by memory bandwidth,
network or algorithms, and there is room to investigate and to improve its performance.

\subsection{Scalability of Simplified Models}

In standard thermal simulations, various properties should be stored, such as porosity, density of each phase,
viscosity, enthalpy, internal energy, saturations, temperature, and mole fractions.
The memory consumption is huge, which makes hard to benchmark larger models unless a larger supercomputer is
available, such as Summit from Oak Ridge National Laboratory and Blue Waters from the National Center for
Supercomputing Applications (NCSA) at the University of Illinois at Urbana-Champaign. Here simplified models
are designed to benchmark model with more grid blocks. However, grid generation, grid load balancing, data
management, distributed matrix and vector, linear solver and preconditioner are all tested.
All simulated problems are run on the Cray XC30 supercomputer.

\begin{example}
    \normalfont
    \label{sca-sp-43b}
    This example studies a simplified problem with a grid dimension of 42.8 billion grid blocks and 1024 nodes
    are empolyed.  The linear solver is BICGSTAB and the preconditioner is the RAS method. The RAS method has
    good scalability, since the communications are local, and only one matrix is required, which is not the
    case as for AMG solver, which may have many coarser matrices. Table \ref{table-sca-sp-43b} presents running time and memory used.
\end{example}

\begin{table}[!htb]
    \centering
    \begin{tabular}{|c|c|c|c|c|c|}
        \hline
        CPU cores &  Total time (s)  &   Solver time (s) & Solver speedup  & Overall speedup & Memory (GB)  \\
        \hline
        1024      & 744.31           & 381.40            & 1.00 (100\%)    & 1.00 (100\%)    & 47,869.86    \\
        2048      & 364.30           & 183.82            & 2.07 (103.5\%)  & 2.04 (102\%)    & 48,792.74    \\
        3072      & 264.48           & 137.74            & 2.76 (92.0\%)   & 2.81 (93.7)     & 49,591.08    \\
        6144      & 131.82           & 68.92             & 5.53 (92.2\%)   & 5.64  (94.0\%)  & 53,875.08    \\
        12288     & 72.61            & 37.95             & 10.05 (83.8\%)  & 10.25 (85.0\%)  & 55,398.00    \\
        24576     & 41.65            & 23.72             & 16.07 (67.0\%)  & 17.87 (74.0\%)  & 56,178.96    \\
        \hline
    \end{tabular}
    \caption{Summary of Example \ref{sca-sp-43b}}
    \label{table-sca-sp-43b}
\end{table}

\begin{figure}[!htb]
    \centering
    \includegraphics[width=0.53\linewidth, angle=270]{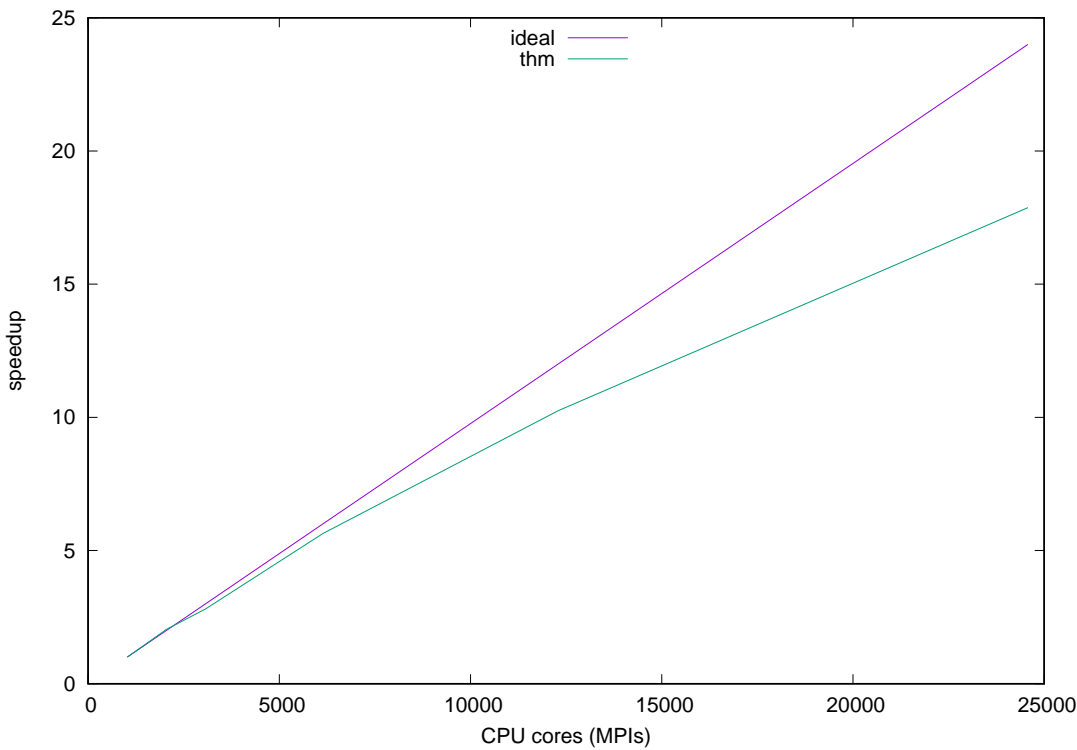}
    \caption{Example \ref{sca-sp-43b}: scalability curve}
    \label{fig-sca-sp-43b}
\end{figure}

\begin{figure}[!htb]
    \centering
    \includegraphics[width=0.53\linewidth, angle=270]{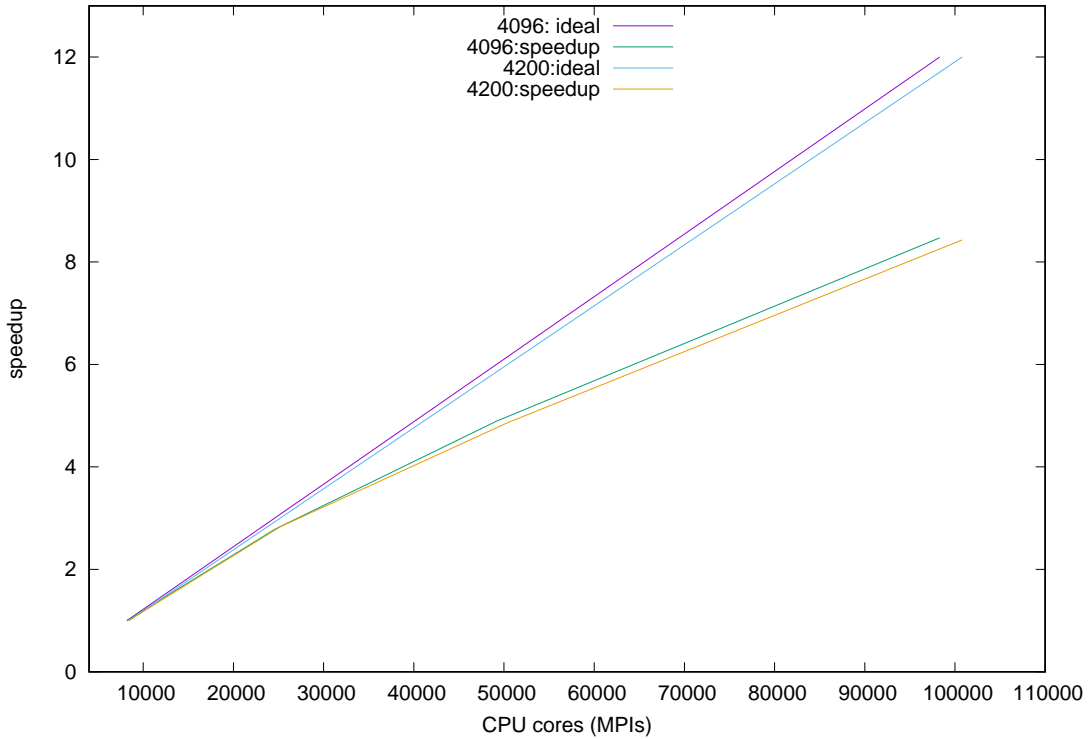}
    \caption{Example \ref{sca-sp-216b}: scalability curve}
    \label{fig-sca-sp-216b}
\end{figure}

\begin{example}
    \normalfont
    \label{sca-sp-216b}
    This example studies a simplified problem with a grid dimension of 216 billion grid blocks.
    The linear solver is BICGSTAB and the preconditioner is the RAS method.
    Table \ref{table-sca-sp-216b} presents running time and memory used for 4096 computation nodes.
    Table \ref{table-sca-sp-216b2} presents running time and memory used for 4200 computation nodes.
    Each node uses 2, 4, 6, 12 and 24 cores. Figure \ref{fig-sca-sp-216b} is the speedup curve.
\end{example}

\begin{table}[!htb]
    \centering
    \begin{tabular}{|c|c|c|c|c|c|}
        \hline
        CPU cores &  Total time (s)  &   Solver time (s) & Solver speedup & Overall speedup & Memory (GB)   \\
        \hline
        8192      & 472.64           & 240.15            & 1.00 (100\%)   & 1.00 (100\%)    & 252,194.32    \\
        16384     & 249.68           & 130.52            & 1.84 (92.0\%)  & 1.89 (94.5\%)   & 255,154.24    \\
        24576     & 169.84           & 88.58             & 2.71 (90.4\%)  & 2.78 (92.7\%)   & 258,981.60    \\
        49152     & 96.53            & 51.97             & 4.62 (77.0\%)  & 4.89 (81.5\%)   & 269,806.56    \\
        98304     & 55.77            & 31.89             & 7.53 (62.8\%)  & 8.47 (70.6\%)   & 289,981.44    \\
        \hline
    \end{tabular}
    \caption{Summary of Example \ref{sca-sp-216b}}
    \label{table-sca-sp-216b}
\end{table}

\begin{table}[!htb]
    \centering
    \begin{tabular}{|c|c|c|c|c|c|}
        \hline
        CPU cores &  Total time (s)  &   Solver time (s) & Solver speedup & Overall speedup & Memory (GB)   \\
        \hline
        8400      & 456.68           &  229.65           &  1.00 (100\%)  & 1.00 (100\%)    & 252,915.14     \\
        16800     & 237.28           &  121.62           &  1.89 (94.4\%) & 1.92 (96.s\%)   & 255,850.05     \\
        25200     & 161.45           &  82.23            &  2.79 (93\%)   & 2.83 (94.1\%)   & 259,589.53     \\
        50400     & 93.97            &  50.54            &  4.54 (75.7\%) & 4.86 (80.9\%)   & 270,134.15     \\
        100800    & 54.16            &  30.94            &  7.42 (61.8\%) & 8.43 (70.2\%)   & 290,869.03     \\
        \hline
    \end{tabular}
    \caption{Summary of Example \ref{sca-sp-216b}}
    \label{table-sca-sp-216b2}
\end{table}

\section{Conclusions}

This paper introduces a parallel thermal simulator on distribted-memory parallel computers, where MPI is
employed for communications. The simulator is designed to handle giant models with billions even trillions of
grid blocks using hundreds of thousands of CPU cores. Its mathematical models and numerical methods are
presented. Numerical experiments are carried out to verify the methods and implementations, which show
that our simulator can match commercial software and it has excellent scalability, and it can handle
extremely large-scale reservoir models.

\section*{Acknowledgements}
The support of Department of Chemical and Petroleum Engineering,
University of Calgary and Reservoir Simulation  Research Group is gratefully
acknowledged. The research is partly supported by NSERC/AIEES/Foundation
CMG, AITF iCore, IBM Thomas J. Watson Research Center, and the Frank
and Sarah Meyer FCMG Collaboration Centre for Visualization and Simulation.
The research is also enabled in part by support
provided by WestGrid (www.westgrid.ca), SciNet (www.scinethpc.ca)
and Compute Canada Calcul Canada (www.computecanada.ca).

\end{document}